\documentclass{pasj00}

\def\commenta{$^*$}
\def\commentb{$^\dagger$}
\def\commentc{$^\ddagger$}
\def\commentd{$^\S$}
\def\commente{$^\|$}

\def\submitted{submitted}
\def\inpress{in press}

\def\arxiv#1{ (arXiv astro-ph/#1)}

\DeclareAbbreviation\AAHam{Astron. Abh. Hamburg. Sternw.}
\DeclareAbbreviation\AARv{Astron. Astrophys. Rev.}
\DeclareAbbreviation\AAS{American Astron. Soc. Meeting Abstracts}
\DeclareAbbreviation\AcA{Acta Astron.}
\DeclareAbbreviation\actaa{Acta Astron.}
\DeclareAbbreviation\actaapS{Acta Astrophs. Sin.}
\DeclareAbbreviation\Afz{Astrofizika}
\DeclareAbbreviation\AGAb{Astronomische Gesellschaft Abstract Ser.}
\DeclareAbbreviation\an{Astron. Nachr.}
\DeclareAbbreviation\AnAp{Annales d'Astrophysique}
\DeclareAbbreviation\AnTok{Tokyo Astron. Obs. Annals, Sec. Ser.}
\DeclareAbbreviation\Ap{Astrophysics}
\DeclareAbbreviation\ARep{Astron. Rep.}
\DeclareAbbreviation\AstBu{Astrophys. Bull.}
\DeclareAbbreviation\ATel{Astron. Telegram}
\DeclareAbbreviation\ATsir{Astron. Tsirk.}
\DeclareAbbreviation\AcApS{Acta Astrophys. Sinica}
\DeclareAbbreviation\AstL{Astron. Lett.}
\DeclareAbbreviation\BaltA{Baltic Astron.}
\DeclareAbbreviation\BANS{Bull. of the Astron. Institutes of the Netherlands Suppl. Ser.}
\DeclareAbbreviation\BASI{Bull. Astron. Soc. India}
\DeclareAbbreviation\BeSN{Be Newslett.}
\DeclareAbbreviation\BHarO{Harvard Coll. Obs. Bull.}
\DeclareAbbreviation\CBET{Cent. Bur. Electron. Telegrams}
\DeclareAbbreviation\CEAB{Central European Astrophys. Bull.}
\DeclareAbbreviation\ChJAA{Chinese J. of Astron. and Astrophys.}
\DeclareAbbreviation\caa{Chinese J. of Astron. and Astrophys.}
\DeclareAbbreviation\CoAsi{Asiago Contr.}
\DeclareAbbreviation\CoSka{Contributions of the Astronomical Observatory Skalnat\'e Pleso}
\DeclareAbbreviation\GCN{GRB Coord. Netw. Circ.}
\DeclareAbbreviation\ErgAN{Erg. Astron. Nachr.}
\DeclareAbbreviation\ibvs{IBVS}
\DeclareAbbreviation\IEEEP{IEEE Proc.}
\DeclareAbbreviation\JAD{J. Astron. Data}
\DeclareAbbreviation\JApA{J. of Astrophys. and Astron.}
\DeclareAbbreviation\JAVSO{J. American Assoc. Variable Star Obs.}
\DeclareAbbreviation\JBAA{J. Br. Astron. Assoc.}
\DeclareAbbreviation\JPhCS{J. of Physics Conference Series}
\DeclareAbbreviation\JPSJ{J. Phys. Soc. Japan}
\DeclareAbbreviation\JSARA{J. of the Southeastern Assoc. for Research in Astron.}
\DeclareAbbreviation\LowOB{Lowell Obs. Bull.}
\DeclareAbbreviation\MitAG{Mitteil. der Astronom. Gesell. Hamburg}
\DeclareAbbreviation\MitVS{Mitteil. Ver\"{a}nderl. Sterne}
\DeclareAbbreviation\MmSAI{Mem. Soc. Astron. Ital.}
\DeclareAbbreviation\memsai{Mem. Soc. Astron. Ital.}
\DeclareAbbreviation\Msngr{Messenger}
\DeclareAbbreviation\NewA{New Astron.}
\DeclareAbbreviation\na{New Astron.}
\DeclareAbbreviation\NewAR{New Astron. Rev.}
\DeclareAbbreviation\nar{New Astron. Rev.}
\DeclareAbbreviation\NInfo{Nauchnye Informatsii}
\DeclareAbbreviation\NPhS{Nature Physical Science}
\DeclareAbbreviation\OAP{Odessa Astron. Publ.}
\DeclareAbbreviation\Obs{Observatory}
\DeclareAbbreviation\OEJV{Open Eur. J. on Variable Stars}
\DeclareAbbreviation\PASA{Publ. Astron. Soc. Australia}
\DeclareAbbreviation\PASAu{Publ. Astron. Soc. Australia}
\DeclareAbbreviation\PAZh{Pis'ma AZh}
\DeclareAbbreviation\PJAB{Proc. Japan Acad. Ser. B}
\DeclareAbbreviation\POBeo{Publ. de l'Observatoire Astronomique de Beograd}
\DeclareAbbreviation\PCCP{Phys. Chem. Chem. Phys.}
\DeclareAbbreviation\PhR{Phys. Rep.}
\DeclareAbbreviation\PVSS{Publ. Variable Stars Sect. R. Astron. Soc. New Zealand}
\DeclareAbbreviation\PZ{Perem. Zvezdy}
\DeclareAbbreviation\PZP{Perem. Zvezdy, Prilozh.}
\DeclareAbbreviation\QJRAS{QJRAS}
\DeclareAbbreviation\RA{Ricerche Astronomiche}
\DeclareAbbreviation\RMxAA{Rev. Mexicana Astron. Astrof.}
\DeclareAbbreviation\RvMA{Reviews of Modern Astron.}
\DeclareAbbreviation\RvMP{Reviews of Modern Phys.}
\DeclareAbbreviation\SASS{Society for Astronom. Sciences Ann. Symp.}
\DeclareAbbreviation\Sci{Science}
\DeclareAbbreviation\SPIE{SPIE Proc.}
\DeclareAbbreviation\SvA{Soviet Astronomy}
\DeclareAbbreviation\SvAL{Soviet Astronomy Letters}
\DeclareAbbreviation\VeSon{Ver\"{o}ff. Sternw. Sonneberg}
\DeclareAbbreviation\VSOLJBul{VSOLJ Variable Star Bull.}
\DeclareAbbreviation\yCat{VizieR Online Data Catalog}
\DeclareAbbreviation\ZA{Z. Astrophys.}

\def\ASPConf#1#2{ASP Conf. Ser. #1, #2}

\def\PublisherCambridge{Cambridge: Cambridge University Press}

\def\PublisherASP{San Francisco: ASP}

\newcounter{author}
\def\authorcount#1#2{\refstepcounter{author}\label{#1}
                     \altaffiltext{\ref{#1}}{#2}}

\begin{document}
\SetRunningHead{T. Kato et al.}{Period Variations in SU UMa-Type Dwarf Novae V}

\Received{201X/XX/XX}
\Accepted{201X/XX/XX}

\title{Survey of Period Variations of Superhumps in SU UMa-Type Dwarf Novae.
    V: The Fifth Year (2012--2013)}

\author{Taichi~\textsc{Kato},\altaffilmark{\ref{affil:Kyoto}*}
        Franz-Josef~\textsc{Hambsch},\altaffilmark{\ref{affil:GEOS}}$^,$\altaffilmark{\ref{affil:BAV}}$^,$\altaffilmark{\ref{affil:Hambsch}}
        Hiroyuki~\textsc{Maehara},\altaffilmark{\ref{affil:Kiso}}$^,$\altaffilmark{\ref{affil:HidaKwasan}}
        Gianluca~\textsc{Masi},\altaffilmark{\ref{affil:Masi}}
        Francesca~\textsc{Nocentini},\altaffilmark{\ref{affil:Masi}}
        Pavol~A.~\textsc{Dubovsky},\altaffilmark{\ref{affil:Dubovsky}}
        Igor~\textsc{Kudzej},\altaffilmark{\ref{affil:Dubovsky}}
        Kazuyoshi~\textsc{Imamura},\altaffilmark{\ref{affil:OUS}}
        Minako~\textsc{Ogi},\altaffilmark{\ref{affil:OUS}}
        Kenji~\textsc{Tanabe},\altaffilmark{\ref{affil:OUS}}
        Hidehiko~\textsc{Akazawa},\altaffilmark{\ref{affil:OUS}}
        Thomas~\textsc{Krajci},\altaffilmark{\ref{affil:Krajci}}
        Ian~\textsc{Miller},\altaffilmark{\ref{affil:Miller}}
        Enrique~de~\textsc{Miguel},\altaffilmark{\ref{affil:Miguel}}$^,$\altaffilmark{\ref{affil:Miguel2}}
        Arne~\textsc{Henden},\altaffilmark{\ref{affil:AAVSO}}
        Ryo~\textsc{Noguchi},\altaffilmark{\ref{affil:OKU}}
        Takehiro~\textsc{Ishibashi},\altaffilmark{\ref{affil:OKU}}
        Rikako~\textsc{Ono},\altaffilmark{\ref{affil:OKU}}
        Miho~\textsc{Kawabata},\altaffilmark{\ref{affil:OKU}}
        Hiroshi~\textsc{Kobayashi},\altaffilmark{\ref{affil:OKU}}
        Daisuke~\textsc{Sakai},\altaffilmark{\ref{affil:OKU}}
        Hirochika~\textsc{Nishino},\altaffilmark{\ref{affil:OKU}}
        Hisami~\textsc{Furukawa},\altaffilmark{\ref{affil:OKU}}
        Kazunari~\textsc{Masumoto},\altaffilmark{\ref{affil:OKU}}
        Katsura~\textsc{Matsumoto},\altaffilmark{\ref{affil:OKU}}
        Colin~\textsc{Littlefield},\altaffilmark{\ref{affil:LCO}}
        Tomohito~\textsc{Ohshima},\altaffilmark{\ref{affil:Kyoto}}
        Chikako~\textsc{Nakata},\altaffilmark{\ref{affil:Kyoto}}
        Satoshi~\textsc{Honda},\altaffilmark{\ref{affil:Gunma}}$^,$\altaffilmark{\ref{affil:NHAO}}
        Kenzo~\textsc{Kinugasa},\altaffilmark{\ref{affil:Gunma}}$^,$\altaffilmark{\ref{affil:Nobeyama}}
        Osamu~\textsc{Hashimoto},\altaffilmark{\ref{affil:Gunma}}
        William~\textsc{Stein},\altaffilmark{\ref{affil:Stein}}
        Roger~D.~\textsc{Pickard},\altaffilmark{\ref{affil:BAAVSS}}$^,$\altaffilmark{\ref{affil:Pickard}}
        Seiichiro~\textsc{Kiyota},\altaffilmark{\ref{affil:Kis}}
        Elena~P.~\textsc{Pavlenko},\altaffilmark{\ref{affil:CrAO}}
        Oksana~I.~\textsc{Antonyuk},\altaffilmark{\ref{affil:CrAO}}
        Aleksei~V.~\textsc{Baklanov},\altaffilmark{\ref{affil:CrAO}}
        Kirill~\textsc{Antonyuk},\altaffilmark{\ref{affil:CrAO}}
        Denis~\textsc{Samsonov},\altaffilmark{\ref{affil:CrAO}}
        Nikolaj~\textsc{Pit},\altaffilmark{\ref{affil:CrAO}}
        Aleksei~\textsc{Sosnovskij},\altaffilmark{\ref{affil:CrAO}}
        Arto~\textsc{Oksanen},\altaffilmark{\ref{affil:Nyrola}}
        Caisey~\textsc{Harlingten},\altaffilmark{\ref{affil:Harlingten}}
        Jenni~\textsc{Tyysk\"a},\altaffilmark{\ref{affil:Tyyska}}
        Berto~\textsc{Monard},\altaffilmark{\ref{affil:Monard}}$^,$\altaffilmark{\ref{affil:Monard2}}
        Sergey~Yu.~\textsc{Shugarov},\altaffilmark{\ref{affil:Sternberg}}$^,$\altaffilmark{\ref{affil:Slovak}}
        Drahomir~\textsc{Chochol},\altaffilmark{\ref{affil:Slovak}}
        Kiyoshi~\textsc{Kasai},\altaffilmark{\ref{affil:Kai}}
        Yutaka~\textsc{Maeda},\altaffilmark{\ref{affil:Mdy}}
        Kenji~\textsc{Hirosawa},\altaffilmark{\ref{affil:Hsk}}
        Hiroshi~\textsc{Itoh},\altaffilmark{\ref{affil:Ioh}}
        Richard~\textsc{Sabo},\altaffilmark{\ref{affil:Sabo}}
        Joseph~\textsc{Ulowetz},\altaffilmark{\ref{affil:Ulowetz}}
        Etienne~\textsc{Morelle},\altaffilmark{\ref{affil:Morelle}}
        Ra\'ul~\textsc{Michel},\altaffilmark{\ref{affil:UNAM}}
        Genaro~\textsc{Su\'arez},\altaffilmark{\ref{affil:UNAM}}
        Nick~\textsc{James},\altaffilmark{\ref{affil:James}}
        Shawn~\textsc{Dvorak},\altaffilmark{\ref{affil:Dvorak}}
        Irina~B.~\textsc{Voloshina},\altaffilmark{\ref{affil:Sternberg}}
        Michael~\textsc{Richmond},\altaffilmark{\ref{affil:RIT}}
        Bart~\textsc{Staels},\altaffilmark{\ref{affil:AAVSO}}$^,$\altaffilmark{\ref{affil:Staels}}
        David~\textsc{Boyd},\altaffilmark{\ref{affil:DavidBoyd}}
        Maksim~V.~\textsc{Andreev},\altaffilmark{\ref{affil:Terskol}}$^,$\altaffilmark{\ref{affil:ICUkraine}}
        Nikolai~\textsc{Parakhin},\altaffilmark{\ref{affil:Terskol}}
        Natalia~\textsc{Katysheva},\altaffilmark{\ref{affil:Sternberg}}
        Atsushi~\textsc{Miyashita},\altaffilmark{\ref{affil:Seikei}}
        Kazuhiro~\textsc{Nakajima},\altaffilmark{\ref{affil:Njh}}
        Greg~\textsc{Bolt},\altaffilmark{\ref{affil:Bolt}}
        Stefano~\textsc{Padovan},\altaffilmark{\ref{affil:AAVSO}}
        Peter~\textsc{Nelson},\altaffilmark{\ref{affil:Nelson}}
        Donn~R.~\textsc{Starkey},\altaffilmark{\ref{affil:Starkey}}
        Denis~\textsc{Buczynski},\altaffilmark{\ref{affil:Buczynski}}
        Peter~\textsc{Starr},\altaffilmark{\ref{affil:Starr}}
        William~N.~\textsc{Goff},\altaffilmark{\ref{affil:Goff}}
        Denis~\textsc{Denisenko},\altaffilmark{\ref{affil:Denisenko}}
        Christopher~S.~\textsc{Kochanek},\altaffilmark{\ref{affil:Ohio}}
        Benjamin~\textsc{Shappee},\altaffilmark{\ref{affil:Ohio}}
        Krzysztof~Z.~\textsc{Stanek},\altaffilmark{\ref{affil:Ohio}}
        Jos\'e~L.~\textsc{Prieto},\altaffilmark{\ref{affil:Princeton}}
        Koh-ichi~\textsc{Itagaki},\altaffilmark{\ref{affil:Itagaki}}
        Shizuo~\textsc{Kaneko},\altaffilmark{\ref{affil:Kaneko}}
        Rod~\textsc{Stubbings},\altaffilmark{\ref{affil:Stubbings}}
        Eddy~\textsc{Muyllaert},\altaffilmark{\ref{affil:VVSBelgium}}
        Jeremy~\textsc{Shears},\altaffilmark{\ref{affil:Shears}}
        Patrick~\textsc{Schmeer},\altaffilmark{\ref{affil:Schmeer}}
        Gary~\textsc{Poyner},\altaffilmark{\ref{affil:Poyner}}
        Miguel~\textsc{Rodr\'{\i}guez Marco},\altaffilmark{\ref{affil:MRO}}
}

\authorcount{affil:Kyoto}{
     Department of Astronomy, Kyoto University, Kyoto 606-8502}
\email{$^*$tkato@kusastro.kyoto-u.ac.jp}

\authorcount{affil:GEOS}{
     Groupe Europ\'een d'Observations Stellaires (GEOS),
     23 Parc de Levesville, 28300 Bailleau l'Ev\^eque, France}

\authorcount{affil:BAV}{
     Bundesdeutsche Arbeitsgemeinschaft f\"ur Ver\"anderliche Sterne
     (BAV), Munsterdamm 90, 12169 Berlin, Germany}

\authorcount{affil:Hambsch}{
     Vereniging Voor Sterrenkunde (VVS), Oude Bleken 12, 2400 Mol, Belgium}

\authorcount{affil:Kiso}{
     Kiso Observatory, Institute of Astronomy, School of Science, 
     The University of Tokyo 10762-30, Mitake, Kiso-machi, Kiso-gun,
     Nagano 397-0101}

\authorcount{affil:HidaKwasan}{
     Kwasan and Hida Observatories, Kyoto University, Yamashina,
     Kyoto 607-8471}

\authorcount{affil:Masi}{
     The Virtual Telescope Project, Via Madonna del Loco 47, 03023
     Ceccano (FR), Italy}

\authorcount{affil:Dubovsky}{
     Vihorlat Observatory, Mierova 4, Humenne, Slovakia}

\authorcount{affil:OUS}{
     Department of Biosphere-Geosphere System Science, Faculty of Informatics,
     Okayama University of Science, 1-1 Ridai-cho, Okayama, Okayama 700-0005}

\authorcount{affil:Krajci}{
     Astrokolkhoz Observatory,
     Center for Backyard Astrophysics New Mexico, PO Box 1351 Cloudcroft,
     New Mexico 83117, USA}

\authorcount{affil:Miller}{
     Furzehill House, Ilston, Swansea, SA2 7LE, UK}

\authorcount{affil:Miguel}{
     Departamento de F\'isica Aplicada, Facultad de Ciencias
     Experimentales, Universidad de Huelva,
     21071 Huelva, Spain}

\authorcount{affil:Miguel2}{
     Center for Backyard Astrophysics, Observatorio del CIECEM,
     Parque Dunar, Matalasca\~nas, 21760 Almonte, Huelva, Spain}

\authorcount{affil:AAVSO}{
     American Association of Variable Star Observers, 49 Bay State Rd.,
     Cambridge, MA 02138, USA}

\authorcount{affil:OKU}{
     Osaka Kyoiku University, 4-698-1 Asahigaoka, Osaka 582-8582}

\authorcount{affil:LCO}{
     Department of Physics, University of Notre Dame, Notre Dame,
     Indiana 46556, USA}

\authorcount{affil:Gunma}{
     Gunma Astronomical Observatory, 6860-86 Nakayama, Takayama, 
     Agatsuma, Gunma 377-0702}

\authorcount{affil:NHAO}{
     Center for Astronomy, University of Hyogo, 407-2, Nishigaichi,
     Sayo-cho, Sayo, Hyogo 679-5313}

\authorcount{affil:Nobeyama}{
     Nobeyama Radio Observatory, NAOJ, 462-2 Nobeyama, Minamimaki,
     Minamisaku, Nagano 384-1305}

\authorcount{affil:Stein}{
     6025 Calle Paraiso, Las Cruces, New Mexico 88012, USA}

\authorcount{affil:BAAVSS}{
     The British Astronomical Association, Variable Star Section (BAA VSS),
     Burlington House, Piccadilly, London, W1J 0DU, UK}

\authorcount{affil:Pickard}{
     3 The Birches, Shobdon, Leominster, Herefordshire, HR6 9NG, UK}

\authorcount{affil:Kis}{
     Variable Star Observers League in Japan (VSOLJ),
     7-1 Kitahatsutomi, Kamagaya, Chiba 273-0126}

\authorcount{affil:CrAO}{
     Crimean Astrophysical Observatory, Kyiv Shevchenko 
     National University, 98409, Nauchny, Crimea, Ukraine}

\authorcount{affil:Nyrola}{
     Hankasalmi observatory, Jyvaskylan Sirius ry, Vertaalantie
     419, FI-40270 Palokka, Finland}

\authorcount{affil:Harlingten}{
     Searchlight Observatory Network, The Grange , Erpingham,
     Norfolk, NR11 7QX, UK}

\authorcount{affil:Tyyska}{
     Jyv\"askyl\"an Lyseo upper secondary school, IB section,
     Yliopistonkatu 13, FI-40101 Jyv\"askyl\"a, Finland}

\authorcount{affil:Monard}{
     Bronberg Observatory, Center for Backyard Astronomy Pretoria,
     PO Box 11426, Tiegerpoort 0056, South Africa}

\authorcount{affil:Monard2}{
     Kleinkaroo Observatory, Center for Backyard Astronomy Kleinkaroo,
     Sint Helena 1B, PO Box 281, Calitzdorp 6660, South Africa}

\authorcount{affil:Sternberg}{
     Sternberg Astronomical Institute, Lomonosov Moscow University, 
     Universitetsky Ave., 13, Moscow 119992, Russia}

\authorcount{affil:Slovak}{
     Astronomical Institute of the Slovak Academy of Sciences, 05960,
     Tatranska Lomnica, the Slovak Republic}

\authorcount{affil:Kai}{
     Baselstrasse 133D, CH-4132 Muttenz, Switzerland}

\authorcount{affil:Mdy}{
     Kaminishiyamamachi 12-14, Nagasaki, Nagasaki 850-0006}

\authorcount{affil:Hsk}{
     216-4 Maeda, Inazawa-cho, Inazawa-shi, Aichi 492-8217}

\authorcount{affil:Ioh}{
     VSOLJ, 1001-105 Nishiterakata, Hachioji, Tokyo 192-0153}

\authorcount{affil:Sabo}{
     2336 Trailcrest Dr., Bozeman, Montana 59718, USA}

\authorcount{affil:Ulowetz}{
     Center for Backyard Astrophysics Illinois,
     Northbrook Meadow Observatory, 855 Fair Ln, Northbrook,
     Illinois 60062, USA}

\authorcount{affil:Morelle}{
     9 rue Vasco de GAMA, 59553 Lauwin Planque, France}

\authorcount{affil:UNAM}{
     Instituto de Astronom\'{\i}a UNAM, Apartado Postal 877, 22800 Ensenada
     B.C., M\'{e}xico}

\authorcount{affil:James}{
     11 Tavistock Road, Chelmsford, Essex CM1 6JL, UK}

\authorcount{affil:Dvorak}{
     Rolling Hills Observatory, 1643 Nightfall Drive,
     Clermont, Florida 34711, USA}

\authorcount{affil:RIT}{
     Physics Department, Rochester Institute of Technology, Rochester,
     New York 14623, USA}

\authorcount{affil:Staels}{
     Center for Backyard Astrophysics (Flanders),
     American Association of Variable Star Observers (AAVSO),
     Alan Guth Observatory, Koningshofbaan 51, Hofstade, Aalst, Belgium}

\authorcount{affil:DavidBoyd}{
     Silver Lane, West Challow, Wantage, OX12 9TX, UK}

\authorcount{affil:Terskol}{
     Institute of Astronomy, Russian Academy of Sciences, 361605 Peak Terskol,
     Kabardino-Balkaria, Russia}

\authorcount{affil:ICUkraine}{
     International Center for Astronomical, Medical and Ecological Research
     of NASU, Ukraine 27 Akademika Zabolotnoho Str. 03680 Kyiv,
     Ukraine}

\authorcount{affil:Seikei}{
     Seikei Meteorological Observatory, Seikei High School,
     3-3-1, Kichijoji-Kitamachi, Musashino-shi, Tokyo 180-8633}

\authorcount{affil:Njh}{
     Variable Star Observers League in Japan (VSOLJ),
     124 Isatotyo, Teradani, Kumano, Mie 519-4673}

\authorcount{affil:Bolt}{
     Camberwarra Drive, Craigie, Western Australia 6025, Australia}

\authorcount{affil:Nelson}{
     1105 Hazeldean Rd, Ellinbank 3820, Australia}

\authorcount{affil:Starkey}{
     DeKalb Observatory, H63, 2507 County Road 60, Auburn, Indiana 46706, USA}

\authorcount{affil:Buczynski}{
     Conder Brow Observatory, Fell Acre, Conder Brow, Little Fell Lane,
     Scotforth, Lancs LA2 0RQ, England}

\authorcount{affil:Starr}{
     Warrumbungle Observatory, Tenby, 841 Timor Rd,
     Coonabarabran NSW 2357, Australia}

\authorcount{affil:Goff}{
     13508 Monitor Ln., Sutter Creek, California 95685, USA}

\authorcount{affil:Denisenko}{
     Space Research Institute (IKI), Russian Academy of Sciences, Moscow,
     Russia}

\authorcount{affil:Ohio}{
     Department of Astronomy, the Ohio State University, Columbia,
     OH 43210, USA}

\authorcount{affil:Princeton}{
     Department of Astrophysical Sciences, Princeton University,
     NJ 08544, USA}

\authorcount{affil:Itagaki}{
     Itagaki Astronomical Observatory, Teppo-cho, Yamagata 990-2492}

\authorcount{affil:Kaneko}{
     14-7 Kami-Yashiki, Kakegawa, Shizuoka 436-0049}

\authorcount{affil:Stubbings}{
     Tetoora Observatory, Tetoora Road, Victoria, Australia}

\authorcount{affil:VVSBelgium}{
     Vereniging Voor Sterrenkunde (VVS), Moffelstraat 13 3370
     Boutersem, Belgium}

\authorcount{affil:Shears}{
     ``Pemberton'', School Lane, Bunbury, Tarporley, Cheshire, CW6 9NR, UK}

\authorcount{affil:Schmeer}{
     Bischmisheim, Am Probstbaum 10, 66132 Saarbr\"{u}cken, Germany}

\authorcount{affil:Poyner}{
     BAA Variable Star Section, 67 Ellerton Road, Kingstanding,
     Birmingham B44 0QE, UK}

\authorcount{affil:MRO}{
     Alberdi, 42, 2F 28029 Madrid, Spain}



\maketitle

\begin{abstract}
   Continuing the project described by \citet{Pdot}, we collected
times of superhump maxima for SU UMa-type dwarf novae 
mainly observed during the 2012--2013 season.  We found
three objects (V444 Peg, CSS J203937 and MASTER J212624)
having strongly positive period derivatives despite the long
orbital period ($P_{\rm orb}$). 
By using the period of growing stage (stage A)
superhumps, we obtained mass ratios for six objects.
We characterized nine new WZ Sge-type dwarf novae.
We made a pilot survey of the decline rate of slowly fading
part of SU UMa-type and WZ Sge-type outbursts.
The decline time scale was found to generally follow the expected
$P_{\rm orb}^{1/4}$ dependence and WZ Sge-type outbursts
also generally follow this trend.
There are some objects which show slower
decline rates, and we consider these objects good candidates
for period bouncers.  We also studied unusual behavior
in some objects, including BK Lyn which made a transition
from an ER UMa-type state to the novalike (standstill) state
in 2013 and unusually frequent occurrence of superoutbursts
in NY Ser and CR Boo.  We applied least absolute 
shrinkage and selection operator (Lasso) power spectral
analysis, which has been proven to be very effective in
analyzing the Kepler data, to ground-based photometry of BK Lyn
and detected the dramatic disappearance of the signal of 
negative superhumps in 2013.  We suggested that
the mass-transfer rates did not vary strongly between
the ER UMa-type state and novalike state in BK Lyn, and
this transition was less likely caused by a systematic
variation of the mass-transfer rate.
\end{abstract}

\section{Introduction}

   Cataclysmic variables (CVs) are close binary systems
transferring matter from a low-mass dwarf secondary to
a white dwarf.  The transferred matter forms an accretion
disk.  In dwarf novae, a class of CVs, the instability
in the accretion disk produces outbursts.  SU UMa-type
dwarf novae, a class of DNe, show superhumps during
their long outbursts (superoutbursts), whose period is
generally a few percent longer than the orbital period.
It is generally considered that the tidal instability
in the accretion disk caused by the 3:1 resonance
\citet{whi88tidal} is responsible for the superhump
and superoutburst phenomenon (\cite{osa89suuma};
\cite{osa96review}).
[for general information of CVs, DNe, SU UMa-type dwarf novae and 
superhumps, see e.g. \citet{war95book}].

   In a series of papers \citet{Pdot}, \citet{Pdot2}, \citet{Pdot3}
and \citet{Pdot4}, we systematically surveyed SU UMa-type dwarf novae 
particularly laying emphasis on period variations of superhumps 
(positive superhumps;
superhumps having periods longer than the orbital period),
which has been the main theme since \citet{Pdot}.
The change in the superhump period reflects the precession
angular velocity of the eccentric (or flexing) disk, and
would be an excellent probe for studying the structure
of the accretion disk during dwarf nova outbursts.
In addition to the systematic survey of period variations of 
superhumps, we have studied properties of newly discovered 
WZ Sge-type dwarf novae [for WZ Sge-type dwarf novae, see e.g. 
\citet{bai79wzsge}; \citet{dow90wxcet}; \citet{kat01hvvir}].

   In the meantime, there has been epoch-making progress
in understanding the SU UMa-type phenomenon and interpreting
the superhump period, and we review the history shortly.

   An anticipation of this new progress started with the
analysis of Kepler data of V344 Lyr \citep{sti10v344lyr},
who detected likely persistent negative superhumps in quiescence.
\citet{can10v344lyr} studied the systematics of outburst
in V344 Lyr.  \citet{woo11v344lyr} analyzed the positive and
negative superhumps in the same object.

   In \citet{Pdot3}, we analyzed the Kepler data of 
V344 Lyr and V1504 Cyg in the same way we have
used and compared them with other SU UMa-type dwarf novae.
This work led to two works, \citet{osa13v1504cygKepler},
\citet{osa13v344lyrv1504cyg}, in which the thermal-tidal
instability (TTI) model \citep{osa89suuma} has been proven
to be the best explanation of the Kepler observation.
In particular, \citet{osa13v1504cygKepler} used the frequency
of negative superhumps (superhumps having periods shorter
than the orbital period) which are believed to arise from
a tilted accretion disk (e.g. \cite{har95v503cyg};
\cite{pat97v603aql}; \cite{woo07negSH})
to derive the variation of the disk radius and 
directly confirmed the prediction of the TTI model.
\citet{osa13v344lyrv1504cyg} further
analyzed the frequency variation of the co-existing negative and 
positive superhumps, and confirmed that the variation of 
the disk radius predicted by the TTI model gives
a consistent explanation of the variation of the negative and 
positive superhumps.

   \citet{osa13v344lyrv1504cyg} also introduced the pressure
effect interpreting the periods of positive superhumps.
Although the pressure effect has been long known
(\cite{lub92SH}; \cite{hir93SHperiod};
\cite{mur98SH}; \cite{mon01SH}; \cite{pea06SH}),
there have been no direct applications to interpretation
of period variation of (positive) superhumps during
the superoutburst.  The treatment of the pressure effect
is difficult, and \citet{osa13v344lyrv1504cyg} partly succeeded
in interpreting the variation of the superhump period
during the initial growing stage (stage A) and fully developed 
stage (early stage B) [see \citet{Pdot} for the notation
of stages A-B-C of superhumps].
This interpretation led to an important
consequence: the precession frequency of the superhumps
during the growing stage (stage A) corresponds to 
the dynamical precession rate at the radius of the
3:1 resonance.  This interpretation enabled us to dynamically
determine the binary mass-ratios ($q = M_2/M_1$) only from
superhump observations and the orbital period.
This method has been indeed shown to be effective
by comparisons with $q$ values by eclipse observations
or radial-velocity studies \citep{kat13qfromstageA}.
This method has been successfully used to characterize
the WZ Sge-type dwarf novae with multiple rebrightenings
\citep{nak13j2112j2037} and identifying the elusive
period bouncers [CVs evolved beyond the minimum orbital period 
during their secular evolution; for a recent review of CV
evolution, see e.g. \citet{kni11CVdonor}] \citep{kat13j1222}.
We should also note that our initial approximation
in \citet{Pdot} did not include the pressure effect and 
the disk radius was normalized to the radius of 3:1 resonance
for the early phase of the stage B superhumps, and 
the resultant radii were systematically estimated
larger in \citet{Pdot}.  Since the qualitative estimation 
of the pressure effect during the superoutburst is still 
a difficult task, we leave the understanding of 
the period variation of superhumps to future works.

   In \citet{Pdot3}, we also made a pilot survey of
variation of superhump amplitudes and indicated that 
the amplitudes of superhumps are strongly correlated
with orbital periods, and the dependence on the inclination is
weak in systems with inclinations smaller than 80$^{\circ}$.
In \citet{Pdot4}, we systematically studied ER UMa-type 
dwarf novae [see e.g. \citet{kat95eruma}; \citet{rob95eruma}],
which has recently become a hot topic in the field of 
cataclysmic variable since the discovery of negative
superhump even during the superoutburst \citep{ohs12eruma}
and the possible identification of BK Lyn -- 
an object recently changed from a novalike variable
to an ER UMa-type dwarf nova -- with an ancient
classical nova \citep{pat13bklyn}, shedding light
on the evidence of the long-sought transition from 
classical novae to dwarf novae [the hibernation model,
see e.g. \citet{liv87hibernation}].

   \citet{can12v344lyr}, \citet{bar12j1939}, \citet{ram12v447lyr}
also studied Kepler data for dwarf novae V344 Lyr, V1504 Cyg,
the background dwarf nova of KIC 4378554 and V447 Lyr.
The work by \citet{can12v344lyr} tried to reproduce the outburst
morphology by the pure thermal instability model
[see \citet{osa13v1504cygv344lyrpaper3} for a discussion
on the difficulty of this model].  \citet{ram12v447lyr}
and \citet{can12ugemLC} indicated that the presence of 
a ``shoulder'' in long outbursts of an SS Cyg-type dwarf nova,
which \citet{ram12v447lyr} and \citet{can12ugemLC} considered
to be analogous to the precursor outburst in SU UMa-type
dwarf novae.  There was also a rapid progress in interpretation
and numerical simulation of negative and positive superhumps, 
a part of which is associated with Kepler observations 
(\cite{mon10negSH}; \cite{mon10disktilt}; \cite{mon12negSHSPH}; 
\cite{mon12negposSH}).

   Keeping the brand-new progress in dwarf novae in mind, 
let's go onto the new observations.

   This paper is structured as follows.  Section \ref{sec:obs}
describes the observation and analysis technique, section
\ref{sec:individual} deals with individual objects
we observed, section \ref{sec:generaldiscuss} discusses 
the general properties following \citet{Pdot} and \citet{Pdot3},
section \ref{sec:fadingrate} provides a new pilot survey
of the decline rate during the superoutburst, which has 
led to a new promising method for identifying period bouncers,
section \ref{sec:OGLEDNe} examined the recently reported
new OGLE dwarf novae, section \ref{sec:topics} presents
topics on some objects which show new types of behavior
in dwarf novae, and finally section \ref{sec:summary}
is for the summary.

\section{Observation and Analysis}\label{sec:obs}

   The data were obtained under campaigns led by 
the VSNET Collaboration \citep{VSNET}.
In some objects, we used the public data from 
the AAVSO International Database\footnote{
   $<$http://www.aavso.org/data-download$>$.
}.

   The majority of the data were acquired
by time-resolved CCD photometry by using 30 cm-class telescopes, 
whose observational details will be presented in
future papers dealing with analysis and discussion on
individual objects of interest.
The list of outbursts and observers is summarized in 
table \ref{tab:outobs}.
The data analysis was performed just in the same way described
in \citet{Pdot} and \citet{Pdot3}.  
The times of all observations are expressed in 
Barycentric Julian Dates (BJD).

   We also used the same abbreviations: $P_{\rm orb}$ for
the orbital period and $\varepsilon \equiv P_{\rm SH}/P_{\rm orb}-1$ for 
the fractional superhump excess.   After \citet{osa13v1504cygKepler},
the alternative fractional superhump excess in the frequency unit
$\varepsilon^* \equiv 1-P_{\rm orb}/P_{\rm SH}-1 = \varepsilon/(1+\varepsilon)$
has been introduced since this fractional superhump excess
can be directly compared to the precession rate.  We therefore
used $\varepsilon^*$ when referring the precession rate.

   We used phase dispersion minimization (PDM; \cite{PDM})
for period analysis and 1$\sigma$ errors for the PDM analysis
was estimated by the methods of \citet{fer89error} and \citet{Pdot2}.
In \citet{Pdot4}, we introduced least absolute 
shrinkage and selection operator (Lasso) method 
(\cite{lasso}; \cite{kat12perlasso}), which has been proven to be 
very effective in separating closely spaced periods.
We have further extended this Lasso analysis to two-dimensional
power spectra (\cite{kat13j1924}; \cite{osa13v344lyrv1504cyg};
\cite{kat13j1939v585lyrv516lyr}).  These two-dimensional
Lasso power spectra have been sometimes helpful in detecting
negative superhumps (cf. \cite{osa13v344lyrv1504cyg}) as well as
superhumps with varying frequencies (cf. \cite{kat13j1924}).

   The derived $P_{\rm SH}$, $P_{\rm dot}$ and other parameters
are listed in table \ref{tab:perlist} in same format as in
\citet{Pdot}.  The definitions of parameters $P_1, P_2, E_1, E_2$
and $P_{\rm dot}$ are the same as in \citet{Pdot}.
We also presented comparisons of $O-C$ diagrams between different
superoutbursts since this has been one of the motivations of
these surveys (cf. \cite{uem05tvcrv}).

   We used the same terminology of superhumps summarized in
\citet{Pdot3}.  We especially call reader's attention to
the term ``late superhumps''.  We only used ``traditional''
late superhumps when an $\sim$0.5 phase shift is confirmed
[see also table 1 in \citet{Pdot3} for various types
of superhumps; the lack of an $\sim$0.5 
phase shift in V585 Lyr in the Kepler data has been also
confirmed \citep{kat13j1939v585lyrv516lyr}].
Early superhumps are double-wave humps seen during the early stages
of WZ Sge-type dwarf novae, and have period close to the orbital
periods (\cite{kat96alcom}; \cite{kat02wzsgeESH}; 
\cite{osa02wzsgehump}).
We used the period of early superhumps as approximate
orbital period, since their periods only differ by 
less than 0.1\% (\cite{ish02wzsgeletter}; \cite{kat02wzsgeESH}).

   As in \citet{Pdot}, we have used coordinate-based 
optical transient (OT) designations for some objects, such as 
Catalina Real-time Transient Survey (CRTS; \cite{CRTS})\footnote{
   $<$http://nesssi.cacr.caltech.edu/catalina/$>$.
   For the information of the individual Catalina CVs, see
   $<$http://nesssi.cacr.caltech.edu/catalina/AllCV.html$>$.
} transients
and listed the original identifiers in table \ref{tab:outobs}.
The CRTS team has recently provided the public data
release\footnote{
  $<$http://nesssi.cacr.caltech.edu/DataRelease/$>$.
}
and provided the International Astronomical Union (IAU)
designations for the cataloged objects.
We used these IAU designations whenever available
starting from the present paper.

\begin{table*}
\caption{List of Superoutbursts.}\label{tab:outobs}
\begin{center}
\begin{tabular}{ccccl}
\hline
Subsection & Object & Year & Observers or references\commenta & ID\commentb \\
\hline
\ref{obj:kxaql}    & KX Aql       & 2012 & Ioh, KU, SRI, AAVSO, Hsk & \\
\ref{obj:nncam}    & NN Cam       & 2012 & OUS & \\
\ref{obj:v485cen}  & V485 Cen     & 2013 & OkC, HaC & \\
\ref{obj:zcha}     & Z Cha        & 2013 & SPE & \\
\ref{obj:yzcnc}    & YZ Cnc       & 2011 & SWI, HMB, DKS, MEV, Boy, & \\
                   &              &      & BSt, UJH, Nyr, AAVSO & \\
\ref{obj:gzcnc}    & GZ Cnc       & 2013 & Kis, Hsk & \\
\ref{obj:v503cyg}  & V503 Cyg     & 2012 & SWI, MEV, AAVSO, Boy & \\
                   &              & 2012b & RPc, IMi & \\
\ref{obj:ovdra}    & OV Dra       & 2013 & LCO & \\
\ref{obj:aqeri}    & AQ Eri       & 2012 & OUS, Kis, Aka, Hsk & \\
\ref{obj:v660her}  & V660 Her     & 2012 & IMi & \\
                   &              & 2013 & DPV, NDJ & \\
\ref{obj:v1227her} & V1227 Her    & 2012a & IMi, RPc, Mhh, OKU, deM & \\ 
                   &              & 2012b & SRI, IMi, CRI & \\
                   &              & 2013 & RPc & \\
\ref{obj:mmhya}    & MM Hya       & 2013 & IMi, Mdy & \\
\ref{obj:abnor}    & AB Nor       & 2013 & MLF, HaC & \\
\ref{obj:dtoct}    & DT Oct       & 2013 & OkC & \\
\ref{obj:grori}    & GR Ori       & 2013 & AKz, Ioh, MLF, SWI, Mdy, Aka, Nel, & \\
                   &              &      & deM, AAVSO, KU, RPc, DKS, DRS, SAc, & \\
                   &              &      & DPV, HaC, PSD, SRI, Buc & \\
\ref{obj:v444peg}  & V444 Peg     & 2012 & Mhh, HaC, Hsk, IMi, AAVSO & \\
\ref{obj:v521peg}  & V521 Peg     & 2012 & Hsk & \\
\ref{obj:v368per}  & V368 Per     & 2012 & AAVSO, IMi & \\
\ref{obj:typsa}    & TY PsA       & 2012 & HaC, OUS, Kis, Aka & \\
\ref{obj:qwser}    & QW Ser       & 2009 & Ioh, Njh & \\
                   &              & 2013 & HaC, Aka & \\
\ref{obj:v493ser}  & V493 Ser     & 2013 & Aka, Mic, HaC & \\
\ref{obj:awsge}    & AW Sge       & 2012 & Kai, BSt, HMB, Mas, DPV, & \\
                   &              &      & Mhh, OUS, IMi, AAVSO & \\
\ref{obj:v1212tau} & V1212 Tau    & 2013 & RPc & \\
\ref{obj:bzuma}    & BZ UMa       & 2012 & Aka & \\
\ref{obj:ciuma}    & CI UMa       & 2013 & CRI & \\
\ref{obj:cyuma}    & CY UMa       & 2013 & DPV & \\
\ref{obj:mruma}    & MR UMa       & 2013 & DPV, AAVSO & \\
\ref{obj:asassn13ao} & ASAS SN-13ao & 2013 & KU, DPV & \\
\ref{obj:asassn13as} & ASAS SN-13as & 2013 & DPV & \\
\ref{obj:asassn13ax} & ASAS SN-13ax & 2013 & deM, DPV, IMi, Shu, NKa & \\
\ref{obj:asassn13bj} & ASAS SN-13bj & 2013 & LCO & \\
\ref{obj:asassn13bm} & ASAS SN-13bm & 2013 & Kai, NDJ & \\
\ref{obj:asassn13bp} & ASAS SN-13bp & 2013 & NDJ, Mdy & \\
\ref{obj:asassn13br} & ASAS SN-13br & 2013 & DPV, Kai, Shu, Mic & \\
\ref{obj:j015051}  & CSS J015051  & 2012 & KU, Mhh, LCO & CSS111006:015052$+$332622 \\
\ref{obj:j015321}  & CSS J015321  & 2012 & AKz, Mas & CSS081026:015321$+$340857 \\
\hline
  \multicolumn{5}{l}{\parbox{530pt}{\commenta Key to observers:
Aka (H. Akazawa, OUS),
AKz (Astrokolkhoz Obs.),
BSt (B. Staels),
Boy\commentc (D. Boyd),
Buc (D. Buczynski),
CRI (Crimean Astrophys. Obs.),
deM (E. de Miguel),
DKS\commentc (S. Dvorak),
DPV (P. Dubovsky),
DRS\commentc (D. Starkey),
GBo (G. Bolt),
GFB\commentc (W. Goff),
HaC (F.-J. Hambsch, remote obs. in Chile)
HMB (F.-J. Hambsch),
Hsk (K. Hirosawa),
IMi\commentc (I. Miller),
Ioh (H. Itoh),
Kai (K. Kasai),
Kis (S. Kiyota),
Kra (T. Krajci),
KU (Kyoto U., campus obs.),
LCO\commentc (C. Littlefield),
MEV\commentc (E. Morelle),
MLF (B. Monard),
Mas (G. Masi),
Mdy (Y. Maeda),
Mhh (H. Maehara),
Mic (R. Michel-Murillo),
NDJ\commentc (N. James),
Nel\commentc (P. Nelson),
NKa (N. Katysheva),
Nyr\commentc (Nyrola and Hankasalmi Obs.),
OkC\commentc (A. Oksanen, remote obs. in Chile)
OKU (Osaya Kyoiku U.),
OUS (Okayama U. of Science),
PSD\commentc (S. Padovan),
RIT (M. Richmond),
RPc\commentc (R. Pickard),
SAc (Seikei High School),
Shu (S. Shugarov),
SPE\commentc (P. Starr),
SRI\commentc (R. Sabo),
SWI\commentc (W. Stein),
Ter (Terskol Obs.),
UJH\commentc (J. Ulowetz),
Vol (I. Voloshina),
AAVSO (AAVSO database)
}} \\
  \multicolumn{5}{l}{\commentb Original identifications, discoverers or data source.} \\
  \multicolumn{5}{l}{\commentc Inclusive of observations from the AAVSO database.} \\
\end{tabular}
\end{center}
\end{table*}

\addtocounter{table}{-1}
\begin{table*}
\caption{List of Superoutbursts (continued).}
\begin{center}
\begin{tabular}{ccccl}
\hline
Subsection & Object & Year & Observers or references\commenta & ID\commentb \\
\hline
\ref{obj:j102842}  & CSS J102842  & 2013 & MLF, deM & CSS090331:102843$-$081927 \\
\ref{obj:j105835}  & CSS J105835  & 2012 & Kis, HaC, OKU, Nyr & CSS081025:105835$+$054706 \\
\ref{obj:j150904}  & CSS J150904  & 2013 & AKz & CSS130324:150904$+$465057 \\
--                 & CSS J174033  & 2013 & Ohshima et al. in prep. & CSS130418:174033$+$414756 \\
\ref{obj:j203937}  & CSS J203937  & 2012 & MLF, deM, RIT, HaC, LCO, & CSS120813:203938$-$042908\\
                   &              &      & Mas, UJH, AAVSO & \\
\ref{obj:j214934}  & CSS J214934  & 2012 & Mas & CSS120922:214934$-$121908 \\
\ref{obj:dde26}    & DDE 26       & 2012 & LCO, CRI & \\
\ref{obj:j000820}  & MASTER J000820 & 2012 & Mas & MASTER OT J000820.50$+$773119.1 \\
\ref{obj:j001952}  & MASTER J001952 & 2012 & OKU, Mas & MASTER OT J001952.31$+$464933.0 \\
\ref{obj:j030128}  & MASTER J030128 & 2012 & deM, Mas, RPc, Kra & MASTER OT J030128.77$+$401104.9 \\
\ref{obj:j042609}  & MASTER J042609 & 2012 & AKz, deM, MEV, Mhh, HaC, & MASTER OT J042609.34$+$354144.8 \\
                   &                &      & OKU, LCO, SWI, Vol & \\
\ref{obj:j054317}  & MASTER J054317 & 2012 & Mhh, HaC, OKU, AKz, Mas & MASTER OT J054317.95$+$093114.8 \\
\ref{obj:j064725}  & MASTER J064725 & 2013 & SWI, Nyr, LCO, OKU, DPV & MASTER OT J064725.70$+$491543.9 \\
\ref{obj:j073418}  & MASTER J073418 & 2013 & IMi, RIT & MASTER OT J073418.66$+$271310.5 \\
\ref{obj:j081110}  & MASTER J081110 & 2012 & AKz, Mas, Mhh & MASTER OT J081110.46$+$660008.5 \\
\ref{obj:j094759}  & MASTER J094759 & 2013 & OKU, CRI, KU, & MASTER OT J094759.83$+$061044.4 \\
                   &                &      & HaC, deM, Mdy & \\
\ref{obj:j105025}  & MASTER J105025 & 2012 & Kra, Mas & MASTER OT J105025.99$+$332811.4 \\
\ref{obj:j111759}  & MASTER J111759 & 2013 & deM, DPV & MASTER OT J111759.87$+$765131.6 \\
\ref{obj:j165236}  & MASTER J165236 & 2013 & KU, Kra, DPV, AKz & MASTER OT J165236.22$+$460513.2 \\
\ref{obj:j174902}  & MASTER J174902 & 2013 & DPV & MASTER OT J174902.10$+$191331.2 \\
\ref{obj:j181953}  & MASTER J181953 & 2013 & OKU, Kai, KU, Mic, CRI, Mdy & MASTER OT J181953.76$+$361356.5 \\
--                 & MASTER J203749 & 2012 & \citet{nak13j2112j2037} & MASTER OT J203749.39$+$552210.3 \\
--                 & MASTER J211258 & 2013 & \citet{nak13j2112j2037} & MASTER OT J211258.65$+$242145.4 \\
\ref{obj:j212624}  & MASTER J212624 & 2013 & Shu, DPV, Kai & MASTER OT J212624.16$+$253827.2 \\
\ref{obj:j112619}  & OT J112619   & 2013 & SWI, UJH, deM, GFB, & CSS130106:112619$+$084651 \\
                   &              &      & Mas, DKS, SRI, IMi &  \\
\ref{obj:j191443}  & OT J191443   & 2012 & Mas & Itagaki \citep{yam08j1914cbet1535} \\
\ref{obj:j205146}  & OT J205146   & 2012 & Mhh, Vol, Shu & CSS121004:205146$-$035827 \\
\ref{obj:j220641}  & OT J220641   & 2012 & Kra, Mas & CSS 110921:220641$+$301436 \\
\ref{obj:j232727}  & OT J232727   & 2012 & Mhh, LCO, CRI, MEV, Mas, & \citet{ita12j2327cbet3228} \\
                   &              &      & Ter, OKU, Kis, Shu, RPc & \\
\ref{obj:j062703}  & PNV J062703  & 2013 & Nyr, deM, Ioh, DPV, & PNV J06270375$+$3952504 (Kaneko) \\
                   &              &      & Aka, OUS, AAVSO & \\
\ref{obj:j075107}  & SDSS J075107 & 2013 & IMi & SDSS J075107.50$+$300628.4 \\
\ref{obj:j080033}  & SDSS J080033 & 2012 & Mas & SDSS J080033.86$+$192416.5 \\
\ref{obj:j162520}  & SDSS J162520 & 2010 & \citet{Pdot2} & SDSS J162520.29$+$120308.7 (Wils) \\
\ref{obj:j122221}  & SSS J122221 & 2013 & \citet{kat13j1222} & SSS J122221.7$-$311523 \\
\ref{obj:j224739}  & SSS J224739  & 2012 & HaC, GBo & SSS J224739.7$-$362253 \\
\ref{obj:j153756}  & TCP J153756  & 2013 & MLF, Kis, SWI, HaC, UJH & TCP J15375685$-$2440136 (Itagaki) \\
\ref{obj:j175219}  & TCP J175219  & 2012 & Mas & TCP J17521907$+$5001155 (Mikuz) \\
\hline
\end{tabular}
\end{center}
\end{table*}

\begin{table*}
\caption{Superhump Periods and Period Derivatives}\label{tab:perlist}
\begin{center}
\begin{tabular}{c@{\hspace{7pt}}c@{\hspace{7pt}}c@{\hspace{7pt}}c@{\hspace{7pt}}c@{\hspace{7pt}}c@{\hspace{7pt}}c@{\hspace{7pt}}c@{\hspace{7pt}}c@{\hspace{7pt}}c@{\hspace{7pt}}c@{\hspace{7pt}}c@{\hspace{7pt}}c@{\hspace{7pt}}c}
\hline
Object & Year & $P_1$ (d) & err & \multicolumn{2}{c}{$E_1$\commenta} & $P_{\rm dot}$\commentb & err\commentb & $P_2$ (d) & err & \multicolumn{2}{c}{$E_2$\commenta} & $P_{\rm orb}$ (d)\commentc & Q\commentd \\
\hline
NN Cam & 2012 & 0.074140 & 0.000057 & 0 & 41 & -- & -- & -- & -- & -- & -- & 0.0717 & C \\
V485 Cen & 2013 & 0.042136 & 0.000013 & 0 & 95 & 4.2 & 1.5 & -- & -- & -- & -- & 0.040995 & C \\
YZ Cnc & 2011 & 0.090648 & 0.000096 & 18 & 34 & -- & -- & 0.090379 & 0.000036 & 40 & 130 & 0.0868 & B \\
GZ Cnc & 2013 & 0.092842 & 0.000084 & 0 & 76 & -- & -- & -- & -- & -- & -- & 0.08825 & C2 \\
V503 Cyg & 2012 & 0.081446 & 0.000096 & 21 & 51 & -- & -- & 0.081121 & 0.000026 & 49 & 123 & 0.077759 & CM \\
V503 Cyg & 2012b & 0.081232 & 0.000165 & 0 & 24 & -- & -- & -- & -- & -- & -- & 0.077759 & CG \\
AQ Eri & 2012 & 0.062398 & 0.000097 & 0 & 63 & 29.5 & 13.4 & -- & -- & -- & -- & 0.06094 & C \\
V660 Her & 2012 & 0.080891 & 0.000164 & 0 & 26 & -- & -- & -- & -- & -- & -- & -- & CG \\
V660 Her & 2013 & 0.081081 & 0.000040 & 0 & 50 & -- & -- & 0.080568 & 0.000128 & 99 & 126 & -- & C \\
V1227 Her & 2012a & 0.065032 & 0.000083 & 0 & 29 & -- & -- & 0.064839 & 0.000136 & 92 & 123 & -- & CP \\
V1227 Her & 2012b & 0.065150 & 0.000027 & 0 & 122 & 7.5 & 2.2 & -- & -- & -- & -- & -- & BP \\
V1227 Her & 2013 & 0.065083 & 0.000046 & 0 & 19 & -- & -- & -- & -- & -- & -- & -- & CP \\
MM Hya & 2013 & -- & -- & -- & -- & -- & -- & 0.058633 & 0.000103 & 0 & 29 & 0.057590 & C \\
AB Nor & 2013 & 0.079756 & 0.000027 & 0 & 30 & -- & -- & 0.079413 & 0.000032 & 28 & 105 & -- & C \\
GR Ori & 2013 & 0.058333 & 0.000021 & 11 & 142 & 6.4 & 1.5 & -- & -- & -- & -- & -- & B \\
V444 Peg & 2012 & 0.097645 & 0.000052 & 10 & 61 & 14.5 & 6.2 & -- & -- & -- & -- & -- & B \\
V521 Peg & 2012 & 0.0603 & 0.0002 & 0 & 3 & -- & -- & -- & -- & -- & -- & -- & C \\
V368 Per & 2012 & -- & -- & -- & -- & -- & -- & 0.078946 & 0.000050 & 0 & 27 & -- & C \\
TY PsA & 2012 & 0.087809 & 0.000019 & 19 & 91 & $-$1.2 & 2.6 & 0.087655 & 0.000032 & 98 & 179 & 0.08423 & A \\
QW Ser & 2009 & 0.076858 & 0.000014 & 0 & 132 & $-$2.0 & 1.3 & -- & -- & -- & -- & 0.074572 & CG \\
QW Ser & 2013 & 0.077088 & 0.000051 & 0 & 34 & -- & -- & 0.076496 & 0.000016 & 46 & 86 & 0.074572 & C \\
V493 Ser & 2013 & 0.082917 & 0.000062 & 0 & 54 & -- & -- & 0.082618 & 0.000030 & 53 & 132 & 0.08001 & C \\
AW Sge & 2012 & 0.074733 & 0.000030 & 16 & 59 & $-$3.1 & 7.5 & 0.074312 & 0.000064 & 57 & 97 & -- & B \\
CI UMa & 2013 & 0.062381 & 0.000084 & 0 & 32 & -- & -- & -- & -- & -- & -- & -- & C \\
MR UMa & 2013 & 0.065336 & 0.000069 & 0 & 34 & -- & -- & 0.064615 & 0.000035 & 47 & 110 & -- & C \\
ASAS SN-13as & 2013 & 0.072522 & 0.000095 & 0 & 42 & -- & -- & -- & -- & -- & -- & -- & CG2 \\
ASAS SN-13ax & 2013 & 0.056155 & 0.000010 & 52 & 177 & 4.5 & 0.6 & -- & -- & -- & -- & -- & A \\
ASAS SN-13bm & 2013 & 0.069015 & 0.000026 & 0 & 61 & -- & -- & -- & -- & -- & -- & -- & C \\
ASAS SN-13bp & 2013 & 0.06828 & 0.00016 & 0 & 5 & -- & -- & -- & -- & -- & -- & -- & C \\
ASAS SN-13br & 2013 & 0.065337 & 0.000020 & 0 & 79 & 9.6 & 1.6 & 0.065008 & 0.000029 & 76 & 125 & -- & B \\
DDE 26 & 2012 & 0.089320 & 0.000063 & 0 & 31 & $-$27.4 & 10.6 & -- & -- & -- & -- & -- & CG \\
CSS J015051 & 2012 & 0.072706 & 0.000032 & 0 & 74 & 1.9 & 5.9 & 0.072293 & -- & 74 & 116 & -- & C \\
CSS J015321 & 2012 & 0.096658 & 0.000096 & 0 & 23 & 74.0 & 27.0 & -- & -- & -- & -- & -- & C \\
CSS J102842 & 2013 & 0.038197 & 0.000006 & 154 & 315 & 2.6 & 0.5 & -- & -- & -- & -- & -- & C \\
CSS J105835 & 2012 & 0.057882 & 0.000025 & 33 & 117 & 6.5 & 3.0 & -- & -- & -- & -- & -- & C \\
CSS J150904 & 2013 & 0.069860 & 0.000043 & 0 & 18 & -- & -- & -- & -- & -- & -- & 0.06844 & C \\
CSS J174033 & 2013 & 0.045548 & 0.000003 & 60 & 385 & 1.6 & 0.1 & -- & -- & -- & -- & 0.045048 & AE \\
CSS J203937 & 2012 & 0.111210 & 0.000042 & 18 & 98 & 9.0 & 3.3 & -- & -- & -- & -- & 0.10572 & B \\
MASTER J000820 & 2012 & 0.082697 & 0.000087 & 0 & 13 & -- & -- & -- & -- & -- & -- & -- & C \\
MASTER J001952 & 2012 & 0.060955 & 0.000035 & 15 & 113 & 10.4 & 2.7 & -- & -- & -- & -- & -- & C \\
MASTER J030128 & 2012 & 0.062831 & 0.000166 & 96 & 114 & -- & -- & -- & -- & -- & -- & -- & C2 \\
MASTER J042609 & 2012 & 0.067557 & 0.000029 & 0 & 40 & -- & -- & 0.066904 & 0.000052 & 81 & 158 & 0.065502 & C \\
MASTER J054317 & 2012 & 0.075949 & 0.000029 & 39 & 131 & 6.5 & 3.4 & 0.075608 & 0.000044 & 123 & 211 & -- & B \\
MASTER J064725 & 2013 & 0.067774 & 0.000041 & 0 & 82 & -- & -- & 0.067337 & 0.000031 & 82 & 157 & -- & C \\
MASTER J073418 & 2013 & -- & -- & -- & -- & -- & -- & 0.061999 & 0.000055 & 0 & 36 & -- & C \\
MASTER J081110 & 2012 & 0.058137 & 0.000011 & 40 & 246 & 4.0 & 0.3 & -- & -- & -- & -- & -- & B \\
\hline
  \multicolumn{13}{l}{\commenta Interval used for calculating the period (corresponding to $E$ in section \ref{sec:individual}).} \\
  \multicolumn{13}{l}{\commentb Unit $10^{-5}$.} \\
  \multicolumn{13}{l}{\parbox{500pt}{\commentc References: 
NN Cam (Denisenko, D. 2007, vsnet-alert 9557);
V485 Cen \citep{aug96v485cen};
YZ Cnc \citep{sha88yzcnc};
GZ Cnc \citep{she07CVspec};
V503 Cyg (this work);
AQ Eri \citep{tho96Porb};
MM Hya \citep{pat03suumas};
TY PsA, QW Ser (this work);
V493 Ser \citet{Pdot};
CSS J150904 (this work);
CSS J174033 (T.~Ohshima et al. in preparation);
CSS J203937, MASTER J042609, MASTER J094759, MASTER J181953 (this work);
MASTER J203749, MASTER J211258 \citep{nak13j2112j2037};
OT J112619, OT J232727, PNV J062703, TCP J153756 (this work).
}}\\
  \multicolumn{13}{l}{\parbox{500pt}{\commentd Data quality and comments. A: excellent, B: partial coverage or slightly low quality, C: insufficient coverage or observations with large scatter, G: $P_{\rm dot}$ denotes global $P_{\rm dot}$, M: observational gap in middle stage, 2: late-stage coverage, the listed period may refer to $P_2$, E: $P_{\rm orb}$ refers to the period of early superhumps, P: $P_{\rm orb}$ refers to a shorter stable periodicity recorded in outburst.}} \\
\end{tabular}
\end{center}
\end{table*}

\addtocounter{table}{-1}
\begin{table*}
\caption{Superhump Periods and Period Derivatives (continued)}
\begin{center}
\begin{tabular}{c@{\hspace{7pt}}c@{\hspace{7pt}}c@{\hspace{7pt}}c@{\hspace{7pt}}c@{\hspace{7pt}}c@{\hspace{7pt}}c@{\hspace{7pt}}c@{\hspace{7pt}}c@{\hspace{7pt}}c@{\hspace{7pt}}c@{\hspace{7pt}}c@{\hspace{7pt}}c@{\hspace{7pt}}c}
\hline
Object & Year & $P_1$ & err & \multicolumn{2}{c}{$E_1$} & $P_{\rm dot}$ & err & $P_2$ & err & \multicolumn{2}{c}{$E_2$} & $P_{\rm orb}$ & Q \\
\hline
MASTER J094759 & 2013 & 0.056121 & 0.000020 & 45 & 214 & 3.0 & 1.1 & -- & -- & -- & -- & 0.05588 & BE \\
MASTER J111759 & 2013 & -- & -- & -- & -- & -- & -- & 0.069721 & 0.000029 & 100 & 172 & -- & C \\
MASTER J165236 & 2013 & 0.084732 & 0.000085 & 11 & 32 & -- & -- & -- & -- & -- & -- & -- & C \\
MASTER J174902 & 2013 & 0.101908 & 0.000040 & 0 & 40 & -- & -- & -- & -- & -- & -- & -- & C \\
MASTER J181953 & 2013 & 0.057519 & 0.000010 & 35 & 157 & 2.6 & 1.1 & -- & -- & -- & -- & 0.05684 & BE \\
MASTER J203749 & 2012 & 0.061307 & 0.000009 & 36 & 157 & 2.9 & 1.0 & -- & -- & -- & -- & 0.06062 & CE \\
MASTER J211258 & 2012 & 0.060227 & 0.000008 & 50 & 155 & 0.8 & 1.0 & -- & -- & -- & -- & 0.059732 & BE \\
MASTER J212624 & 2013 & 0.091281 & 0.000073 & 9 & 65 & 28.6 & 4.3 & -- & -- & -- & -- & -- & C \\
OT J112619 & 2013 & 0.054886 & 0.000010 & 55 & 260 & 3.6 & 0.4 & -- & -- & -- & -- & 0.05423 & BE \\
OT J191443 & 2012 & 0.071331 & 0.000033 & 0 & 13 & -- & -- & -- & -- & -- & -- & -- & C \\
OT J205146 & 2012 & 0.057245 & 0.000035 & 0 & 93 & 12.3 & 2.5 & 0.056799 & 0.000057 & 92 & 216 & -- & C \\
OT J220641 & 2012 & 0.071152 & 0.000106 & 0 & 9 & -- & -- & -- & -- & -- & -- & -- & C \\
OT J232727 & 2012 & 0.053438 & 0.000012 & 41 & 217 & 4.0 & 1.1 & -- & -- & -- & -- & 0.05277 & BE \\
PNV J062703 & 2013 & 0.059026 & 0.000026 & 0 & 108 & 6.3 & 1.3 & -- & -- & -- & -- & 0.05787 & CE \\
SDSS J075107 & 2013 & 0.057980 & 0.000016 & 0 & 51 & $-$2.9 & 4.1 & -- & -- & -- & -- & -- & C2 \\
SDSS J080033 & 2012 & 0.080421 & 0.000034 & 0 & 26 & -- & -- & -- & -- & -- & -- & -- & C \\
SSS J122221 & 2013 & 0.076486 & 0.000013 & 203 & 362 & $-$1.1 & 0.7 & -- & -- & -- & -- & -- & B \\
TCP J153756 & 2013 & 0.061899 & 0.000023 & -- & -- & -- & -- & -- & -- & -- & -- & 0.06101 & CE \\
TCP J175219 & 2012 & 0.066925 & 0.000091 & 0 & 13 & -- & -- & -- & -- & -- & -- & -- & C \\
\hline
\end{tabular}
\end{center}
\end{table*}

\section{Individual Objects}\label{sec:individual}

\subsection{KX Aquilae}\label{obj:kxaql}

   KX Aql had long been known as a dwarf nova with very low
outburst frequency (cf. \cite{gar79kxaql}).  Based on the
long bright outburst in 1980, this object was suspected
to be an SU UMa-type dwarf nova (see \cite{Pdot2} for more
history).  \citet{tap01kxaql} first presented a spectrum of
this object clearly showing the low mass-transfer rate.
The long-awaited superoutburst finally occurred in 2010 and
superhumps were detected \citep{Pdot2}.

\begin{figure}
  \begin{center}
    \FigureFile(88mm,70mm){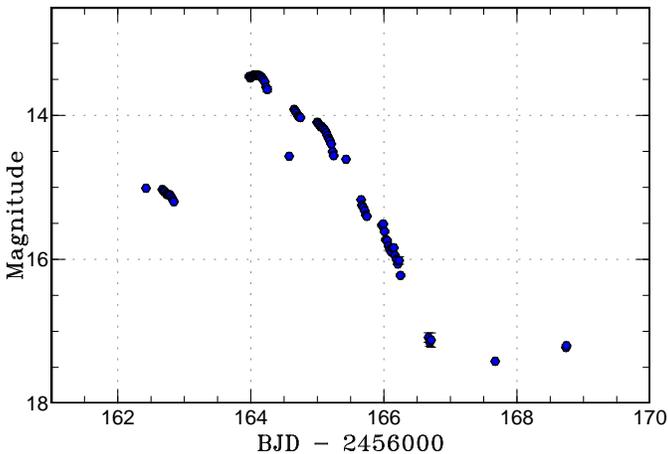}
  \end{center}
  \caption{Light curve of KX Aql (2012).  The data were binned
  to 0.02~d.
  }
  \label{fig:kxaql2012lc}
\end{figure}

   Although the 2012 outburst was not a superoutburst,
we report on it here because it showed an interesting phenomenon.
The outburst was reported on 2012 August 22.907 UT at an unfiltered
CCD magnitude of 16.0 (BAAVSS alert 2997).  The object further
brightened relatively slowly for an SU UMa-type dwarf nova.
Furthermore, there was a precursor-like fading during the
rising stage of the outburst (figure \ref{fig:kxaql2012lc}).
Such precursor-like phenomenon in a normal outburst has been
seen in frequently outbursting objects such as V1504 Cyg
\citep{Pdot3} or V516 Lyr \citep{kat13j1939v585lyrv516lyr}.
In V1504 Cyg, negative superhumps [likely ``impulsive negative superhump'',
\citet{osa13v344lyrv1504cyg}] appeared.  In V516 Lyr,
inside-out outburst may by responsible for the double precursor
event \citep{kat13j1939v585lyrv516lyr}.  In the case of KX Aql,
the slow rise may be interpreted as a signature of
an inside-out outburst, and such a case is rare in low mass-transfer
systems.  In the present data, we could not detect either
positive or negative superhumps during the fading part of
the outburst.

\subsection{NN Camelopardalis}\label{obj:nncam}

   NN Cam = NSV 1485 is a recently recognized dwarf nova
(cf. \cite{khr05nsv1485}).  The first-ever recorded
superoutburst, preceded by a distinct precursor, was observed 
in 2007 \citep{Pdot}.
We reported on two more superoutbursts in 2009 \citep{Pdot2}
and 2011 \citep{Pdot4}.  We observed another
superoutburst in 2012.  The times of superhump maxima are listed
in table \ref{tab:nncamoc2012}.  The resultant period suggests
that the observation recorded stage B superhumps.

\begin{table}
\caption{Superhump maxima of NN Cam (2012)}\label{tab:nncamoc2012}
\begin{center}
\begin{tabular}{ccccc}
\hline
$E$ & max\commenta & error & $O-C$\commentb & $N$\commentc \\
\hline
0 & 56241.1152 & 0.0005 & $-$0.0012 & 63 \\
1 & 56241.1915 & 0.0007 & 0.0011 & 58 \\
39 & 56244.0111 & 0.0078 & 0.0032 & 40 \\
40 & 56244.0803 & 0.0013 & $-$0.0016 & 74 \\
41 & 56244.1546 & 0.0013 & $-$0.0015 & 71 \\
\hline
  \multicolumn{5}{l}{\commenta BJD$-$2400000.} \\
  \multicolumn{5}{l}{\commentb Against max $= 2456241.1163 + 0.074140 E$.} \\
  \multicolumn{5}{l}{\commentc Number of points used to determine the maximum.} \\
\end{tabular}
\end{center}
\end{table}

\subsection{V485 Centauri}\label{obj:v485cen}

   V485 Cen is one of the prototypes of a small group of
dwarf novae having orbital periods below the period minimum
while displaying hydrogen-rich spectra (\cite{aug93v485cen};
\cite{aug96v485cen}; see also T. Ohshima in prep. for
the discussion of this group of objects).  \citet{ole97v485cen}
reported on a positive period derivative.  We reported
re-analysis of \citet{ole97v485cen} and two additional
superoutbursts in 2001 and 2004 in \citet{Pdot}.

   We observed the 2013 April superoutburst of this ultracompact
binary.  The times of superhump maxima are listed in table
\ref{tab:v485cenoc2013}.  The resultant $P_{\rm dot}$ was
similar to those obtained in the past observations
(\cite{ole97v485cen}; \cite{Pdot} see also this reference
for the correction of \cite{ole97v485cen}).
The $O-C$ diagram also showed the behavior similar to that
in the past superoutbursts.

\begin{figure}
  \begin{center}
    \FigureFile(88mm,70mm){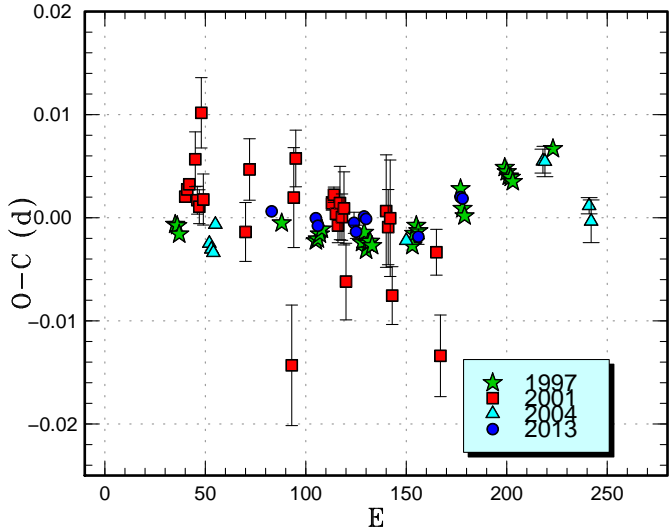}
  \end{center}
  \caption{Comparison of $O-C$ diagrams of V485 Cen between different
  superoutbursts.  A period of 0.04212~d was used to draw this figure.
  Approximate cycle counts ($E$) after the start of the superoutburst
  were used.  Since the start of the 2013 superoutburst
  was not well constrained, we shifted the $O-C$ diagram
  to best fit the others.
  }
  \label{fig:v485cencomp2}
\end{figure}

\begin{table}
\caption{Superhump maxima of V485 Cen (2013)}\label{tab:v485cenoc2013}
\begin{center}
\begin{tabular}{ccccc}
\hline
$E$ & max\commenta & error & $O-C$\commentb & $N$\commentc \\
\hline
0 & 56399.8162 & 0.0003 & 0.0014 & 42 \\
22 & 56400.7422 & 0.0002 & 0.0004 & 42 \\
23 & 56400.7836 & 0.0002 & $-$0.0004 & 38 \\
41 & 56401.5421 & 0.0010 & $-$0.0004 & 14 \\
42 & 56401.5833 & 0.0006 & $-$0.0013 & 15 \\
46 & 56401.7533 & 0.0003 & 0.0002 & 41 \\
47 & 56401.7951 & 0.0004 & $-$0.0001 & 41 \\
73 & 56402.8885 & 0.0007 & $-$0.0022 & 33 \\
94 & 56403.7769 & 0.0004 & 0.0013 & 41 \\
95 & 56403.8189 & 0.0004 & 0.0011 & 31 \\
\hline
  \multicolumn{5}{l}{\commenta BJD$-$2400000.} \\
  \multicolumn{5}{l}{\commentb Against max $= 2456399.8149 + 0.042136 E$.} \\
  \multicolumn{5}{l}{\commentc Number of points used to determine the maximum.} \\
\end{tabular}
\end{center}
\end{table}

\subsection{Z Chameleontis}\label{obj:zcha}

   Only single-night observation for the 2013 February superoutburst
of this well-known dwarf nova was available.
Two superhump maxima were recorded:
BJD 2456341.1272(10) ($E=46$), 2456341.2042(5) ($E=46$).

\subsection{YZ Cancri}\label{obj:yzcnc}

   This is also a well-known SU UMa-type dwarf nova
(cf. the light curve in \cite{szk84AAVSO}).  The reported
superhump period \citep{pat79SH} long remained incorrect
until a new measurement became available \citep{Pdot}.

\begin{figure}
  \begin{center}
    \FigureFile(88mm,100mm){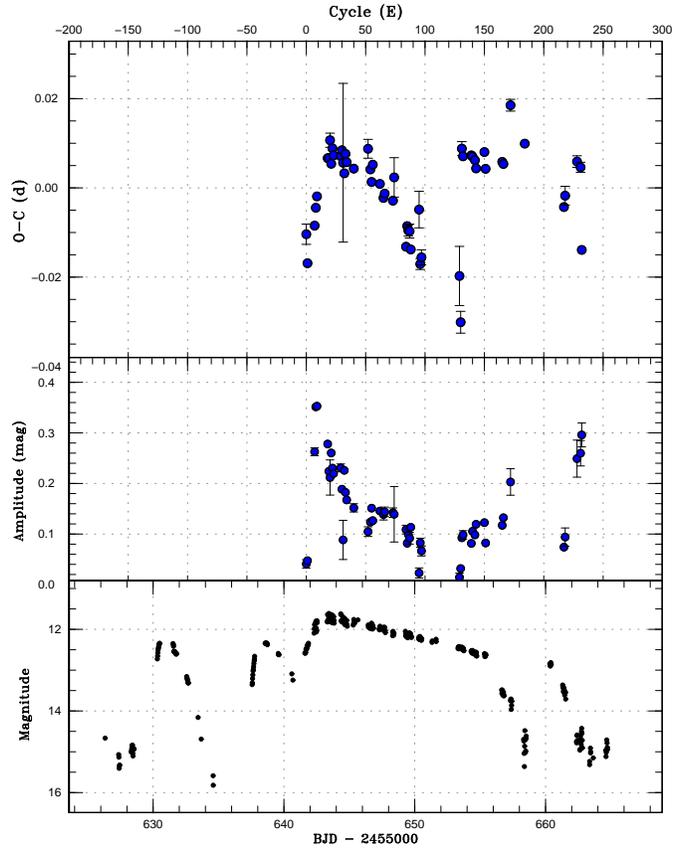}
  \end{center}
  \caption{$O-C$ diagram of superhumps in YZ Cnc (2011).
     (Upper:) $O-C$ diagram.  The maxima after $E=301$
     were traditional late superhumps with an $\sim$0.5 phase jump.
     We used a period of 0.090768~d for calculating the $O-C$ residuals.
     (Middle:) Amplitudes of superhumps.  The growing superhumps
     during stage A were clearly detected,
     (Lower:) Light curve.  The superoutburst had a precursor
     outburst and a dip following it.
  }
  \label{fig:yzcnc2011humpall}
\end{figure}

   A superoutburst in 2011 March was observed.  This outburst
showed a precursor 4~d before the maximum, and the rising
stage to the maximum and growing superhumps were partly
observed (figure \ref{fig:yzcnc2011humpall}). 
There was a deep dip following the precursor, 
as in the superoutburst of V344 Lyr around BJD 2456190 in Kepler 
data \citep{osa13v1504cygv344lyrpaper3}.  Unlike the Kepler data,
our observations were unable to detect superhumps evolving
during the dip phase.
The times of superhump maxima are listed in table 
\ref{tab:yzcncoc2011}.  The maxima after $E=301$
were traditional late superhumps with an $\sim$0.5 phase
jump.  This transition to traditional late superhumps
occurred $\sim$4~d before the rapid decline from the
superoutburst plateau.  These late superhumps persisted
during the post-superoutburst stage and survived during
the next normal outburst. 

   This observation made the first detection of stage A
superhumps in YZ Cnc.  Since YZ Cnc is known as one of
the most frequently outbursting SU UMa-type dwarf novae,
this detection confirmed the wide existence of stage A superhumps
in SU UMa-type dwarf novae other than ER UMa-type dwarf novae
(see a discussion in \cite{kat13qfromstageA}).
The nominal $q$ value from this measured period of stage A
superhumps was 0.17, which appears to be slightly too small.
This was probably due to the rapid growth of the pressure
effect in high-$q$ systems, resulting a slowing the precession
rate even during the later part of stage A.  If we could have
recorded growing superhump during the dip following the
precursor, we would have obtained a better $q$ estimate,
which could become a target for the next observation.

   The rapid growth of the superhump amplitude when
the object was rising to the maximum (figure
\ref{fig:yzcnc2011humpall}) agrees well to the consequence
of the TTI model (\cite{osa13v1504cygKepler};
\cite{osa13v344lyrv1504cyg}).

\begin{table}
\caption{Superhump maxima of YZ Cnc (2011)}\label{tab:yzcncoc2011}
\begin{center}
\begin{tabular}{ccccc}
\hline
$E$ & max\commenta & error & $O-C$\commentb & $N$\commentc \\
\hline
0 & 55641.6988 & 0.0023 & $-$0.0104 & 42 \\
1 & 55641.7831 & 0.0008 & $-$0.0169 & 112 \\
7 & 55642.3361 & 0.0003 & $-$0.0085 & 191 \\
8 & 55642.4309 & 0.0002 & $-$0.0045 & 193 \\
9 & 55642.5242 & 0.0003 & $-$0.0019 & 66 \\
18 & 55643.3497 & 0.0003 & 0.0066 & 95 \\
19 & 55643.4405 & 0.0003 & 0.0066 & 95 \\
20 & 55643.5352 & 0.0016 & 0.0107 & 21 \\
21 & 55643.6207 & 0.0003 & 0.0054 & 136 \\
22 & 55643.7150 & 0.0002 & 0.0088 & 239 \\
23 & 55643.8041 & 0.0002 & 0.0073 & 239 \\
29 & 55644.3486 & 0.0005 & 0.0071 & 46 \\
30 & 55644.4407 & 0.0004 & 0.0084 & 94 \\
31 & 55644.5286 & 0.0178 & 0.0056 & 16 \\
32 & 55644.6170 & 0.0005 & 0.0032 & 135 \\
33 & 55644.7122 & 0.0003 & 0.0076 & 245 \\
34 & 55644.8011 & 0.0003 & 0.0057 & 245 \\
40 & 55645.3442 & 0.0006 & 0.0043 & 91 \\
52 & 55646.4379 & 0.0021 & 0.0087 & 33 \\
54 & 55646.6148 & 0.0006 & 0.0041 & 89 \\
55 & 55646.7028 & 0.0005 & 0.0013 & 270 \\
56 & 55646.7974 & 0.0004 & 0.0052 & 300 \\
62 & 55647.3377 & 0.0004 & 0.0008 & 96 \\
65 & 55647.6069 & 0.0008 & $-$0.0023 & 53 \\
66 & 55647.6986 & 0.0008 & $-$0.0013 & 74 \\
73 & 55648.3324 & 0.0006 & $-$0.0029 & 81 \\
74 & 55648.4284 & 0.0044 & 0.0023 & 18 \\
84 & 55649.3205 & 0.0007 & $-$0.0132 & 72 \\
85 & 55649.4159 & 0.0008 & $-$0.0086 & 94 \\
86 & 55649.5057 & 0.0012 & $-$0.0096 & 93 \\
87 & 55649.5963 & 0.0016 & $-$0.0097 & 69 \\
88 & 55649.6830 & 0.0005 & $-$0.0139 & 249 \\
95 & 55650.3273 & 0.0041 & $-$0.0049 & 81 \\
96 & 55650.4059 & 0.0012 & $-$0.0171 & 86 \\
97 & 55650.4981 & 0.0017 & $-$0.0156 & 92 \\
129 & 55653.3985 & 0.0066 & $-$0.0198 & 101 \\
130 & 55653.4789 & 0.0024 & $-$0.0302 & 95 \\
131 & 55653.6086 & 0.0016 & 0.0087 & 80 \\
132 & 55653.6976 & 0.0008 & 0.0070 & 77 \\
139 & 55654.3332 & 0.0006 & 0.0072 & 98 \\
140 & 55654.4237 & 0.0005 & 0.0069 & 105 \\
142 & 55654.6045 & 0.0006 & 0.0061 & 98 \\
143 & 55654.6934 & 0.0005 & 0.0043 & 123 \\
150 & 55655.3324 & 0.0009 & 0.0079 & 124 \\
151 & 55655.4194 & 0.0004 & 0.0042 & 151 \\
165 & 55656.6917 & 0.0004 & 0.0057 & 240 \\
166 & 55656.7820 & 0.0004 & 0.0053 & 243 \\
172 & 55657.3398 & 0.0013 & 0.0184 & 103 \\
184 & 55658.4204 & 0.0004 & 0.0098 & 178 \\
217 & 55661.4016 & 0.0009 & $-$0.0044 & 181 \\
218 & 55661.4949 & 0.0021 & $-$0.0019 & 99 \\
228 & 55662.4102 & 0.0013 & 0.0058 & 85 \\
231 & 55662.6812 & 0.0011 & 0.0045 & 95 \\
232 & 55662.7535 & 0.0008 & $-$0.0140 & 95 \\
\hline
  \multicolumn{5}{l}{\commenta BJD$-$2400000.} \\
  \multicolumn{5}{l}{\commentb Against max $= 2455641.7092 + 0.090768 E$.} \\
  \multicolumn{5}{l}{\commentc Number of points used to determine the maximum.} \\
\end{tabular}
\end{center}
\end{table}

\subsection{GZ Cancri}\label{obj:gzcnc}

   GZ Cnc is a variable star discovered by Takamizawa (TmzV34),
which later turned out to be a dwarf nova \citep{kat01gzcnc}.
\citet{kat02gzcncnsv10934} reported on a high number of short
outbursts in this system.  \citet{tap03gzcnc} finally clarified
that the object is an object near the lower edge of the period gap.
Although the orbital period [0.08825(28)~d] would suggest
an SU UMa-type dwarf nova, no secure superoutburst had been
recorded until 2010 (there was a long outburst in 2007, which
was only recognized retrospectively).

\begin{figure}
  \begin{center}
    \FigureFile(88mm,70mm){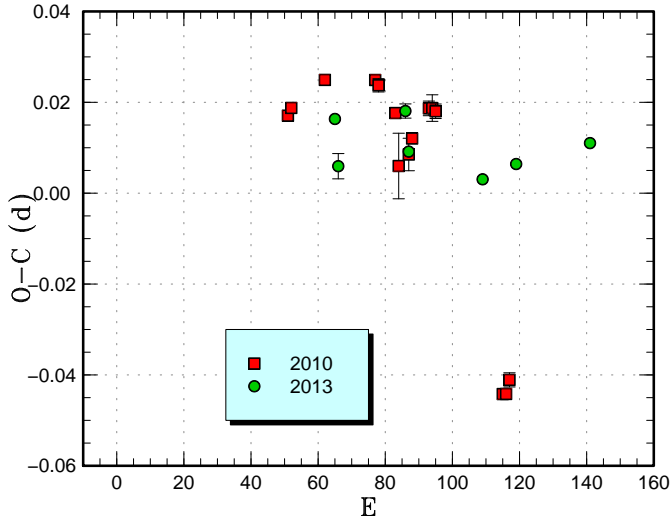}
  \end{center}
  \caption{Comparison of $O-C$ diagrams of GZ Cnc between different
  superoutbursts.  A period of 0.09290~d was used to draw this figure.
  Approximate cycle counts ($E$) after the start of the superoutburst
  were used.
  }
  \label{fig:gzcnccomp}
\end{figure}

   In 2010, the long-awaited superoutburst of this object was
recorded \citep{Pdot2}.  There was another superoutburst in 2013
February and we observed it.  The AAVSO data indicate that this
outburst started with a precursor outburst.  The times of
superhump maxima are listed in table \ref{tab:gzcncoc2013}.
Although our observations covered only the late stage of 
the superoutburst, the resultant period was close to that
of stage B superhumps in \citet{Pdot2}.  A PDM analysis also
yielded a period of 0.09291(2)~d.  This period
appears to be incompatible with an assumption of a large period
change in \citet{Pdot2}.  We could not, however,
find a solution to smoothly express both the 2010 and 2013 
observations mainly due to the limited coverage 
in both data sets.
It looks like that a phase jump occurred in the 2010 superoutburst,
while it did not occur in the 2013 one (figure \ref{fig:gzcnccomp}).
Future observations are absolutely needed to solve this issue.
According to the AAVSO data, an outburst just preceding
this superoutburst was longer (3--4~d) in duration,
and this outburst may have been a failed superoutburst
similar to the one seen in the Kepler data of V1504 Cyg
\citep{Pdot3} and in NY Ser (subsection \ref{sec:nyser}).

\begin{table}
\caption{Superhump maxima of GZ Cnc (2013)}\label{tab:gzcncoc2013}
\begin{center}
\begin{tabular}{ccccc}
\hline
$E$ & max\commenta & error & $O-C$\commentb & $N$\commentc \\
\hline
0 & 56329.9777 & 0.0009 & 0.0045 & 55 \\
1 & 56330.0602 & 0.0028 & $-$0.0058 & 36 \\
21 & 56331.9304 & 0.0016 & 0.0075 & 96 \\
22 & 56332.0144 & 0.0004 & $-$0.0014 & 166 \\
44 & 56334.0521 & 0.0008 & $-$0.0062 & 84 \\
54 & 56334.9844 & 0.0008 & $-$0.0023 & 158 \\
76 & 56337.0328 & 0.0012 & 0.0036 & 105 \\
\hline
  \multicolumn{5}{l}{\commenta BJD$-$2400000.} \\
  \multicolumn{5}{l}{\commentb Against max $= 2456329.9732 + 0.092842 E$.} \\
  \multicolumn{5}{l}{\commentc Number of points used to determine the maximum.} \\
\end{tabular}
\end{center}
\end{table}

\subsection{V503 Cygni}\label{obj:v503cyg}

   V503 Cyg is one of the representative dwarf novae which
exhibited negative superhumps \citep{har95v503cyg}.
\citet{kat02v503cyg} noticed a dramatic variation of the number
of normal outbursts and suggested that normal outbursts may be
suppressed in certain conditions.  \citet{osa13v1504cygKepler}
proposed that the disk tilt, which is supposed to cause
negative superhumps, is responsible for the suppression
of normal outbursts.  In recent years, V503 Cyg did not show
distinct negative superhumps (\cite{Pdot4}; \cite{pav12v503cyg}).

\begin{figure}
  \begin{center}
    \FigureFile(88mm,70mm){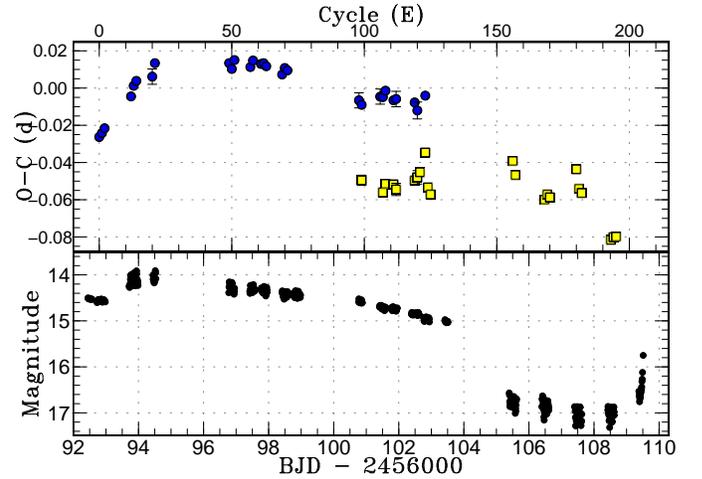}
  \end{center}
  \caption{$O-C$ diagram of superhumps in V503 Cyg (2012).
     (Upper): $O-C$ diagram.  The filled circles represent primary
     maxima of superhumps.  The filled squares represent secondary maxima
     of superhumps and persisting superhumps.  A period of 0.08146~d
     was used to draw this figure.
     (Lower): Light curve.  The observations were binned to 0.008~d.}
  \label{fig:v503cyghumpall}
\end{figure}

   We observed the 2012 June superoutburst, including
the quiescent segments before and after this superoutburst.
We also partly observed the 2012 September superoutburst.
The times of superhump maxima in the 2012 June superoutburst
are listed in table \ref{tab:v503cygoc2012}.
After BJD 2456101 ($E \le 106$),
the profile became double waves, and we listed in this table
one of the peaks which are on the smooth extension from
the earlier epochs of superhump maxima.  The other peak
became stronger relative to this peak in the later course
of the superoutburst.  This behavior is very similar to V344 Lyr
(\cite{woo11v344lyr}; \cite{Pdot3}) and ER UMa \citep{Pdot}.
(see figure \ref{fig:v503cyghumpall} for the $O-C$ behavior).
The times of the secondary maxima (during the late plateau phase)
and times of the post-superoutburst superhumps are listed in
table \ref{tab:v503cygoc2012sec}.  The exact identification
of the post-superoutburst superhumps is not perfectly clear:
on the first three night (BJD 2456105--2456107), the period 
was close to the superhump period before this phase, while 
the period was much shorter [0.0795(1)~d]
on the last two nights (BJD 2456107--2456109).  The nature
of the superhump may have changed during this period.

   We can see a very clear stage A--B transition in the
early part of this superoutburst.  This part followed
a precursor outburst.  Stage A superhumps were for the
first time detected in V503 Cyg.  The $\varepsilon^*$ value of
stage A superhump was 6.88(12)\%, which corresponds to $q=0.218(5)$.

   The times of superhump maxima during the 2012 September
superoutburst are listed in table \ref{tab:v503cygoc2012b}.

\begin{figure}
  \begin{center}
    \FigureFile(88mm,70mm){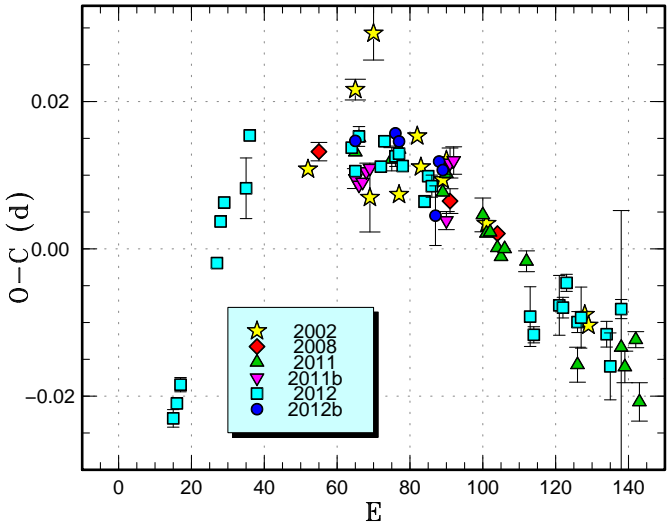}
  \end{center}
  \caption{Comparison of $O-C$ diagrams of V503 Cyg between different
  superoutbursts.  A period of 0.08152~d was used to draw this figure.
  Approximate cycle counts ($E$) after the start of the superoutburst
  were used.  Since the starts of the superoutbursts other than
  the 2012 one were not well constrained, we shifted the $O-C$ diagrams
  to best fit the best-recorded 2012 one.
  }
  \label{fig:v503cygcomp}
\end{figure}

   A comparison of $O-C$ diagrams between different superoutbursts
is shown in figure \ref{fig:v503cygcomp}.

   We also analyzed quiescent segments after removing the first
quiescent interval following the superoutburst, when superhump
still persisted.  The four quiescent segments in 2012 did not
show a sign of negative superhumps, and this behavior was similar
to the 2011 one \citep{Pdot4}.  The orbital period determined
from the 2012 quiescent data was 0.077755(1)~d.  Similarly,
the 2011 quiescent data \citep{Pdot4} yielded 0.077766(5)~d.
A combined analysis of the 2011 and 2012 data yielded
0.0777591(2)~d, which we adopted as the refined orbital period.
This period is in good agreement with 0.077760(3)~d
reported in \citet{pav12v503cyg}.

\begin{table}
\caption{Superhump maxima of V503 Cyg (2012)}\label{tab:v503cygoc2012}
\begin{center}
\begin{tabular}{ccccc}
\hline
$E$ & max\commenta & error & $O-C$\commentb & $N$\commentc \\
\hline
0 & 56092.7677 & 0.0012 & $-$0.0267 & 76 \\
1 & 56092.8513 & 0.0008 & $-$0.0246 & 77 \\
2 & 56092.9353 & 0.0010 & $-$0.0220 & 75 \\
12 & 56093.7670 & 0.0003 & $-$0.0049 & 141 \\
13 & 56093.8542 & 0.0003 & 0.0008 & 234 \\
14 & 56093.9383 & 0.0003 & 0.0035 & 178 \\
20 & 56094.4293 & 0.0041 & 0.0057 & 15 \\
21 & 56094.5180 & 0.0004 & 0.0130 & 37 \\
49 & 56096.7989 & 0.0006 & 0.0129 & 124 \\
50 & 56096.8773 & 0.0003 & 0.0098 & 159 \\
51 & 56096.9635 & 0.0013 & 0.0145 & 50 \\
57 & 56097.4485 & 0.0005 & 0.0108 & 109 \\
58 & 56097.5335 & 0.0005 & 0.0143 & 57 \\
61 & 56097.7760 & 0.0014 & 0.0124 & 108 \\
62 & 56097.8579 & 0.0006 & 0.0128 & 159 \\
63 & 56097.9377 & 0.0003 & 0.0112 & 115 \\
69 & 56098.4220 & 0.0008 & 0.0067 & 57 \\
70 & 56098.5070 & 0.0003 & 0.0102 & 126 \\
71 & 56098.5872 & 0.0015 & 0.0089 & 18 \\
98 & 56100.7705 & 0.0041 & $-$0.0072 & 52 \\
99 & 56100.8495 & 0.0011 & $-$0.0097 & 76 \\
106 & 56101.4242 & 0.0041 & $-$0.0053 & 68 \\
107 & 56101.5054 & 0.0014 & $-$0.0055 & 69 \\
108 & 56101.5903 & 0.0012 & $-$0.0021 & 81 \\
111 & 56101.8295 & 0.0014 & $-$0.0072 & 85 \\
112 & 56101.9116 & 0.0041 & $-$0.0066 & 86 \\
119 & 56102.4800 & 0.0017 & $-$0.0084 & 90 \\
120 & 56102.5572 & 0.0045 & $-$0.0127 & 87 \\
123 & 56102.8095 & 0.0013 & $-$0.0048 & 155 \\
\hline
  \multicolumn{5}{l}{\commenta BJD$-$2400000.} \\
  \multicolumn{5}{l}{\commentb Against max $= 2456092.7944 + 0.081463 E$.} \\
  \multicolumn{5}{l}{\commentc Number of points used to determine the maximum.} \\
\end{tabular}
\end{center}
\end{table}

\begin{table}
\caption{Superhump maxima (secondary and post-superoutburst maximum) of
V503 Cyg (2012)}\label{tab:v503cygoc2012sec}
\begin{center}
\begin{tabular}{ccccc}
\hline
$E$ & max\commenta & error & $O-C$\commentb & $N$\commentc \\
\hline
0 & 56100.8090 & 0.0027 & $-$0.0043 & 86 \\
8 & 56101.4542 & 0.0019 & $-$0.0092 & 86 \\
9 & 56101.5402 & 0.0015 & $-$0.0043 & 70 \\
12 & 56101.7843 & 0.0015 & $-$0.0040 & 66 \\
13 & 56101.8631 & 0.0033 & $-$0.0065 & 85 \\
20 & 56102.4381 & 0.0020 & $-$0.0003 & 80 \\
21 & 56102.5212 & 0.0036 & 0.0016 & 84 \\
22 & 56102.6055 & 0.0006 & 0.0046 & 62 \\
24 & 56102.7789 & 0.0005 & 0.0156 & 103 \\
25 & 56102.8417 & 0.0005 & $-$0.0030 & 159 \\
26 & 56102.9193 & 0.0012 & $-$0.0066 & 112 \\
57 & 56105.4626 & 0.0010 & 0.0178 & 65 \\
58 & 56105.5365 & 0.0007 & 0.0104 & 72 \\
69 & 56106.4193 & 0.0009 & $-$0.0006 & 60 \\
70 & 56106.5036 & 0.0009 & 0.0025 & 73 \\
71 & 56106.5835 & 0.0008 & 0.0011 & 73 \\
81 & 56107.4132 & 0.0008 & 0.0183 & 40 \\
82 & 56107.4842 & 0.0008 & 0.0080 & 68 \\
83 & 56107.5633 & 0.0009 & 0.0059 & 67 \\
94 & 56108.4343 & 0.0009 & $-$0.0170 & 48 \\
95 & 56108.5171 & 0.0009 & $-$0.0153 & 68 \\
96 & 56108.5989 & 0.0015 & $-$0.0149 & 59 \\
\hline
  \multicolumn{5}{l}{\commenta BJD$-$2400000.} \\
  \multicolumn{5}{l}{\commentb Against max $= 2456100.8133 + 0.081255 E$.} \\
  \multicolumn{5}{l}{\commentc Number of points used to determine the maximum.} \\
\end{tabular}
\end{center}
\end{table}

\begin{table}
\caption{Superhump maxima of V503 Cyg (2012b)}\label{tab:v503cygoc2012b}
\begin{center}
\begin{tabular}{ccccc}
\hline
$E$ & max\commenta & error & $O-C$\commentb & $N$\commentc \\
\hline
0 & 56177.6011 & 0.0005 & $-$0.0018 & 85 \\
11 & 56178.4989 & 0.0007 & 0.0024 & 88 \\
12 & 56178.5793 & 0.0006 & 0.0016 & 90 \\
22 & 56179.3844 & 0.0040 & $-$0.0056 & 27 \\
23 & 56179.4733 & 0.0005 & 0.0021 & 90 \\
24 & 56179.5536 & 0.0005 & 0.0012 & 90 \\
\hline
  \multicolumn{5}{l}{\commenta BJD$-$2400000.} \\
  \multicolumn{5}{l}{\commentb Against max $= 2456177.6029 + 0.081232 E$.} \\
  \multicolumn{5}{l}{\commentc Number of points used to determine the maximum.} \\
\end{tabular}
\end{center}
\end{table}

\subsection{OV Draconis}\label{obj:ovdra}

   This object (=SDSS J125023.85$+$665525.5) is a CV 
selected during the course of 
the Sloan Digital Sky Survey (SDSS) \citep{szk03SDSSCV2}.
\citet{dil08SDSSCV} confirmed that this is a deeply eclipsing CV.
The 2008 and 2009 superoutbursts were reported in \citet{Pdot2} and
another one in 2011 was reported in \citet{Pdot3}.

   Only single-night observation for the 2013 February superoutburst
of this eclipsing dwarf nova (=SDSS J125023.85$+$665525.5)
was available.  Three superhump maxima were recorded:
BJD 2456341.7864(12) ($E=56$), 2456341.8432(10) ($E=34$),
2456341.9070(11) ($E=60$).  The epoch of the eclipse minimum
(average of four eclipse observations) was BJD 2456341.8182(1)
determined by the Markov-chain Monte Carlo (MCMC) analysis introduced
in \citet{Pdot4}.  We also updated the eclipse ephemeris using our
2008--2013 observations:
\begin{equation}
{\rm Min(BJD)} = 2453407.5600(1) + 0.058735677(4) E
\label{equ:ovdraecl}.
\end{equation}

\subsection{AQ Eridani}\label{obj:aqeri}

   AQ Eri is a well-known dwarf nova whose SU UMa-type nature
was first clarified by \citet{kat89aqeri}.  \citet{kat91aqeri} and
\citet{kat01aqeri} reported observation of superhumps with
limited coverage and \citet{kat99aqeri} reported a photometric
study of a normal outburst.  The spectroscopic study and the
measurement of the orbital period was performed by
\citet{men93wxcetaqericuvel}.  The first well-observed superoutburst
occurred in 2008 \citet{Pdot}.  Two more superoutbursts in
2010 and 2011 were also reported in \citet{Pdot2} and
\citet{Pdot4}, respectively.

\begin{figure}
  \begin{center}
    \FigureFile(88mm,70mm){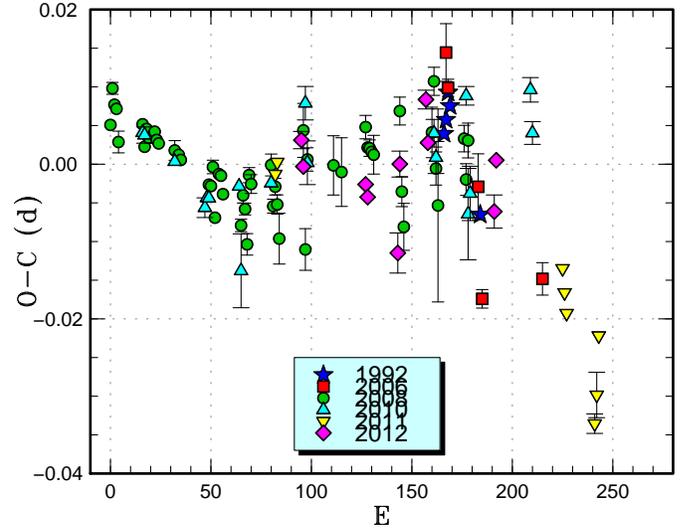}
  \end{center}
  \caption{Comparison of $O-C$ diagrams of AQ Eri between different
  superoutbursts.  A period of 0.06238~d was used to draw this figure.
  Approximate cycle counts ($E$) after the start of the superoutburst
  were used.  Since the start of the 2012 superoutburst
  was not well constrained, we shifted the $O-C$ diagram
  to best fit the best-recorded 2008 one.
  }
  \label{fig:aqericomp3}
\end{figure}

   We observed the 2012 superoutburst during its (presumably) 
later part.  The times of superhump maxima are listed in table
\ref{tab:aqerioc2012}.  Stages B and C can be recognized.
A comparison of $O-C$ diagrams of AQ Eri between different
superoutbursts is shown in figure \ref{fig:aqericomp3}.
The object shows long-lasting stage C as in QZ Vir (\cite{Pdot};
\cite{ohs11qzvir}).

\begin{table}
\caption{Superhump maxima of AQ Eri (2012)}\label{tab:aqerioc2012}
\begin{center}
\begin{tabular}{ccccc}
\hline
$E$ & max\commenta & error & $O-C$\commentb & $N$\commentc \\
\hline
0 & 56213.2582 & 0.0011 & 0.0035 & 25 \\
1 & 56213.3172 & 0.0006 & 0.0001 & 58 \\
32 & 56215.2486 & 0.0005 & $-$0.0018 & 264 \\
33 & 56215.3094 & 0.0003 & $-$0.0035 & 319 \\
48 & 56216.2379 & 0.0026 & $-$0.0105 & 67 \\
49 & 56216.3117 & 0.0017 & 0.0010 & 68 \\
62 & 56217.1310 & 0.0012 & 0.0095 & 111 \\
63 & 56217.1878 & 0.0008 & 0.0040 & 111 \\
96 & 56219.2375 & 0.0021 & $-$0.0045 & 46 \\
97 & 56219.3065 & 0.0004 & 0.0022 & 123 \\
\hline
  \multicolumn{5}{l}{\commenta BJD$-$2400000.} \\
  \multicolumn{5}{l}{\commentb Against max $= 2456213.2548 + 0.062366 E$.} \\
  \multicolumn{5}{l}{\commentc Number of points used to determine the maximum.} \\
\end{tabular}
\end{center}
\end{table}

\subsection{V660 Herculis}\label{obj:v660her}

   V660 Her was discovered as a dwarf nova by \citet{shu75v660her}.
\citet{spo98alcomv544herv660herv516cygdxand} reported color
variations during an outburst.  \citet{liu99CVspec1} reported
a spectrum confirming the dwarf nova-type nature and suggested
that the object has a low mass-transfer rate and a short
orbital period.  \citet{tho03kxaqlftcampucmav660herdmlyr}
conducted a radial-velocity study and confirmed the short
[0.07826(8)~d] orbital period.
\citet{ole05v660her} confirmed the SU UMa-type nature
during the 2004 superoutburst.
We also reported the 2009 superoutburst in \citep{Pdot2}.

\begin{figure}
  \begin{center}
    \FigureFile(88mm,70mm){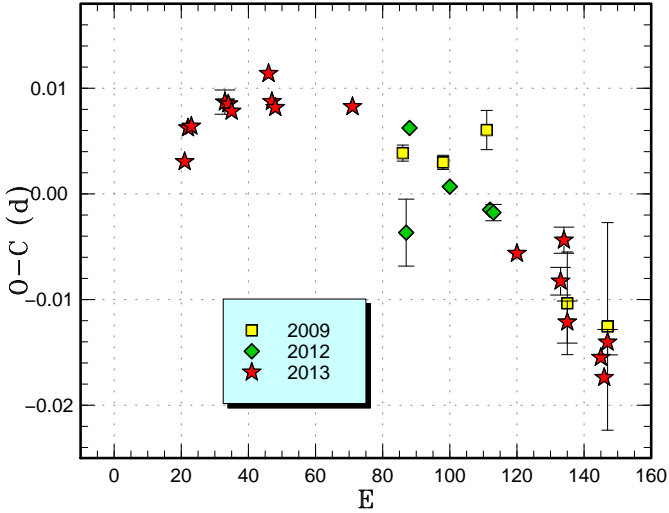}
  \end{center}
  \caption{Comparison of $O-C$ diagrams of V660 Her between different
  superoutbursts.  A period of 0.08100~d was used to draw this figure.
  Approximate cycle counts ($E$) after the start of the superoutburst
  were used.  Since the start of the 2012 superoutburst
  was not well constrained, we shifted the $O-C$ diagram
  to best fit the rest.}
  \label{fig:v660hercomp}
\end{figure}

   The 2012 superoutburst was detected by G. Poyner visually.
The times of superhump maxima are listed in table
\ref{tab:v660heroc2012}.
The result for the 2013 superoutburst, which was detected by
M. Rodr\'{\i}guez and contained a precursor (vsnet-alert 15930,
15945), is shown in table \ref{tab:v660heroc2013}.
The data showed both stage B and C superhumps.
A comparison of the $O-C$ diagrams (figure \ref{fig:v660hercomp})
suggests that the 2012 observation recorded early stage C.
The initial part of the 2013 observation may have recorded
the terminal part of stage A.

\begin{table}
\caption{Superhump maxima of V660 Her (2012)}\label{tab:v660heroc2012}
\begin{center}
\begin{tabular}{ccccc}
\hline
$E$ & max\commenta & error & $O-C$\commentb & $N$\commentc \\
\hline
0 & 56177.3446 & 0.0032 & $-$0.0051 & 30 \\
1 & 56177.4355 & 0.0006 & 0.0049 & 86 \\
13 & 56178.4019 & 0.0006 & 0.0007 & 77 \\
25 & 56179.3717 & 0.0005 & $-$0.0002 & 84 \\
26 & 56179.4525 & 0.0008 & $-$0.0004 & 77 \\
\hline
  \multicolumn{5}{l}{\commenta BJD$-$2400000.} \\
  \multicolumn{5}{l}{\commentb Against max $= 2456177.3496 + 0.080891 E$.} \\
  \multicolumn{5}{l}{\commentc Number of points used to determine the maximum.} \\
\end{tabular}
\end{center}
\end{table}

\begin{table}
\caption{Superhump maxima of V660 Her (2013)}\label{tab:v660heroc2013}
\begin{center}
\begin{tabular}{ccccc}
\hline
$E$ & max\commenta & error & $O-C$\commentb & $N$\commentc \\
\hline
0 & 56482.3481 & 0.0006 & $-$0.0073 & 34 \\
1 & 56482.4323 & 0.0002 & $-$0.0039 & 84 \\
2 & 56482.5134 & 0.0002 & $-$0.0036 & 81 \\
12 & 56483.3257 & 0.0011 & 0.0005 & 23 \\
13 & 56483.4065 & 0.0004 & 0.0005 & 79 \\
14 & 56483.4868 & 0.0003 & $-$0.0000 & 169 \\
25 & 56484.3814 & 0.0006 & 0.0055 & 54 \\
26 & 56484.4598 & 0.0004 & 0.0030 & 83 \\
27 & 56484.5402 & 0.0007 & 0.0027 & 45 \\
50 & 56486.4033 & 0.0004 & 0.0069 & 82 \\
99 & 56490.3584 & 0.0006 & 0.0017 & 36 \\
112 & 56491.4088 & 0.0013 & 0.0014 & 29 \\
113 & 56491.4936 & 0.0012 & 0.0055 & 87 \\
114 & 56491.5669 & 0.0020 & $-$0.0021 & 44 \\
124 & 56492.3735 & 0.0008 & $-$0.0036 & 41 \\
125 & 56492.4526 & 0.0008 & $-$0.0054 & 43 \\
126 & 56492.5370 & 0.0012 & $-$0.0018 & 34 \\
\hline
  \multicolumn{5}{l}{\commenta BJD$-$2400000.} \\
  \multicolumn{5}{l}{\commentb Against max $= 2456482.3554 + 0.080821 E$.} \\
  \multicolumn{5}{l}{\commentc Number of points used to determine the maximum.} \\
\end{tabular}
\end{center}
\end{table}

\subsection{V1227 Herculis}\label{obj:v1227her}

   This object (=SDSS J165359.06$+$201010.4) was selected
as a CV during the course of the SDSS \citep{szk06SDSSCV5},
who detected superhumps with a period of 1.58 hr during 
one of its superoutburst.  Two superoutbursts have been
observed in detail in 2010 \citep{Pdot2} and 2012 May
(\cite{Pdot4}; \cite{she13v1227her}).

\begin{figure}
  \begin{center}
    \FigureFile(88mm,70mm){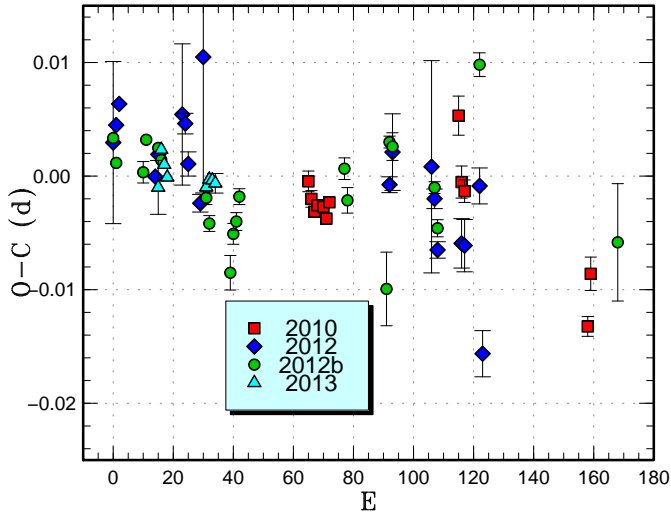}
  \end{center}
  \caption{Comparison of $O-C$ diagrams of V1227 Her between different
  superoutbursts.  A period of 0.06520~d was used to draw this figure.
  Approximate cycle counts ($E$) after the start of the superoutburst
  were used.  Since the start of the 2010 and 2013 superoutbursts
  were not well constrained, we shifted the $O-C$ diagrams
  to best fit the best-recorded 2012b one.
  }
  \label{fig:v1227hercomp2}
\end{figure}

   The object again underwent two superoutbursts in 2012 September
(detection by R. Sabo and relayed by J. Shears, BAAVSS alert 3017)
and 2013 May (detection by J. Shears, BAAVSS alert 3288).
Since new data became available after the publication of
\citet{Pdot4}, we have updated the table of superhump maxima
in the 2012 May superoutburst (table \ref{tab:v1227heroc2012})
and listed the revised periods in table \ref{tab:perlist}.
The times of superhump maxima for the new superoutbursts 
are listed in tables
\ref{tab:v1227heroc2012b} and \ref{tab:v1227heroc2013}.
Although $E=168$ in the 2012 September observation likely
correspond to a stage C superhump, we could not determine
the period of stage C superhumps.
We have been able to determine $P_{\rm dot}$ of stage B superhumps
in this object for the first time.
In figure \ref{fig:v1227hercomp2}, we show a comparison
of $O-C$ diagrams of V1227 Her between different superoutbursts.
In \citet{Pdot4}, we assumed that the 2010 observation was started
soon after the start of the outburst.  It appears less likely
because the object started rapid fading 9~d after the detection,
and because the $O-C$ diagram does not match the present observation.
We therefore shifted the 2010 observation for 65 cycles
to best match the present observation.

\begin{figure}
  \begin{center}
    \FigureFile(88mm,110mm){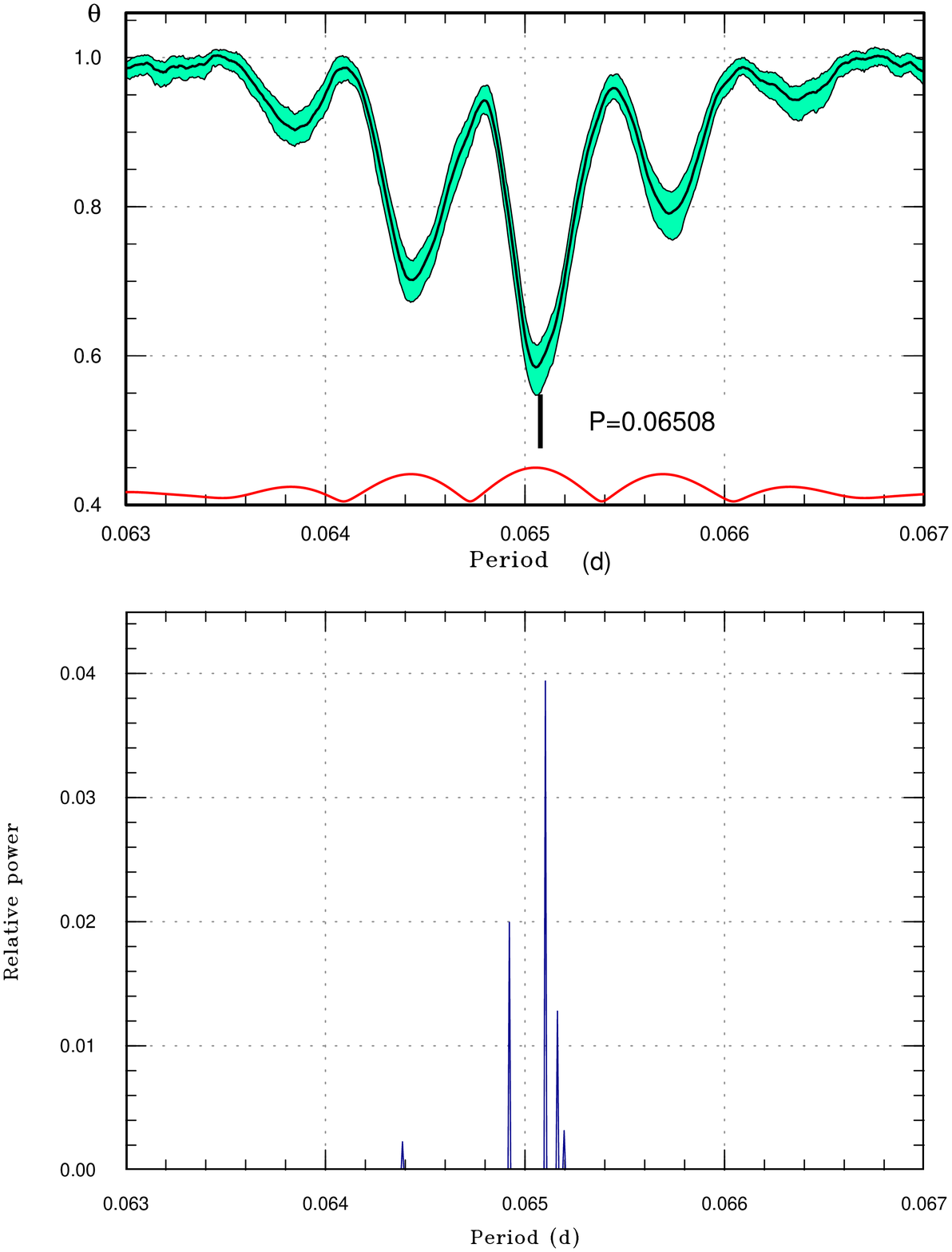}
  \end{center}
  \caption{Period analysis in V1227 Her (2012 May).
     (Upper): PDM analysis.  The curve at the bottom
     refers to the window function.
     (Lower): Lasso analysis ($\log \lambda=-2.11$).
     The main signal of the superhump is split due to the
     variation in the period.}
  \label{fig:v1227her2012pdm}
\end{figure}

\begin{figure}
  \begin{center}
    \FigureFile(88mm,110mm){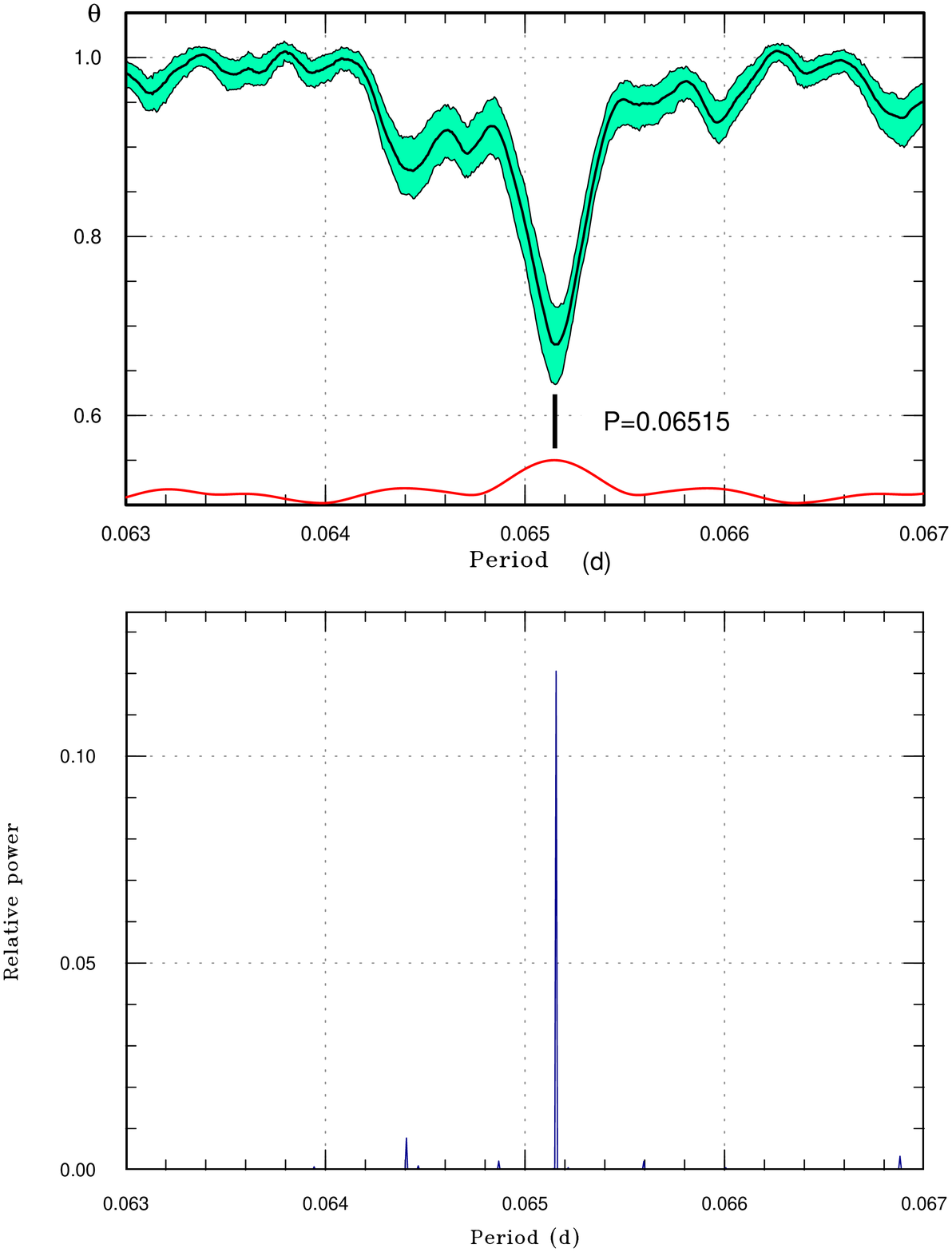}
  \end{center}
  \caption{Period analysis in V1227 Her (2012 September).
     (Upper): PDM analysis.  The curve at the bottom
     refers to the window function.
     (Lower): Lasso analysis ($\log \lambda=-2.07$).}
  \label{fig:v1227her2012bpdm}
\end{figure}

\begin{figure}
  \begin{center}
    \FigureFile(88mm,110mm){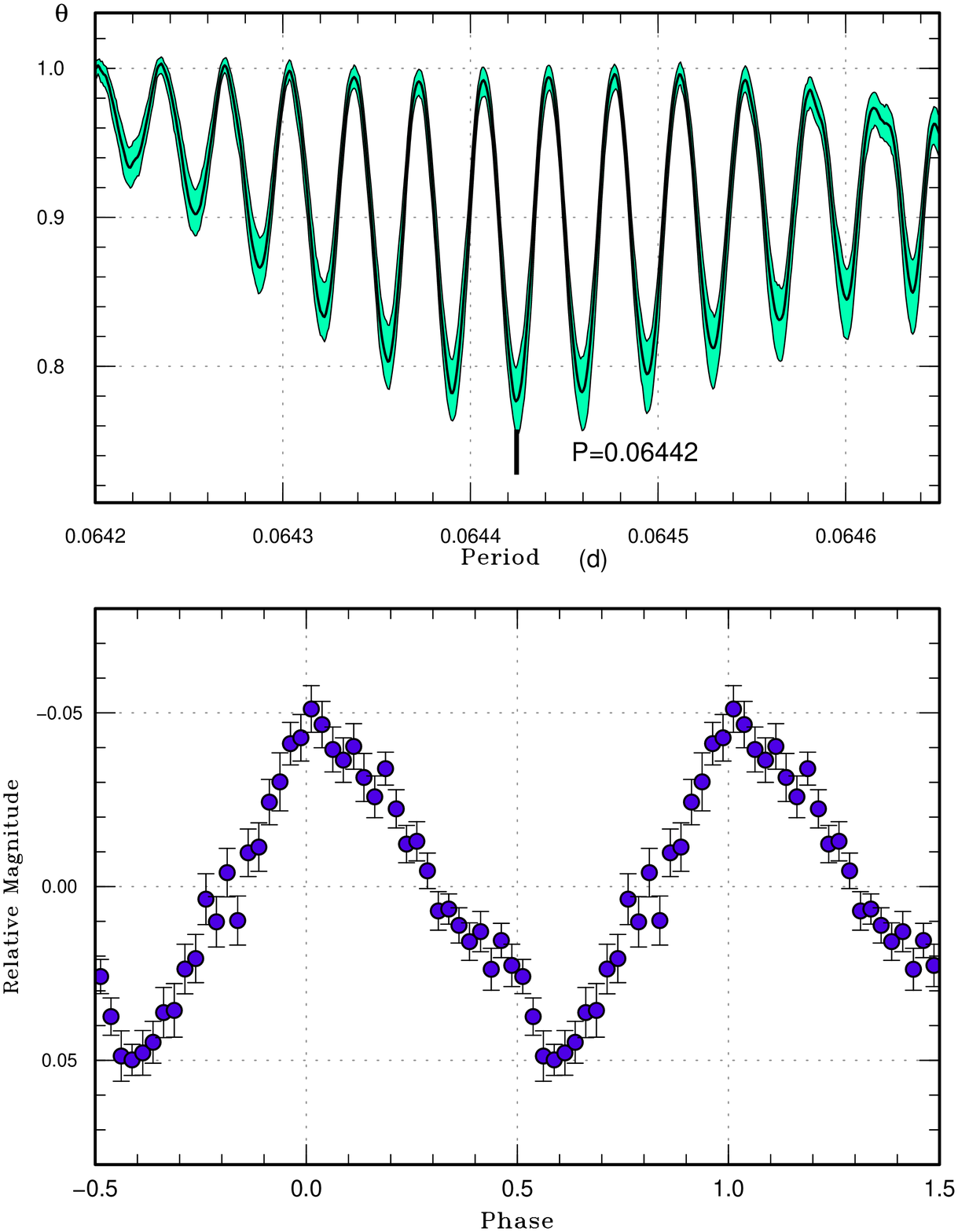}
  \end{center}
  \caption{Possible orbital signal of V1227 Her from the 2012 observation.
     (Upper): PDM analysis.
     (Lower): Phase-averaged profile.}
  \label{fig:v1227herporbshpdm}
\end{figure}

   The 2012 May and September superoutbursts were also studied
by \citet{she13v1227her} using the data parly overlapping with
the present data.  \citet{she13v1227her} reported the analysis of 
superhumps and also suggested the possible orbital period.
The orbital period in \citet{she13v1227her} was suggested
by the possible presence of eclipse-like phenomena.
Since this period is close to the side lobe of the main superhump
signal (e.g. see the window function in figure \ref{fig:v1227her2012pdm}),
we made both the PDM and Lasso analyses (figures
\ref{fig:v1227her2012pdm}, \ref{fig:v1227her2012bpdm})
It appears that the signal around 0.0644~d is commonly
present in these data.  By combining these data together,
a PDM analysis yielded a period of 0.0644246(5)~d
(figure \ref{fig:v1227herporbshpdm}, the selection of the period
was made by comparing with the periods obtained
from individual superoutbursts).
Following \citet{she13v1227her}, we adopted this period as
the candidate orbital period.  The profile is not, however,
eclipse-like as reported in \citet{she13v1227her}.
As described in \citet{she13v1227her}
this $P_{\rm orb}$ gives a relatively small $\varepsilon$.
The identity of the period needs to be confirmed by
further observations.

\begin{table}
\caption{Superhump maxima of V1227 Her (2012 May)}\label{tab:v1227heroc2012}
\begin{center}
\begin{tabular}{ccccc}
\hline
$E$ & max\commenta & error & $O-C$\commentb & $N$\commentc \\
\hline
0 & 56062.5237 & 0.0071 & $-$0.0020 & 8 \\
1 & 56062.5904 & 0.0004 & $-$0.0004 & 67 \\
2 & 56062.6574 & 0.0004 & 0.0015 & 52 \\
14 & 56063.4328 & 0.0004 & $-$0.0039 & 105 \\
15 & 56063.4999 & 0.0004 & $-$0.0018 & 116 \\
23 & 56064.0246 & 0.0062 & 0.0023 & 19 \\
24 & 56064.0890 & 0.0009 & 0.0016 & 70 \\
25 & 56064.1506 & 0.0011 & $-$0.0019 & 52 \\
29 & 56064.4077 & 0.0008 & $-$0.0050 & 47 \\
30 & 56064.4857 & 0.0120 & 0.0080 & 21 \\
92 & 56068.5138 & 0.0007 & 0.0019 & 64 \\
93 & 56068.5818 & 0.0034 & 0.0048 & 15 \\
106 & 56069.4275 & 0.0093 & 0.0046 & 26 \\
107 & 56069.4898 & 0.0009 & 0.0019 & 57 \\
108 & 56069.5504 & 0.0007 & $-$0.0026 & 55 \\
116 & 56070.0722 & 0.0022 & $-$0.0013 & 132 \\
117 & 56070.1372 & 0.0023 & $-$0.0014 & 82 \\
122 & 56070.4682 & 0.0016 & 0.0042 & 119 \\
123 & 56070.5185 & 0.0020 & $-$0.0105 & 80 \\
\hline
  \multicolumn{5}{l}{\commenta BJD$-$2400000.} \\
  \multicolumn{5}{l}{\commentb Against max $= 2456062.5237 + 0.065067 E$.} \\
  \multicolumn{5}{l}{\commentc Number of points used to determine the maximum.} \\
\end{tabular}
\end{center}
\end{table}

\begin{table}
\caption{Superhump maxima of V1227 Her (2012 September)}\label{tab:v1227heroc2012b}
\begin{center}
\begin{tabular}{ccccc}
\hline
$E$ & max\commenta & error & $O-C$\commentb & $N$\commentc \\
\hline
0 & 56182.6775 & 0.0003 & 0.0036 & 56 \\
1 & 56182.7404 & 0.0003 & 0.0014 & 57 \\
10 & 56183.3260 & 0.0009 & 0.0007 & 33 \\
11 & 56183.3940 & 0.0003 & 0.0036 & 67 \\
15 & 56183.6539 & 0.0003 & 0.0029 & 54 \\
16 & 56183.7179 & 0.0003 & 0.0019 & 56 \\
31 & 56184.6919 & 0.0006 & $-$0.0013 & 56 \\
32 & 56184.7547 & 0.0007 & $-$0.0035 & 51 \\
39 & 56185.2065 & 0.0015 & $-$0.0077 & 32 \\
40 & 56185.2750 & 0.0009 & $-$0.0043 & 47 \\
41 & 56185.3413 & 0.0008 & $-$0.0032 & 69 \\
42 & 56185.4086 & 0.0007 & $-$0.0010 & 51 \\
77 & 56187.6913 & 0.0010 & 0.0019 & 56 \\
78 & 56187.7537 & 0.0011 & $-$0.0009 & 56 \\
91 & 56188.5928 & 0.0032 & $-$0.0085 & 19 \\
92 & 56188.6709 & 0.0005 & 0.0044 & 56 \\
93 & 56188.7357 & 0.0012 & 0.0040 & 34 \\
107 & 56189.6442 & 0.0005 & 0.0006 & 55 \\
108 & 56189.7057 & 0.0008 & $-$0.0029 & 54 \\
122 & 56190.6322 & 0.0010 & 0.0116 & 31 \\
168 & 56193.6135 & 0.0052 & $-$0.0034 & 23 \\
\hline
  \multicolumn{5}{l}{\commenta BJD$-$2400000.} \\
  \multicolumn{5}{l}{\commentb Against max $= 2456182.6739 + 0.065137 E$.} \\
  \multicolumn{5}{l}{\commentc Number of points used to determine the maximum.} \\
\end{tabular}
\end{center}
\end{table}

\begin{table}
\caption{Superhump maxima of V1227 Her (2013)}\label{tab:v1227heroc2013}
\begin{center}
\begin{tabular}{ccccc}
\hline
$E$ & max\commenta & error & $O-C$\commentb & $N$\commentc \\
\hline
0 & 56418.4412 & 0.0022 & $-$0.0012 & 30 \\
1 & 56418.5091 & 0.0002 & 0.0016 & 72 \\
2 & 56418.5730 & 0.0001 & 0.0003 & 72 \\
3 & 56418.6370 & 0.0002 & $-$0.0007 & 72 \\
16 & 56419.4838 & 0.0003 & $-$0.0005 & 71 \\
17 & 56419.5497 & 0.0002 & 0.0003 & 72 \\
18 & 56419.6148 & 0.0002 & 0.0003 & 72 \\
19 & 56419.6794 & 0.0008 & $-$0.0002 & 37 \\
\hline
  \multicolumn{5}{l}{\commenta BJD$-$2400000.} \\
  \multicolumn{5}{l}{\commentb Against max $= 2456418.4424 + 0.065115 E$.} \\
  \multicolumn{5}{l}{\commentc Number of points used to determine the maximum.} \\
\end{tabular}
\end{center}
\end{table}

\subsection{MM Hydrae}\label{obj:mmhya}

   MM Hya was originally selected as a CV by the Palomer-Green 
survey \citep{gre82PGsurveyCV}.  \citet{mis95PGCV} suggested
that it may be a WZ Sge-type dwarf nova based on the short
orbital period.  \citet{pat03suumas} reported a mean $P_{\rm SH}$ 
of 0.05868 d during the 1998 superoutburst.  Not much details
were known about this object except for the mean supercycle
of $\sim$380~d, which disqualifies the expectation by
\citet{mis95PGCV}.  In 2011, a first well-observed superoutburst
occurred and we reported on it in \citet{Pdot3}.  Information
on other (rather poorly studied) superoutbursts was also reported
in \citet{Pdot}, \citet{Pdot3} and \citet{Pdot4}.

   We observed the late plateau phase and the declining phase
of the 2013 superoutburst.  The times of maxima are listed in
table \ref{tab:mmhyaoc2013}.  These superhumps are stage C
superhumps.

\begin{table}
\caption{Superhump maxima of MM Hya (2013)}\label{tab:mmhyaoc2013}
\begin{center}
\begin{tabular}{ccccc}
\hline
$E$ & max\commenta & error & $O-C$\commentb & $N$\commentc \\
\hline
0 & 56364.4629 & 0.0008 & 0.0025 & 61 \\
1 & 56364.5176 & 0.0009 & $-$0.0014 & 56 \\
2 & 56364.5766 & 0.0010 & $-$0.0010 & 45 \\
17 & 56365.4592 & 0.0011 & 0.0021 & 57 \\
18 & 56365.5132 & 0.0011 & $-$0.0026 & 58 \\
28 & 56366.0984 & 0.0119 & $-$0.0037 & 68 \\
29 & 56366.1648 & 0.0015 & 0.0040 & 74 \\
\hline
  \multicolumn{5}{l}{\commenta BJD$-$2400000.} \\
  \multicolumn{5}{l}{\commentb Against max $= 2456364.4604 + 0.058633 E$.} \\
  \multicolumn{5}{l}{\commentc Number of points used to determine the maximum.} \\
\end{tabular}
\end{center}
\end{table}

\subsection{AB Normae}\label{obj:abnor}

   AB Nor was discovered by \citet{swo30abnorbrlup} during
the photographic survey of the southern Milky Way.  Only
little had been known before 1997, when regular monitoring
by R. Stubbings started.  During the 2000 outburst,
which followed by a precursor outburst, superhumps were
detected by W. S. G. Walker (vsnet-alert 4589; see
\cite{kat04nsv10934mmscoabnorcal86} for more history).
\citet{kat04nsv10934mmscoabnorcal86} reported on the 2002
superoutburst and established the superhump period
[since this work was before the establishment of superhump
stages, the updated interpretation was reported in
\citet{Pdot}].

\begin{figure}
  \begin{center}
    \FigureFile(88mm,70mm){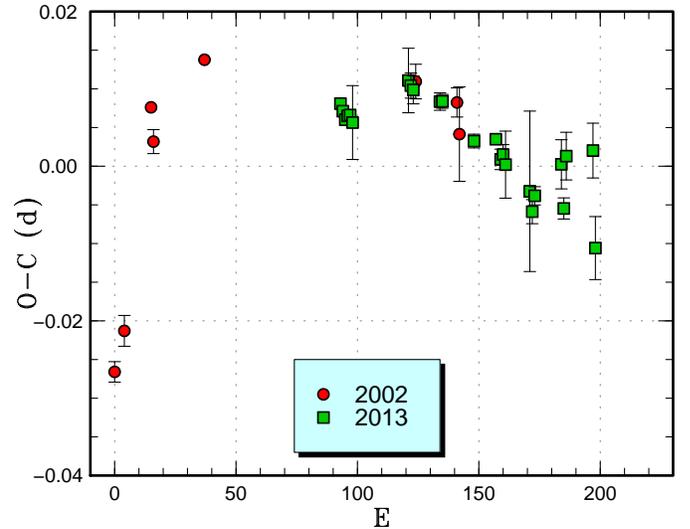}
  \end{center}
  \caption{Comparison of $O-C$ diagrams of AB Nor between different
  superoutbursts.  A period of 0.07962~d was used to draw this figure.
  Since the start of the 2013 superoutburst
  was not well constrained, we shifted the $O-C$ diagram
  to best fit the better-defined 2002 one.
  }
  \label{fig:abnorcomp}
\end{figure}

   We observed the 2013 superoutburst.
The times of superhump maxima are listed in table
\ref{tab:abnoroc2013}.  The final part of stage B and stage C
were recorded.  A combined $O-C$ diagram (figure \ref{fig:abnorcomp})
suggests that stage B may be relatively long for this
$P_{\rm SH}$.  Since the middle of stage B was not well
observed, observations of this stage is needed.

\begin{table}
\caption{Superhump maxima of AB Nor (2013)}\label{tab:abnoroc2013}
\begin{center}
\begin{tabular}{ccccc}
\hline
$E$ & max\commenta & error & $O-C$\commentb & $N$\commentc \\
\hline
0 & 56436.2565 & 0.0006 & $-$0.0012 & 184 \\
1 & 56436.3351 & 0.0005 & $-$0.0021 & 182 \\
2 & 56436.4136 & 0.0005 & $-$0.0031 & 183 \\
3 & 56436.4938 & 0.0006 & $-$0.0024 & 184 \\
4 & 56436.5735 & 0.0006 & $-$0.0022 & 183 \\
5 & 56436.6521 & 0.0048 & $-$0.0031 & 61 \\
28 & 56438.4888 & 0.0042 & 0.0052 & 15 \\
29 & 56438.5678 & 0.0016 & 0.0047 & 19 \\
30 & 56438.6469 & 0.0018 & 0.0042 & 13 \\
41 & 56439.5212 & 0.0011 & 0.0041 & 18 \\
42 & 56439.6008 & 0.0009 & 0.0043 & 20 \\
55 & 56440.6307 & 0.0009 & 0.0007 & 17 \\
64 & 56441.3476 & 0.0005 & 0.0021 & 184 \\
66 & 56441.5042 & 0.0013 & $-$0.0003 & 19 \\
67 & 56441.5844 & 0.0013 & 0.0005 & 18 \\
68 & 56441.6628 & 0.0043 & $-$0.0007 & 7 \\
78 & 56442.4555 & 0.0104 & $-$0.0029 & 7 \\
79 & 56442.5325 & 0.0016 & $-$0.0054 & 17 \\
80 & 56442.6142 & 0.0012 & $-$0.0033 & 20 \\
91 & 56443.4941 & 0.0032 & 0.0022 & 17 \\
92 & 56443.5680 & 0.0014 & $-$0.0034 & 17 \\
93 & 56443.6544 & 0.0031 & 0.0035 & 9 \\
104 & 56444.5309 & 0.0035 & 0.0056 & 18 \\
105 & 56444.5979 & 0.0041 & $-$0.0069 & 20 \\
\hline
  \multicolumn{5}{l}{\commenta BJD$-$2400000.} \\
  \multicolumn{5}{l}{\commentb Against max $= 2456436.2577 + 0.079496 E$.} \\
  \multicolumn{5}{l}{\commentc Number of points used to determine the maximum.} \\
\end{tabular}
\end{center}
\end{table}

\subsection{DT Octantis}\label{obj:dtoct}

   DT Oct = NSV 10934 was originally discovered as a large-amplitude 
suspected variable star of unknown classification.
\citet{kat02gzcncnsv10934} suggested an X-ray identification
and reported the detection of multiple outbursts, leading to
a dwarf nova-type classification.  \citet{kat04nsv10934mmscoabnorcal86}
reported the first detection of superhumps, whose result was
updated in \citet{Pdot}.  Two other superoutbursts in 2003 (second
one in 2003) and 2008 with poorer coverage were also reported
in \citet{Pdot}.

   Only the final stage of the 2013 March--April superoutburst
was observed.  A single superhump maximum BJD 2456376.8560(6)
($N=74$) was measured.

\subsection{GR Orionis}\label{obj:grori}

   As introduced in \citet{kat12DNSDSS},\footnote{
   There was an incorrect citation in \citet{kat12DNSDSS}.
   GR Ori already faded
   to 13.0 on February 8 \citep{thi16grori}.  Note that
   the magnitude system in \citet{thi16grori} was probably
   1--2 mag brighter than the present one.
} GR Ori was initially recorded as a nova, which was later suspected 
to be a dwarf nova.  The second known outburst was detected
on 2013 February 11 at a visual magnitude of 13.0 by
R. Stubbings (vsnet-outburst 15096).
Astrometric measurement by D. Buczynski during the outburst confirmed 
the suggested quiescent identification (\cite{rob00oldnova};
\cite{kat12DNSDSS}).  \citet{ara13groriatel4811} confirmed 
the dwarf nova-type nature by spectroscopy.
Although early observations recorded some variations, no definite 
period of early superhumps was obtained.  The object is likely to have
a low orbital inclination.

\begin{figure}
  \begin{center}
    \FigureFile(88mm,110mm){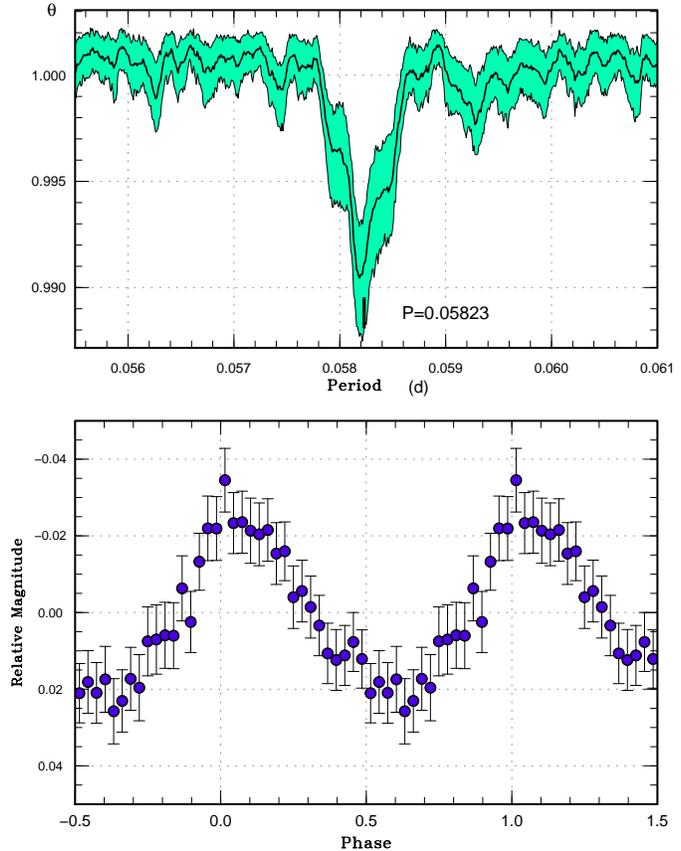}
  \end{center}
  \caption{Superhumps in GR Ori (2013). (Upper): PDM analysis.
     (Lower): Phase-averaged profile.}
  \label{fig:grorishpdm}
\end{figure}

\begin{figure}
  \begin{center}
    \FigureFile(88mm,70mm){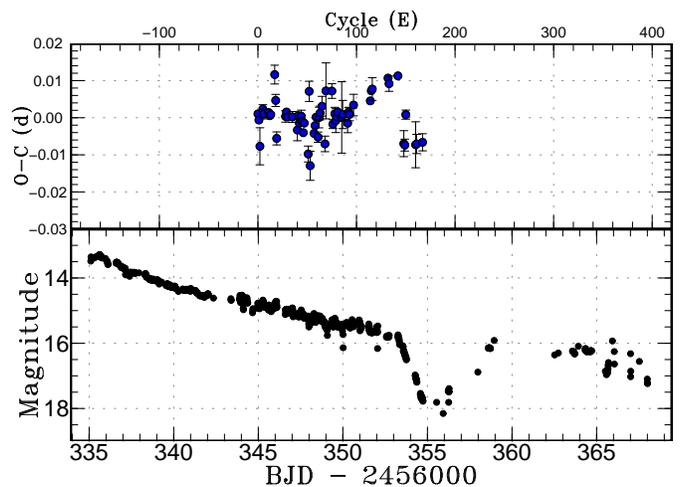}
  \end{center}
  \caption{$O-C$ diagram of superhumps in GR Ori (2013).
     (Upper): $O-C$ diagram.  A period of 0.05827~d
     was used to draw this figure.
     (Lower): Light curve.  The observations were binned to 0.012~d.}
  \label{fig:grorihumpall}
\end{figure}

   Ordinary superhumps appeared on February 20--21
(vsnet-alert\footnote{
  The vsnet-alert archive can be seen at
  $<$http://ooruri.kusastro.kyoto-u.ac.jp/pipermail/\\vsnet-alert/$>$.
} 15430, 15432, 15434; figure \ref{fig:grorishpdm}).
The times of superhump maxima are listed in table
\ref{tab:grorioc2013}.  Due to the faintness of the object, the errors 
were relatively large despite fair coverage.
During $11 \le E \le 142$, we obtained a positive $P_{\rm dot}$ of
$+6.4(1.5) \times 10^{-5}$ for stage B (figure \ref{fig:grorihumpall}).
Although there was a hint of stage A in the earliest part of 
the observation, the growing stage of superhumps was unfortunately 
missed.  The was a possible phase jump after $E=142$, corresponding to 
the time of the rapid decline.

   The object underwent a long-lasting rebrightening (vsnet-alert
15479, 15485, 15508; figure \ref{fig:grorishpdm}) similar to
those of WZ Sge (\cite{ish02wzsgeletter}; \cite{pat02wzsge}; 
\cite{Pdot}), AL Com (\cite{pat96alcom}; \cite{nog97alcom};
\cite{ish02wzsgeletter}) and OT J012059.6$+$325545 \citep{Pdot3}.
The overall behavior was not atypical for a short $P_{\rm orb}$ 
WZ Sge-type dwarf nova.

\begin{table}
\caption{Superhump maxima of GR Ori (2013)}\label{tab:grorioc2013}
\begin{center}
\begin{tabular}{ccccc}
\hline
$E$ & max\commenta & error & $O-C$\commentb & $N$\commentc \\
\hline
0 & 56344.9799 & 0.0009 & 0.0010 & 143 \\
1 & 56345.0365 & 0.0009 & $-$0.0006 & 137 \\
2 & 56345.0877 & 0.0050 & $-$0.0077 & 118 \\
5 & 56345.2726 & 0.0012 & 0.0023 & 97 \\
6 & 56345.3295 & 0.0013 & 0.0010 & 129 \\
11 & 56345.6213 & 0.0002 & 0.0014 & 61 \\
12 & 56345.6787 & 0.0003 & 0.0005 & 55 \\
13 & 56345.7373 & 0.0005 & 0.0008 & 61 \\
17 & 56345.9811 & 0.0025 & 0.0116 & 104 \\
18 & 56346.0324 & 0.0016 & 0.0046 & 102 \\
19 & 56346.0805 & 0.0018 & $-$0.0056 & 83 \\
28 & 56346.6108 & 0.0015 & 0.0003 & 47 \\
29 & 56346.6703 & 0.0004 & 0.0015 & 62 \\
30 & 56346.7272 & 0.0005 & 0.0001 & 61 \\
35 & 56347.0186 & 0.0015 & 0.0001 & 44 \\
40 & 56347.3064 & 0.0029 & $-$0.0034 & 28 \\
41 & 56347.3685 & 0.0006 & 0.0005 & 69 \\
42 & 56347.4248 & 0.0008 & $-$0.0015 & 46 \\
44 & 56347.5432 & 0.0016 & 0.0003 & 22 \\
45 & 56347.6000 & 0.0007 & $-$0.0011 & 178 \\
46 & 56347.6553 & 0.0008 & $-$0.0041 & 175 \\
47 & 56347.7161 & 0.0004 & $-$0.0015 & 117 \\
51 & 56347.9409 & 0.0021 & $-$0.0099 & 98 \\
52 & 56348.0160 & 0.0027 & 0.0070 & 88 \\
53 & 56348.0543 & 0.0039 & $-$0.0131 & 97 \\
57 & 56348.2961 & 0.0007 & $-$0.0043 & 139 \\
58 & 56348.3564 & 0.0007 & $-$0.0022 & 192 \\
59 & 56348.4170 & 0.0008 & 0.0000 & 79 \\
61 & 56348.5281 & 0.0014 & $-$0.0054 & 19 \\
62 & 56348.5917 & 0.0028 & $-$0.0001 & 48 \\
63 & 56348.6512 & 0.0019 & 0.0011 & 40 \\
65 & 56348.7695 & 0.0027 & 0.0030 & 37 \\
68 & 56348.9342 & 0.0021 & $-$0.0072 & 99 \\
69 & 56349.0068 & 0.0075 & 0.0071 & 136 \\
75 & 56349.3563 & 0.0020 & 0.0070 & 98 \\
76 & 56349.4057 & 0.0012 & $-$0.0019 & 124 \\
78 & 56349.5250 & 0.0018 & 0.0009 & 16 \\
79 & 56349.5814 & 0.0032 & $-$0.0009 & 62 \\
80 & 56349.6408 & 0.0006 & 0.0001 & 70 \\
81 & 56349.7003 & 0.0005 & 0.0013 & 55 \\
85 & 56349.9319 & 0.0096 & $-$0.0001 & 61 \\
86 & 56349.9910 & 0.0038 & 0.0007 & 61 \\
91 & 56350.2800 & 0.0026 & $-$0.0017 & 50 \\
92 & 56350.3412 & 0.0013 & 0.0013 & 108 \\
93 & 56350.3989 & 0.0017 & 0.0007 & 33 \\
97 & 56350.6345 & 0.0029 & 0.0032 & 16 \\
114 & 56351.6262 & 0.0010 & 0.0043 & 79 \\
115 & 56351.6872 & 0.0009 & 0.0070 & 61 \\
116 & 56351.7459 & 0.0031 & 0.0075 & 25 \\
132 & 56352.6812 & 0.0010 & 0.0104 & 56 \\
133 & 56352.7380 & 0.0021 & 0.0089 & 29 \\
142 & 56353.2645 & 0.0007 & 0.0110 & 111 \\
148 & 56353.5959 & 0.0035 & $-$0.0073 & 141 \\
149 & 56353.6538 & 0.0016 & $-$0.0077 & 118 \\
150 & 56353.7202 & 0.0012 & 0.0005 & 95 \\
160 & 56354.2948 & 0.0062 & $-$0.0076 & 134 \\
161 & 56354.3533 & 0.0026 & $-$0.0074 & 134 \\
167 & 56354.7034 & 0.0023 & $-$0.0069 & 55 \\
\hline
  \multicolumn{5}{l}{\commenta BJD$-$2400000.} \\
  \multicolumn{5}{l}{\commentb Against max $= 2456344.9789 + 0.058272 E$.} \\
  \multicolumn{5}{l}{\commentc Number of points used to determine the maximum.} \\
\end{tabular}
\end{center}
\end{table}

\subsection{V444 Pegasi}\label{obj:v444peg}

\begin{figure}
  \begin{center}
    \FigureFile(88mm,50mm){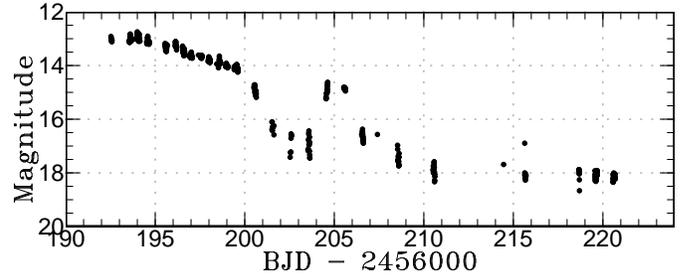}
  \end{center}
  \caption{Light curve of V444 Peg (2013).  A post-superoutburst
  rebrightening occurred immediately after the superoutburst.
  The data were binned to 0.005~d.
  }
  \label{fig:v444peglc}
\end{figure}

   This object (=OT J213701.8$+$071446) is a dwarf nova
discovered by K. Itagaki (cf. \cite{Pdot}).  The analysis of
the 2008 superoutburst was reported in \citet{Pdot}.
CRTS detected another bright outburst (12.9 mag on 2012
September 21, communicated by E. Muyllaert, cvnet-outburst 4943).
The times of superhump maxima are listed in table
\ref{tab:v444pegoc2012}.  Despite the long $P_{\rm SH}$,
the period did not decrease with time.  Using the likely stage B
superhumps, we obtained a significantly positive $P_{\rm dot}$.
Similar large positive $P_{\rm dot}$ in long-period systems
were seen in GX Cas \citep{Pdot3} and SDSS J170213.26$+$322954.1
(hereafter SDSS J170213, \cite{Pdot4}).
The object is also unusual in that it showed a post-superoutburst
rebrightening only 5~d after the rapid fading of the superoutburst
(figure \ref{fig:v444peglc}).
Such a phenomenon immediately following the superoutburst
is usually seen only in WZ Sge-type dwarf novae
[``dip'' phenomenon in WZ Sge (\cite{ish02wzsgeletter};
\cite{pat02wzsge}; \cite{Pdot}), AL Com (\cite{ish02wzsgeletter};
\cite{pat96alcom}), OT J012059.6$+$325545 (\cite{Pdot3})].

\begin{figure}
  \begin{center}
    \FigureFile(88mm,70mm){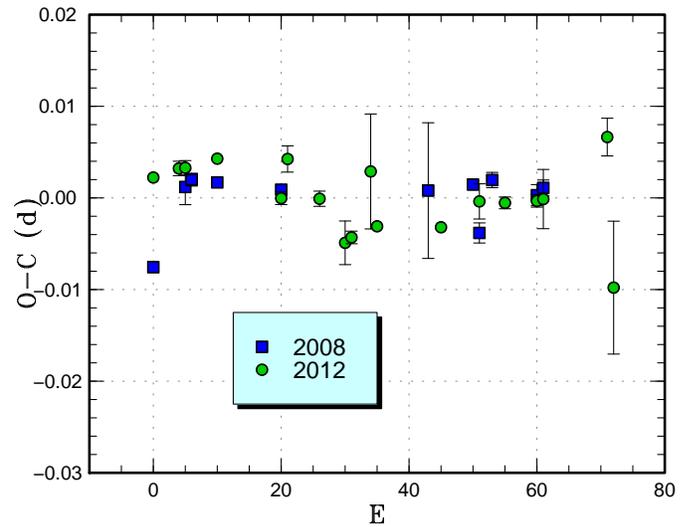}
  \end{center}
  \caption{Comparison of $O-C$ diagrams of V444 Peg between different
  superoutbursts.  A period of 0.09770~d was used to draw this figure.
  Since the starts of the outbursts were not well constrained,
  we used cycle counts ($E$) after the start of the observation.
  }
  \label{fig:v444pegcomp}
\end{figure}

   A comparison of the $O-C$ diagrams of V444 Peg between different
superoutbursts is given in figure \ref{fig:v444pegcomp}.
It looks like that the 2008 observation recorded stages A and B,
rather than stages B and C, as identified in \citep{Pdot}.

\begin{table}
\caption{Superhump maxima of V444 Peg (2012)}\label{tab:v444pegoc2012}
\begin{center}
\begin{tabular}{ccccc}
\hline
$E$ & max\commenta & error & $O-C$\commentb & $N$\commentc \\
\hline
0 & 56193.6015 & 0.0005 & $-$0.0001 & 42 \\
4 & 56193.9933 & 0.0008 & 0.0012 & 67 \\
5 & 56194.0911 & 0.0008 & 0.0013 & 68 \\
10 & 56194.5806 & 0.0005 & 0.0026 & 42 \\
20 & 56195.5533 & 0.0007 & $-$0.0010 & 24 \\
21 & 56195.6552 & 0.0014 & 0.0033 & 27 \\
26 & 56196.1394 & 0.0008 & $-$0.0007 & 87 \\
30 & 56196.5254 & 0.0024 & $-$0.0052 & 24 \\
31 & 56196.6237 & 0.0007 & $-$0.0046 & 38 \\
34 & 56196.9240 & 0.0063 & 0.0028 & 120 \\
35 & 56197.0157 & 0.0003 & $-$0.0031 & 279 \\
45 & 56197.9926 & 0.0004 & $-$0.0026 & 268 \\
51 & 56198.5816 & 0.0019 & 0.0007 & 38 \\
55 & 56198.9723 & 0.0006 & 0.0008 & 291 \\
60 & 56199.4610 & 0.0007 & 0.0013 & 80 \\
61 & 56199.5589 & 0.0032 & 0.0016 & 17 \\
71 & 56200.5426 & 0.0021 & 0.0090 & 46 \\
72 & 56200.6239 & 0.0073 & $-$0.0074 & 39 \\
\hline
  \multicolumn{5}{l}{\commenta BJD$-$2400000.} \\
  \multicolumn{5}{l}{\commentb Against max $= 2456193.6016 + 0.097635 E$.} \\
  \multicolumn{5}{l}{\commentc Number of points used to determine the maximum.} \\
\end{tabular}
\end{center}
\end{table}

\subsection{V521 Pegasi}\label{obj:v521peg}

\begin{figure}
  \begin{center}
    \FigureFile(88mm,110mm){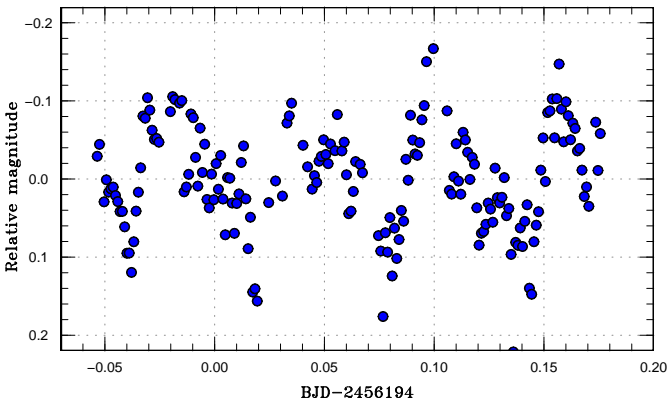}
  \end{center}
  \caption{Superhumps in V521 Peg (2012).}
  \label{fig:v521peglc}
\end{figure}

   This object (=HS 2219$+$1824) is a dwarf nova reported in
\citet{rod05hs2219}.  \citet{rod05hs2219} reported the detection
of superhumps with a period of 0.06184~d and orbital modulations
with a period of 0.0599~d.  Since then, the object underwent
superoutbursts in unfavorable condition and new period measurements
have not been available.  The 2012 superoutburst was detected
by P. Schmeer visually (vsnet-alert 14926).  Only single-night
observation was available (vsnet-alert 14959, figure \ref{fig:v521peglc}).
The times of superhump maxima arr listed in table
\ref{tab:v521pegoc2012}.  The period with the PDM method
was 0.0603(2)~d.  It is not known in which stage this object
was observed.  The period is too different from the one by
\citet{rod05hs2219} to be considered as a consequence of
a stage B--C transition.  Since \citet{rod05hs2219} observed
the very early stage of the outburst, their data may have been
contaminated by stage A superhumps.  We need better observations
to establish the basic periods in this system.

\begin{table}
\caption{Superhump maxima of V521 Peg (2012)}\label{tab:v521pegoc2012}
\begin{center}
\begin{tabular}{ccccc}
\hline
$E$ & max\commenta & error & $O-C$\commentb & $N$\commentc \\
\hline
0 & 56193.9798 & 0.0009 & $-$0.0004 & 42 \\
1 & 56194.0411 & 0.0017 & 0.0008 & 29 \\
2 & 56194.1002 & 0.0007 & $-$0.0002 & 39 \\
3 & 56194.1604 & 0.0009 & $-$0.0001 & 36 \\
\hline
  \multicolumn{5}{l}{\commenta BJD$-$2400000.} \\
  \multicolumn{5}{l}{\commentb Against max $= 2456193.9802 + 0.060095 E$.} \\
  \multicolumn{5}{l}{\commentc Number of points used to determine the maximum.} \\
\end{tabular}
\end{center}
\end{table}

\subsection{V368 Persei}\label{obj:v368per}

   V368 Per was discovered by \citet{ric69v1233aql}, who recorded
one outburst (JD 2440152.4) lasting for $\sim$2~d.
\citet{kur73v368per} studied this object on
photographic plates and recorded two additional outbursts.
The second outburst (JD 2441599.4) lasted for 7~d and faded
from 15.1 mag to 15.7 mag.  Based on these observations
\citet{kur73v368per} suspected the object to be a Z Cam-type
dwarf nova.  This suggestion was probably due to the long
duration of his second outburst, not based on any signature
of a standstill.  \citet{bus79VS17} further studied this object
on Sonneberg plates and found seven more outbursts.
One of these outburst (JD 2440151--2440153) was recorded for
three consecutive nights and rapid rise (1 mag in 0.24~d)
and fading (1 mag in 1~d) were recorded.  No long outburst
was recorded.  \citet{bus79VS17} classified this object
as a dwarf nova.

\begin{figure}
  \begin{center}
    \FigureFile(88mm,110mm){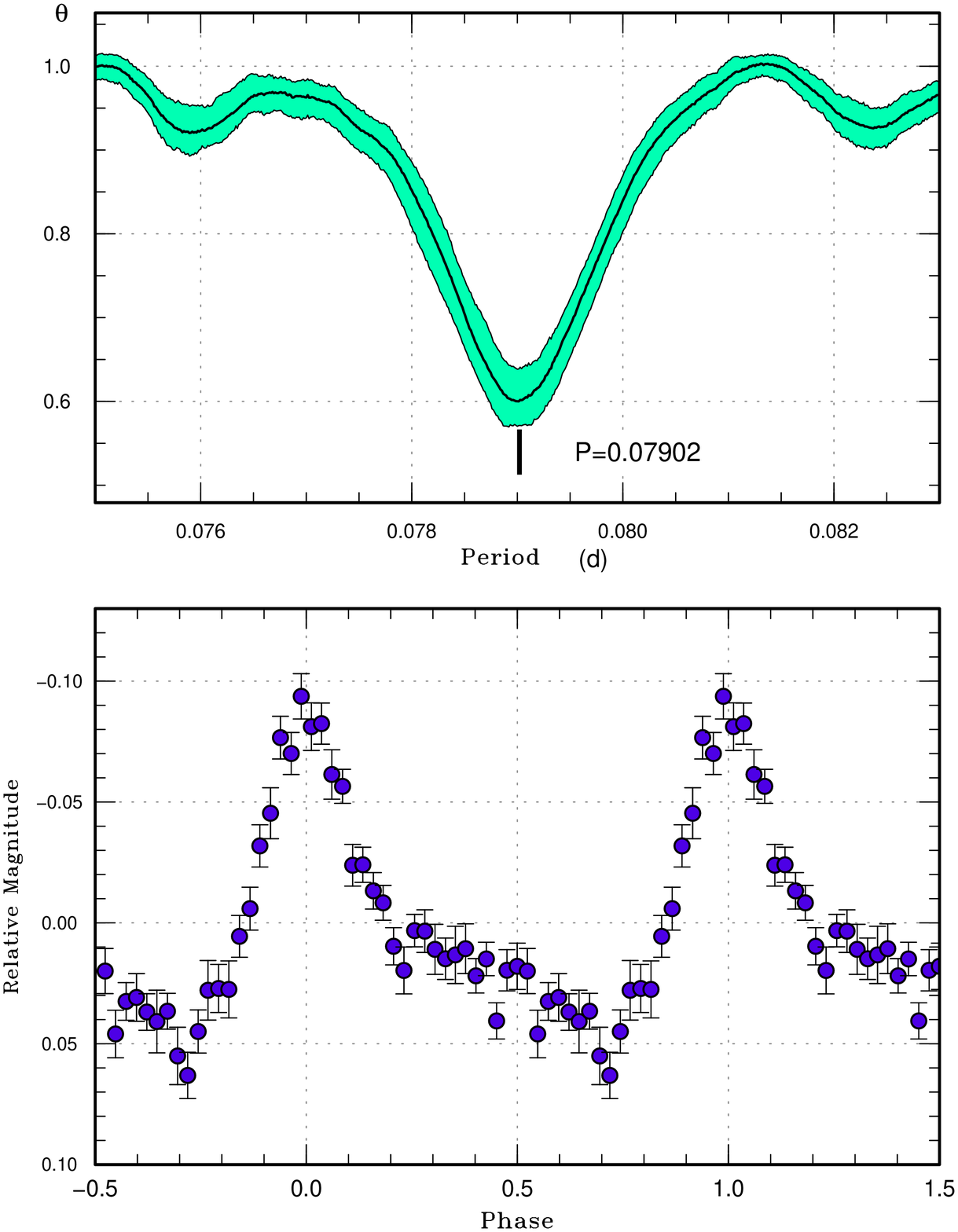}
  \end{center}
  \caption{Superhumps in V368 Per (2012). (Upper): PDM analysis.
     (Lower): Phase-averaged profile.}
  \label{fig:v368pershpdm}
\end{figure}

   The object has been largely neglected until S. Brady started
systematically monitoring it with a CCD.  Although he detected 
an outburst on 2007 October 13 and detected superhump-like
modulations, this finding was not broadly communicated.
During an outburst in 2012 December, I. Miller reported
the detection of superhumps (BAAVSS alert 3113).
Follow-up observations confirmed this finding (figure
\ref{fig:v368pershpdm}.
The times of superhump maxima are listed in table
\ref{tab:v368peroc2012}.  Since the object was observed only
during the final part of the superoutburst, we identified
these superhumps as stage C superhumps.  Although there was
observation 7~d earlier than the first superhump maximum
in table \ref{tab:v368peroc2012}, only one superhump minimum
was recorded.  We did not include this observation because
no superhump maximum was fully recorded.  The superhump amplitude,
however, was as large as 0.4 mag.

   The rapidly fading outburst in \citet{bus79VS17} was most likely
a normal outburst, and the long-lasting one in \citet{kur73v368per}
was a superoutburst.  This object is yet another example
in which a superoutburst was confused as a standstill
[see e.g. AQ Eri \citep{kat89aqeri},
TT Boo (\cite{mei66ttboo}; \cite{GCVS3})].

\begin{table}
\caption{Superhump maxima of V368 Per (2012)}\label{tab:v368peroc2012}
\begin{center}
\begin{tabular}{ccccc}
\hline
$E$ & max\commenta & error & $O-C$\commentb & $N$\commentc \\
\hline
0 & 56269.4514 & 0.0008 & $-$0.0007 & 68 \\
1 & 56269.5301 & 0.0009 & $-$0.0010 & 82 \\
2 & 56269.6103 & 0.0010 & 0.0003 & 70 \\
11 & 56270.3214 & 0.0007 & 0.0009 & 174 \\
12 & 56270.4013 & 0.0011 & 0.0018 & 109 \\
23 & 56271.2661 & 0.0090 & $-$0.0018 & 36 \\
24 & 56271.3486 & 0.0014 & 0.0018 & 56 \\
25 & 56271.4269 & 0.0023 & 0.0011 & 69 \\
26 & 56271.5018 & 0.0021 & $-$0.0029 & 67 \\
27 & 56271.5842 & 0.0021 & 0.0006 & 77 \\
\hline
  \multicolumn{5}{l}{\commenta BJD$-$2400000.} \\
  \multicolumn{5}{l}{\commentb Against max $= 2456269.4521 + 0.078946 E$.} \\
  \multicolumn{5}{l}{\commentc Number of points used to determine the maximum.} \\
\end{tabular}
\end{center}
\end{table}

\subsection{TY Piscis Austrini}\label{obj:typsa}

   TY PsA is a well known SU UMa-type dwarf nova since
\citet{bar82typsa}.  The reported observations since then, however,
were rather restricted to high-speed photometry to detect
dwarf nova oscillations \citep{war89typsa} and spectroscopy
\citep{odo92typsa}.  Until our report on the 2008 superoutburst
\citep{Pdot}, the superhump period with a limited accuracy
by \citet{war89typsa} had long been used.

\begin{figure}
  \begin{center}
    \FigureFile(88mm,70mm){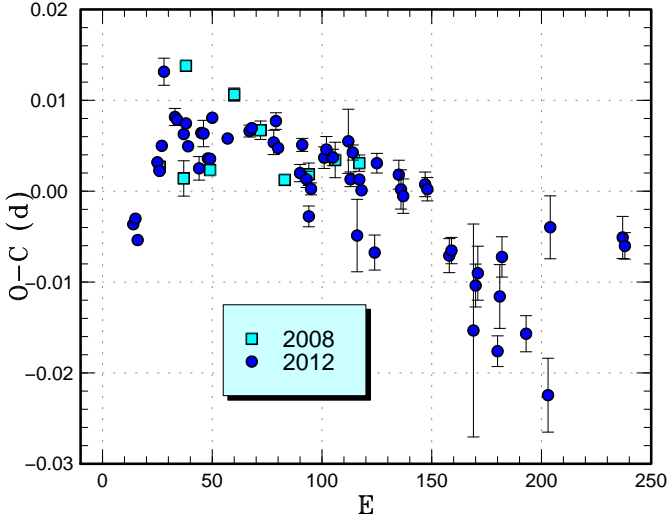}
  \end{center}
  \caption{Comparison of $O-C$ diagrams of TY PsA between different
  superoutbursts.  A period of 0.08787~d was used to draw this figure.
  Approximate cycle counts ($E$) after the start of the superoutburst
  were used.
  }
  \label{fig:typsacomp}
\end{figure}

   We could obtain a very good coverage of the 2012 superoutburst
and its post-superoutburst state until the second normal outburst
following this superoutburst.  The times of superhump maxima
during the superoutburst and the initial part of the post-superoutburst
period are listed in table \ref{tab:typsaoc2012}.  The rapid
decline from the superoutburst occurred between $E=123$ and $E=133$
and there was no evidence of a phase 0.5 jump as seen in
VW Hyi \citep{Pdot4}.  In other words, there was no evidence
of traditional late superhumps.  As is common in relatively
long-$P_{\rm orb}$ systems (e.g. V344 Lyr: \cite{Pdot3}),
the stage B--C transition was relatively smooth.
The superhumps in the post-superoutburst stage can be interpreted
as a smooth extension of stage C superhumps, and this makes
a clear contrast to VW Hyi.
A comparison of the $O-C$ between the different superoutburst
is shown in figure \ref{fig:typsacomp}.

   Using the post-superoutburst data, we could determine
the orbital period to be 0.08423(1)~d.

\begin{figure}
  \begin{center}
    \FigureFile(88mm,95mm){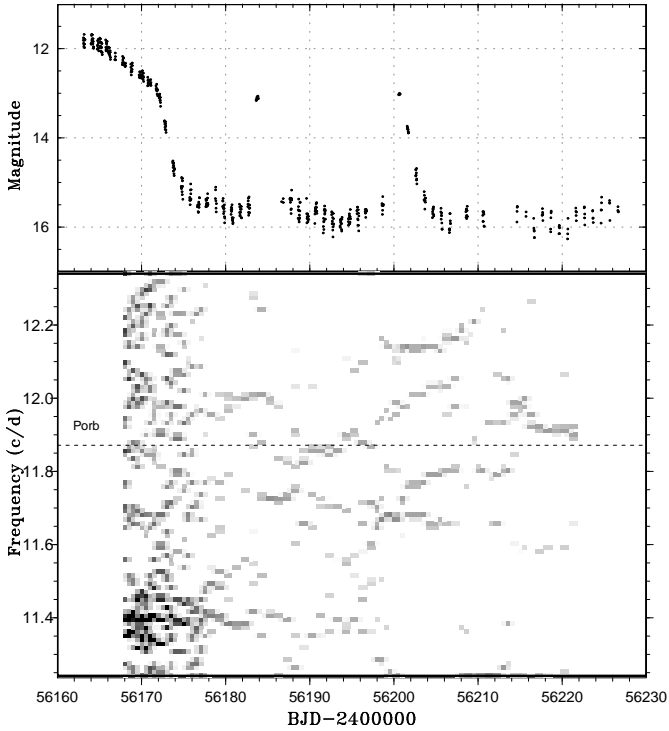}
  \end{center}
  \caption{Two-dimensional Lasso period analysis of TY PsA.
     (Upper): Light curve (binned to 0.01~d).
     (Lower): Two-dimensional Lasso analysis (10~d window,
     0.5~d shift and $\log \lambda=-6.5$).}
  \label{fig:typsalasso}
\end{figure}

\begin{figure}
  \begin{center}
    \FigureFile(88mm,110mm){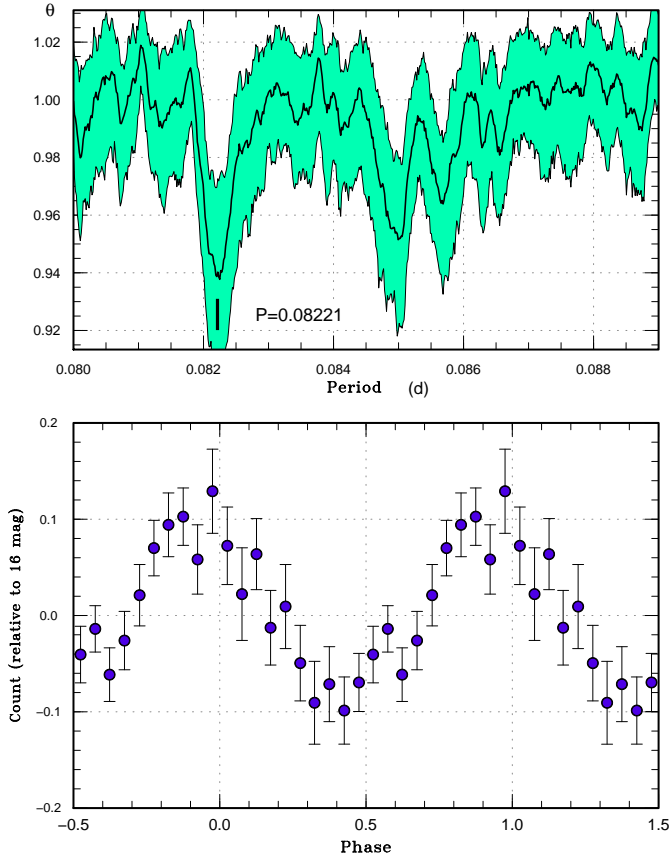}
  \end{center}
  \caption{Negative superhumps in TY PsA around the second normal
     outburst following the superoutburst. (Upper): PDM analysis.
     (Lower): Phase-averaged profile.}
  \label{fig:typsansh}
\end{figure}

   We applied two-dimensional Lasso analysis
(figure \ref{fig:typsalasso}).  The superhumps were detected
as a rather broad band due to the non-sinusoidal profile.
The main signal during the quiescent period (BJD 2456186--2456200)
following the first normal outburst after the superoutburst
was the orbital period.  During and after the second normal
outburst, a signal around 12.15~c/d appeared.  This signal
cannot be explained by a one-day alias of the superhump period,
and we consider it to be the signal of negative superhumps.
A PDM analysis (range BJD 2456198--2456210) also supports
this identification (figure \ref{fig:typsansh}).
The $\varepsilon$ for this negative superhump was $-2.4$\%
(or $\varepsilon^*=-2.5$\%), a value similar to that obtained
in the Kepler data of V344 Lyr and V1504 Cyg 
(cf. \cite{osa13v344lyrv1504cyg}).  Becuase of the relatively
sparse observation after the second normal outburst, it is not
clear whether this signal of negative superhumps persisted
or not.  It has been shown that a state with negative superhumps
(likely arising from a tilted disk) tends to suppress
normal outbursts (\cite{osa13v1504cygKepler};
\cite{ohs12eruma}; \cite{zem13eruma}).  TY PsA did not show
a normal outburst for more than 25~d and this may be a result
of such suppression of outbursts.

   Stage A superhumps were also observed.  The $\varepsilon^*$
for stage A superhumps was 4.86(12)\%, which corresponds to
$q=$0.142(4).

\begin{table}
\caption{Superhump maxima of TY PsA (2012)}\label{tab:typsaoc2012}
\begin{center}
\begin{tabular}{ccccc}
\hline
$E$ & max\commenta & error & $O-C$\commentb & $N$\commentc \\
\hline
0 & 56163.0866 & 0.0004 & $-$0.0110 & 105 \\
1 & 56163.1751 & 0.0004 & $-$0.0103 & 390 \\
2 & 56163.2606 & 0.0005 & $-$0.0126 & 268 \\
11 & 56164.0600 & 0.0004 & $-$0.0033 & 141 \\
12 & 56164.1469 & 0.0005 & $-$0.0042 & 111 \\
13 & 56164.2375 & 0.0004 & $-$0.0013 & 121 \\
14 & 56164.3336 & 0.0015 & 0.0069 & 41 \\
19 & 56164.7679 & 0.0009 & 0.0024 & 28 \\
20 & 56164.8555 & 0.0005 & 0.0021 & 27 \\
23 & 56165.1175 & 0.0003 & 0.0008 & 152 \\
24 & 56165.2066 & 0.0002 & 0.0021 & 105 \\
25 & 56165.2919 & 0.0005 & $-$0.0004 & 101 \\
30 & 56165.7289 & 0.0013 & $-$0.0024 & 19 \\
31 & 56165.8206 & 0.0006 & 0.0016 & 29 \\
32 & 56165.9084 & 0.0015 & 0.0016 & 16 \\
34 & 56166.0814 & 0.0003 & $-$0.0009 & 285 \\
35 & 56166.1693 & 0.0003 & $-$0.0009 & 540 \\
36 & 56166.2616 & 0.0004 & 0.0037 & 343 \\
43 & 56166.8745 & 0.0006 & 0.0020 & 30 \\
53 & 56167.7540 & 0.0007 & 0.0037 & 30 \\
54 & 56167.8422 & 0.0006 & 0.0041 & 31 \\
64 & 56168.7193 & 0.0013 & 0.0034 & 15 \\
65 & 56168.8095 & 0.0009 & 0.0058 & 24 \\
66 & 56168.8944 & 0.0006 & 0.0029 & 19 \\
76 & 56169.7704 & 0.0009 & 0.0010 & 25 \\
77 & 56169.8613 & 0.0007 & 0.0042 & 25 \\
79 & 56170.0333 & 0.0009 & 0.0006 & 154 \\
80 & 56170.1171 & 0.0011 & $-$0.0034 & 118 \\
81 & 56170.2080 & 0.0007 & $-$0.0003 & 94 \\
87 & 56170.7386 & 0.0012 & 0.0036 & 24 \\
88 & 56170.8274 & 0.0014 & 0.0046 & 25 \\
91 & 56171.0901 & 0.0009 & 0.0040 & 139 \\
98 & 56171.7070 & 0.0035 & 0.0064 & 18 \\
99 & 56171.7907 & 0.0008 & 0.0023 & 33 \\
100 & 56171.8815 & 0.0008 & 0.0053 & 30 \\
102 & 56172.0481 & 0.0040 & $-$0.0037 & 51 \\
103 & 56172.1421 & 0.0004 & 0.0026 & 217 \\
104 & 56172.2288 & 0.0005 & 0.0015 & 231 \\
110 & 56172.7492 & 0.0019 & $-$0.0049 & 33 \\
111 & 56172.8469 & 0.0011 & 0.0050 & 35 \\
121 & 56173.7243 & 0.0016 & 0.0046 & 30 \\
122 & 56173.8106 & 0.0022 & 0.0031 & 34 \\
123 & 56173.8977 & 0.0019 & 0.0024 & 23 \\
133 & 56174.7777 & 0.0014 & 0.0046 & 44 \\
134 & 56174.8650 & 0.0013 & 0.0041 & 45 \\
144 & 56175.7364 & 0.0019 & $-$0.0023 & 34 \\
145 & 56175.8249 & 0.0014 & $-$0.0017 & 45 \\
155 & 56176.6948 & 0.0117 & $-$0.0097 & 44 \\
156 & 56176.7876 & 0.0024 & $-$0.0046 & 45 \\
157 & 56176.8768 & 0.0030 & $-$0.0032 & 35 \\
166 & 56177.6591 & 0.0017 & $-$0.0110 & 47 \\
167 & 56177.7529 & 0.0035 & $-$0.0049 & 39 \\
168 & 56177.8452 & 0.0022 & $-$0.0005 & 33 \\
179 & 56178.8033 & 0.0020 & $-$0.0080 & 30 \\
189 & 56179.6752 & 0.0041 & $-$0.0139 & 45 \\
190 & 56179.7816 & 0.0035 & 0.0047 & 30 \\
223 & 56182.6802 & 0.0023 & 0.0063 & 41 \\
224 & 56182.7671 & 0.0015 & 0.0055 & 30 \\
\hline
  \multicolumn{5}{l}{\commenta BJD$-$2400000.} \\
  \multicolumn{5}{l}{\commentb Against max $= 2456163.0977 + 0.087786 E$.} \\
  \multicolumn{5}{l}{\commentc Number of points used to determine the maximum.} \\
\end{tabular}
\end{center}
\end{table}

\subsection{QW Serpentis}\label{obj:qwser}

   This object was originally discovered as
a Mira-type or a dwarf nova (TmzV46) \citep{tak98qwser}.
\citet{kat99qwser} reported CCD photometry of the brightening
in 1999 October, confirming the dwarf nova-type classification.
The object was further studied during the 2000 outburst, when 
superhumps were first detected (\cite{pat03suumas}; \cite{nog04qwser}).
The 2003 superoutburst was one of the best observed ones
among SU UMa-type dwarf novae (\cite{ole03qwser}; \cite{nog04qwser}).
The object was also selected as a CV in the SDSS
\citep{szk09SDSSCV7}.

   We observed the 2009 and 2013 superoutbursts.  The times of superhump
maxima are listed in table \ref{tab:qwseroc2013}.
The later part of stage B and stage C were observed.
Around the start of the rapid fading ($E \ge 98$), the period 
appeared to increase.  We consider this as a result of the decrease 
in the pressure effect as discussed in \citet{nak13j2112j2037} and 
did not include them for determining the period of stage C.
The period of this phase [0.07661(9)~d] corresponds to
$q=0.21$ if we assume a disk radius of 0.30$A$ and $q=0.13$
for a disk radius of 0.38$A$, where $A$ is the binary separation.

\begin{figure}
  \begin{center}
    \FigureFile(88mm,70mm){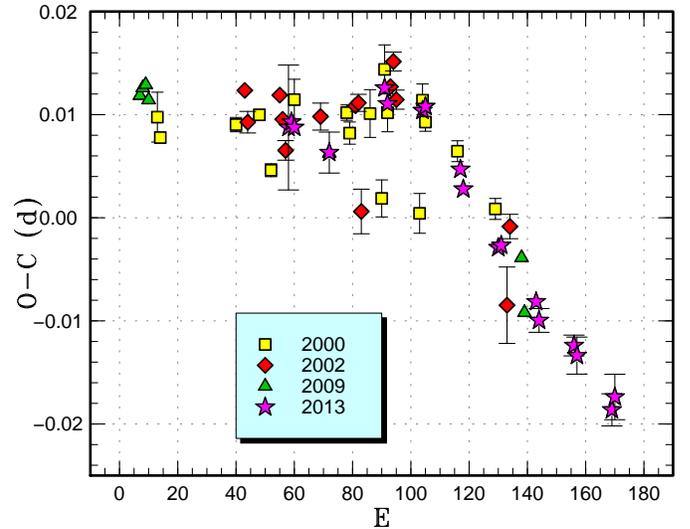}
  \end{center}
  \caption{Comparison of $O-C$ diagrams of QW Ser between different
  superoutbursts.  A period of 0.07700~d was used to draw this figure.
  Approximate cycle counts ($E$) after the start of the superoutburst
  were used.}
  \label{fig:qwsercomp}
\end{figure}

   A comparison of $O-C$ diagrams between different superoutbursts
is shown in figure \ref{fig:qwsercomp}.  All superoutbursts
followed a similar pattern of stage B--C transition.

   We have also updated the orbital period to be 0.074572(1)~d
using the CRTS data.  This value is in good agreement with
those in \citet{ole03qwser} (photometric); \citet{pat03suumas}
(radial-velocity study).

\begin{table}
\caption{Superhump maxima of QW Ser (2009)}\label{tab:qwseroc2009}
\begin{center}
\begin{tabular}{ccccc}
\hline
$E$ & max\commenta & error & $O-C$\commentb & $N$\commentc \\
\hline
0 & 54983.9862 & 0.0004 & $-$0.0009 & 102 \\
1 & 54984.0639 & 0.0003 & $-$0.0000 & 148 \\
2 & 54984.1413 & 0.0002 & 0.0004 & 226 \\
3 & 54984.2168 & 0.0002 & $-$0.0009 & 260 \\
65 & 54988.9856 & 0.0003 & 0.0027 & 136 \\
131 & 54994.0575 & 0.0005 & 0.0019 & 128 \\
132 & 54994.1291 & 0.0005 & $-$0.0032 & 88 \\
\hline
  \multicolumn{5}{l}{\commenta BJD$-$2400000.} \\
  \multicolumn{5}{l}{\commentb Against max $= 2454983.9871 + 0.076858 E$.} \\
  \multicolumn{5}{l}{\commentc Number of points used to determine the maximum.} \\
\end{tabular}
\end{center}
\end{table}

\begin{table}
\caption{Superhump maxima of QW Ser (2013)}\label{tab:qwseroc2013}
\begin{center}
\begin{tabular}{ccccc}
\hline
$E$ & max\commenta & error & $O-C$\commentb & $N$\commentc \\
\hline
0 & 56482.9942 & 0.0061 & $-$0.0059 & 15 \\
1 & 56483.0717 & 0.0006 & $-$0.0051 & 85 \\
2 & 56483.1482 & 0.0006 & $-$0.0054 & 75 \\
14 & 56484.0697 & 0.0020 & $-$0.0048 & 84 \\
33 & 56485.5390 & 0.0008 & 0.0064 & 43 \\
34 & 56485.6145 & 0.0007 & 0.0051 & 48 \\
46 & 56486.5378 & 0.0006 & 0.0075 & 43 \\
47 & 56486.6152 & 0.0006 & 0.0081 & 47 \\
59 & 56487.5331 & 0.0008 & 0.0051 & 23 \\
60 & 56487.6082 & 0.0009 & 0.0035 & 22 \\
72 & 56488.5265 & 0.0008 & 0.0009 & 22 \\
73 & 56488.6037 & 0.0009 & 0.0013 & 22 \\
85 & 56489.5223 & 0.0009 & $-$0.0011 & 19 \\
86 & 56489.5975 & 0.0012 & $-$0.0026 & 19 \\
98 & 56490.5190 & 0.0010 & $-$0.0020 & 20 \\
99 & 56490.5950 & 0.0018 & $-$0.0027 & 19 \\
111 & 56491.5138 & 0.0016 & $-$0.0049 & 20 \\
112 & 56491.5920 & 0.0022 & $-$0.0034 & 19 \\
\hline
  \multicolumn{5}{l}{\commenta BJD$-$2400000.} \\
  \multicolumn{5}{l}{\commentb Against max $= 2456483.0001 + 0.076744 E$.} \\
  \multicolumn{5}{l}{\commentc Number of points used to determine the maximum.} \\
\end{tabular}
\end{center}
\end{table}

\subsection{V493 Serpentis}\label{obj:v493ser}

   This object (=SDSS J155644.24$-$000950.2) was selected as
a dwarf nova by SDSS \citep{szk02SDSSCVs}.
\citet{wou04CV4} obtained 0.07408(1)~d from quiescent orbital humps.
Although H. Maehara detected superhumps during the 2006 superoutburst,
it was impossible to determine an accurate period.  The 2007
superoutburst was well-observed and was summarized in \citet{Pdot}.

\begin{figure}
  \begin{center}
    \FigureFile(88mm,70mm){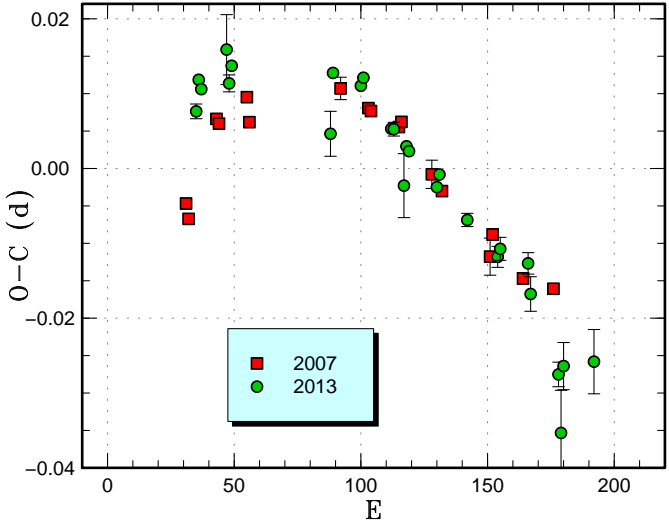}
  \end{center}
  \caption{Comparison of $O-C$ diagrams of V493 Ser between different
  superoutbursts.  A period of 0.08296~d was used to draw this figure.
  Approximate cycle counts ($E$) after the start of the superoutburst
  were used.  Since the start of the 2013 superoutburst
  was not well constrained, we shifted the $O-C$ diagram
  to best fit the better-recorded 2007 one.
  }
  \label{fig:v493sercomp}
\end{figure}

   We observed the 2013 superoutburst.  The times of superhump maxima
are listed in table \ref{tab:v493seroc2013}.  Although $P_{\rm dot}$
for stage B superhumps was not measured, there was a clear stage B--C
transition.  Following the rapid fading, there was a likely
phase jump ($E \ge 143$).  A PDM analysis of the post-superoutburst
part yielded a period of 0.08285(6)~d, not very different from
that of stage C superhumps.  As in QW Ser, the disk radius
0.30$A$ and 0.38$A$ correspond to $q$=0.26 and $q$=0.16, respectively.
A comparison of $O-C$ diagrams between 2007 and 2013 superoutburst
is shown in table \ref{fig:v493sercomp}.

\begin{table}
\caption{Superhump maxima of V493 Ser (2013)}\label{tab:v493seroc2013}
\begin{center}
\begin{tabular}{ccccc}
\hline
$E$ & max\commenta & error & $O-C$\commentb & $N$\commentc \\
\hline
0 & 56478.6582 & 0.0010 & $-$0.0113 & 60 \\
1 & 56478.7454 & 0.0002 & $-$0.0069 & 130 \\
2 & 56478.8271 & 0.0002 & $-$0.0078 & 129 \\
12 & 56479.6620 & 0.0047 & 0.0001 & 17 \\
13 & 56479.7404 & 0.0011 & $-$0.0042 & 46 \\
14 & 56479.8257 & 0.0002 & $-$0.0016 & 136 \\
53 & 56483.0521 & 0.0030 & $-$0.0004 & 50 \\
54 & 56483.1432 & 0.0008 & 0.0080 & 150 \\
65 & 56484.0540 & 0.0008 & 0.0092 & 99 \\
66 & 56484.1380 & 0.0007 & 0.0105 & 153 \\
77 & 56485.0438 & 0.0006 & 0.0066 & 128 \\
78 & 56485.1267 & 0.0009 & 0.0068 & 139 \\
82 & 56485.4510 & 0.0043 & 0.0003 & 13 \\
83 & 56485.5392 & 0.0004 & 0.0058 & 47 \\
84 & 56485.6215 & 0.0003 & 0.0054 & 51 \\
95 & 56486.5293 & 0.0005 & 0.0035 & 46 \\
96 & 56486.6139 & 0.0005 & 0.0055 & 51 \\
107 & 56487.5204 & 0.0009 & 0.0023 & 19 \\
119 & 56488.5110 & 0.0014 & 0.0005 & 25 \\
120 & 56488.5950 & 0.0015 & 0.0019 & 24 \\
131 & 56489.5056 & 0.0014 & 0.0028 & 21 \\
132 & 56489.5845 & 0.0023 & $-$0.0010 & 21 \\
143 & 56490.4863 & 0.0016 & $-$0.0089 & 17 \\
144 & 56490.5614 & 0.0057 & $-$0.0164 & 20 \\
145 & 56490.6533 & 0.0032 & $-$0.0072 & 14 \\
157 & 56491.6494 & 0.0043 & $-$0.0035 & 14 \\
\hline
  \multicolumn{5}{l}{\commenta BJD$-$2400000.} \\
  \multicolumn{5}{l}{\commentb Against max $= 2456478.6695 + 0.082697 E$.} \\
  \multicolumn{5}{l}{\commentc Number of points used to determine the maximum.} \\
\end{tabular}
\end{center}
\end{table}

\subsection{AW Sagittae}\label{obj:awsge}

   AW Sge was discovered as a dwarf nova early in the history 
\citep{wol06awsge}.  Although very little were known since then,
the SU UMa-type nature was clarified during the 2000 superoutburst.
The 2000 and 2006 superoutbursts were reported in \citet{Pdot}.

   The 2012 superoutburst was timely detected by R. Stubbings
(vsnet-alert 14768) and subsequent observations managed to
record the evolving stage of superhumps (vsnet-alert 14770,
14776, 14777, 14789).  This outburst was the best observed
one in this object.

\begin{figure}
  \begin{center}
    \FigureFile(88mm,70mm){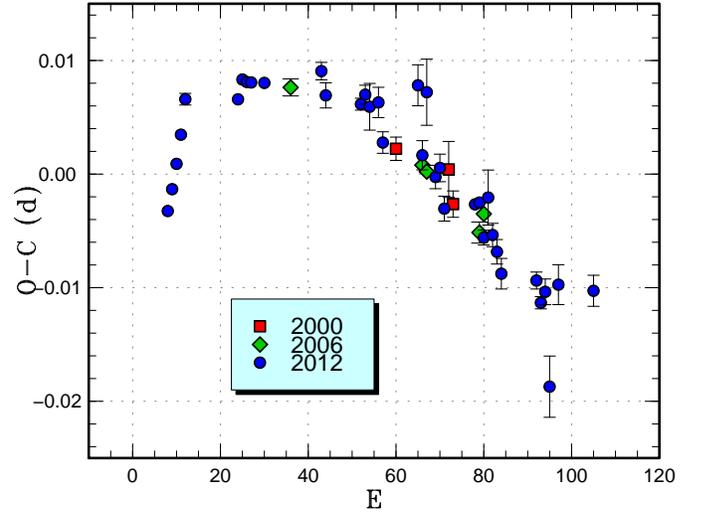}
  \end{center}
  \caption{Comparison of $O-C$ diagrams of AW Sge between different
  superoutbursts.  A period of 0.07480~d was used to draw this figure.
  Approximate cycle counts ($E$) after the start of the superoutburst
  were used.  Since the start of the 2006 superoutburst
  was not well constrained, we shifted the $O-C$ diagram
  to best fit the best-recorded 2012 one.
  }
  \label{fig:awsgecomp}
\end{figure}

   The times of superhump maxima are listed in table \ref{tab:awsgeoc2012}.
The object showed clear stages A ($E \le 4$), B and C.
The stage B--C transition occurred at around $E=$57--59.
A comparison of the $O-C$ between the different superoutburst
is shown in figure \ref{fig:awsgecomp}.

\begin{table}
\caption{Superhump maxima of AW Sge (2012)}\label{tab:awsgeoc2012}
\begin{center}
\begin{tabular}{ccccc}
\hline
$E$ & max\commenta & error & $O-C$\commentb & $N$\commentc \\
\hline
0 & 56127.1419 & 0.0004 & $-$0.0119 & 155 \\
1 & 56127.2186 & 0.0004 & $-$0.0098 & 142 \\
2 & 56127.2957 & 0.0004 & $-$0.0074 & 53 \\
3 & 56127.3730 & 0.0004 & $-$0.0047 & 126 \\
4 & 56127.4510 & 0.0005 & $-$0.0014 & 73 \\
16 & 56128.3485 & 0.0003 & 0.0007 & 51 \\
17 & 56128.4251 & 0.0002 & 0.0026 & 78 \\
18 & 56128.4997 & 0.0002 & 0.0026 & 74 \\
19 & 56128.5744 & 0.0003 & 0.0027 & 79 \\
22 & 56128.7988 & 0.0004 & 0.0032 & 31 \\
35 & 56129.7722 & 0.0008 & 0.0065 & 35 \\
36 & 56129.8449 & 0.0011 & 0.0045 & 15 \\
44 & 56130.4425 & 0.0005 & 0.0051 & 80 \\
45 & 56130.5182 & 0.0008 & 0.0061 & 87 \\
46 & 56130.5919 & 0.0021 & 0.0052 & 39 \\
48 & 56130.7419 & 0.0013 & 0.0059 & 13 \\
49 & 56130.8131 & 0.0009 & 0.0026 & 19 \\
57 & 56131.4166 & 0.0018 & 0.0090 & 28 \\
58 & 56131.4852 & 0.0013 & 0.0030 & 28 \\
59 & 56131.5656 & 0.0029 & 0.0087 & 46 \\
61 & 56131.7077 & 0.0010 & 0.0016 & 11 \\
62 & 56131.7833 & 0.0012 & 0.0026 & 19 \\
63 & 56131.8545 & 0.0011 & $-$0.0008 & 13 \\
70 & 56132.3785 & 0.0004 & 0.0008 & 168 \\
71 & 56132.4535 & 0.0004 & 0.0011 & 203 \\
72 & 56132.5252 & 0.0007 & $-$0.0018 & 148 \\
73 & 56132.6035 & 0.0024 & 0.0019 & 44 \\
74 & 56132.6750 & 0.0010 & $-$0.0012 & 14 \\
75 & 56132.7483 & 0.0011 & $-$0.0025 & 13 \\
76 & 56132.8212 & 0.0013 & $-$0.0043 & 21 \\
84 & 56133.4190 & 0.0007 & $-$0.0035 & 167 \\
85 & 56133.4918 & 0.0005 & $-$0.0053 & 178 \\
86 & 56133.5676 & 0.0011 & $-$0.0042 & 58 \\
87 & 56133.6340 & 0.0027 & $-$0.0123 & 24 \\
89 & 56133.7926 & 0.0018 & $-$0.0030 & 20 \\
97 & 56134.3905 & 0.0014 & $-$0.0022 & 106 \\
\hline
  \multicolumn{5}{l}{\commenta BJD$-$2400000.} \\
  \multicolumn{5}{l}{\commentb Against max $= 2456127.1538 + 0.074627 E$.} \\
  \multicolumn{5}{l}{\commentc Number of points used to determine the maximum.} \\
\end{tabular}
\end{center}
\end{table}

\subsection{V1212 Tauri}\label{obj:v1212tau}

   V1212 Tau was discovered as an eruptive object near M45
\citep{par83v1212tau}.  Although this object was not named
for a long time, possibly due to the confusion as being a possible
Pleiades flare star, members of the Variable Star Observers 
League in Japan (VSOLJ) suspected the
dwarf nova-type nature of this object and started monitoring
since 1987.  The first secure outburst was reported in
2007 by G. Gualdoni (vsnet-alert 9190).  During this outburst,
J. Patterson reported the detection of superhumps (see
\cite{Pdot3} for more history).  The well-observed 2011
superoutburst was reported in \citet{Pdot3} and another
superoutburst in 2011 was also recorded \citet{Pdot4}.

   The 2013 superoutburst was observed on a single night.
We obtained a single superhump maximum of BJD 2456330.3366(13)
($N=56$).

\subsection{BZ Ursae Majoris}\label{obj:bzuma}

   This object was discovered as a dwarf nova by \citet{mar68bzuma},
and was relatively well monitored by the AAVSO since 1968.
Up to 1984, secure outbursts were recorded at a typical rate
of $\sim$1 per year.  Despite this, \citet{wen82bzuma} suggested
that the object is an intermediate object between U Gem star
and WZ Sge-type star based on the analysis of the photographic
plates.  Queerly, no secure large-amplitude outbursts were recorded
by the AAVSO between 1985 and 1990.  The object was also
identified as a CV by the PG Survey \citet{gre82PGsurveyCV}.
\citet{kal86bzumaiauc} reported a fading ($B=17.8$) episode
in 1986 December.  Around these years, the object may have spent
a period of decreased mass-transfer rate, and several detections
of (supposedly rare) outbursts were reported in IAU Circulars
(e.g. \cite{sch90bzumaiauc}).
\citet{rin90bzuma} was the first to report
the orbital period of 0.0679~d [see \citet{jur94bzuma};
\citet{rin94czoriv1193oribzuma} for more details].
Although the orbital period  suggested an SU UMa-type
dwarf nova, no superoutburst was definitely identified
\citep{jur94bzuma}.  \citet{rin94czoriv1193oribzuma} also reported
anomalous low-frequency variations in the radial-velocity data.
\citet{kat99bzuma}, \citet{pri04bzuma} and
\citet{jia10v1159oribzuma} reported time-series observations 
during short (normal) outbursts.
\citet{gan03eipscv396hyabzumaeycyg} reported
UV observations by the HST and \citet{neu06bzuma} reported
spectroscopic observations during a normal outburst.
Since 1992, bright (10--11 mag) outbursts were more frequently
seen than before.  The object was also selected
as a CV in the SDSS \citep{szk03SDSSCV2}.
The long-awaited superoutburst was finally recorded in
2007 (\cite{pri09bzuma}; \cite{Pdot}).

   The object again underwent
a superoutburst on 2012 November 27 (H. Maehara, vsnet-alert 15148).
The outburst was not well observed due to the delay in
the announcement of the superoutburst.  Only a part of the
rapid fading stage was recorded.  A PDM analysis yielded a period
of 0.06982(4)~d, which is consistent with the period of stage C
superhumps \citep{Pdot}.  Due to the large irregularities,
we did not attempt to obtain $O-C$ values.  According to
the AAVSO data, the object was already in full outburst
on November 26 at a visual magnitude of 11.3.  The object was
still in quiescence 2~d before.  It was likely that the rise
was faster than in 2007, when a slow start of the outburst
(likely an inside-out outburst) was recorded \citep{Pdot}.

\subsection{CI Ursae Majoris}\label{obj:ciuma}

   CI UMa was discovered as a dwarf nova by \citet{gor72ciuma}.
There was a report of photographic records of outbursts
in \citet{kol79cpdraciuma}.
The SU UMa-type nature was established by the detection of
superhumps in \citet{nog97ciuma}.  In \citet{Pdot}, we reported
on superoutbursts in 2001, 2003 amd 2006.  \citet{par06ciuma}
also reported the detection of superhumps in the 2006 superoutburst.
There was another moderately well-observed superoutburst in
2011 \citep{Pdot3}.

   We observed the final part of the 2013 April superoutburst.
The times of superhumps are listed in table
\ref{tab:ciumaoc2013}.  The observed superhumps are most likely
stage C superhumps.

\begin{table}
\caption{Superhump maxima of CI UMa (2013)}\label{tab:ciumaoc2013}
\begin{center}
\begin{tabular}{ccccc}
\hline
$E$ & max\commenta & error & $O-C$\commentb & $N$\commentc \\
\hline
0 & 56385.3116 & 0.0013 & $-$0.0040 & 11 \\
1 & 56385.3778 & 0.0012 & $-$0.0001 & 18 \\
2 & 56385.4414 & 0.0002 & 0.0012 & 195 \\
3 & 56385.5034 & 0.0002 & 0.0007 & 228 \\
4 & 56385.5681 & 0.0003 & 0.0030 & 228 \\
5 & 56385.6270 & 0.0007 & $-$0.0004 & 53 \\
32 & 56387.3113 & 0.0007 & $-$0.0004 & 145 \\
\hline
  \multicolumn{5}{l}{\commenta BJD$-$2400000.} \\
  \multicolumn{5}{l}{\commentb Against max $= 2456385.3155 + 0.062381 E$.} \\
  \multicolumn{5}{l}{\commentc Number of points used to determine the maximum.} \\
\end{tabular}
\end{center}
\end{table}

\subsection{CY Ursae Majoris}\label{obj:cyuma}

   CY UMa was discovered as a dwarf nova by \citet{gor77cyuma}.
The VSOLJ members started monitoring this object since 1987,
and recorded a long outburst in 1988 January.  This outburst
turned out to be a superoutburst (\citet{kat88cyuma}; the reported
superhump period was incorrect, see \cite{kat97cyuma}).
There was a report on its outburst activity \citep{wat89cyuma}.
\citet{szk92CVspec} reported a spectrum characteristic to
a dwarf nova.
\citet{har95cyuma} was the first to establish the superhump period
during the 1995 superoutburst.  \citet{kat95cyuma} also reported 
observations of the two superoutburst in 1991--1992 and in 1993.
The orbital period was determined by \citet{mar95cyuma} and
\citet{tho96Porb}.  \citet{kat99cyuma} reported observations
of superhumps during the 1999 superoutburst.  A summary of
these observations and the 2009 superoutburst were reported
in \citet{Pdot}.  The object was also selected
as a CV in the SDSS \citep{szk05SDSSCV4}.

   Only single-night observation was obtained during the
2013 April superoutburst.  Only one superhump maximum of
BJD 2456397.4024(8) ($N=106$) was recorded.

\subsection{MR Ursae Majoris}\label{obj:mruma}

   MR UMa = 1RXP J113123$+$4322.5 is an ROSAT-selected CV
\citep{wei97mruma}.  Its first superoutburst was recorded
in 2002 (vsnet-alert 7221).  Superhumps during this superoutburst
was recorded by different groups including us 
(cf. \cite{hol02mruma}).  The results for the 2001 and 2003
superoutbursts were reported in \citet{Pdot}.  The 2007 and
2010 superoutbursts were reported in \citet{Pdot2} and
\citet{Pdot4}, respectively.  The object was also selected
as a CV in the SDSS \citep{szk06SDSSCV5}.

\begin{figure}
  \begin{center}
    \FigureFile(88mm,70mm){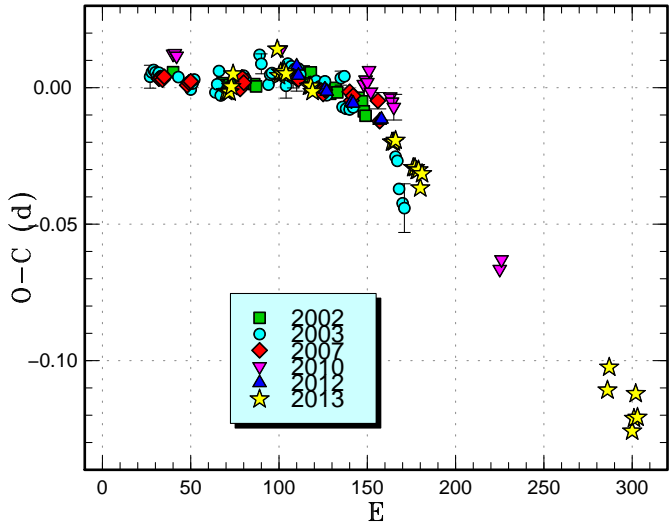}
  \end{center}
  \caption{Comparison of $O-C$ diagrams of MR UMa between different
  superoutbursts.  A period of 0.06512~d was used to draw this figure.
  Approximate cycle counts ($E$) after the start of the superoutburst
  were used.  Since the starts of the superoutbursts other than
  the 2013 one were not well constrained, we shifted the $O-C$ diagrams
  to best fit the others.  The cycle counts of the last two maxima 
  in the 2010 superoutburst were shifted by one from \citet{Pdot2}.
  }
  \label{fig:mrumacomp4}
\end{figure}

   We observed the later stage of the 2013 superoutburst.
The times of superhump maxima are listed in table
\ref{tab:mrumaoc2013}.  Both stages B and C can be identified.
Although post-superoutburst superhumps were likely recorded
($E \ge 216$), we did not include these maxima for deriving
the period of stage C superhumps because these maxima were not
on a smooth extension of the times of stage C superhump
during the late plateau phase.  The combined $O-C$ curve,
however, suggest that they are a part of the persisting
stage C superhumps during the post-superoutburst stage
(figure \ref{fig:mrumacomp4}).  The $O-C$ behavior is very
similar to that of QZ Vir (figure 160 in \cite{Pdot}).

\begin{table}
\caption{Superhump maxima of MR UMa (2013)}\label{tab:mrumaoc2013}
\begin{center}
\begin{tabular}{ccccc}
\hline
$E$ & max\commenta & error & $O-C$\commentb & $N$\commentc \\
\hline
0 & 56354.4000 & 0.0003 & $-$0.0214 & 132 \\
1 & 56354.4652 & 0.0004 & $-$0.0208 & 121 \\
2 & 56354.5301 & 0.0005 & $-$0.0205 & 105 \\
3 & 56354.5972 & 0.0003 & $-$0.0179 & 131 \\
4 & 56354.6669 & 0.0006 & $-$0.0127 & 35 \\
29 & 56356.3040 & 0.0020 & 0.0104 & 80 \\
30 & 56356.3594 & 0.0003 & 0.0012 & 113 \\
31 & 56356.4259 & 0.0007 & 0.0031 & 97 \\
32 & 56356.4904 & 0.0003 & 0.0031 & 128 \\
33 & 56356.5563 & 0.0005 & 0.0044 & 130 \\
34 & 56356.6206 & 0.0003 & 0.0042 & 136 \\
47 & 56357.4626 & 0.0004 & 0.0069 & 65 \\
48 & 56357.5278 & 0.0003 & 0.0075 & 58 \\
49 & 56357.5909 & 0.0003 & 0.0060 & 59 \\
94 & 56360.5028 & 0.0006 & 0.0127 & 72 \\
95 & 56360.5679 & 0.0007 & 0.0134 & 86 \\
96 & 56360.6337 & 0.0008 & 0.0145 & 66 \\
106 & 56361.2748 & 0.0011 & 0.0101 & 84 \\
107 & 56361.3400 & 0.0018 & 0.0107 & 95 \\
108 & 56361.4040 & 0.0011 & 0.0102 & 95 \\
109 & 56361.4694 & 0.0010 & 0.0109 & 95 \\
110 & 56361.5279 & 0.0018 & 0.0049 & 95 \\
111 & 56361.5982 & 0.0012 & 0.0107 & 95 \\
216 & 56368.3566 & 0.0009 & $-$0.0097 & 70 \\
217 & 56368.4301 & 0.0016 & $-$0.0008 & 70 \\
230 & 56369.2532 & 0.0022 & $-$0.0169 & 52 \\
231 & 56369.3230 & 0.0010 & $-$0.0117 & 70 \\
232 & 56369.3972 & 0.0007 & $-$0.0021 & 69 \\
233 & 56369.4536 & 0.0010 & $-$0.0102 & 64 \\
\hline
  \multicolumn{5}{l}{\commenta BJD$-$2400000.} \\
  \multicolumn{5}{l}{\commentb Against max $= 2456354.4214 + 0.064560 E$.} \\
  \multicolumn{5}{l}{\commentc Number of points used to determine the maximum.} \\
\end{tabular}
\end{center}
\end{table}

\subsection{ASASSN-13ao}\label{obj:asassn13ao}

   This object was discovered by ASAS-SN survey \citep{ASASSN}
on 2013 June 8 \citep{sta13asassn13aoatel5118}.  The coordinates are
\timeform{12h 43m 12.05s}, \timeform{+43D 31' 59.9''}.\footnote{
   All the coordinates in this paper are J2000.0 ones.
}
\citet{sta13asassn13aoatel5118} indicated a $g$=21.2-mag SDSS 
counterpart and suggested that this outburst 
has a large outburst amplitude.
The SDSS colors also suggested a short orbital period
(vsnet-alert 15827).  Subsequent observations detected
superhumps (vsnet-alert 15837).  Only two superhump maxima
were recorded: BJD 2456457.0258(8) ($N$=165) and
2456458.0085(16) ($N$=126).  The object already faded
to 18.6 mag only 3~d later.  It was most likely that
we only observed the terminal stage of the superoutburst.
A PDM analysis favored a period of 0.0895(1)~d, although
a longer one-day alias is still possible.

\subsection{ASASSN-13as}\label{obj:asassn13as}

   This object was discovered by ASAS-SN survey on 2013
June 26 \citep{sta13asassn13asatel5168}.  The coordinates are
\timeform{17h 23m 06.3s}, \timeform{+17D 57' 55.9''}.
Follow-up observations detected superhumps (vsnet-alert 15992).
The times of superhump maxima are listed in table
\ref{tab:asassn13asoc2013}.  There was a decreasing trend
of the superhump period.  There may have been a stage A-B
or stage B-C transition between BJD 2456475 and 2456476.

\begin{figure}
  \begin{center}
    \FigureFile(88mm,110mm){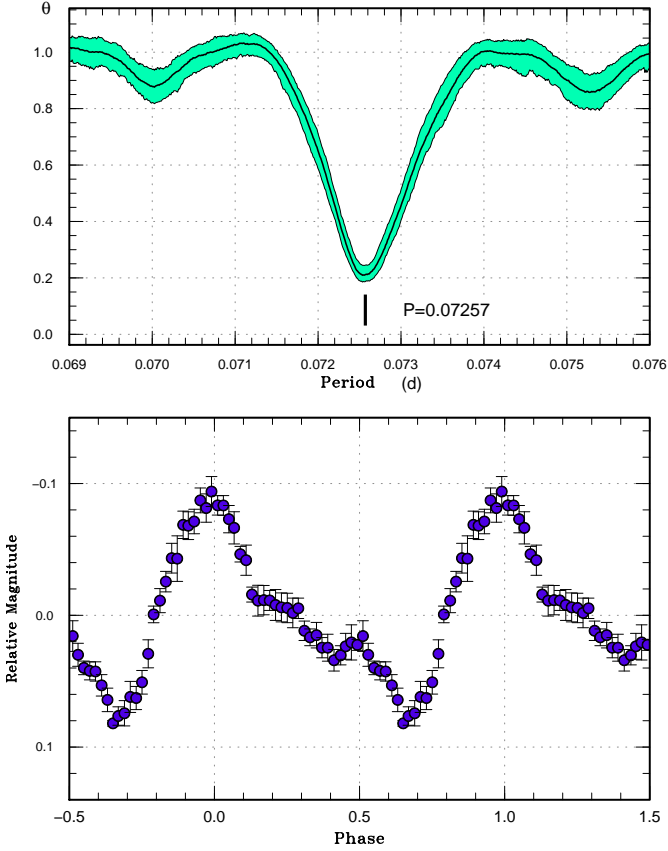}
  \end{center}
  \caption{Superhumps in ASASSN-13as (2013). (Upper): PDM analysis.
     (Lower): Phase-averaged profile.}
  \label{fig:asassn13asshpdm}
\end{figure}

\begin{table}
\caption{Superhump maxima of ASASSN-13as}\label{tab:asassn13asoc2013}
\begin{center}
\begin{tabular}{ccccc}
\hline
$E$ & max\commenta & error & $O-C$\commentb & $N$\commentc \\
\hline
0 & 56475.3829 & 0.0008 & $-$0.0022 & 40 \\
1 & 56475.4562 & 0.0005 & $-$0.0014 & 36 \\
2 & 56475.5293 & 0.0007 & $-$0.0009 & 28 \\
14 & 56476.4009 & 0.0006 & 0.0006 & 39 \\
15 & 56476.4733 & 0.0006 & 0.0004 & 34 \\
27 & 56477.3523 & 0.0065 & 0.0091 & 15 \\
28 & 56477.4163 & 0.0007 & 0.0006 & 36 \\
29 & 56477.4899 & 0.0009 & 0.0016 & 40 \\
41 & 56478.3502 & 0.0057 & $-$0.0083 & 14 \\
42 & 56478.4315 & 0.0028 & 0.0005 & 16 \\
\hline
  \multicolumn{5}{l}{\commenta BJD$-$2400000.} \\
  \multicolumn{5}{l}{\commentb Against max $= 2456475.3851 + 0.072522 E$.} \\
  \multicolumn{5}{l}{\commentc Number of points used to determine the maximum.} \\
\end{tabular}
\end{center}
\end{table}

\subsection{ASASSN-13ax}\label{obj:asassn13ax}

\begin{figure}
  \begin{center}
    \FigureFile(88mm,110mm){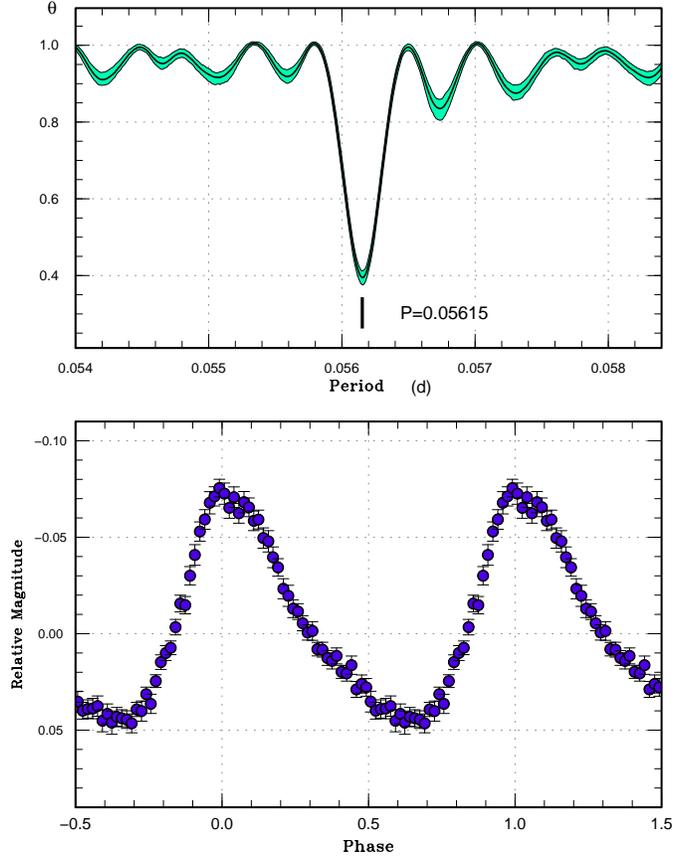}
  \end{center}
  \caption{Ordinary superhumps in ASASSN-13ax (2013).
     Stage A superhumps were excluded from the analysis. 
     (Upper): PDM analysis.
     (Lower): Phase-averaged profile.}
  \label{fig:asassn13axshpdm}
\end{figure}

\begin{figure}
  \begin{center}
    \FigureFile(88mm,70mm){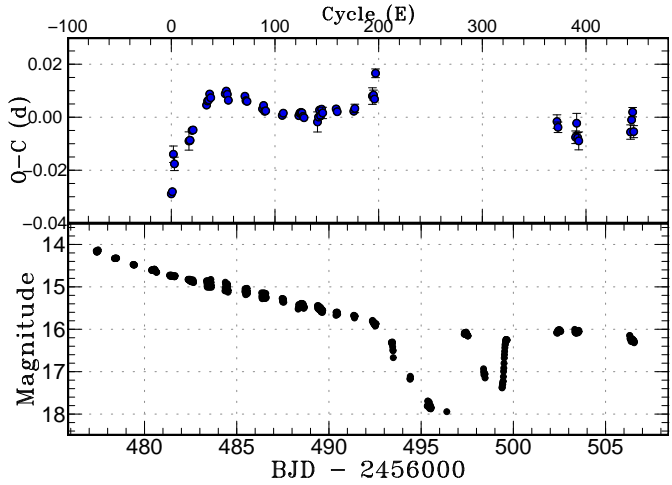}
  \end{center}
  \caption{$O-C$ diagram of superhumps in ASASSN-13ax (2013).
     (Upper): $O-C$ diagram.  A period of 0.056212~d
     was used to draw this figure.
     (Lower): Light curve.  The observations were binned to 0.011~d.}
  \label{fig:asassn13axhumpall}
\end{figure}

   This object was discovered by ASAS-SN survey on 2013
June 1 \citep{cop13asassn13axatel5195}.  The coordinates are
\timeform{18h 00m 05.8s}, \timeform{+52D 56' 33.7''}.
The SDSS quiescent counterpart has a magnitude of $g=21.2$,
making the object as a large-amplitude ($\sim$7.7 mag)
WZ Sge-type dwarf nova.  Subsequent observations detected
possible early superhumps (vsnet-alert 15944), whose period
was not convincingly determined due to the low amplitudes
and short observing runs.
Ordinary superhumps started to grow 7~d after the discovery
(vsnet-alert 15951, 15964, 15986, 15995, 16007, 16013,
16032; figure \ref{fig:asassn13axshpdm}).
The object then entered a temporary dip
19~d after the discovery (vsnet-alert 16034
(figure \ref{fig:asassn13axhumpall}).
The dip lasted for $\sim$3~d and the object then entered
the rebrightening phase.  The rebrightening started
with a precursor-like outburst (BJD 2456497, July 23,
vsnet-alert 16060),
which was followed by a shallow dip (BJD 2456499, July 25), 
then by a plateau-type rebrightening (vsnet-alert 16076,
16082).

\begin{figure}
  \begin{center}
    \FigureFile(88mm,110mm){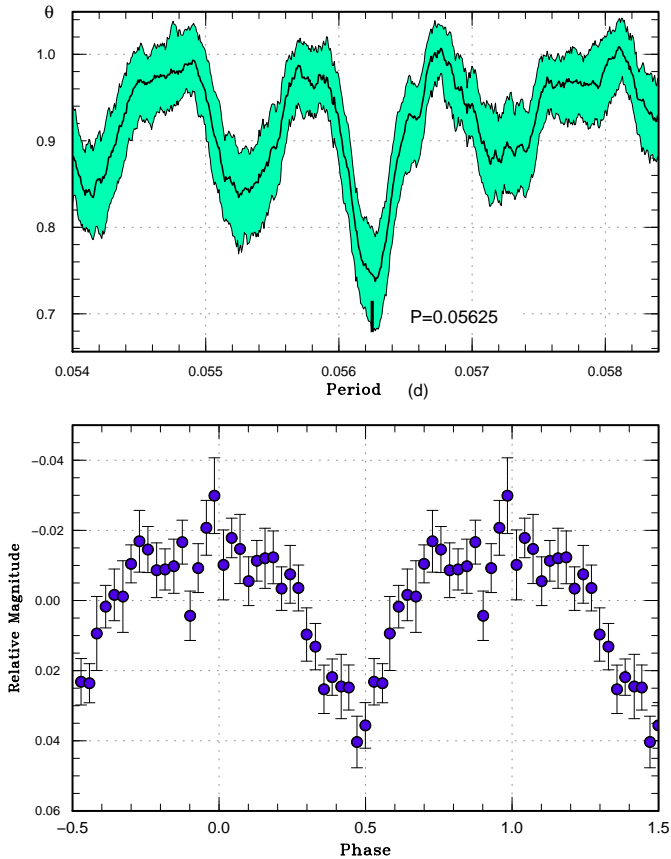}
  \end{center}
  \caption{Superhumps in ASASSN-13ax during the rebrightening
     phase (2013).
     (Upper): PDM analysis.
     (Lower): Phase-averaged profile.}
  \label{fig:asassn13axshrebpdm}
\end{figure}

   The times of (ordinary) superhump maxima are listed in
table \ref{tab:asassn13axoc2013}.  The times for $E \ge 372$
correspond to the superhumps recorded during the rebrightening
phase.  The cycle count may not be continuous with the earlier
part of this figure.
During the dip and the precursor-like outburst,
the superhump signal was weak and we could not measure
the times of maxima.  In the early part, clear 
stages A and B can be recognized
(figure \ref{fig:asassn13axhumpall})
As in most WZ Sge-type
dwarf novae, the object entered the rapid fading stage
without a transition to stage C.  The $P_{\rm dot}$ for
stage B superhumps was determined excluding $E=$88--91,
when an increase of the period was probably due to the reduction 
of the pressure effect \citep{nak13j2112j2037}.
The period of stage A superhump was determined by the PDM
method using the segment of BJD 2456481.40--2456483.63.
The mean superhump period during the rebrightening stage
was 0.05625(2)~d (PDM method), which showed a sharper
minimum than maximum, and the amplitude was low
(figure \ref{fig:asassn13axshrebpdm}).  Since the amplitudes
of the superhumps decrease around the dip and showed
a regrowth preceded by a precursor-like outburst,
the rebrightening bore some resemblance to
an ordinary superoutburst of an SU UMa-type dwarf nova,
rather than a long rebrightening consisting of
repetitive rebrightenings as in WZ Sge (2001)
(\cite{ish02wzsgeletter}; \cite{pat02wzsge}; \cite{Pdot}).
A similar rebrightening with a precursor-like outburst
was seen in AL Com (1995) \citep{nog97alcom}.

\begin{table}
\caption{Superhump maxima of ASASSN-13ax (2013)}\label{tab:asassn13axoc2013}
\begin{center}
\begin{tabular}{ccccc}
\hline
$E$ & max\commenta & error & $O-C$\commentb & $N$\commentc \\
\hline
0 & 56481.4244 & 0.0012 & $-$0.0289 & 91 \\
1 & 56481.4814 & 0.0012 & $-$0.0281 & 92 \\
2 & 56481.5518 & 0.0031 & $-$0.0140 & 78 \\
3 & 56481.6043 & 0.0025 & $-$0.0176 & 63 \\
17 & 56482.3999 & 0.0034 & $-$0.0090 & 77 \\
18 & 56482.4563 & 0.0007 & $-$0.0088 & 149 \\
18 & 56482.4564 & 0.0007 & $-$0.0087 & 148 \\
20 & 56482.5726 & 0.0006 & $-$0.0049 & 107 \\
21 & 56482.6289 & 0.0010 & $-$0.0048 & 64 \\
34 & 56483.3691 & 0.0007 & 0.0046 & 49 \\
35 & 56483.4268 & 0.0002 & 0.0061 & 117 \\
36 & 56483.4834 & 0.0002 & 0.0065 & 100 \\
37 & 56483.5419 & 0.0002 & 0.0088 & 98 \\
38 & 56483.5967 & 0.0003 & 0.0074 & 70 \\
52 & 56484.3852 & 0.0002 & 0.0089 & 29 \\
53 & 56484.4424 & 0.0002 & 0.0099 & 48 \\
54 & 56484.4974 & 0.0003 & 0.0087 & 30 \\
55 & 56484.5514 & 0.0012 & 0.0064 & 8 \\
71 & 56485.4523 & 0.0003 & 0.0080 & 50 \\
72 & 56485.5068 & 0.0003 & 0.0063 & 60 \\
73 & 56485.5627 & 0.0002 & 0.0060 & 59 \\
88 & 56486.4031 & 0.0003 & 0.0032 & 31 \\
89 & 56486.4606 & 0.0005 & 0.0045 & 71 \\
90 & 56486.5146 & 0.0003 & 0.0023 & 59 \\
91 & 56486.5710 & 0.0003 & 0.0024 & 55 \\
107 & 56487.4687 & 0.0005 & 0.0007 & 39 \\
108 & 56487.5257 & 0.0004 & 0.0016 & 57 \\
123 & 56488.3680 & 0.0007 & 0.0006 & 51 \\
124 & 56488.4251 & 0.0004 & 0.0016 & 104 \\
125 & 56488.4816 & 0.0005 & 0.0019 & 133 \\
126 & 56488.5377 & 0.0004 & 0.0017 & 109 \\
127 & 56488.5925 & 0.0005 & 0.0003 & 80 \\
128 & 56488.6482 & 0.0009 & $-$0.0001 & 47 \\
141 & 56489.3774 & 0.0038 & $-$0.0017 & 27 \\
142 & 56489.4355 & 0.0004 & 0.0001 & 173 \\
143 & 56489.4943 & 0.0005 & 0.0028 & 148 \\
144 & 56489.5483 & 0.0004 & 0.0006 & 136 \\
145 & 56489.6071 & 0.0007 & 0.0031 & 52 \\
146 & 56489.6619 & 0.0021 & 0.0017 & 32 \\
159 & 56490.3941 & 0.0010 & 0.0032 & 47 \\
160 & 56490.4493 & 0.0006 & 0.0021 & 54 \\
176 & 56491.3489 & 0.0012 & 0.0023 & 24 \\
177 & 56491.4061 & 0.0016 & 0.0034 & 19 \\
194 & 56492.3664 & 0.0031 & 0.0080 & 25 \\
195 & 56492.4230 & 0.0019 & 0.0084 & 30 \\
196 & 56492.4777 & 0.0013 & 0.0069 & 28 \\
197 & 56492.5436 & 0.0016 & 0.0166 & 25 \\
372 & 56502.3625 & 0.0025 & $-$0.0015 & 29 \\
373 & 56502.4166 & 0.0020 & $-$0.0037 & 29 \\
390 & 56503.3684 & 0.0024 & $-$0.0075 & 28 \\
391 & 56503.4299 & 0.0037 & $-$0.0022 & 30 \\
392 & 56503.4807 & 0.0020 & $-$0.0076 & 29 \\
393 & 56503.5357 & 0.0034 & $-$0.0088 & 29 \\
443 & 56506.3496 & 0.0027 & $-$0.0055 & 28 \\
444 & 56506.4104 & 0.0013 & $-$0.0009 & 31 \\
445 & 56506.4696 & 0.0017 & 0.0021 & 25 \\
446 & 56506.5184 & 0.0023 & $-$0.0053 & 30 \\
\hline
  \multicolumn{5}{l}{\commenta BJD$-$2400000.} \\
  \multicolumn{5}{l}{\commentb Against max $= 2456481.4533 + 0.056212 E$.} \\
  \multicolumn{5}{l}{\commentc Number of points used to determine the maximum.} \\
\end{tabular}
\end{center}
\end{table}

\subsection{ASASSN-13bj}\label{obj:asassn13bj}

\begin{figure}
  \begin{center}
    \FigureFile(88mm,110mm){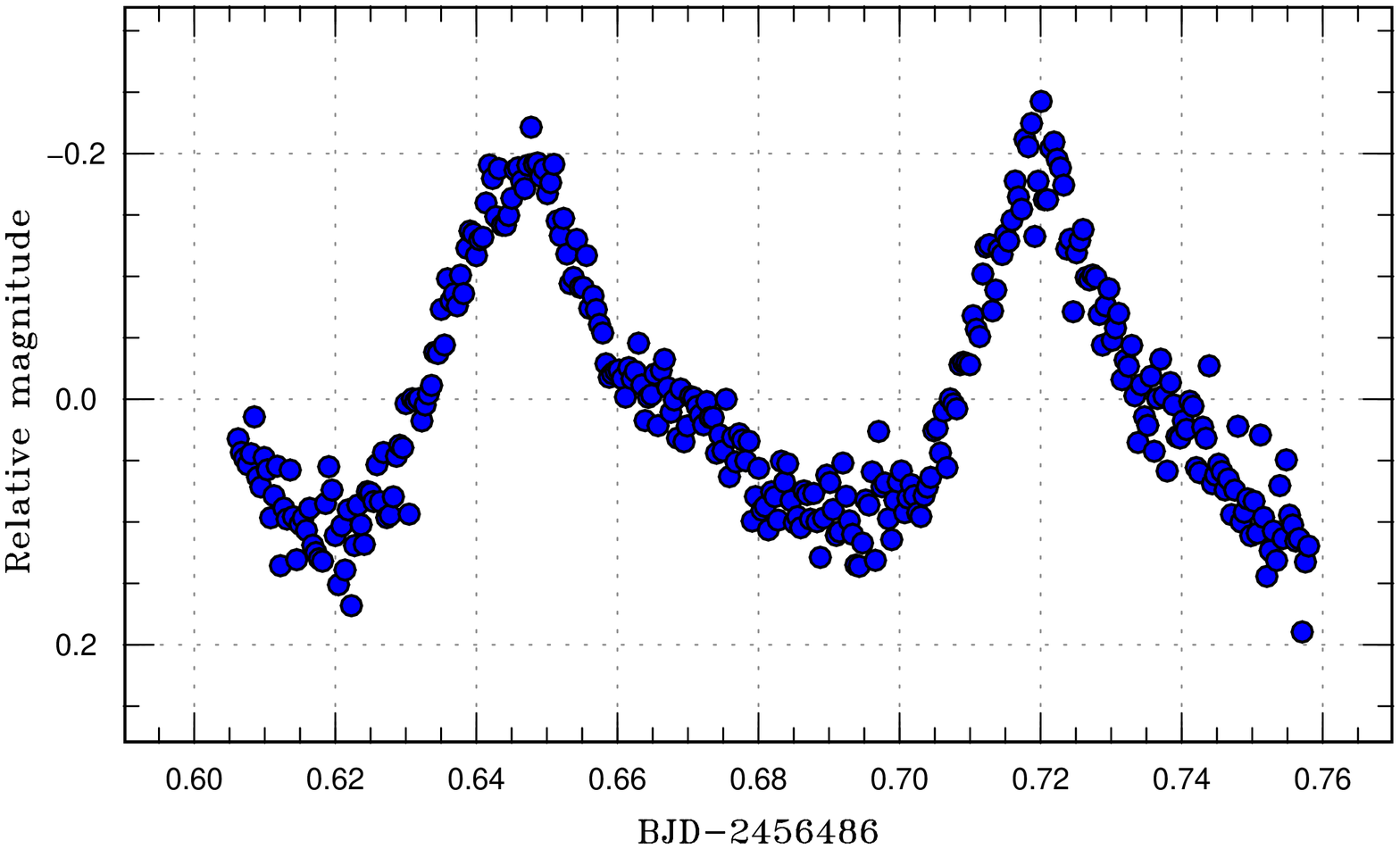}
  \end{center}
  \caption{Superhumps in ASASSN-13bj (2013).}
  \label{fig:asassn13bj}
\end{figure}

   This object was discovered by ASAS-SN survey on 2013
July 10.  The coordinates are
\timeform{16h 00m 20.39s}, \timeform{+70D 50' 09.4''}.
The object has an ROSAT counterpart (1RXS J160017.2$+$705029,
vsnet-alert 15960).  Only single-night observation
was available, which recorded two superhump maxima
(figure \ref{fig:asassn13bj}):
BJD 2456486.6477(2) $N=128$ and 2456486.7212(3) $N=128$.
The period by the PDM method is 0.0731(3)~d.

\subsection{ASASSN-13bm}\label{obj:asassn13bm}

\begin{figure}
  \begin{center}
    \FigureFile(88mm,110mm){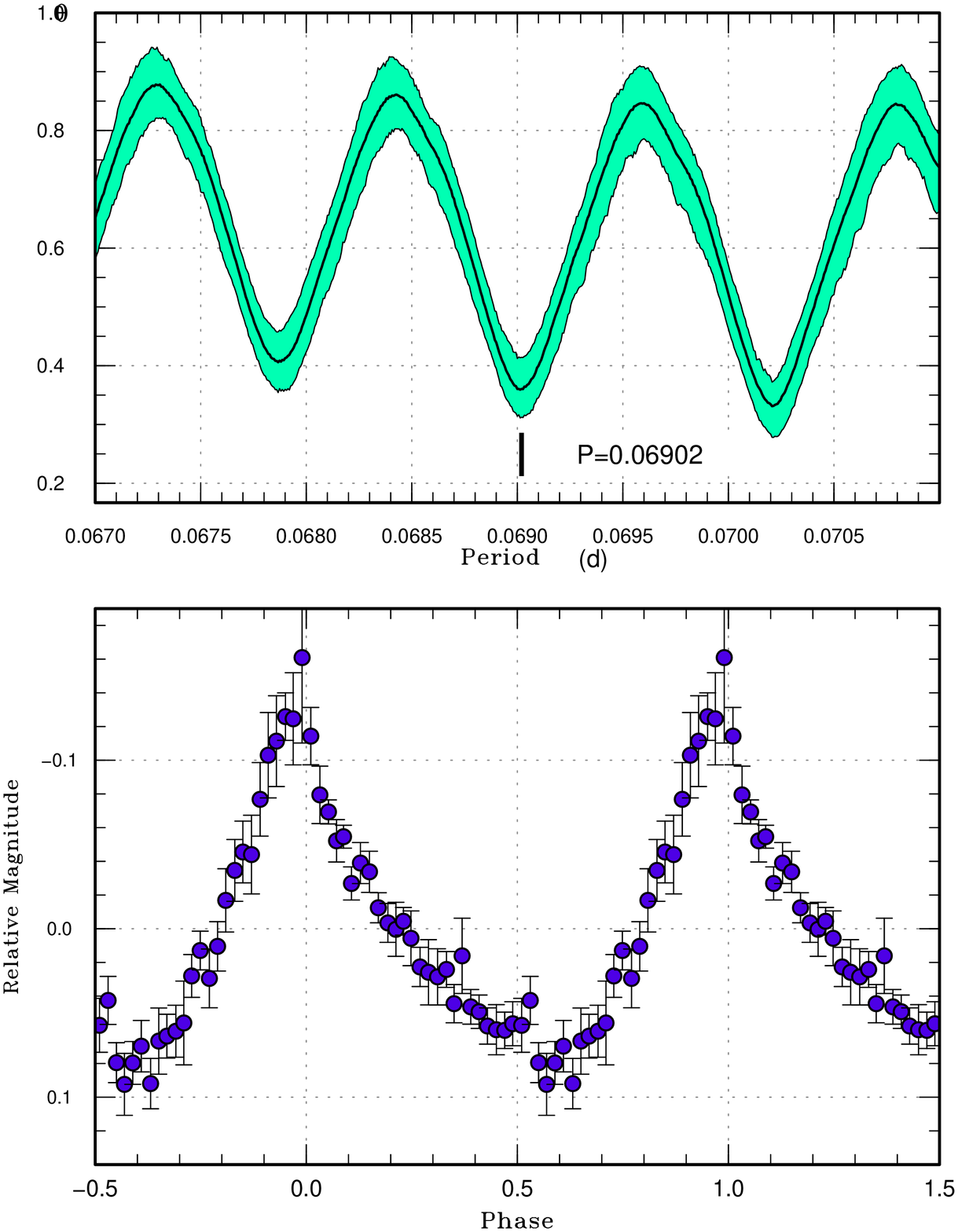}
  \end{center}
  \caption{Superhumps in ASASSN-13bm (2013). (Upper): PDM analysis.
     The alias was selected based on the continuous run on BJD 2456489.
     (Lower): Phase-averaged profile.}
  \label{fig:asassn13bmshpdm}
\end{figure}

   This object was discovered by ASAS-SN survey on 2013
July 12.  The coordinates are
\timeform{19h 25m 14.46s}, \timeform{+68D 51' 22.1''}.
Subsequent observations recorded superhumps
(vsnet-alert 16021; figure \ref{fig:asassn13bmshpdm}).
The times of superhump maxima are listed in table
\ref{tab:asassn13bmoc2013}.  The stage of the superhumps
is not known.

\begin{table}
\caption{Superhump maxima of ASASSN-13bm (2013)}\label{tab:asassn13bmoc2013}
\begin{center}
\begin{tabular}{ccccc}
\hline
$E$ & max\commenta & error & $O-C$\commentb & $N$\commentc \\
\hline
0 & 56489.3507 & 0.0031 & 0.0023 & 33 \\
1 & 56489.4160 & 0.0004 & $-$0.0015 & 73 \\
2 & 56489.4864 & 0.0010 & $-$0.0001 & 62 \\
3 & 56489.5547 & 0.0005 & $-$0.0008 & 60 \\
4 & 56489.6246 & 0.0023 & 0.0001 & 34 \\
61 & 56493.5584 & 0.0006 & 0.0001 & 74 \\
\hline
  \multicolumn{5}{l}{\commenta BJD$-$2400000.} \\
  \multicolumn{5}{l}{\commentb Against max $= 2456489.3484 + 0.069015 E$.} \\
  \multicolumn{5}{l}{\commentc Number of points used to determine the maximum.} \\
\end{tabular}
\end{center}
\end{table}

\subsection{ASASSN-13bp}\label{obj:asassn13bp}

\begin{figure}
  \begin{center}
    \FigureFile(88mm,110mm){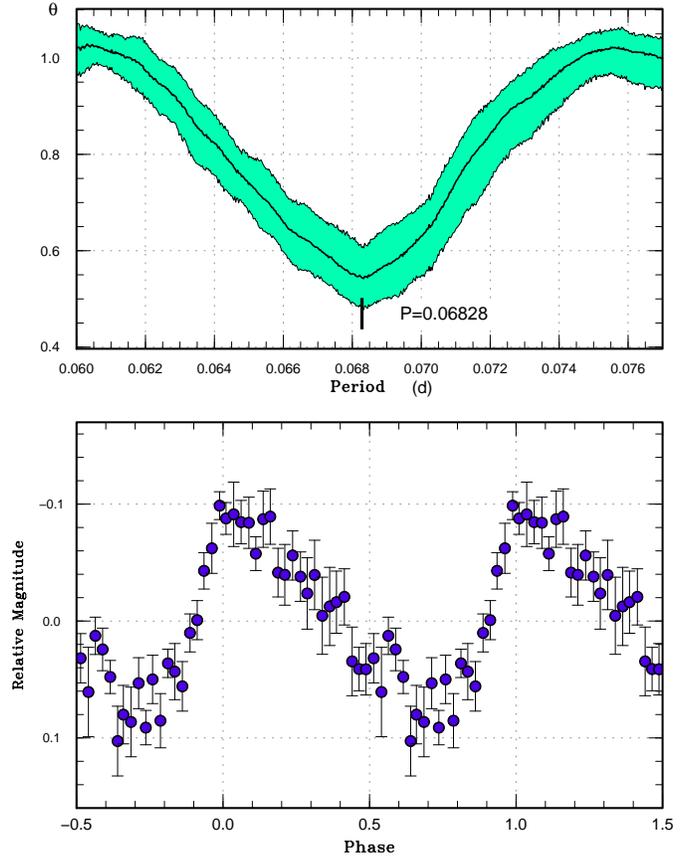}
  \end{center}
  \caption{Superhumps in ASASSN-13bp (2013). (Upper): PDM analysis.
     (Lower): Phase-averaged profile.}
  \label{fig:asassn13bpshpdm}
\end{figure}

   This object was discovered by ASAS-SN survey on 2013
July 15.  The coordinates are
\timeform{22h 53m 50.51s}, \timeform{+33D 30' 32.5''}.
The object showed three outbursts in the CRTS data
on BJD 2454299, 2454358 and 2454382.  The SDSS colors
suggested an short orbital period based on the method
of \citet{kat12DNSDSS} (vsnet-alert 15997).
Subsequent observations confirmed the presence of
superhumps (vsnet-alert 16018, 16019; 
figure \ref{fig:asassn13bpshpdm}).
The times of superhump maxima are listed in table
\ref{tab:asassn13bpoc2013}.  The superhump period by the PDM
method was 0.06828(16)~d.

\begin{table}
\caption{Superhump maxima of ASASSN-13bp (2013)}\label{tab:asassn13bpoc2013}
\begin{center}
\begin{tabular}{ccccc}
\hline
$E$ & max\commenta & error & $O-C$\commentb & $N$\commentc \\
\hline
0 & 56492.1678 & 0.0016 & 0.0016 & 61 \\
1 & 56492.2318 & 0.0025 & $-$0.0022 & 29 \\
4 & 56492.4382 & 0.0008 & 0.0008 & 74 \\
5 & 56492.5050 & 0.0007 & $-$0.0002 & 74 \\
\hline
  \multicolumn{5}{l}{\commenta BJD$-$2400000.} \\
  \multicolumn{5}{l}{\commentb Against max $= 2456492.1662 + 0.067805 E$.} \\
  \multicolumn{5}{l}{\commentc Number of points used to determine the maximum.} \\
\end{tabular}
\end{center}
\end{table}

\subsection{ASASSN-13br}\label{obj:asassn13br}

   This object was discovered by ASAS-SN survey on 2013
July 19.  The coordinates are
\timeform{17h 03m 43.01s}, \timeform{+66D 07' 45.5''}.
Although there is a $g=17.78$ SDSS star close to this location,
a close inspection of the SDSS image and the analysis of
the coordinates indicated that the true quiescent counterpart
was a much fainter ($^sim$21.5 mag) close companion to this
SDSS object.  The actual outburst amplitude was estimated
to be $\sim$7 mag (vsnet-alert 16026).

\begin{figure}
  \begin{center}
    \FigureFile(88mm,110mm){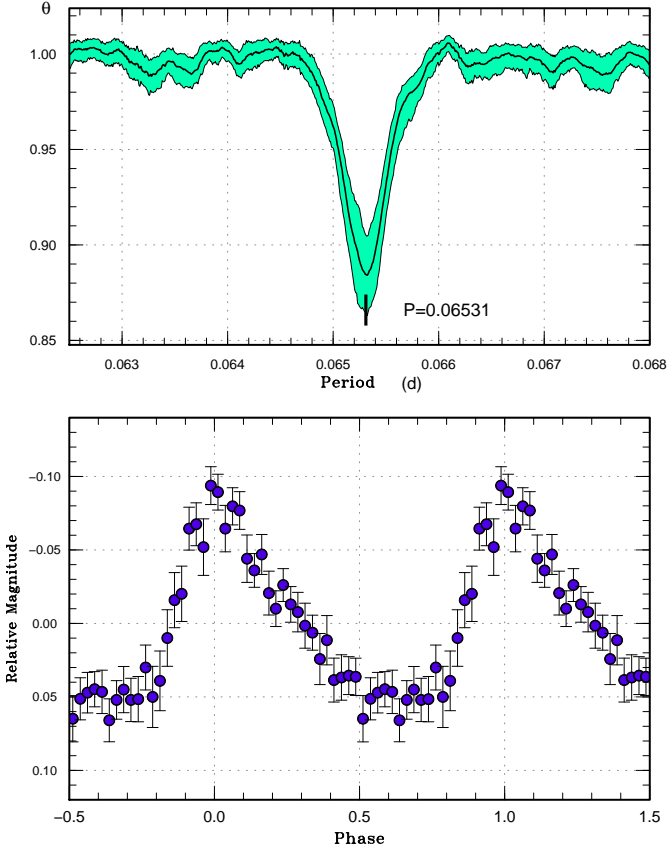}
  \end{center}
  \caption{Superhumps in ASASSN-13br (2013). (Upper): PDM analysis.
     (Lower): Phase-averaged profile.}
  \label{fig:asassn13brshpdm}
\end{figure}

   Subsequent observations detected superhumps (vsnet-alert
16037, 16045, 16049, 16051, 16055, 16071, 16077, 16083,
16100; figure \ref{fig:asassn13brshpdm}).
The object entered the rapid fading stage 12~d
after the outburst detection (vsnet-alert 16105).
The object is more likely an ordinary SU UMa-type dwarf nova
rather than an extreme WZ Sge-type dwarf nova.

   The times of superhump maxima are listed in table
\ref{tab:asassn13broc2013}.  Stages B and C are clearly seen,
and stage B had a clearly positive $P_{\rm dot}$ of 
$+9.6(1.6) \times 10^{-5}$.
The epoch $E=153$ corresponds to a post-superoutburst
superhump, and we disregarded it in estimating the period
of stage C superhumps.

\begin{table}
\caption{Superhump maxima of ASASSN-13br (2013)}\label{tab:asassn13broc2013}
\begin{center}
\begin{tabular}{ccccc}
\hline
$E$ & max\commenta & error & $O-C$\commentb & $N$\commentc \\
\hline
0 & 56495.3688 & 0.0005 & $-$0.0032 & 78 \\
1 & 56495.4368 & 0.0003 & $-$0.0003 & 106 \\
2 & 56495.5014 & 0.0004 & $-$0.0010 & 103 \\
3 & 56495.5666 & 0.0007 & $-$0.0010 & 71 \\
16 & 56496.4139 & 0.0005 & $-$0.0015 & 48 \\
17 & 56496.4780 & 0.0003 & $-$0.0025 & 63 \\
18 & 56496.5434 & 0.0005 & $-$0.0024 & 58 \\
30 & 56497.3239 & 0.0012 & $-$0.0044 & 19 \\
31 & 56497.3912 & 0.0007 & $-$0.0023 & 35 \\
32 & 56497.4556 & 0.0004 & $-$0.0031 & 34 \\
33 & 56497.5205 & 0.0007 & $-$0.0034 & 113 \\
34 & 56497.5865 & 0.0045 & $-$0.0026 & 24 \\
36 & 56497.7215 & 0.0005 & 0.0020 & 49 \\
37 & 56497.7834 & 0.0005 & $-$0.0013 & 50 \\
38 & 56497.8496 & 0.0004 & $-$0.0003 & 49 \\
46 & 56498.3716 & 0.0017 & 0.0000 & 43 \\
47 & 56498.4345 & 0.0008 & $-$0.0023 & 102 \\
48 & 56498.5011 & 0.0008 & $-$0.0010 & 141 \\
49 & 56498.5662 & 0.0014 & $-$0.0011 & 84 \\
62 & 56499.4166 & 0.0011 & 0.0016 & 105 \\
63 & 56499.4856 & 0.0007 & 0.0054 & 153 \\
64 & 56499.5503 & 0.0011 & 0.0049 & 173 \\
65 & 56499.6187 & 0.0077 & 0.0081 & 29 \\
76 & 56500.3353 & 0.0005 & 0.0074 & 30 \\
77 & 56500.4008 & 0.0006 & 0.0077 & 35 \\
78 & 56500.4654 & 0.0007 & 0.0070 & 34 \\
79 & 56500.5307 & 0.0009 & 0.0071 & 33 \\
107 & 56502.3498 & 0.0011 & 0.0004 & 25 \\
108 & 56502.4168 & 0.0009 & 0.0021 & 34 \\
109 & 56502.4824 & 0.0011 & 0.0025 & 25 \\
110 & 56502.5478 & 0.0018 & 0.0027 & 27 \\
122 & 56503.3224 & 0.0020 & $-$0.0052 & 18 \\
123 & 56503.3908 & 0.0007 & $-$0.0020 & 34 \\
124 & 56503.4598 & 0.0014 & 0.0018 & 32 \\
125 & 56503.5192 & 0.0013 & $-$0.0040 & 33 \\
153 & 56505.3335 & 0.0040 & $-$0.0156 & 11 \\
\hline
  \multicolumn{5}{l}{\commenta BJD$-$2400000.} \\
  \multicolumn{5}{l}{\commentb Against max $= 2456495.3720 + 0.065210 E$.} \\
  \multicolumn{5}{l}{\commentc Number of points used to determine the maximum.} \\
\end{tabular}
\end{center}
\end{table}

\subsection{CSS J015051.7$+$332621}\label{obj:j015051}

\begin{figure}
  \begin{center}
    \FigureFile(88mm,110mm){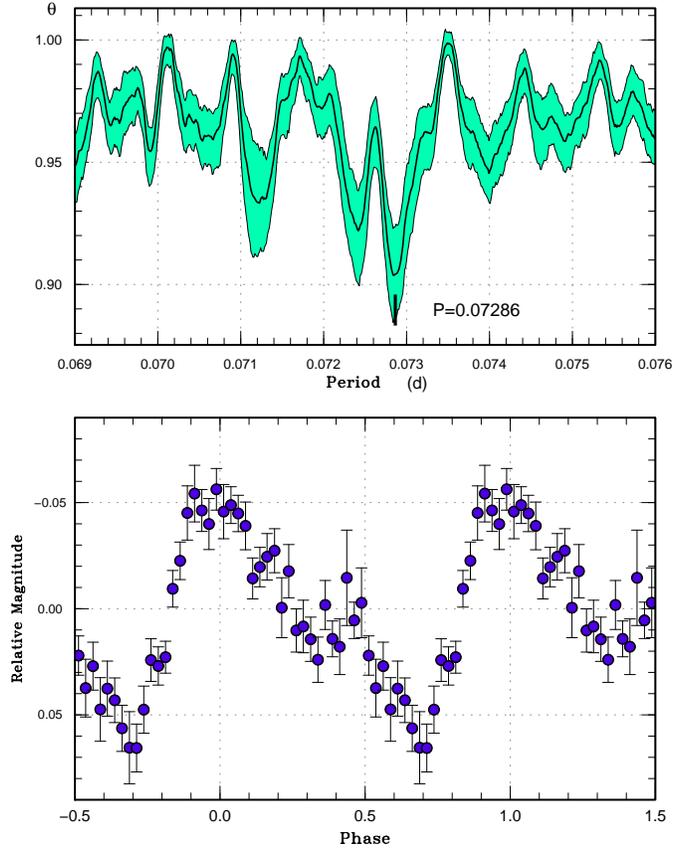}
  \end{center}
  \caption{Superhumps in CSS J015051 (2012). (Upper): PDM analysis.
     (Lower): Phase-averaged profile.}
  \label{fig:j015051shpdm}
\end{figure}

   This object was detected as a transient by CRTS
(=CSS111006:015052$+$332622, hereafter CSS J015051) on
2011 October 6.  The quiescent SDSS color suggested an orbital
period of 0.069~d \citep{kat12DNSDSS}.  Another outburst was
detected by the MASTER network \citep{MASTER}
reaching 14.5 mag on 2012 October 10
(\cite{den12j0150atel4475}; vsnet-alert 14993).
Subsequent observations detected superhumps
(vsnet-alert 15001, 15011, 15026; figure \ref{fig:j015051shpdm}).
The times of superhump maxima are listed in table
\ref{tab:j015051oc2012}.  Although the coverage was not
sufficient, stages B and C can be recognized (the nature of
$E=144$ hump is unclear, and was disregarded in analysis).

\begin{table}
\caption{Superhump maxima of CSS J015051 (2012)}\label{tab:j015051oc2012}
\begin{center}
\begin{tabular}{ccccc}
\hline
$E$ & max\commenta & error & $O-C$\commentb & $N$\commentc \\
\hline
0 & 56211.6952 & 0.0018 & $-$0.0104 & 58 \\
19 & 56213.0713 & 0.0023 & $-$0.0094 & 52 \\
20 & 56213.1487 & 0.0004 & $-$0.0043 & 118 \\
59 & 56215.9835 & 0.0008 & 0.0078 & 127 \\
60 & 56216.0583 & 0.0006 & 0.0103 & 297 \\
61 & 56216.1284 & 0.0006 & 0.0080 & 275 \\
74 & 56217.0727 & 0.0014 & 0.0114 & 141 \\
116 & 56220.1091 & 0.0041 & 0.0080 & 293 \\
144 & 56222.1062 & 0.0067 & $-$0.0213 & 150 \\
\hline
  \multicolumn{5}{l}{\commenta BJD$-$2400000.} \\
  \multicolumn{5}{l}{\commentb Against max $= 2456211.7056 + 0.072374 E$.} \\
  \multicolumn{5}{l}{\commentc Number of points used to determine the maximum.} \\
\end{tabular}
\end{center}
\end{table}

\subsection{CSS J015321.3$+$340855}\label{obj:j015321}

   This object was discovered by CRTS (=CSS081026:015321$+$340857,
hereafter CSS J015321) on 2008 October 26.  Seven outbursts
were recorded in the CRTS data between 2004 November and
2012 January.  Another outburst was detected by MASTER network
(vsnet-alert 15178) on 2012 December 17.  This is a bright
GALEX UV source and \citet{era11sbs0150} also noted outbursts
between 1972 and 1975 (the object is also known as SBS 0150$+$339)
(vsnet-alert 15178).  Subsequent observations detected superhumps
(vsnet-alert 15180, 15182, 15185, 15187), qualifying
this object as an SU UMa-type dwarf nova in the period gap.
The observation recorded the middle-to-late stage of
the superoutburst.  There was a possible stage B-C transition
between $E=23$ and $E=31$.  Although a large positive $P_{\rm dot}$
was obtained, this value is not very reliable due to the short
observation segment.

\begin{figure}
  \begin{center}
    \FigureFile(88mm,110mm){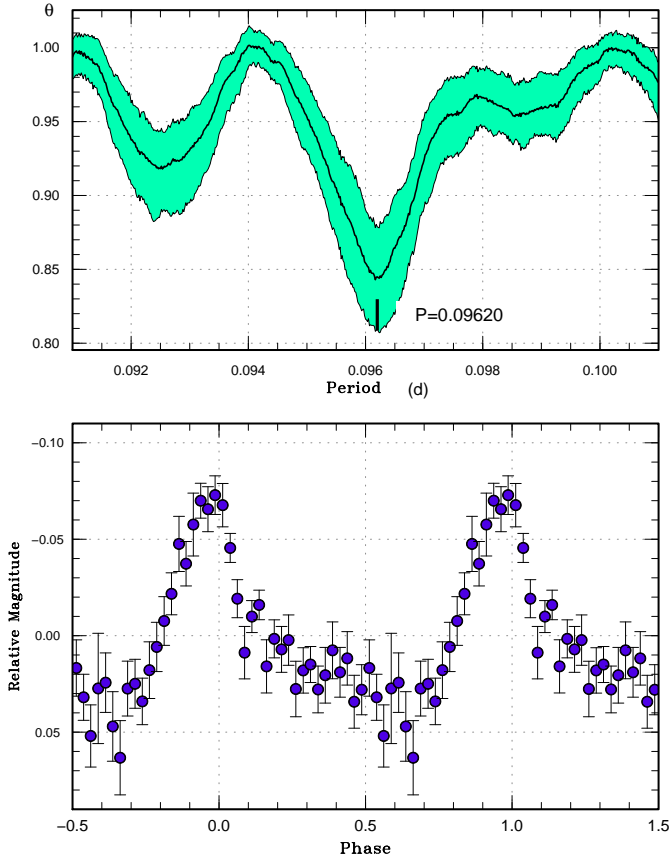}
  \end{center}
  \caption{Superhumps in CSS J015321 (2012). (Upper): PDM analysis.
     (Lower): Phase-averaged profile.}
  \label{fig:j015321shpdm}
\end{figure}

\begin{table}
\caption{Superhump maxima of CSS J015321 (2012)}\label{tab:j015321oc2012}
\begin{center}
\begin{tabular}{ccccc}
\hline
$E$ & max\commenta & error & $O-C$\commentb & $N$\commentc \\
\hline
0 & 56279.5922 & 0.0007 & 0.0013 & 85 \\
1 & 56279.6862 & 0.0008 & $-$0.0012 & 101 \\
2 & 56279.7822 & 0.0010 & $-$0.0015 & 101 \\
7 & 56280.2656 & 0.0011 & $-$0.0003 & 56 \\
8 & 56280.3597 & 0.0020 & $-$0.0027 & 76 \\
17 & 56281.2301 & 0.0016 & $-$0.0001 & 89 \\
18 & 56281.3274 & 0.0015 & 0.0007 & 100 \\
22 & 56281.7200 & 0.0021 & 0.0077 & 100 \\
23 & 56281.8125 & 0.0034 & 0.0038 & 75 \\
31 & 56282.5725 & 0.0066 & $-$0.0077 & 97 \\
\hline
  \multicolumn{5}{l}{\commenta BJD$-$2400000.} \\
  \multicolumn{5}{l}{\commentb Against max $= 2456279.5909 + 0.0964283 E$.} \\
  \multicolumn{5}{l}{\commentc Number of points used to determine the maximum.} \\
\end{tabular}
\end{center}
\end{table}

\subsection{CSS J102842.8$-$081930}\label{obj:j102842}

   This object (=OT J102842.9$-$081927 = CSS090331:102843$-$081927,
hereafter CSS J102842), detected by the CRTS, belongs to
the small group of hydrogen-rich dwarf novae below the period gap 
(see subsection \ref{obj:v485cen}).  \citet{Pdot} reported
superhumps during the 2009 superoutburst.  The 2012 superoutburst
was also partially observed \citet{Pdot4}.

\subsubsection{Photometry}

\begin{figure}
  \begin{center}
    \FigureFile(88mm,70mm){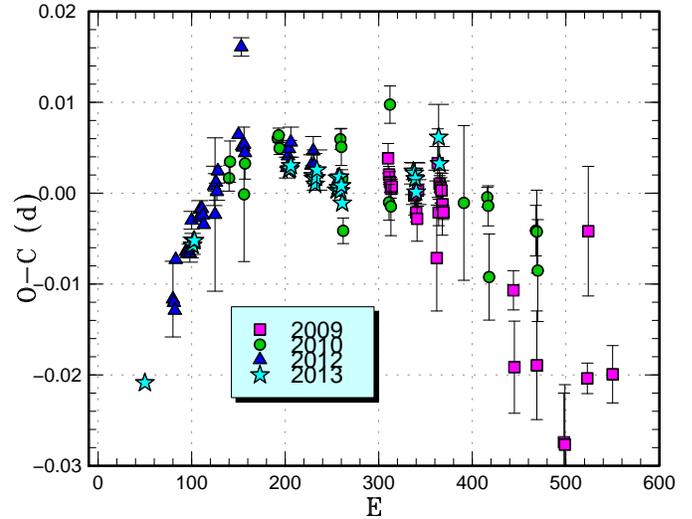}
  \end{center}
  \caption{Comparison of $O-C$ diagrams of CSS J102842 between different
  superoutbursts.  A period of 0.03819~d was used to draw this figure.
  Approximate cycle counts ($E$) after the start of the superoutburst
  were used.  The assumption to draw this figure is very different
  from that in \citet{Pdot4}.  Since all detections were made by CRTS,
  there was unavoidable uncertainty when the outburst started.
  The 2013 outburst, however, was apparently caught in the earlier
  phase than in other outbursts as judged from the brightness and
  long duration of observed stage A.  We therefore assumed
  that the 2013 outburst started ($E=0$) at the time of the CRTS
  detection.  Since the 2012 $O-C$ diagram recorded the similar stage
  of the outburst we could match it to the 2013 one by shifting
  80 cycles.  The starts of other outbursts were more uncertain,
  and we assumed a shift of 310 and 140 cycles for the 2009 and 2010
  outburst, respectively.
  }
  \label{fig:j1028comp2}
\end{figure}

   We observed the 2013 superoutburst.  Both stages A and B were recorded
(table \ref{tab:j102842oc2013}).  The periods of stage A and B
superhumps were similar to those obtained in 2012 \citep{Pdot4}.
A comparison of $O-C$ diagrams between different
superoutbursts is shown in figure \ref{fig:j1028comp2}.
The assumption to draw this figure is very different
from that in \citet{Pdot4}.  Since all detections were made by CRTS,
there was unavoidable uncertainty when the outburst started.
The 2013 outburst, however, was apparently caught in the earlier
phase than in other outbursts as judged from the brightness and
long duration of observed stage A superhumps.  We therefore assumed
that the 2013 outburst started ($E=0$) at the time of the CRTS
detection.  Since the 2012 $O-C$ diagram recorded the similar stage
of the outburst we could match it to the 2013 one by shifting
80 cycles.  The starts of other outbursts were more uncertain,
and we assumed a shift of 310 and 140 cycles for the 2009 and 2010
outburst, respectively.  If this interpretation is correct,
we observed late stage B and stage C superhumps in 2009.

   \citet{wou12SDSSCRTSCVs} reported a possible photometric
period of 52.1(6) min, which they considered to be the orbital
period.  If this is the true orbital period, the $\varepsilon^*$ 
values for stage A superhumps are 6.0\% and 5.9\% for 2013 and 2012
data, respectively.  These values give $q$=0.183--0.178,
and the secondary in CSS J102842 must be very massive for
this orbital period.  The orbital period should be confirmed by 
spectroscopy.

\begin{table}
\caption{Superhump maxima of CSS J102842 (2013)}\label{tab:j102842oc2013}
\begin{center}
\begin{tabular}{ccccc}
\hline
$E$ & max\commenta & error & $O-C$\commentb & $N$\commentc \\
\hline
0 & 56388.3958 & 0.0002 & $-$0.0122 & 30 \\
51 & 56390.3586 & 0.0003 & 0.0006 & 87 \\
52 & 56390.3973 & 0.0004 & 0.0011 & 88 \\
53 & 56390.4355 & 0.0004 & 0.0010 & 86 \\
154 & 56394.3005 & 0.0005 & 0.0042 & 72 \\
155 & 56394.3388 & 0.0003 & 0.0043 & 88 \\
156 & 56394.3773 & 0.0003 & 0.0046 & 88 \\
180 & 56395.2928 & 0.0004 & 0.0024 & 88 \\
181 & 56395.3306 & 0.0004 & 0.0020 & 88 \\
182 & 56395.3682 & 0.0004 & 0.0014 & 88 \\
183 & 56395.4071 & 0.0004 & 0.0021 & 81 \\
184 & 56395.4462 & 0.0022 & 0.0029 & 22 \\
205 & 56396.2459 & 0.0004 & $-$0.0003 & 88 \\
206 & 56396.2855 & 0.0006 & 0.0010 & 88 \\
207 & 56396.3226 & 0.0006 & $-$0.0001 & 95 \\
208 & 56396.3620 & 0.0004 & 0.0010 & 125 \\
209 & 56396.3986 & 0.0005 & $-$0.0006 & 109 \\
210 & 56396.4374 & 0.0007 & $-$0.0001 & 86 \\
211 & 56396.4736 & 0.0009 & $-$0.0020 & 24 \\
287 & 56399.3796 & 0.0010 & $-$0.0020 & 35 \\
288 & 56399.4176 & 0.0012 & $-$0.0022 & 37 \\
289 & 56399.4552 & 0.0017 & $-$0.0029 & 37 \\
290 & 56399.4919 & 0.0014 & $-$0.0044 & 36 \\
314 & 56400.4145 & 0.0036 & 0.0005 & 29 \\
315 & 56400.4497 & 0.0019 & $-$0.0024 & 26 \\
\hline
  \multicolumn{5}{l}{\commenta BJD$-$2400000.} \\
  \multicolumn{5}{l}{\commentb Against max $= 2456388.4080 + 0.038235 E$.} \\
  \multicolumn{5}{l}{\commentc Number of points used to determine the maximum.} \\
\end{tabular}
\end{center}
\end{table}

\subsubsection{Spectroscopy}

   We reported that ``a spectroscopic observation clarified its 
hydrogen-rich nature (vsnet-alert 11166), suggesting that 
the object is similar to V485 Cen and EI Psc'' in \citet{Pdot}.
Since no spectroscopic observation has been reported yet,
we present here our spectrum taken in 2009.

\begin{figure}
  \begin{center}
    \FigureFile(88mm,110mm){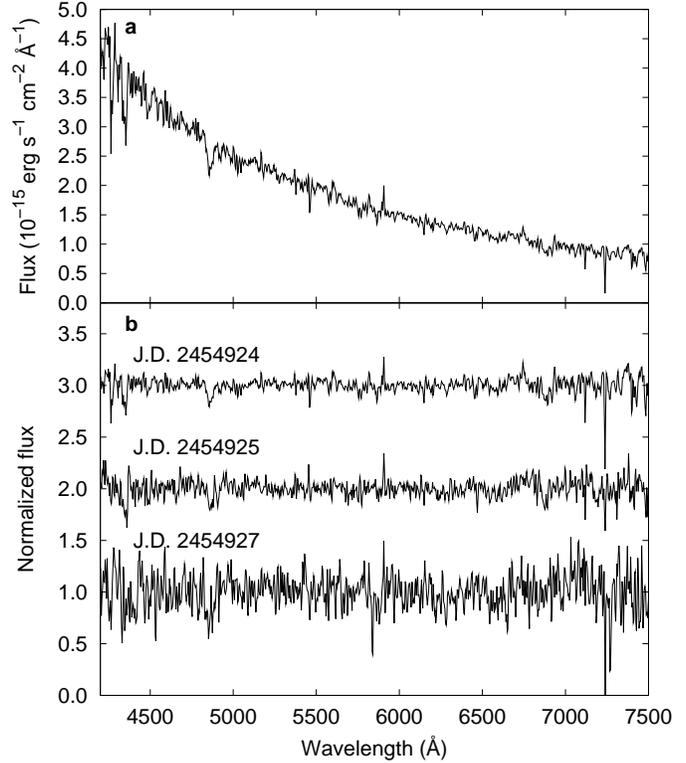}
  \end{center}
  \caption{(Upper:) Optical spectrum of CSS J102842 obtained on
  BJD 2454924.
  (Lower:) Nightly averaged normalized spectra of CSS J102842.}
  \label{fig:j1028spec}
\end{figure}

   We obtained low-resolution optical spectra of the object with
the GLOWS spectrograph attached to the 1.5-m telescope at Gunma 
Astronomical Observatory on 2009 April 2 (35 frames of 180s exposures, 
BJD 2454924.040--.166), April 3 (20 frames, BJD 2454925.090--163)
and April 5 (22 frames, BJD 2454927.067--.147). 
Standard IRAF routines were used for data 
reduction and a flux calibration was performed by using HR 4963.
The optical spectrum of the object on BJD 2454924 is displayed in
the upper panel of figure \ref{fig:j1028spec} and the nightly spectra, 
normalized by the continuum, are presented in the lower panel of
figure \ref{fig:j1028spec}.
These spectra show a smooth blue continuum and the H$\beta$ 
line in absorption.  The equivalent width (EW) of H$\beta$ absorption
line is 12\AA (measured from the summed spectrum).
The H$\alpha$ absorption is much weaker (4\AA),
suggesting that the line has an emission core.
The He\textsc{I} absorption lines are weak in the spectra
(He\textsc{I} 5876 has an EW of 3\AA).
These features indicate that the object is a hydrogen-rich
dwarf nova in outburst, and the weakness of the helium lines
makes a clear contrast against SBS 1108$+$574
(\cite{car13sbs1108}; \cite{lit13sbs1108}) and
CSS J174033.5$+$414756 (\cite{pri13j1740asassn13adatel4999}; 
Ohshima et al. in prep.), which are dwarf novae having
orbital periods below the period gap.  The He\textsc{I} 5876/H$\beta$
ratio is 0.2--0.3, which places the object in a range of
hydrogen-rich CVs within the error of the observation
(see figure 4 of \cite{lit13sbs1108}).

\subsection{CSS J105835.1$+$054703}\label{obj:j105835}

\begin{figure}
  \begin{center}
    \FigureFile(88mm,110mm){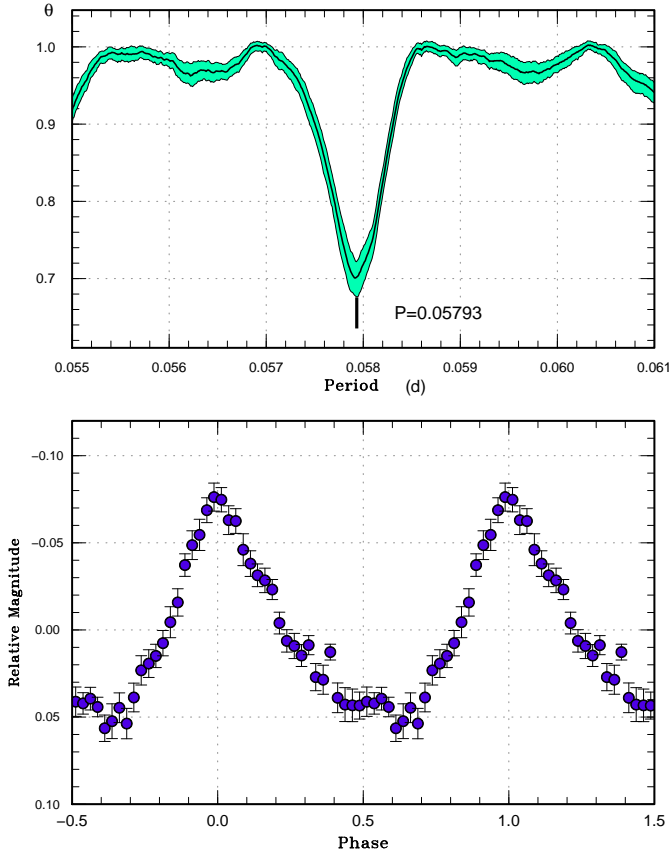}
  \end{center}
  \caption{Superhumps in CSS J105835 (2012). (Upper): PDM analysis.
     (Lower): Phase-averaged profile.}
  \label{fig:j105835shpdm}
\end{figure}

   This object was discovered by CRTS (=CSS081025:105835$+$054706,
hereafter CSS J105835) on 2008 October 25.  There were three
known outburst in 2007 December (15.0 mag), 2008 October (15.8 mag)
and 2012 November--December (14.6 mag).  The 2012 outburst
was detected by CRTS MLS.  Using SDSS colors, \citet{kat12DNSDSS}
estimated an orbital period of 0.07~d.  The 2012 outburst soon
turned out to be a superoutburst by the detection of superhumps
(vsnet-alert 15155, 15166; figure \ref{fig:j105835shpdm}).
The times of superhump maxima are listed in table
\ref{tab:j105835oc2012}.  Although stage A--B transition was
recorded, the data were insufficient to determine the period
of stage A.

\begin{table}
\caption{Superhump maxima of CSS J105835 (2012)}\label{tab:j105835oc2012}
\begin{center}
\begin{tabular}{ccccc}
\hline
$E$ & max\commenta & error & $O-C$\commentb & $N$\commentc \\
\hline
0 & 56268.2217 & 0.0052 & $-$0.0112 & 104 \\
1 & 56268.2872 & 0.0016 & $-$0.0037 & 102 \\
33 & 56270.1551 & 0.0031 & 0.0089 & 60 \\
34 & 56270.2089 & 0.0006 & 0.0048 & 104 \\
35 & 56270.2660 & 0.0004 & 0.0040 & 103 \\
51 & 56271.1940 & 0.0006 & 0.0042 & 100 \\
52 & 56271.2487 & 0.0005 & 0.0010 & 84 \\
68 & 56272.1834 & 0.0011 & 0.0081 & 103 \\
69 & 56272.2331 & 0.0003 & $-$0.0001 & 206 \\
70 & 56272.2893 & 0.0002 & $-$0.0020 & 239 \\
81 & 56272.9270 & 0.0004 & $-$0.0020 & 54 \\
82 & 56272.9847 & 0.0003 & $-$0.0023 & 55 \\
83 & 56273.0451 & 0.0007 & 0.0001 & 31 \\
86 & 56273.2179 & 0.0008 & $-$0.0010 & 91 \\
87 & 56273.2765 & 0.0011 & $-$0.0003 & 89 \\
88 & 56273.3324 & 0.0007 & $-$0.0024 & 47 \\
98 & 56273.9107 & 0.0020 & $-$0.0039 & 24 \\
99 & 56273.9706 & 0.0003 & $-$0.0020 & 55 \\
100 & 56274.0294 & 0.0004 & $-$0.0012 & 49 \\
116 & 56274.9581 & 0.0006 & $-$0.0001 & 55 \\
117 & 56275.0171 & 0.0005 & 0.0009 & 55 \\
\hline
  \multicolumn{5}{l}{\commenta BJD$-$2400000.} \\
  \multicolumn{5}{l}{\commentb Against max $= 2456268.2329 + 0.057977 E$.} \\
  \multicolumn{5}{l}{\commentc Number of points used to determine the maximum.} \\
\end{tabular}
\end{center}
\end{table}

\subsection{CSS J150904.0$+$465057}\label{obj:j150904}

\begin{figure}
  \begin{center}
    \FigureFile(88mm,110mm){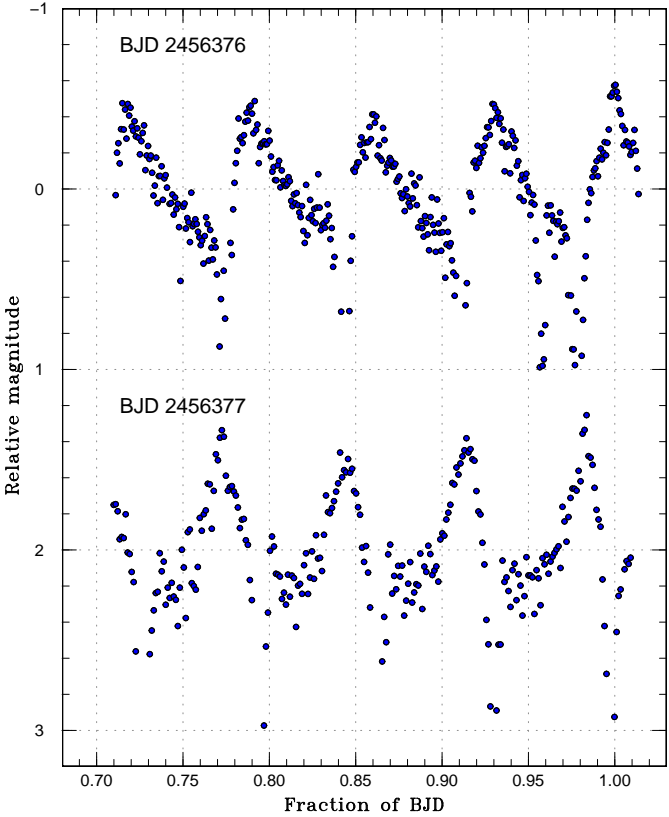}
  \end{center}
  \caption{Light curve of CSS J150904 (2013).  Large-amplitude
    superhumps and shallow eclipses are seen.}
  \label{fig:j150904lc}
\end{figure}

\begin{figure}
  \begin{center}
    \FigureFile(88mm,110mm){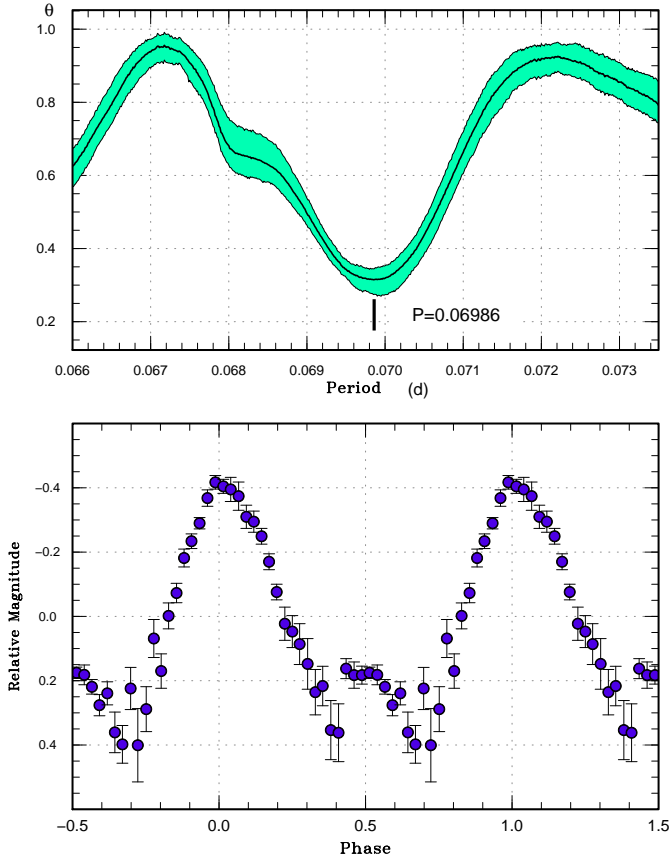}
  \end{center}
  \caption{Superhumps in CSS J150904 (2013). (Upper): PDM analysis.
     (Lower): Phase-averaged profile.}
  \label{fig:j150904shpdm}
\end{figure}

   This object was detected as a transient by CRTS
(=CSS130324:150904$+$465057, hereafter CSS J150904) on
2013 March 24.  Two past outbursts (2007 April, 17.8 mag and 2009 
March, 16.8 mag) were recorded in the CRTS data.
Relatively large intra-night variation was suggestive
of eclipses (vsnet-alert 15543).  Subsequent observations
detected superhumps and shallow eclipses (vsnet-alert 15545,
15548; figures \ref{fig:j150904lc}, \ref{fig:j150904shpdm}).
The times of superhump maxima are listed im table
\ref{tab:j150904oc2013}.
The superhump period given in table \ref{tab:perlist}
was obtained by the PDM analysis.
The MCMC analysis introduced in \citet{Pdot4}
yielded the eclipse ephemeris
of BJD $2456376.9751(6)+0.068440(8) E$, a large systematic
error is, however, expected due to the presence of superhumps.

\begin{table}
\caption{Superhump maxima of CSS J150904 (2013)}\label{tab:j150904oc2013}
\begin{center}
\begin{tabular}{ccccc}
\hline
$E$ & max\commenta & error & $O-C$\commentb & $N$\commentc \\
\hline
0 & 56376.7211 & 0.0008 & $-$0.0016 & 48 \\
1 & 56376.7941 & 0.0006 & 0.0014 & 67 \\
2 & 56376.8634 & 0.0007 & 0.0008 & 63 \\
3 & 56376.9318 & 0.0007 & $-$0.0008 & 64 \\
4 & 56377.0011 & 0.0004 & $-$0.0014 & 49 \\
14 & 56377.7088 & 0.0083 & 0.0065 & 21 \\
15 & 56377.7717 & 0.0010 & $-$0.0006 & 43 \\
16 & 56377.8422 & 0.0010 & $-$0.0000 & 43 \\
17 & 56377.9104 & 0.0013 & $-$0.0018 & 46 \\
18 & 56377.9797 & 0.0014 & $-$0.0025 & 44 \\
\hline
  \multicolumn{5}{l}{\commenta BJD$-$2400000.} \\
  \multicolumn{5}{l}{\commentb Against max $= 2456376.7226 + 0.069977 E$.} \\
  \multicolumn{5}{l}{\commentc Number of points used to determine the maximum.} \\
\end{tabular}
\end{center}
\end{table}

\subsection{CSS J203937.7$-$042907}\label{obj:j203937}

\begin{figure}
  \begin{center}
    \FigureFile(88mm,110mm){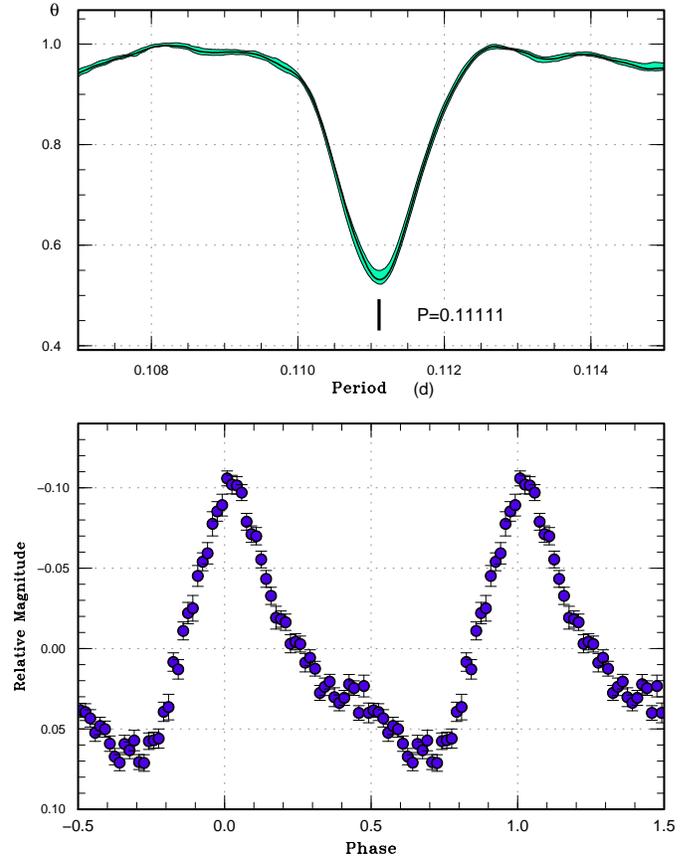}
  \end{center}
  \caption{Superhumps in CSS J203937 (2012). (Upper): PDM analysis.
     (Lower): Phase-averaged profile.}
  \label{fig:j203937shpdm}
\end{figure}

   This object was detected as a transient by CRTS
(=CSS120813:203938$-$042908, hereafter CSS J203937) on
2012 August 13.  Subsequent observation indicated that the
object showed superhumps (figure \ref{fig:j203937shpdm})
and it is a dwarf nova in the period gap (vsnet-alert 14858, 14859).
The times of superhump maxima are listed in table
\ref{tab:j203937oc2012}.  The $O-C$ values indicate that
the period was increasing during the entire observation.
Since the cycle count was somewhat ambiguous in the initial
part, $E=0$ was not used for determining $P_{\rm dot}$
listed in table \ref{tab:perlist}.
If we use this point, $P_{\rm dot}$ becomes $+10.2(2.5) \times 10^{-5}$.

   There have been at least two long-$P_{\rm orb}$ objects
with large positive $P_{\rm dot}$ (GX Cas: \cite{Pdot3};
SDSS J170213: \cite{Pdot}, \cite{Pdot4}).
CSS J203937 appears to be similar to SDSS J170213,
another SU UMa-type dwarf nova in the period gap.
CRTS data recored only one past outburst in 2005 October,
and the frequency of outbursts seems to be as low as
in SDSS J170213 \citep{Pdot4}.

\begin{figure}
  \begin{center}
    \FigureFile(88mm,110mm){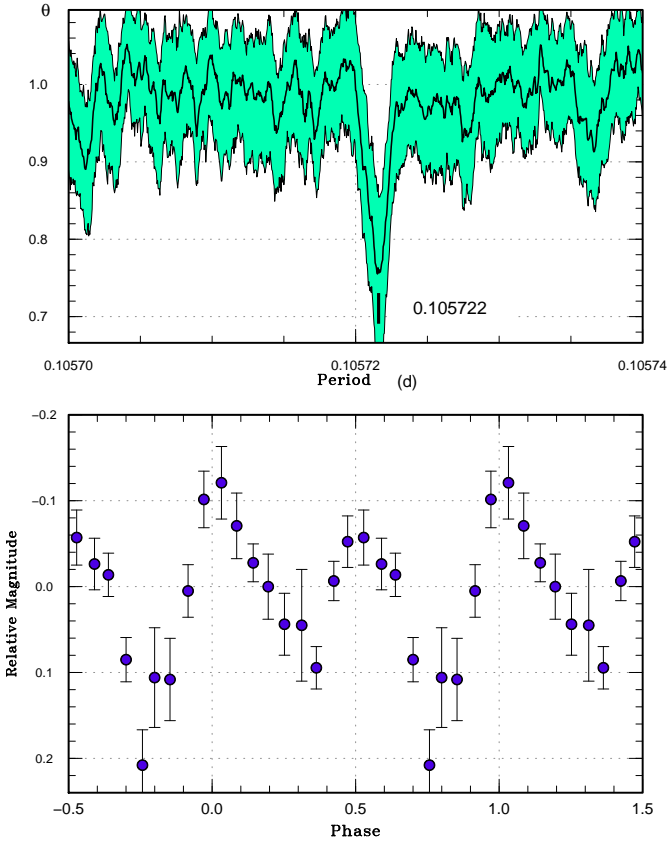}
  \end{center}
  \caption{Orbital variation in CSS J203937 in quiescence.
     (Upper): PDM analysis.
     (Lower): Phase-averaged profile.}
  \label{fig:j203937porbpdm}
\end{figure}

   Using CRTS data in quiescence, we obtained a period of
0.1057216(1)~d (figure \ref{fig:j203937porbpdm}),
which we consider to be the orbital period.
The light curve has double humps, which likely reflect
a combination of the ellipsoidal variation of the secondary
and the orbital hump.
The value of $\varepsilon$ (for mean $P_{\rm SH}$)
amounts to 5.1(1)\%, which is also
similar to that of SDSS J170213 \citep{Pdot4}.

\begin{table}
\caption{Superhump maxima of CSS J203937}\label{tab:j203937oc2012}
\begin{center}
\begin{tabular}{ccccc}
\hline
$E$ & max\commenta & error & $O-C$\commentb & $N$\commentc \\
\hline
0 & 56155.6569 & 0.0005 & 0.0152 & 110 \\
18 & 56157.6427 & 0.0006 & $-$0.0001 & 105 \\
19 & 56157.7531 & 0.0013 & $-$0.0009 & 99 \\
23 & 56158.2058 & 0.0096 & 0.0072 & 74 \\
24 & 56158.3106 & 0.0004 & 0.0008 & 254 \\
25 & 56158.4229 & 0.0003 & 0.0020 & 367 \\
26 & 56158.5327 & 0.0005 & 0.0006 & 159 \\
27 & 56158.6416 & 0.0012 & $-$0.0017 & 38 \\
28 & 56158.7568 & 0.0038 & 0.0023 & 13 \\
34 & 56159.4314 & 0.0066 & 0.0099 & 37 \\
35 & 56159.5265 & 0.0007 & $-$0.0062 & 108 \\
36 & 56159.6397 & 0.0007 & $-$0.0042 & 170 \\
37 & 56159.7562 & 0.0025 & 0.0012 & 75 \\
42 & 56160.3037 & 0.0130 & $-$0.0072 & 29 \\
43 & 56160.4189 & 0.0007 & $-$0.0031 & 129 \\
44 & 56160.5292 & 0.0008 & $-$0.0040 & 108 \\
45 & 56160.6433 & 0.0013 & $-$0.0011 & 39 \\
51 & 56161.3025 & 0.0010 & $-$0.0089 & 167 \\
52 & 56161.4138 & 0.0005 & $-$0.0088 & 299 \\
53 & 56161.5260 & 0.0005 & $-$0.0077 & 342 \\
54 & 56161.6389 & 0.0009 & $-$0.0060 & 106 \\
55 & 56161.7511 & 0.0014 & $-$0.0050 & 63 \\
62 & 56162.5309 & 0.0015 & $-$0.0033 & 76 \\
63 & 56162.6517 & 0.0031 & 0.0062 & 44 \\
69 & 56163.3101 & 0.0007 & $-$0.0023 & 255 \\
70 & 56163.4182 & 0.0006 & $-$0.0055 & 305 \\
71 & 56163.5296 & 0.0013 & $-$0.0052 & 214 \\
72 & 56163.6441 & 0.0012 & $-$0.0019 & 102 \\
73 & 56163.7669 & 0.0073 & 0.0097 & 16 \\
79 & 56164.4243 & 0.0041 & 0.0002 & 57 \\
80 & 56164.5360 & 0.0020 & 0.0007 & 79 \\
81 & 56164.6540 & 0.0027 & 0.0075 & 105 \\
82 & 56164.7604 & 0.0053 & 0.0027 & 87 \\
88 & 56165.4249 & 0.0011 & 0.0003 & 64 \\
89 & 56165.5266 & 0.0038 & $-$0.0093 & 93 \\
90 & 56165.6650 & 0.0049 & 0.0180 & 105 \\
97 & 56166.4328 & 0.0036 & 0.0076 & 59 \\
98 & 56166.5365 & 0.0031 & 0.0001 & 77 \\
\hline
  \multicolumn{5}{l}{\commenta BJD$-$2400000.} \\
  \multicolumn{5}{l}{\commentb Against max $= 2456155.6417 + 0.111170 E$.} \\
  \multicolumn{5}{l}{\commentc Number of points used to determine the maximum.} \\
\end{tabular}
\end{center}
\end{table}

\subsection{CSS J214934.6-121909}\label{obj:j214934}

   This object was detected as a transient by CRTS
(=CSS120922:214934$-$121908, hereafter CSS J214934) on
2012 September 22.  Subsequent observation confirmed the presence
of superhumps (vsnet-alert 14944, 14947).
The times of two superhump maxima are BJD 2456194.3570(4) ($N=66$)
and 2456194.4272(4) ($N=69$).  The period by the PDM method
was 0.0702(5)~d.

\begin{figure}
  \begin{center}
    \FigureFile(88mm,110mm){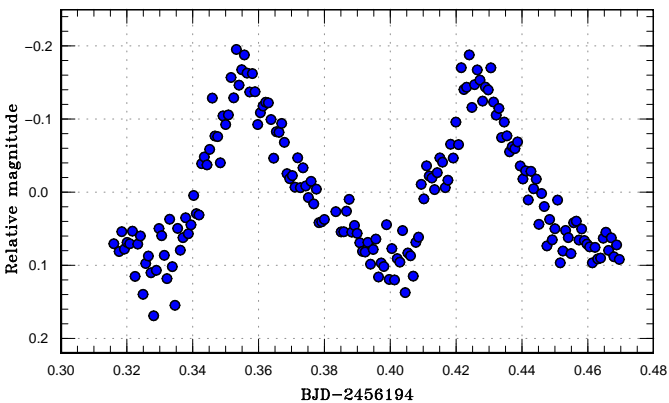}
  \end{center}
  \caption{Superhumps in CSS J214934 (2012).}
  \label{fig:j214934lc}
\end{figure}

\subsection{DDE 26}\label{obj:dde26}

\begin{figure}
  \begin{center}
    \FigureFile(88mm,110mm){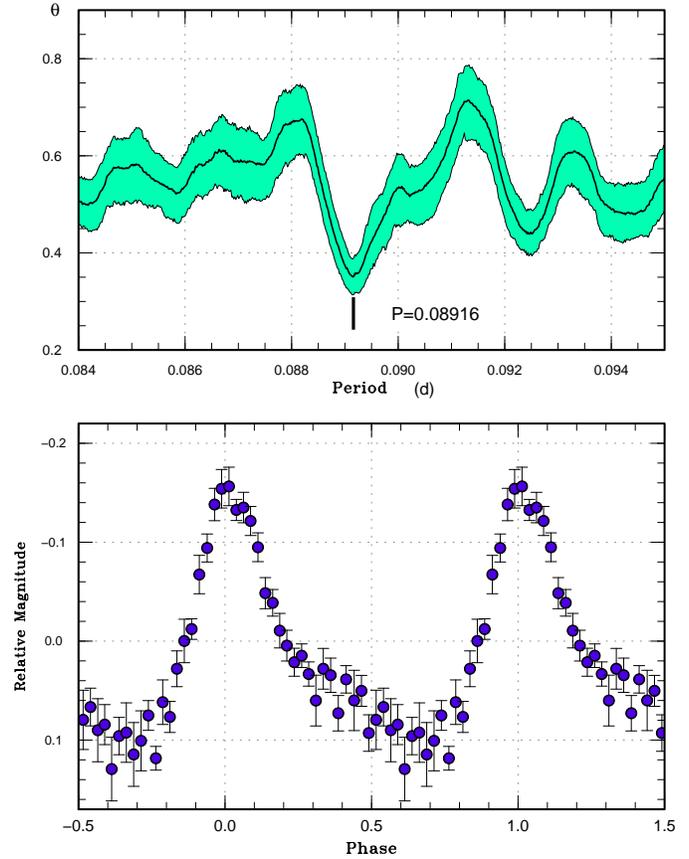}
  \end{center}
  \caption{Superhumps in DDE 26 (2012). (Upper): PDM analysis.
     (Lower): Phase-averaged profile.}
  \label{fig:dde26shpdm}
\end{figure}

   DDE 26 is a dwarf nova discovered by \citep{den12USNOCVs}
and is located at \timeform{22h 03m 28.2s}, \timeform{+30D 56' 37''}.
D. Denisenko detected a bright outburst on 2012 July 25
(vsnet-alert 14792).  Subsequent observations detected
superhumps (vsnet-alert 14799; figure \ref{fig:dde26shpdm}).

   The times of superhump maxima are listed in table \ref{tab:dde2620oc12}.
The $P_{\rm dot}$ was determined globally.  The large negative
$P_{\rm dot}$ indicates that the object belongs to a group
of long-$P_{\rm orb}$ SU UMa-type dwarf novae with large
$P_{\rm SH}$ variations, whose best known member is UV Gem
\citep{Pdot}.

\begin{table}
\caption{Superhump maxima of DDE 26 (2012)}\label{tab:dde2620oc12}
\begin{center}
\begin{tabular}{ccccc}
\hline
$E$ & max\commenta & error & $O-C$\commentb & $N$\commentc \\
\hline
0 & 56134.6787 & 0.0005 & $-$0.0011 & 129 \\
1 & 56134.7696 & 0.0006 & 0.0005 & 152 \\
20 & 56136.4679 & 0.0004 & 0.0017 & 31 \\
31 & 56137.4476 & 0.0011 & $-$0.0011 & 33 \\
\hline
  \multicolumn{5}{l}{\commenta BJD$-$2400000.} \\
  \multicolumn{5}{l}{\commentb Against max $= 2456134.6798 + 0.089320 E$.} \\
  \multicolumn{5}{l}{\commentc Number of points used to determine the maximum.} \\
\end{tabular}
\end{center}
\end{table}

\subsection{MASTER OT J000820.50$+$773119.1}\label{obj:j000820}

\begin{figure}
  \begin{center}
    \FigureFile(88mm,110mm){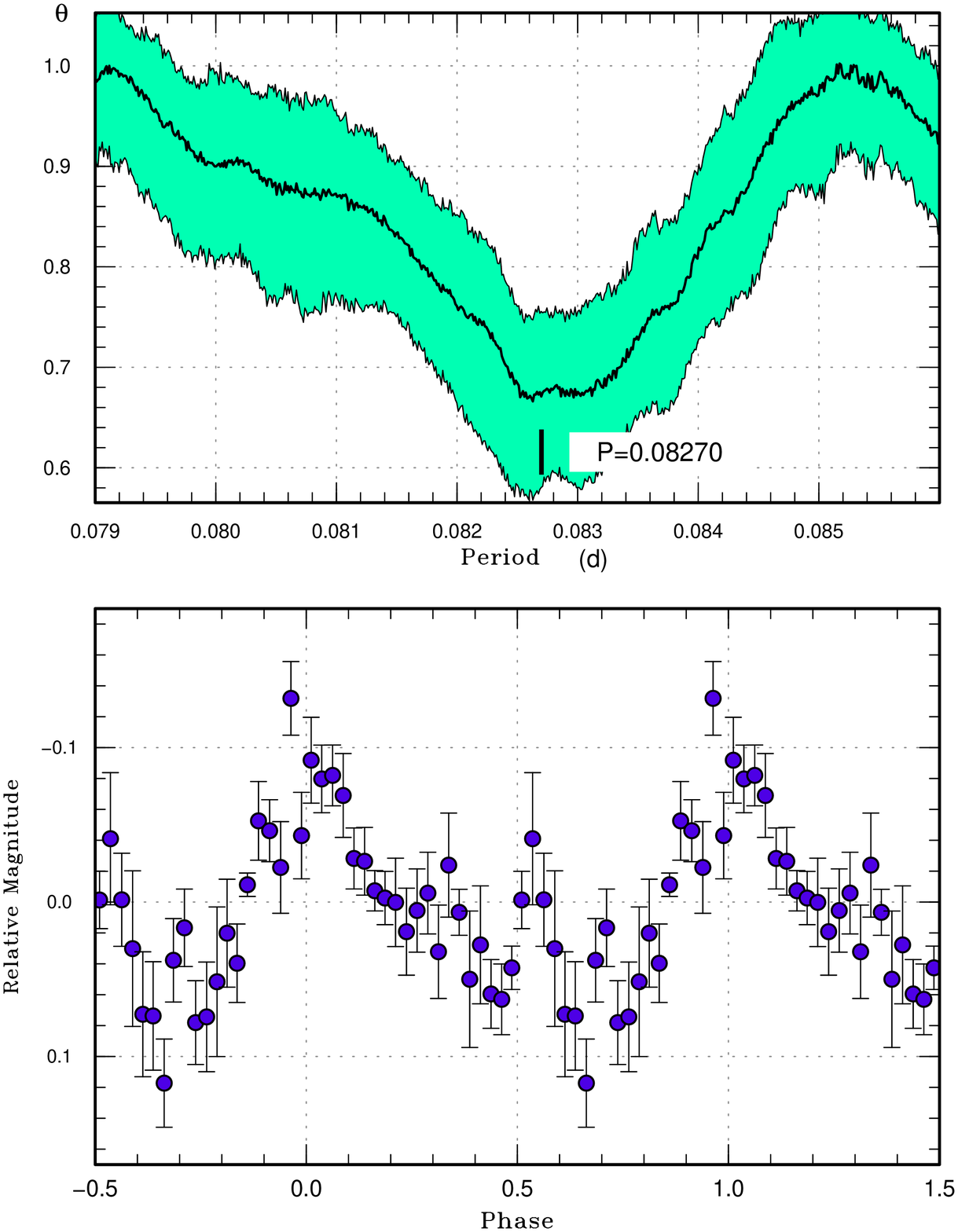}
  \end{center}
  \caption{Superhumps in MASTER J000820 (2012). (Upper): PDM analysis.
     (Lower): Phase-averaged profile.}
  \label{fig:j000820shpdm}
\end{figure}

   This object was detected as a transient by MASTER network
(\cite{den12j0426atel4441}, hereafter MASTER J000820).
Subsequent observations detected superhumps (vsnet-alert 14967,
figure \ref{fig:j000820shpdm}).
The times of superhump maxima are listed in table \ref{tab:j000820oc2012}.
The period listed in table \ref{tab:perlist} was obtained by
the PDM method.

\begin{table}
\caption{Superhump maxima of MASTER J000820}\label{tab:j000820oc2012}
\begin{center}
\begin{tabular}{ccccc}
\hline
$E$ & max\commenta & error & $O-C$\commentb & $N$\commentc \\
\hline
0 & 56203.3346 & 0.0014 & $-$0.0005 & 46 \\
1 & 56203.4187 & 0.0013 & 0.0007 & 41 \\
12 & 56204.3276 & 0.0027 & $-$0.0016 & 45 \\
13 & 56204.4135 & 0.0027 & 0.0014 & 46 \\
\hline
  \multicolumn{5}{l}{\commenta BJD$-$2400000.} \\
  \multicolumn{5}{l}{\commentb Against max $= 2456203.3351 + 0.082841 E$.} \\
  \multicolumn{5}{l}{\commentc Number of points used to determine the maximum.} \\
\end{tabular}
\end{center}
\end{table}

\subsection{MASTER OT J001952.31$+$464933.0}\label{obj:j001952}

\begin{figure}
  \begin{center}
    \FigureFile(88mm,110mm){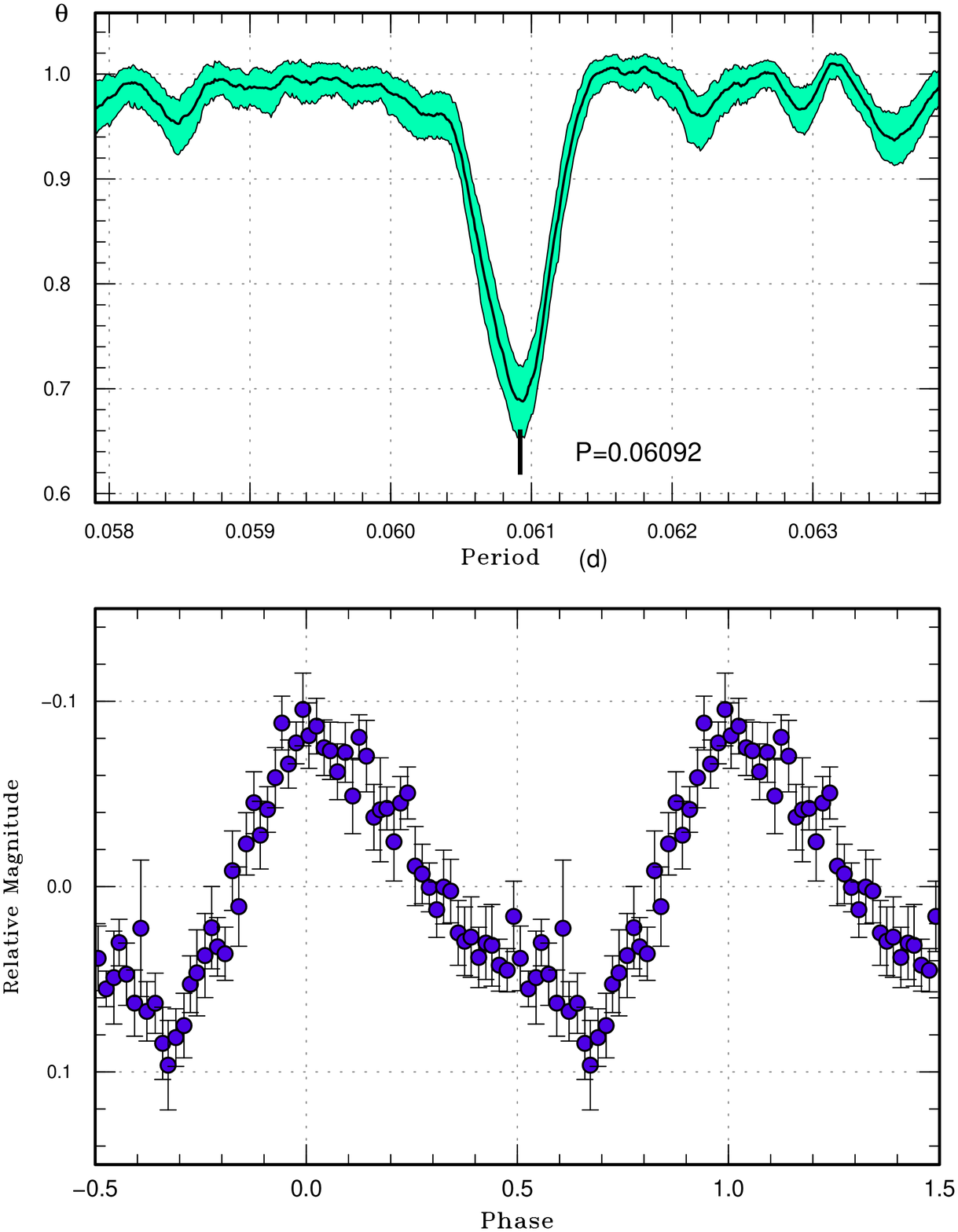}
  \end{center}
  \caption{Superhumps in MASTER J001952 (2012). (Upper): PDM analysis.
     (Lower): Phase-averaged profile.}
  \label{fig:j001952shpdm}
\end{figure}

   This object (hereafter MASTER J001952) was discovered by
MASTER network \citep{den12j0019atel4488} on 2012 October 15.
Follow-up observations detected superhumps (vsnet-alert 15012,
15014, 15015, 15017, 15065; figure \ref{fig:j001952shpdm}).
The times of superhump maxima are listed in table
\ref{tab:j001952oc2012}.  The epoch $E=0$ corresponds to
a stage A superhump.  A positive $P_{\rm dot}$ of 
$+10.4(2.7) \times 10^{-5}$ was observed in stage B superhumps.
This value is typical for this $P_{\rm SH}$.

\begin{table}
\caption{Superhump maxima of MASTER J001952}\label{tab:j001952oc2012}
\begin{center}
\begin{tabular}{ccccc}
\hline
$E$ & max\commenta & error & $O-C$\commentb & $N$\commentc \\
\hline
0 & 56217.3383 & 0.0029 & $-$0.0094 & 34 \\
15 & 56218.2674 & 0.0004 & 0.0050 & 33 \\
16 & 56218.3297 & 0.0005 & 0.0063 & 34 \\
17 & 56218.3889 & 0.0004 & 0.0045 & 34 \\
32 & 56219.3020 & 0.0005 & 0.0028 & 33 \\
33 & 56219.3613 & 0.0005 & 0.0011 & 34 \\
34 & 56219.4268 & 0.0010 & 0.0056 & 12 \\
44 & 56220.0309 & 0.0010 & $-$0.0002 & 84 \\
45 & 56220.0882 & 0.0010 & $-$0.0039 & 82 \\
60 & 56220.9998 & 0.0014 & $-$0.0071 & 47 \\
61 & 56221.0678 & 0.0009 & $-$0.0001 & 82 \\
62 & 56221.1225 & 0.0015 & $-$0.0064 & 82 \\
63 & 56221.1897 & 0.0052 & $-$0.0001 & 18 \\
77 & 56222.0462 & 0.0015 & 0.0026 & 51 \\
78 & 56222.0965 & 0.0017 & $-$0.0081 & 55 \\
79 & 56222.1638 & 0.0050 & $-$0.0018 & 23 \\
111 & 56224.1216 & 0.0022 & 0.0044 & 44 \\
112 & 56224.1840 & 0.0056 & 0.0058 & 39 \\
113 & 56224.2382 & 0.0012 & $-$0.0010 & 44 \\
\hline
  \multicolumn{5}{l}{\commenta BJD$-$2400000.} \\
  \multicolumn{5}{l}{\commentb Against max $= 2456217.3477 + 0.060987 E$.} \\
  \multicolumn{5}{l}{\commentc Number of points used to determine the maximum.} \\
\end{tabular}
\end{center}
\end{table}

\subsection{MASTER OT J030128.77$+$401104.9}\label{obj:j030128}

\begin{figure}
  \begin{center}
    \FigureFile(88mm,110mm){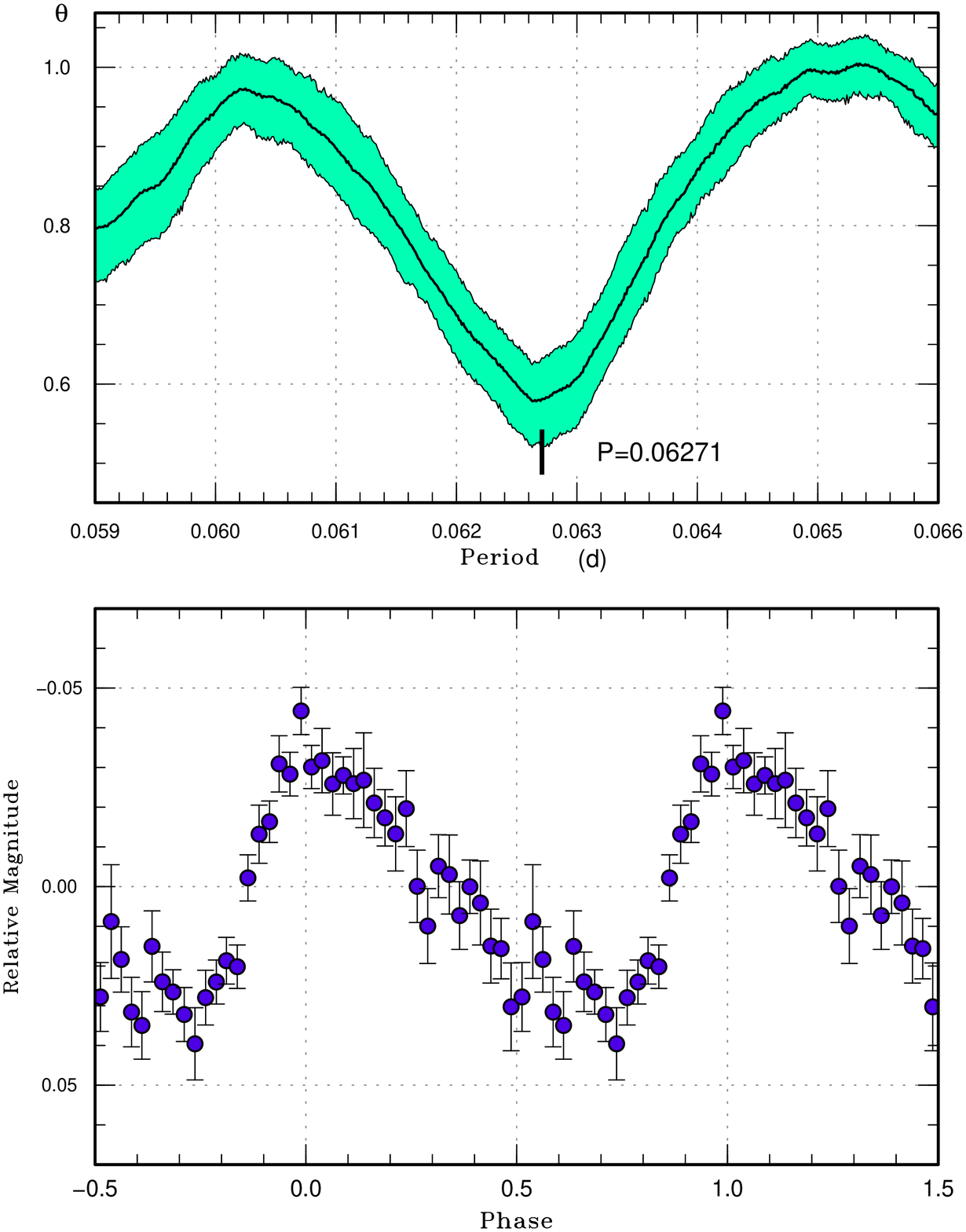}
  \end{center}
  \caption{Superhumps in MASTER J030128 (2012).
     The data fir BJD 2456298--2456300 were used.
     (Upper): PDM analysis.
     (Lower): Phase-averaged profile.}
  \label{fig:j030128shpdm}
\end{figure}

   This object (hereafter MASTER J030128) was discovered by
MASTER network \citep{bal12j0301j1050atel4682} on 2012 December 27.
The outburst amplitude was about 6 mag.
Subsequent observations possibly detected
variations (vsnet-alert 15224), but superhumps clearly appeared
only 8~d later (vsnet-alert 15247; figure \ref{fig:j030128shpdm}). 
Although initial observations possibly recorded the precursor phase 
(vsnet-alert 15247), it is difficult to tell the outburst phase 
due to the gaps in the observation.
The object entered the rapid fading phase
on 2013 January 11.  The times of superhump maxima are listed in
table \ref{tab:j030128oc2012}; the times were well measured only
after $E=96$, and the cycle numbers for $E \le 34$ may not be
correct.  It was not sure whether we observed stage B or C.

\begin{table}
\caption{Superhump maxima of MASTER J030128}\label{tab:j030128oc2012}
\begin{center}
\begin{tabular}{ccccc}
\hline
$E$ & max\commenta & error & $O-C$\commentb & $N$\commentc \\
\hline
0 & 56292.2948 & 0.0213 & 0.0062 & 35 \\
1 & 56292.3560 & 0.0023 & 0.0047 & 35 \\
2 & 56292.4184 & 0.0028 & 0.0044 & 35 \\
34 & 56294.3985 & 0.0018 & $-$0.0217 & 60 \\
96 & 56298.3035 & 0.0015 & $-$0.0035 & 51 \\
97 & 56298.3752 & 0.0010 & 0.0056 & 65 \\
100 & 56298.5580 & 0.0011 & 0.0003 & 49 \\
101 & 56298.6191 & 0.0007 & $-$0.0014 & 62 \\
102 & 56298.6822 & 0.0009 & $-$0.0009 & 46 \\
112 & 56299.3089 & 0.0016 & $-$0.0012 & 42 \\
113 & 56299.3770 & 0.0008 & 0.0043 & 42 \\
114 & 56299.4387 & 0.0030 & 0.0033 & 22 \\
\hline
  \multicolumn{5}{l}{\commenta BJD$-$2400000.} \\
  \multicolumn{5}{l}{\commentb Against max $= 2456292.2886 + 0.062691 E$.} \\
  \multicolumn{5}{l}{\commentc Number of points used to determine the maximum.} \\
\end{tabular}
\end{center}
\end{table}

\subsection{MASTER OT J042609.34$+$354144.8}\label{obj:j042609}

\begin{figure}
  \begin{center}
    \FigureFile(88mm,110mm){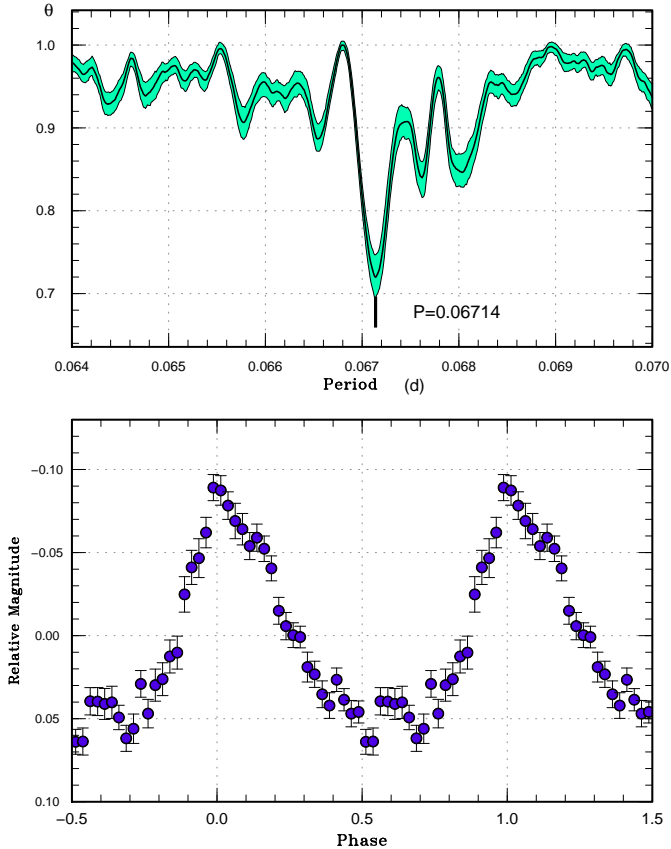}
  \end{center}
  \caption{Superhumps in MASTER J042609 (2012). (Upper): PDM analysis.
     (Lower): Phase-averaged profile.}
  \label{fig:j042609shpdm}
\end{figure}

   This object (hereafter MASTER J042609) was discovered by
MASTER network \citep{den12j0426atel4441}.  The object has
a ROSAT counterpart (1RXS J042608.9$+$354151) and a GALEX
UV source.  There were at least three past outburst detections
in the CRTS data: 2008 November 1 (13.4 mag), 2010 February 19
(14.7 mag), 2011 November 18 (14.4 mag) and the frequency of
outbursts is not particularly low.
Subsequent observations confirmed
superhumps (vsnet-alert 14970, 14979, 14989, 14992;
figure \ref{fig:j042609shpdm}).
The times of superhump maxima are listed in table
\ref{tab:j042609oc2012}.  Since these superhumps were observed 
in the later phase of the superoutburst and
the early post-outburst phase, the period decrease likely
corresponds to a stage B--C transition.
Since only the later part of stage B was observed,
we did not attempt to determine $P_{\rm dot}$.

\begin{figure}
  \begin{center}
    \FigureFile(88mm,110mm){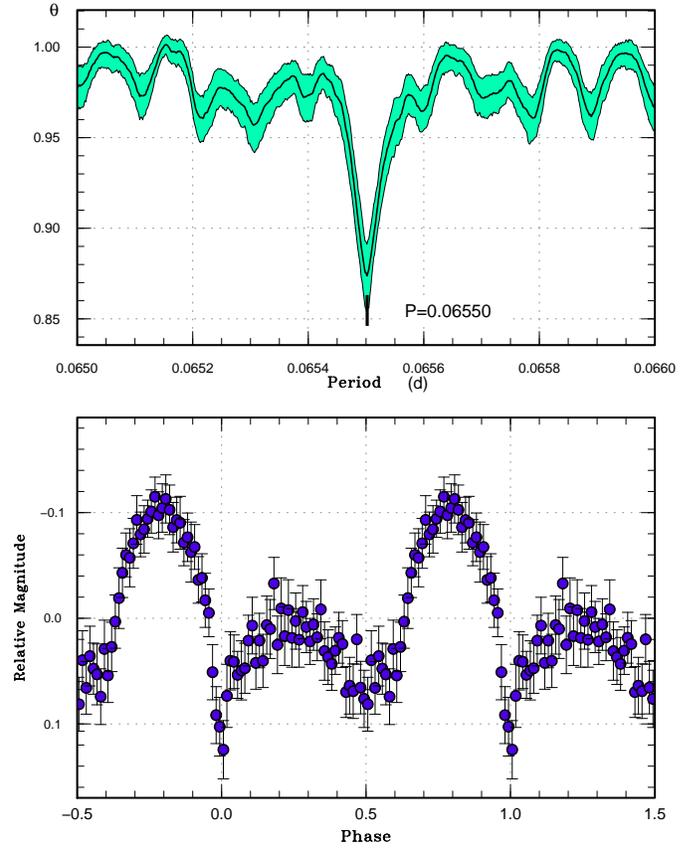}
  \end{center}
  \caption{Orbital modulation in MASTER J042609 in quiescence.
     (Upper): PDM analysis.
     (Lower): Phase-averaged profile.}
  \label{fig:j042609porbpdm}
\end{figure}

   Between 2012 November 20 and 2013 January 5, the object was
observed in quiescence.  A period analysis yielded a light curve
characteristic to a grazing eclipsing system with a prominent
orbital hump (figure \ref{fig:j042609porbpdm}).  The orbital period
determined from this analysis was 0.0655015(17)~d.
By combining with the CRTS data in quiescence, we have obtained
a refined period of 0.0655022(1)~d.  The eclipse minimum in
figure \ref{fig:j042609porbpdm} corresponds to BJD 2456276.6430.
The quiescent orbital profile bears some similarity with
WZ Sge-type dwarf nova, particularly with very low $q$,
in the double-wave modulations
[see e.g. WZ Sge and AL Com \citep{pat96alcom}, V455 And
(\cite{ara05v455and}; \cite{Pdot}), V386 Ser \citep{muk10v386ser},
EZ Lyn (\cite{kat09j0804}; \cite{zha13ezlyn}), BW Scl
(\cite{aug97bwscl}; \cite{Pdot4})].  \citet{zha13ezlyn} interpreted
that these double-wave modulations can be interpreted as a result
of the spiral structure caused by the 2:1 resonance.  Most recently,
however, SDSS J152419.33$+$220920.0, which appears to have a higher
$q$ and less likely achieve the 2:1 resonance, was reported to
show double-wave modulations \citep{mic13j1524}.  MASTER J042609
would add an additional example for a rather ordinary SU UMa-type 
dwarf nova showing double-wave modulations. 
It may be that there are different
mechanisms to produce double-wave modulations in quiescence.  
A classical interpretation assuming a semi-transparent accretion
disk allowing the light from the hot spot to escape in two directions
\citep{ski00wzsge} may be an alternative mechanism.

   The high orbital inclination would explain the relatively small
outburst amplitude for a short-$P_{\rm orb}$ SU UMa-type
dwarf nova.

\begin{table}
\caption{Superhump maxima of MASTER J042609}\label{tab:j042609oc2012}
\begin{center}
\begin{tabular}{ccccc}
\hline
$E$ & max\commenta & error & $O-C$\commentb & $N$\commentc \\
\hline
0 & 56202.6263 & 0.0007 & $-$0.0087 & 67 \\
1 & 56202.6949 & 0.0005 & $-$0.0072 & 65 \\
12 & 56203.4360 & 0.0007 & $-$0.0049 & 93 \\
13 & 56203.5040 & 0.0004 & $-$0.0040 & 124 \\
14 & 56203.5718 & 0.0003 & $-$0.0034 & 191 \\
15 & 56203.6388 & 0.0004 & $-$0.0035 & 162 \\
16 & 56203.7086 & 0.0008 & $-$0.0009 & 63 \\
29 & 56204.5867 & 0.0005 & 0.0040 & 65 \\
31 & 56204.7228 & 0.0007 & 0.0058 & 52 \\
38 & 56205.1918 & 0.0005 & 0.0047 & 119 \\
40 & 56205.3286 & 0.0006 & 0.0071 & 121 \\
81 & 56208.0835 & 0.0004 & 0.0084 & 182 \\
87 & 56208.4849 & 0.0012 & 0.0068 & 39 \\
88 & 56208.5549 & 0.0005 & 0.0096 & 69 \\
89 & 56208.6201 & 0.0004 & 0.0077 & 69 \\
90 & 56208.6899 & 0.0006 & 0.0102 & 63 \\
103 & 56209.5470 & 0.0016 & $-$0.0058 & 46 \\
104 & 56209.6195 & 0.0015 & $-$0.0004 & 45 \\
105 & 56209.6875 & 0.0008 & 0.0004 & 94 \\
106 & 56209.7514 & 0.0009 & $-$0.0028 & 60 \\
107 & 56209.8211 & 0.0011 & $-$0.0003 & 54 \\
120 & 56210.6878 & 0.0016 & $-$0.0067 & 35 \\
148 & 56212.5722 & 0.0004 & $-$0.0029 & 51 \\
158 & 56213.2336 & 0.0018 & $-$0.0131 & 136 \\
\hline
  \multicolumn{5}{l}{\commenta BJD$-$2400000.} \\
  \multicolumn{5}{l}{\commentb Against max $= 2456202.6349 + 0.0671633 E$.} \\
  \multicolumn{5}{l}{\commentc Number of points used to determine the maximum.} \\
\end{tabular}
\end{center}
\end{table}

\subsection{MASTER OT J054317.95$+$093114.8}\label{obj:j054317}

\begin{figure}
  \begin{center}
    \FigureFile(88mm,110mm){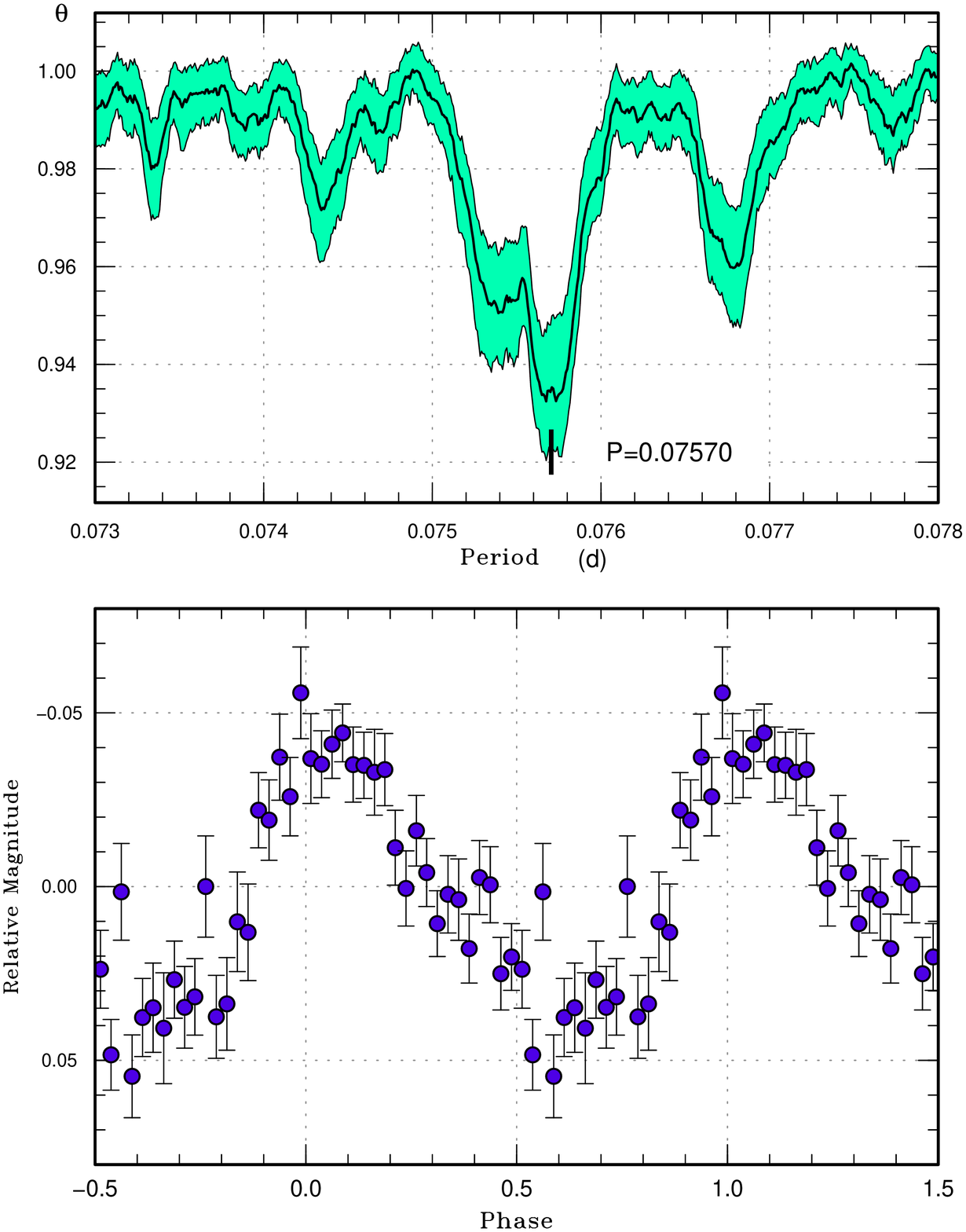}
  \end{center}
  \caption{Superhumps in MASTER J054317 (2012). (Upper): PDM analysis.
     (Lower): Phase-averaged profile.}
  \label{fig:j054317shpdm}
\end{figure}

   This object (hereafter MASTER J054317) was discovered by
MASTER network \citep{bal12j0543atel4446}.  There was no
visible quiescent counterpart in the DSS plates to 21 mag
\citep{bal12j0543atel4446}.  The large outburst amplitude
($\ge$7 mag) caught attention.  For the initial four days,
the object did not show clear superhumps (vsnet-alert 14986).
The object then started to show superhumps (vsnet-alert 14998,
15016, 15018, 15020, figure \ref{fig:j054317shpdm}.
The times of superhump maxima are listed in table
\ref{tab:j054317oc2012}.  Although stage A--C are recognized,
the period of stage A was not determined due to the lack of
observations.  The coverage of stage B was also somewhat
insufficient.

   Despite the large outburst amplitude, the behavior of
superhumps more resembles that of ordinary SU UMa-type dwarf novae
rather than WZ Sge-type dwarf novae.  There are several
long-$P_{\rm orb}$ SU UMa-type dwarf novae with large outburst
amplitudes: V1251 Cyg (\cite{kat95v1251cyg}; \cite{Pdot}),
EF Peg (\cite{how93efpeg}; \cite{kat02efpeg}; \cite{Pdot};
\cite{Pdot2}) and this object may resemble these objects.

\begin{table}
\caption{Superhump maxima of MASTER J054317 (2012)}\label{tab:j054317oc2012}
\begin{center}
\begin{tabular}{ccccc}
\hline
$E$ & max\commenta & error & $O-C$\commentb & $N$\commentc \\
\hline
0 & 56205.8134 & 0.0015 & $-$0.0095 & 20 \\
6 & 56206.2708 & 0.0061 & $-$0.0069 & 33 \\
39 & 56208.7789 & 0.0007 & $-$0.0008 & 24 \\
40 & 56208.8534 & 0.0016 & $-$0.0022 & 17 \\
53 & 56209.8383 & 0.0009 & $-$0.0028 & 32 \\
65 & 56210.7509 & 0.0014 & $-$0.0001 & 23 \\
66 & 56210.8239 & 0.0018 & $-$0.0029 & 24 \\
118 & 56214.7718 & 0.0034 & 0.0026 & 21 \\
123 & 56215.1595 & 0.0007 & 0.0111 & 54 \\
131 & 56215.7654 & 0.0023 & 0.0105 & 24 \\
136 & 56216.1429 & 0.0005 & 0.0090 & 62 \\
137 & 56216.2187 & 0.0010 & 0.0089 & 217 \\
138 & 56216.2931 & 0.0014 & 0.0075 & 154 \\
144 & 56216.7536 & 0.0023 & 0.0131 & 24 \\
145 & 56216.8212 & 0.0052 & 0.0048 & 13 \\
157 & 56217.7311 & 0.0026 & 0.0050 & 23 \\
158 & 56217.8009 & 0.0034 & $-$0.0011 & 24 \\
171 & 56218.7856 & 0.0051 & $-$0.0020 & 24 \\
172 & 56218.8676 & 0.0027 & 0.0042 & 13 \\
182 & 56219.6201 & 0.0006 & $-$0.0014 & 72 \\
183 & 56219.7065 & 0.0199 & 0.0091 & 16 \\
184 & 56219.7751 & 0.0034 & 0.0019 & 24 \\
185 & 56219.8426 & 0.0040 & $-$0.0064 & 22 \\
190 & 56220.2226 & 0.0062 & $-$0.0055 & 147 \\
191 & 56220.2932 & 0.0039 & $-$0.0107 & 101 \\
197 & 56220.7554 & 0.0024 & $-$0.0034 & 24 \\
198 & 56220.8191 & 0.0022 & $-$0.0155 & 24 \\
199 & 56220.9071 & 0.0009 & $-$0.0033 & 70 \\
200 & 56220.9832 & 0.0017 & $-$0.0030 & 70 \\
203 & 56221.2068 & 0.0047 & $-$0.0069 & 145 \\
204 & 56221.2897 & 0.0163 & 0.0001 & 72 \\
210 & 56221.7523 & 0.0195 & 0.0078 & 23 \\
211 & 56221.8091 & 0.0030 & $-$0.0112 & 22 \\
\hline
  \multicolumn{5}{l}{\commenta BJD$-$2400000.} \\
  \multicolumn{5}{l}{\commentb Against max $= 2456205.8229 + 0.075817 E$.} \\
  \multicolumn{5}{l}{\commentc Number of points used to determine the maximum.} \\
\end{tabular}
\end{center}
\end{table}

\subsection{MASTER OT J064725.70$+$491543.9}\label{obj:j064725}

\begin{figure}
  \begin{center}
    \FigureFile(88mm,110mm){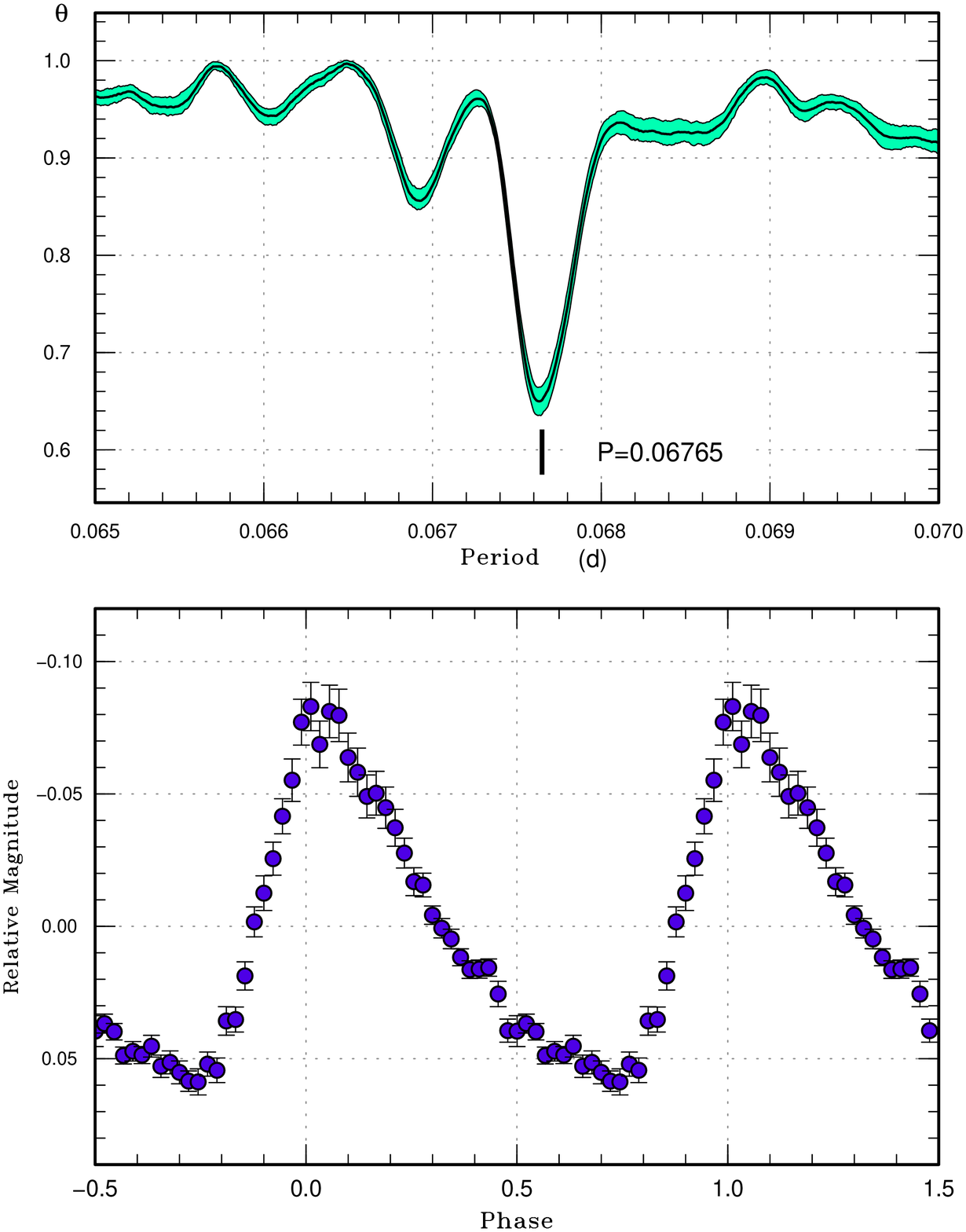}
  \end{center}
  \caption{Superhumps in MASTER J064725 (2013). (Upper): PDM analysis.
     (Lower): Phase-averaged profile.}
  \label{fig:j064725shpdm}
\end{figure}

   This object (hereafter MASTER J064725) was detected as
a dwarf nova by MASTER network on 2013 March 7
\citep{tiu13j0647atel4871}.  Superhumps were immediately
detected (vsnet-alert 15476, 15477, 15495;
figure \ref{fig:j064725shpdm}).
The times of superhump maxima are listed in table
\ref{tab:j064725oc2013}.  Stages B and C can be identified.
In deriving the period of stage C superhumps, we did not
include post-superoutburst maxima ($E \ge 243$) and $E=156$,
which was measured on the rapid declining phase and with a small
amplitude.

\begin{table}
\caption{Superhump maxima of MASTER J064725 (2013)}\label{tab:j064725oc2013}
\begin{center}
\begin{tabular}{ccccc}
\hline
$E$ & max\commenta & error & $O-C$\commentb & $N$\commentc \\
\hline
0 & 56360.0846 & 0.0015 & $-$0.0122 & 37 \\
7 & 56360.5579 & 0.0015 & $-$0.0117 & 126 \\
8 & 56360.6326 & 0.0004 & $-$0.0045 & 182 \\
9 & 56360.6994 & 0.0003 & $-$0.0053 & 190 \\
10 & 56360.7669 & 0.0003 & $-$0.0054 & 174 \\
15 & 56361.1069 & 0.0004 & $-$0.0031 & 30 \\
18 & 56361.3103 & 0.0002 & $-$0.0023 & 101 \\
19 & 56361.3787 & 0.0002 & $-$0.0014 & 137 \\
37 & 56362.6057 & 0.0028 & 0.0097 & 48 \\
38 & 56362.6627 & 0.0002 & $-$0.0008 & 132 \\
39 & 56362.7296 & 0.0002 & $-$0.0015 & 132 \\
42 & 56362.9303 & 0.0007 & $-$0.0034 & 39 \\
43 & 56363.0008 & 0.0005 & $-$0.0005 & 55 \\
57 & 56363.9522 & 0.0008 & 0.0053 & 71 \\
58 & 56364.0180 & 0.0006 & 0.0036 & 63 \\
82 & 56365.6457 & 0.0003 & 0.0101 & 132 \\
83 & 56365.7130 & 0.0005 & 0.0099 & 132 \\
84 & 56365.7793 & 0.0003 & 0.0087 & 132 \\
87 & 56365.9801 & 0.0006 & 0.0068 & 40 \\
88 & 56366.0497 & 0.0008 & 0.0089 & 45 \\
97 & 56366.6535 & 0.0003 & 0.0048 & 132 \\
98 & 56366.7206 & 0.0004 & 0.0044 & 132 \\
99 & 56366.7869 & 0.0003 & 0.0031 & 122 \\
108 & 56367.3919 & 0.0006 & 0.0002 & 75 \\
109 & 56367.4595 & 0.0007 & 0.0002 & 76 \\
110 & 56367.5266 & 0.0010 & $-$0.0003 & 76 \\
111 & 56367.5933 & 0.0010 & $-$0.0011 & 76 \\
123 & 56368.4043 & 0.0011 & $-$0.0006 & 83 \\
124 & 56368.4717 & 0.0010 & $-$0.0008 & 84 \\
125 & 56368.5472 & 0.0020 & 0.0072 & 83 \\
126 & 56368.6069 & 0.0016 & $-$0.0006 & 83 \\
136 & 56369.2885 & 0.0014 & 0.0055 & 38 \\
137 & 56369.3465 & 0.0008 & $-$0.0041 & 80 \\
138 & 56369.4139 & 0.0007 & $-$0.0042 & 118 \\
139 & 56369.4805 & 0.0016 & $-$0.0051 & 45 \\
140 & 56369.5458 & 0.0011 & $-$0.0074 & 40 \\
156 & 56370.6496 & 0.0025 & 0.0157 & 131 \\
157 & 56370.6940 & 0.0017 & $-$0.0075 & 131 \\
243 & 56376.5035 & 0.0016 & $-$0.0069 & 47 \\
244 & 56376.5647 & 0.0018 & $-$0.0132 & 46 \\
\hline
  \multicolumn{5}{l}{\commenta BJD$-$2400000.} \\
  \multicolumn{5}{l}{\commentb Against max $= 2456360.0968 + 0.067546 E$.} \\
  \multicolumn{5}{l}{\commentc Number of points used to determine the maximum.} \\
\end{tabular}
\end{center}
\end{table}

\subsection{MASTER OT J073418.66$+$271310.5}\label{obj:j073418}

   This object (hereafter MASTER J073418) was detected as
a dwarf nova by MASTER network on 2013 February 25
\citep{den12j0811atel4506}.
The 2MASS color using the neural network analysis
\citep{kat12DNSDSS} suggested an orbital period of 0.060~d
(vsnet-alert 15445).  Superhumps were indeed detected
(vsnet-alert 15453, 15461).
The times of superhump maxima are listed in table
\ref{tab:j073418oc2013}.  Since the object faded soon after
these observations, we most likely observed only stage C
superhumps.

\begin{figure}
  \begin{center}
    \FigureFile(88mm,110mm){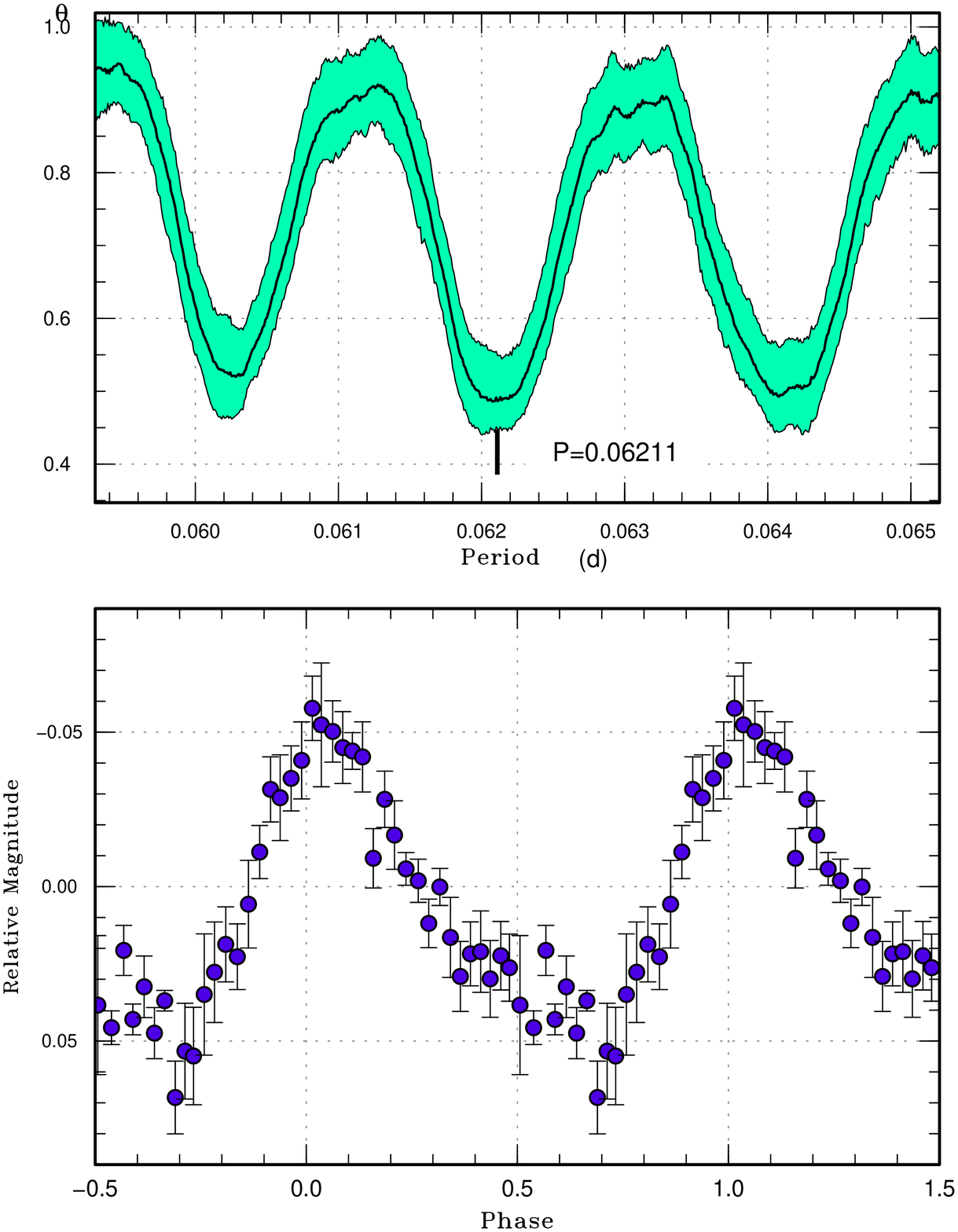}
  \end{center}
  \caption{Superhumps in MASTER J073418 (2013). (Upper): PDM analysis.
     (Lower): Phase-averaged profile.}
  \label{fig:j073418shpdm}
\end{figure}

\begin{table}
\caption{Superhump maxima of MASTER J073418 (2013)}\label{tab:j073418oc2013}
\begin{center}
\begin{tabular}{ccccc}
\hline
$E$ & max\commenta & error & $O-C$\commentb & $N$\commentc \\
\hline
0 & 56352.4162 & 0.0009 & $-$0.0018 & 44 \\
1 & 56352.4803 & 0.0006 & 0.0002 & 61 \\
2 & 56352.5433 & 0.0011 & 0.0013 & 30 \\
32 & 56354.4047 & 0.0009 & 0.0027 & 44 \\
33 & 56354.4662 & 0.0012 & 0.0022 & 45 \\
34 & 56354.5240 & 0.0060 & $-$0.0020 & 24 \\
36 & 56354.6474 & 0.0009 & $-$0.0026 & 27 \\
\hline
  \multicolumn{5}{l}{\commenta BJD$-$2400000.} \\
  \multicolumn{5}{l}{\commentb Against max $= 2456352.4180 + 0.061999 E$.} \\
  \multicolumn{5}{l}{\commentc Number of points used to determine the maximum.} \\
\end{tabular}
\end{center}
\end{table}

\subsection{MASTER OT J081110.46$+$660008.5}\label{obj:j081110}

\begin{figure}
  \begin{center}
    \FigureFile(88mm,110mm){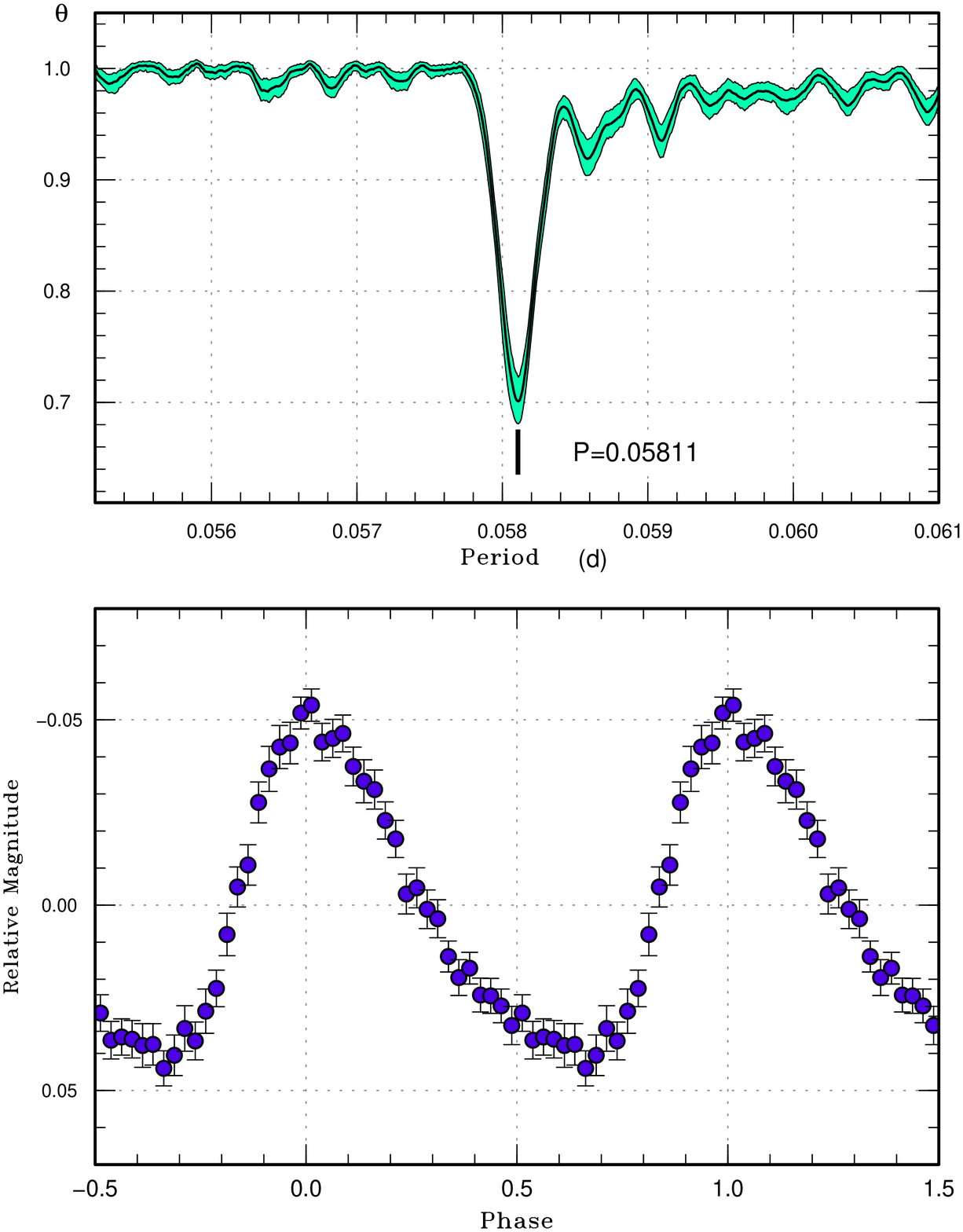}
  \end{center}
  \caption{Superhumps in MASTER J081110 (2012). (Upper): PDM analysis.
     (Lower): Phase-averaged profile.}
  \label{fig:j081110shpdm}
\end{figure}

   This object (hereafter MASTER J081110) was detected as
a large-amplitude transient by MASTER network \citep{den12j0811atel4506}.
The outburst amplitude of $\sim$8 mag and the SDSS color
(vsnet-alert 15037) were suggestive of a WZ Sge-type dwarf nova.
There were modulations resembling early superhumps in the early
follow-up observations (vsnet-alert 15041, 15045, 15049).
Ordinary superhumps grew soon after them (vsnet-alert 15055,
15063, 15066, 15075; figure \ref{fig:j081110shpdm}).
It was likely most of the early superhump phase was missed.
The times of superhump maxima are listed in table
\ref{tab:j081110oc2012}.   Stage A and B are clearly seen.
The $P_{\rm dot}$ was small [$+4.5(0.3) \times 10^{-5}$]
and there was no hint of a transition to stage C, both of
which are usual for a WZ Sge-type dwarf nova.
Although early superhumps were not unambiguously detected,
we consider this object to be a WZ Sge-type dwarf nova.

\begin{table}
\caption{Superhump maxima of MASTER J081110 (2012)}\label{tab:j081110oc2012}
\begin{center}
\begin{tabular}{ccccc}
\hline
$E$ & max\commenta & error & $O-C$\commentb & $N$\commentc \\
\hline
0 & 56224.4837 & 0.0015 & $-$0.0161 & 61 \\
16 & 56225.4249 & 0.0015 & $-$0.0056 & 57 \\
17 & 56225.4763 & 0.0012 & $-$0.0124 & 61 \\
23 & 56225.8338 & 0.0008 & $-$0.0039 & 55 \\
24 & 56225.8911 & 0.0004 & $-$0.0047 & 55 \\
25 & 56225.9512 & 0.0004 & $-$0.0027 & 55 \\
26 & 56226.0105 & 0.0005 & $-$0.0016 & 41 \\
29 & 56226.1872 & 0.0009 & 0.0005 & 96 \\
30 & 56226.2453 & 0.0007 & 0.0004 & 123 \\
31 & 56226.3051 & 0.0008 & 0.0021 & 123 \\
40 & 56226.8354 & 0.0003 & 0.0089 & 55 \\
41 & 56226.8922 & 0.0002 & 0.0075 & 55 \\
42 & 56226.9517 & 0.0002 & 0.0088 & 55 \\
43 & 56227.0103 & 0.0003 & 0.0092 & 43 \\
57 & 56227.8235 & 0.0003 & 0.0081 & 55 \\
58 & 56227.8808 & 0.0003 & 0.0072 & 55 \\
59 & 56227.9395 & 0.0004 & 0.0077 & 54 \\
60 & 56227.9969 & 0.0002 & 0.0070 & 55 \\
74 & 56228.8092 & 0.0004 & 0.0049 & 50 \\
75 & 56228.8670 & 0.0004 & 0.0046 & 55 \\
76 & 56228.9263 & 0.0003 & 0.0056 & 55 \\
77 & 56228.9829 & 0.0003 & 0.0042 & 55 \\
91 & 56229.7963 & 0.0006 & 0.0032 & 39 \\
92 & 56229.8529 & 0.0004 & 0.0015 & 55 \\
93 & 56229.9101 & 0.0004 & 0.0006 & 55 \\
94 & 56229.9674 & 0.0004 & $-$0.0003 & 55 \\
108 & 56230.7794 & 0.0079 & $-$0.0027 & 20 \\
109 & 56230.8396 & 0.0006 & $-$0.0006 & 43 \\
110 & 56230.8972 & 0.0004 & $-$0.0012 & 43 \\
111 & 56230.9548 & 0.0004 & $-$0.0017 & 43 \\
112 & 56231.0125 & 0.0005 & $-$0.0022 & 33 \\
125 & 56231.7705 & 0.0047 & $-$0.0004 & 16 \\
126 & 56231.8253 & 0.0006 & $-$0.0037 & 43 \\
127 & 56231.8824 & 0.0005 & $-$0.0048 & 43 \\
128 & 56231.9405 & 0.0005 & $-$0.0049 & 43 \\
129 & 56231.9987 & 0.0005 & $-$0.0049 & 44 \\
143 & 56232.8146 & 0.0006 & $-$0.0033 & 43 \\
144 & 56232.8731 & 0.0005 & $-$0.0030 & 43 \\
145 & 56232.9296 & 0.0006 & $-$0.0047 & 43 \\
146 & 56232.9889 & 0.0004 & $-$0.0035 & 43 \\
160 & 56233.8046 & 0.0005 & $-$0.0022 & 72 \\
161 & 56233.8604 & 0.0004 & $-$0.0046 & 70 \\
162 & 56233.9171 & 0.0004 & $-$0.0061 & 70 \\
163 & 56233.9759 & 0.0004 & $-$0.0054 & 70 \\
211 & 56236.7737 & 0.0085 & 0.0003 & 24 \\
212 & 56236.8307 & 0.0016 & $-$0.0009 & 36 \\
213 & 56236.8860 & 0.0012 & $-$0.0037 & 35 \\
214 & 56236.9437 & 0.0012 & $-$0.0042 & 35 \\
215 & 56237.0055 & 0.0032 & $-$0.0006 & 35 \\
229 & 56237.8236 & 0.0014 & 0.0031 & 35 \\
230 & 56237.8804 & 0.0010 & 0.0018 & 35 \\
231 & 56237.9414 & 0.0015 & 0.0046 & 35 \\
232 & 56237.9984 & 0.0013 & 0.0035 & 36 \\
246 & 56238.8205 & 0.0044 & 0.0111 & 36 \\
\hline
  \multicolumn{5}{l}{\commenta BJD$-$2400000.} \\
  \multicolumn{5}{l}{\commentb Against max $= 2456224.4998 + 0.058168 E$.} \\
  \multicolumn{5}{l}{\commentc Number of points used to determine the maximum.} \\
\end{tabular}
\end{center}
\end{table}

\subsection{MASTER OT J094759.83$+$061044.4}\label{obj:j094759}

\begin{figure}
  \begin{center}
    \FigureFile(88mm,110mm){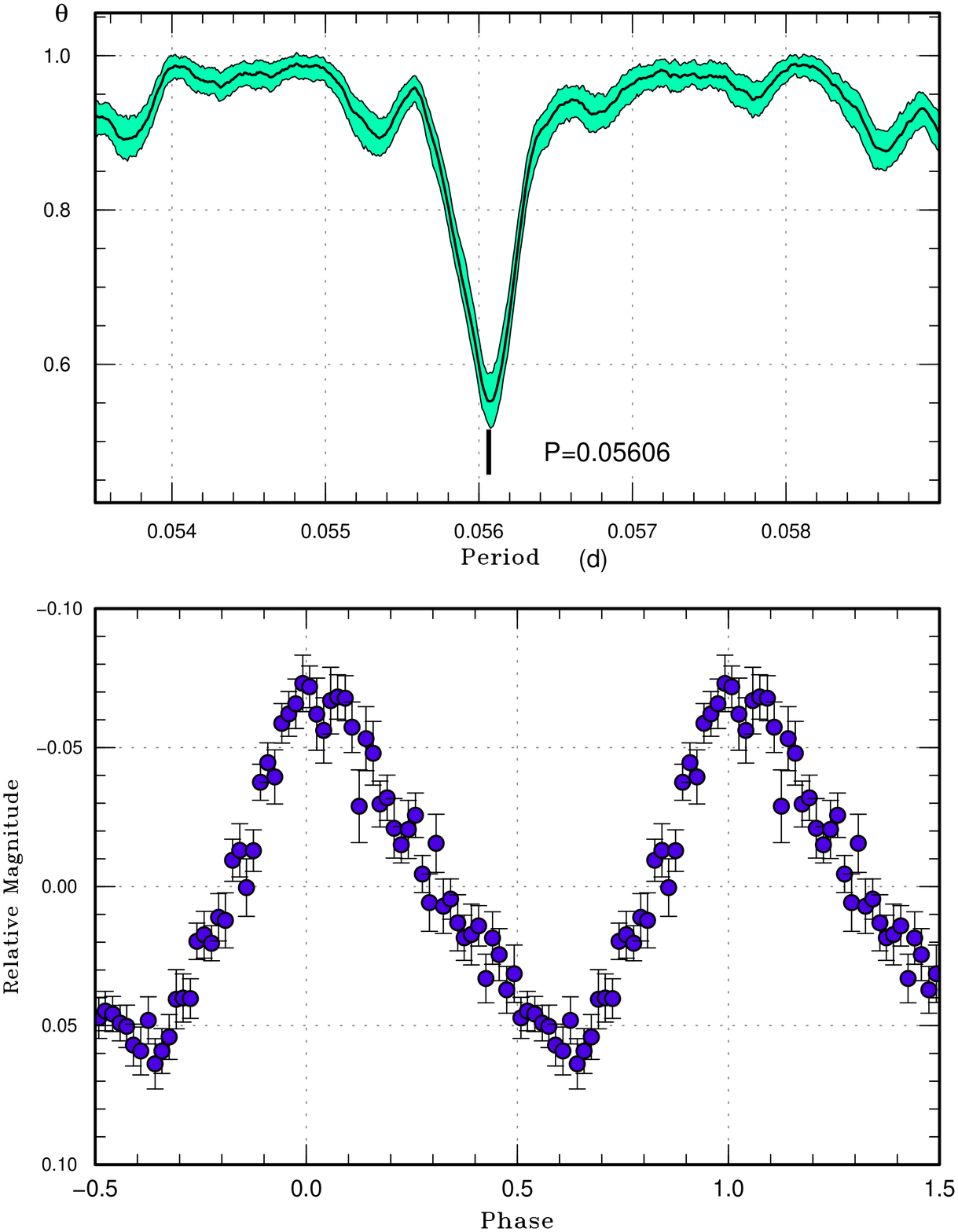}
  \end{center}
  \caption{Superhumps in MASTER J094759 (2013).
     (Upper): PDM analysis.
     (Lower): Phase-averaged profile.}
  \label{fig:j094759shpdm}
\end{figure}

   This object (hereafter MASTER J094759) was discovered by
MASTER network \citep{den13j0947atel5049} on 2013 May 4.
There is an $g=20.4$-mag SDSS counterpart.  There were two
previous fainter outbursts in the CRTS data.
Subsequent observations detected possible early superhumps
(vsnet-alert 15679, 15692).  After $\sim$5~d, ordinary superhumps
started to develop (vsnet-alert 15706; figure \ref{fig:j094759shpdm}).
The times of superhump maxima are listed in table
\ref{tab:j094759oc2013}.

\begin{figure}
  \begin{center}
    \FigureFile(88mm,110mm){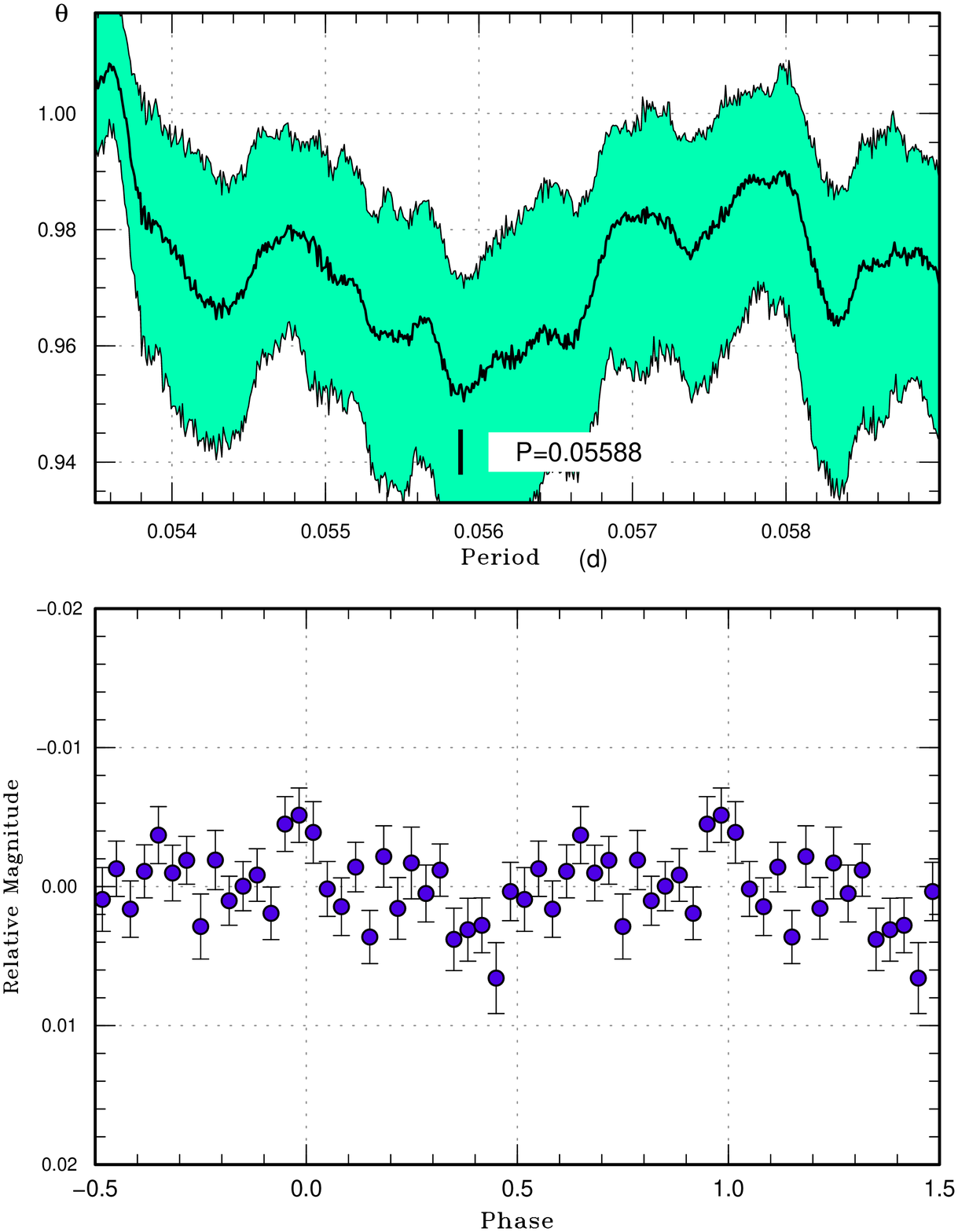}
  \end{center}
  \caption{Possible early superhumps in MASTER J094759 (2013).
     (Upper): PDM analysis.
     (Lower): Phase-averaged profile.}
  \label{fig:j094759eshpdm}
\end{figure}

   The amplitudes of early superhumps were very small and
it was extremely difficult to determine the period.
It was likely the signal was detected on only first two
nights of observations (before BJD 2456419.2;
figure \ref{fig:j094759eshpdm}).  We should bear in mind
that this period [0.05588(9)~d] has a large uncertainty 
due to the short observational baseline.

   The $O-C$ values of ordinary superhumps, however, were
well-defined and showed clear stages A and B.
The period of stage A superhumps was 0.05717(21)~d
and even if we assume a conservative error 0.004~d
in the period of early superhumps, $\epsilon^*$ for
stage A corresponds to 2.3(3)\%, and $q$ is estimated
to be 0.060(8).  The low value appears to be consistent
with the short $P_{\rm orb}$.

\begin{table}
\caption{Superhump maxima of MASTER J094759}\label{tab:j094759oc2013}
\begin{center}
\begin{tabular}{ccccc}
\hline
$E$ & max\commenta & error & $O-C$\commentb & $N$\commentc \\
\hline
0 & 56421.9647 & 0.0010 & $-$0.0154 & 72 \\
1 & 56422.0147 & 0.0009 & $-$0.0218 & 85 \\
6 & 56422.2959 & 0.0012 & $-$0.0218 & 61 \\
10 & 56422.5408 & 0.0020 & $-$0.0018 & 10 \\
11 & 56422.5913 & 0.0057 & $-$0.0076 & 11 \\
27 & 56423.5055 & 0.0007 & 0.0066 & 13 \\
28 & 56423.5595 & 0.0010 & 0.0044 & 13 \\
45 & 56424.5242 & 0.0005 & 0.0128 & 12 \\
46 & 56424.5817 & 0.0005 & 0.0141 & 11 \\
53 & 56424.9734 & 0.0002 & 0.0121 & 104 \\
54 & 56425.0295 & 0.0002 & 0.0119 & 105 \\
55 & 56425.0821 & 0.0011 & 0.0083 & 48 \\
71 & 56425.9817 & 0.0006 & 0.0079 & 144 \\
72 & 56426.0367 & 0.0007 & 0.0065 & 158 \\
73 & 56426.0904 & 0.0019 & 0.0041 & 48 \\
89 & 56426.9959 & 0.0032 & 0.0095 & 47 \\
90 & 56427.0412 & 0.0044 & $-$0.0014 & 28 \\
98 & 56427.4921 & 0.0012 & $-$0.0005 & 12 \\
99 & 56427.5486 & 0.0015 & $-$0.0003 & 13 \\
107 & 56427.9974 & 0.0005 & $-$0.0014 & 66 \\
113 & 56428.3409 & 0.0007 & 0.0045 & 37 \\
124 & 56428.9567 & 0.0026 & 0.0016 & 32 \\
125 & 56429.0049 & 0.0013 & $-$0.0064 & 32 \\
142 & 56429.9634 & 0.0004 & $-$0.0042 & 69 \\
143 & 56430.0169 & 0.0012 & $-$0.0070 & 69 \\
213 & 56433.9599 & 0.0088 & $-$0.0014 & 28 \\
214 & 56434.0042 & 0.0048 & $-$0.0134 & 55 \\
\hline
  \multicolumn{5}{l}{\commenta BJD$-$2400000.} \\
  \multicolumn{5}{l}{\commentb Against max $= 2456421.9802 + 0.056249 E$.} \\
  \multicolumn{5}{l}{\commentc Number of points used to determine the maximum.} \\
\end{tabular}
\end{center}
\end{table}

\subsection{MASTER OT J105025.99$+$332811.4}\label{obj:j105025}

   This object (hereafter MASTER J105025) was discovered by
MASTER network \citep{bal12j0301j1050atel4682} on 2012
December 25.  The object was already in outburst on December 19.
The quiescent SDSS colors suggest an orbital period of 0.07~d
using the method in \citet{kat12DNSDSS} (vsnet-alert 15216).
Although possible superhumps were reported (vsnet-alert 15228),
the duration of the observation was not sufficient to determine
the period.  Another observation 6~d later detected a clearer
superhump at BJD 2456299.0108(15) ($N=67$).  The period was
not determined.

\subsection{MASTER OT J111759.87$+$765131.6}\label{obj:j111759}

\begin{figure}
  \begin{center}
    \FigureFile(88mm,110mm){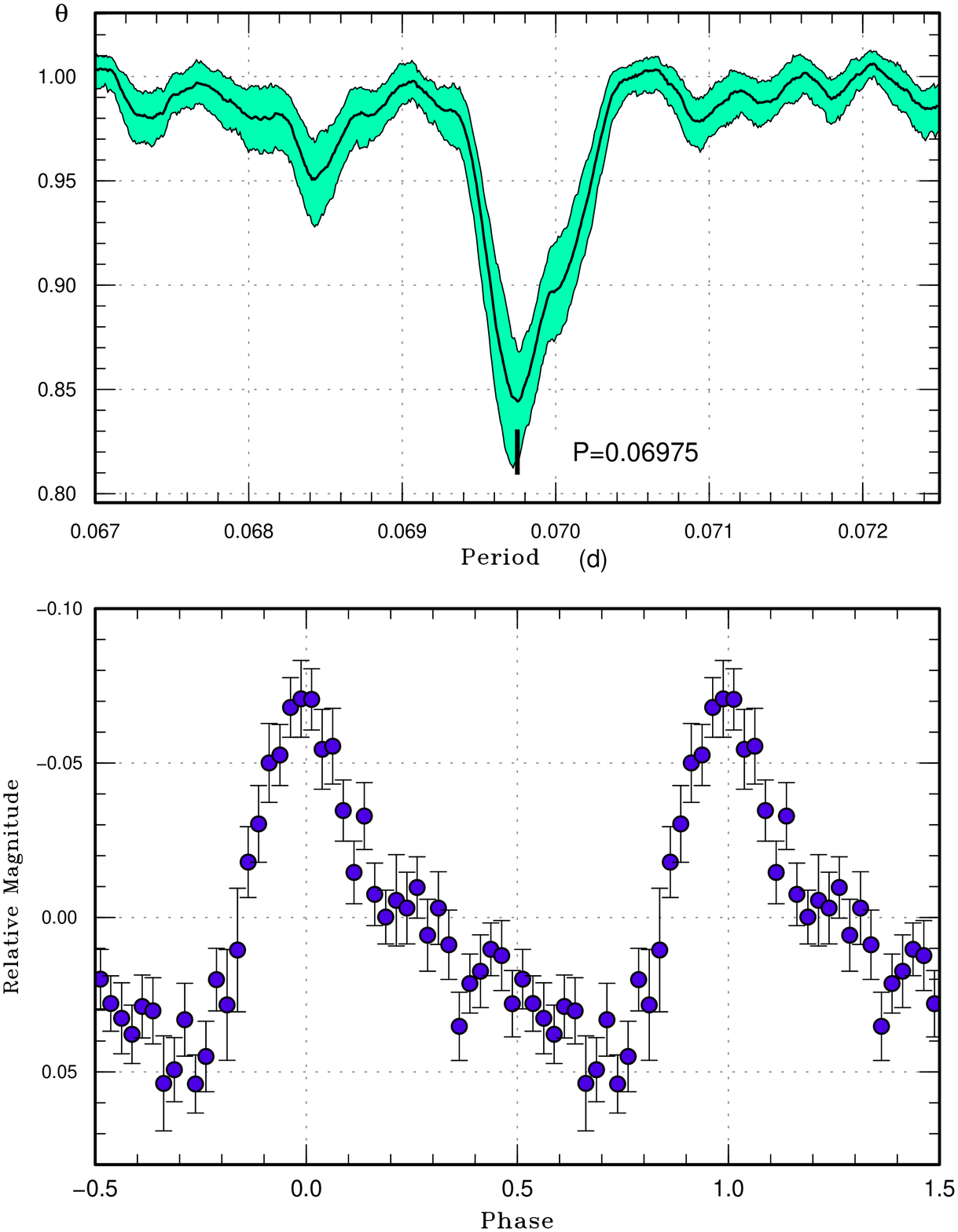}
  \end{center}
  \caption{Superhumps in MASTER J111759 (2013).
     (Upper): PDM analysis.
     (Lower): Phase-averaged profile.}
  \label{fig:j111759shpdm}
\end{figure}

   This object (hereafter MASTER J111759) was discovered by
MASTER network \citep{bal13j1117atel4956} on 2013 April 4.
Subsequent observations confirmed the presence of superhumps
(vsnet-alert 15598, 15610; figure \ref{fig:j111759shpdm}).
The times of superhump maxima are listed in table
\ref{tab:j111759oc2013}.
Most of our observations were obtained between April 12 and 18,
the late stage of the superoutburst.  The superhumps for 
$E \ge 100$ are thus likely stage C superhumps.  Although
$0 \le E \le 2$ may be stage B superhumps, it was impossible
to determine the period.

\begin{table}
\caption{Superhump maxima of MASTER J111759 (2013)}\label{tab:j111759oc2013}
\begin{center}
\begin{tabular}{ccccc}
\hline
$E$ & max\commenta & error & $O-C$\commentb & $N$\commentc \\
\hline
0 & 56388.5388 & 0.0026 & $-$0.0004 & 54 \\
1 & 56388.6119 & 0.0011 & 0.0031 & 84 \\
2 & 56388.6826 & 0.0008 & 0.0041 & 83 \\
100 & 56395.5017 & 0.0007 & $-$0.0016 & 56 \\
101 & 56395.5703 & 0.0006 & $-$0.0027 & 63 \\
102 & 56395.6397 & 0.0005 & $-$0.0030 & 68 \\
115 & 56396.5466 & 0.0006 & $-$0.0014 & 73 \\
116 & 56396.6142 & 0.0007 & $-$0.0034 & 72 \\
117 & 56396.6868 & 0.0013 & $-$0.0005 & 40 \\
127 & 56397.3774 & 0.0012 & $-$0.0062 & 73 \\
128 & 56397.4530 & 0.0009 & $-$0.0003 & 53 \\
129 & 56397.5179 & 0.0024 & $-$0.0051 & 67 \\
130 & 56397.5924 & 0.0008 & $-$0.0002 & 67 \\
131 & 56397.6608 & 0.0013 & $-$0.0014 & 32 \\
155 & 56399.3368 & 0.0013 & 0.0032 & 69 \\
156 & 56399.4032 & 0.0017 & $-$0.0001 & 53 \\
158 & 56399.5456 & 0.0010 & 0.0030 & 140 \\
159 & 56399.6214 & 0.0019 & 0.0092 & 67 \\
169 & 56400.3131 & 0.0020 & 0.0045 & 47 \\
170 & 56400.3793 & 0.0024 & 0.0011 & 69 \\
171 & 56400.4440 & 0.0045 & $-$0.0039 & 71 \\
172 & 56400.5196 & 0.0042 & 0.0020 & 64 \\
\hline
  \multicolumn{5}{l}{\commenta BJD$-$2400000.} \\
  \multicolumn{5}{l}{\commentb Against max $= 2456388.5392 + 0.069642 E$.} \\
  \multicolumn{5}{l}{\commentc Number of points used to determine the maximum.} \\
\end{tabular}
\end{center}
\end{table}

\subsection{MASTER OT J165236.22$+$460513.2}\label{obj:j165236}

\begin{figure}
  \begin{center}
    \FigureFile(88mm,110mm){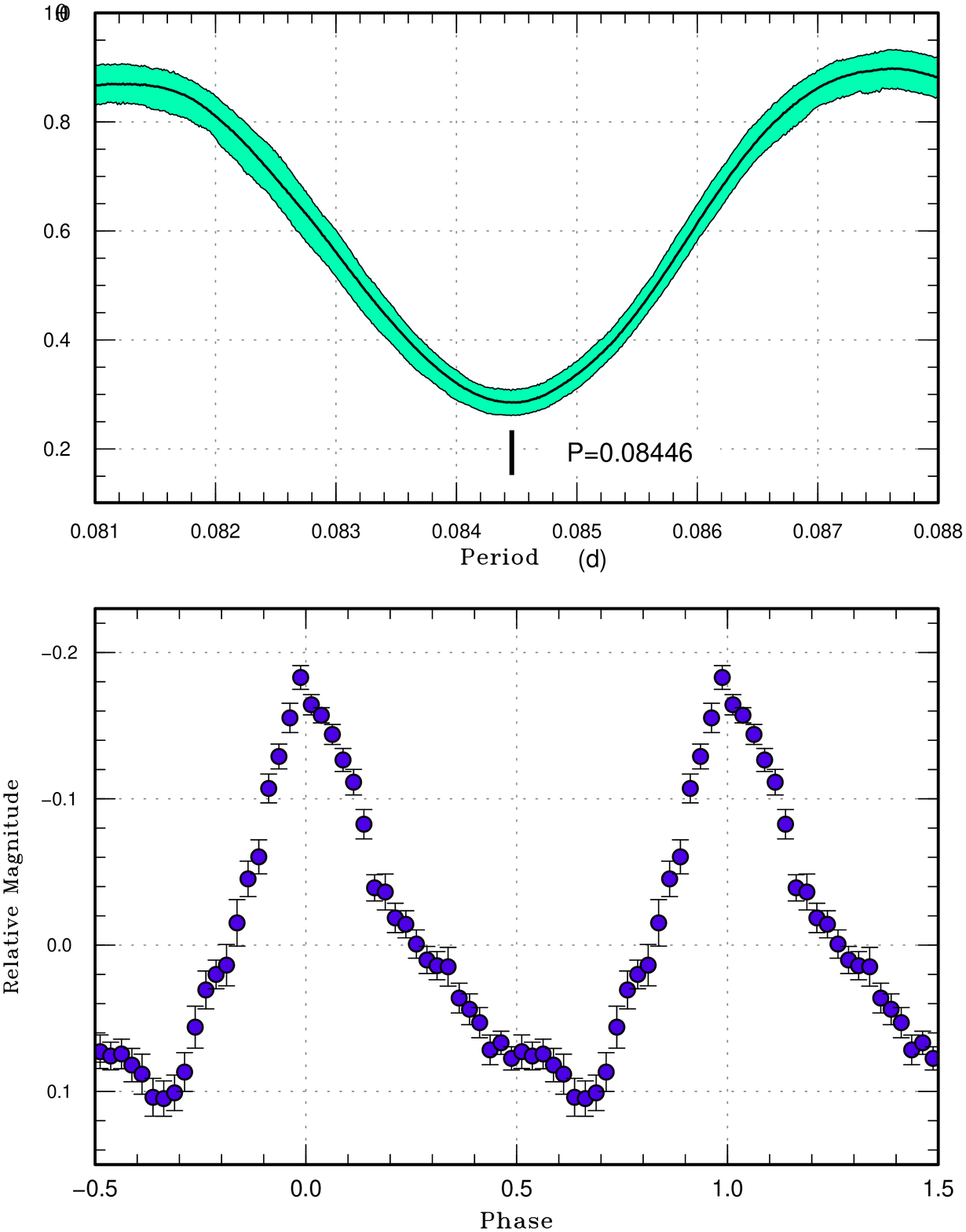}
  \end{center}
  \caption{Superhumps in MASTER J165236 (2013).
     The range of the data was limited to BJD 2456367.2--2456368.7,
     when stage B superhumps were clearly detected.
     (Upper): PDM analysis.
     (Lower): Phase-averaged profile.}
  \label{fig:j165236shpdm}
\end{figure}

   This object (hereafter MASTER J165236) was discovered by
MASTER network \citep{den13j1652atel4881} on 2013 March 12.
The object was suggested to be a WZ Sge-type dwarf nova
due to the large (larger than 7 mag) outburst amplitude
(vsnet-alert 15493).
The quiescent SDSS counterpart ($g=22.1$) has a red color
($g-z=+1.3$), unlike a WZ Sge-type object (vsnet-alert 15496).
The evolution of superhumps was soon detected
(vsnet-alert 15506, 15519, 15535).
The period of superhumps ($\sim$0.084~d) was also unusually
long for a WZ Sge-type dwarf nova.
The times of superhump maxima are listed in table
\ref{tab:j165236oc2013}.  Although the growing stage of the
superhumps was detected, the period of stage A superhumps could
not be determined due to the lack of observations.
The mean profile of stage B superhumps is shown
in figure \ref{fig:j165236shpdm}.
After BJD 2456368, the amplitudes of superhumps became smaller
and the times of maxima were not meaningfully determined.
Although the initial evolution of the superhumps resembled
those of ordinary long-$P_{\rm orb}$ SU UMa-type dwarf novae,
the rapid decay of the superhump amplitude was unusual.
Although the object might be a period bouncer, we do not have
additional data to check this possibility.  Detailed observations
are desired.

\begin{table}
\caption{Superhump maxima of MASTER J165236 (2013)}\label{tab:j165236oc2013}
\begin{center}
\begin{tabular}{ccccc}
\hline
$E$ & max\commenta & error & $O-C$\commentb & $N$\commentc \\
\hline
0 & 56365.8839 & 0.0009 & $-$0.0124 & 64 \\
1 & 56365.9593 & 0.0009 & $-$0.0228 & 77 \\
11 & 56366.8532 & 0.0006 & 0.0134 & 75 \\
12 & 56366.9379 & 0.0007 & 0.0124 & 79 \\
13 & 56367.0282 & 0.0014 & 0.0169 & 32 \\
15 & 56367.1933 & 0.0033 & 0.0105 & 69 \\
16 & 56367.2763 & 0.0005 & 0.0077 & 176 \\
28 & 56368.2952 & 0.0004 & $-$0.0026 & 178 \\
29 & 56368.3774 & 0.0005 & $-$0.0062 & 71 \\
30 & 56368.4646 & 0.0004 & $-$0.0048 & 87 \\
31 & 56368.5473 & 0.0005 & $-$0.0078 & 85 \\
32 & 56368.6367 & 0.0009 & $-$0.0042 & 41 \\
\hline
  \multicolumn{5}{l}{\commenta BJD$-$2400000.} \\
  \multicolumn{5}{l}{\commentb Against max $= 2456365.8963 + 0.085767 E$.} \\
  \multicolumn{5}{l}{\commentc Number of points used to determine the maximum.} \\
\end{tabular}
\end{center}
\end{table}

\subsection{MASTER OT J174902.10$+$191331.2}\label{obj:j174902}

\begin{figure}
  \begin{center}
    \FigureFile(88mm,110mm){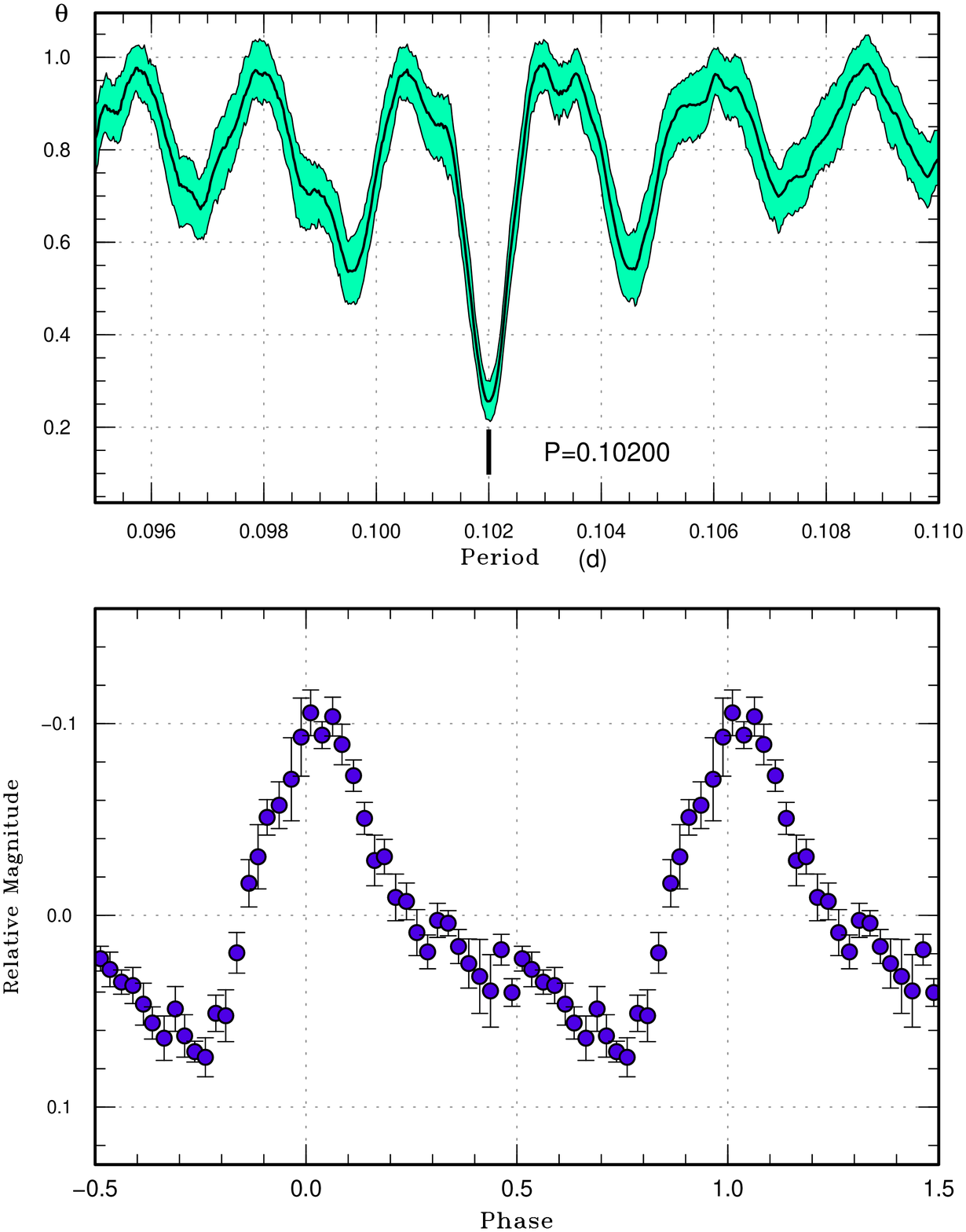}
  \end{center}
  \caption{Ordinary superhumps in MASTER J174902 (2013).
     (Upper): PDM analysis.
     (Lower): Phase-averaged profile.}
  \label{fig:j1749shpdm}
\end{figure}

   This object (hereafter MASTER J174902) was originally discovered by
MASTER network \citep{den12j1749atel4324} on 2012 August 20.
\citet{nes12j1739atel4330} reported a pre-discovery spectrum
on the Digitized First Byurakan Survey and suggested that there
was a past outburst in 1974.  The 2012 outburst was a brief one
and it faded rapidly (R. Pickard, see vsnet-alert 16001).
The object was again detected in outburst on 2013 July 14
by ASAS-SN survey.  The outburst had been detected by C. Chiselbrook
(AAVSO) two days earlier.  Subsequent observations detected
superhumps (vsnet-alert 16044, 16061, 16072; 
figure \ref{fig:j1749shpdm})
and it immediately became apparent that this object is located
in the period gap.
The times of superhump maxima are listed in
table \ref{tab:j191331oc2013}.

\begin{table}
\caption{Superhump maxima of MASTER J174902 (2013)}\label{tab:j174902oc2013}
\begin{center}
\begin{tabular}{ccccc}
\hline
$E$ & max\commenta & error & $O-C$\commentb & $N$\commentc \\
\hline
0 & 56496.4480 & 0.0008 & $-$0.0001 & 49 \\
1 & 56496.5502 & 0.0012 & 0.0002 & 33 \\
30 & 56499.5048 & 0.0017 & $-$0.0005 & 35 \\
38 & 56500.3182 & 0.0042 & $-$0.0023 & 22 \\
39 & 56500.4249 & 0.0006 & 0.0025 & 50 \\
40 & 56500.5244 & 0.0010 & 0.0001 & 48 \\
\hline
  \multicolumn{5}{l}{\commenta BJD$-$2400000.} \\
  \multicolumn{5}{l}{\commentb Against max $= 2456496.4480 + 0.101908 E$.} \\
  \multicolumn{5}{l}{\commentc Number of points used to determine the maximum.} \\
\end{tabular}
\end{center}
\end{table}

\subsection{MASTER OT J181953.76$+$361356.5}\label{obj:j181953}

\begin{figure}
  \begin{center}
    \FigureFile(88mm,110mm){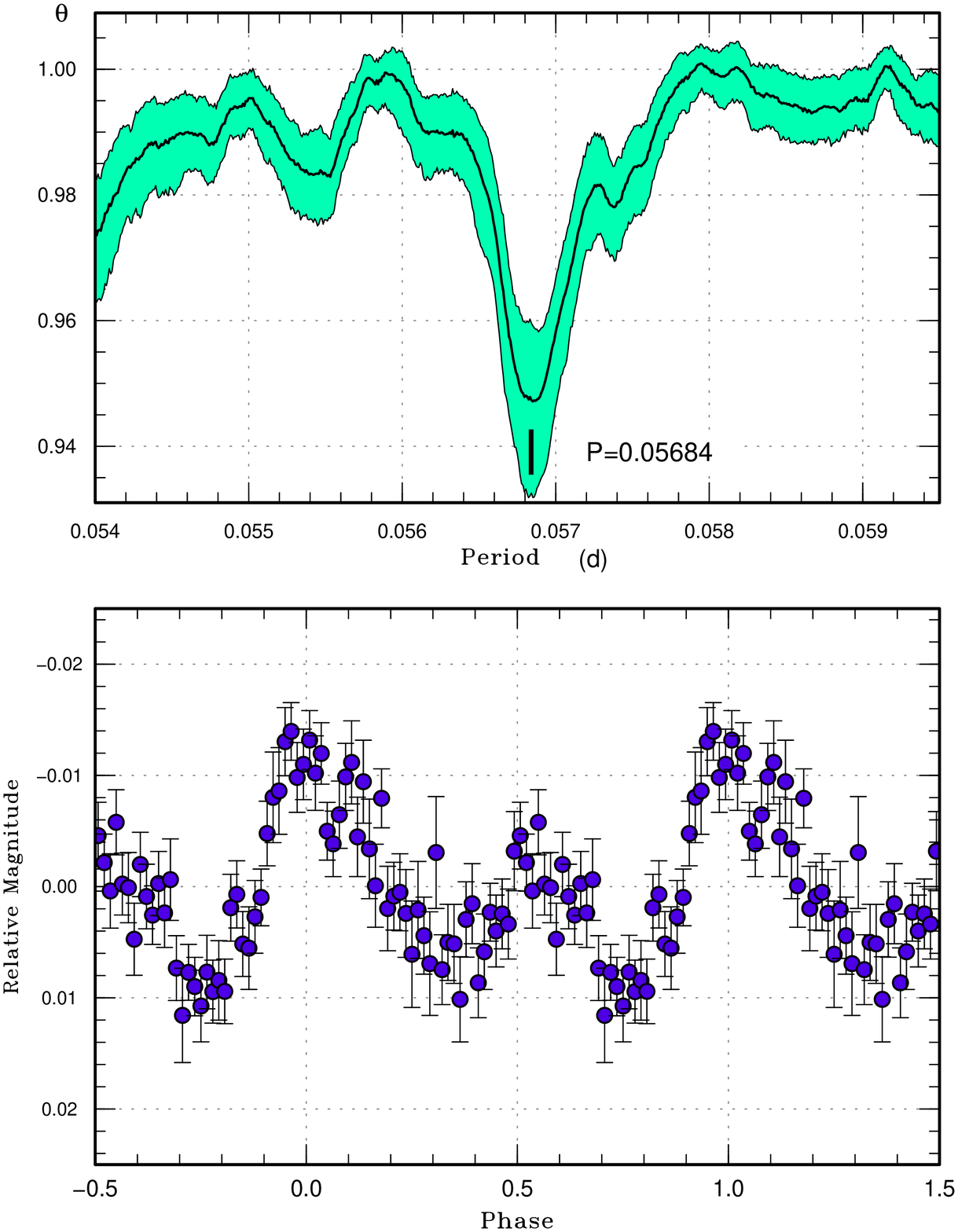}
  \end{center}
  \caption{Early superhumps in MASTER J181953 (2013).
     (Upper): PDM analysis.
     (Lower): Phase-averaged profile.}
  \label{fig:j1819eshpdm}
\end{figure}

\begin{figure}
  \begin{center}
    \FigureFile(88mm,110mm){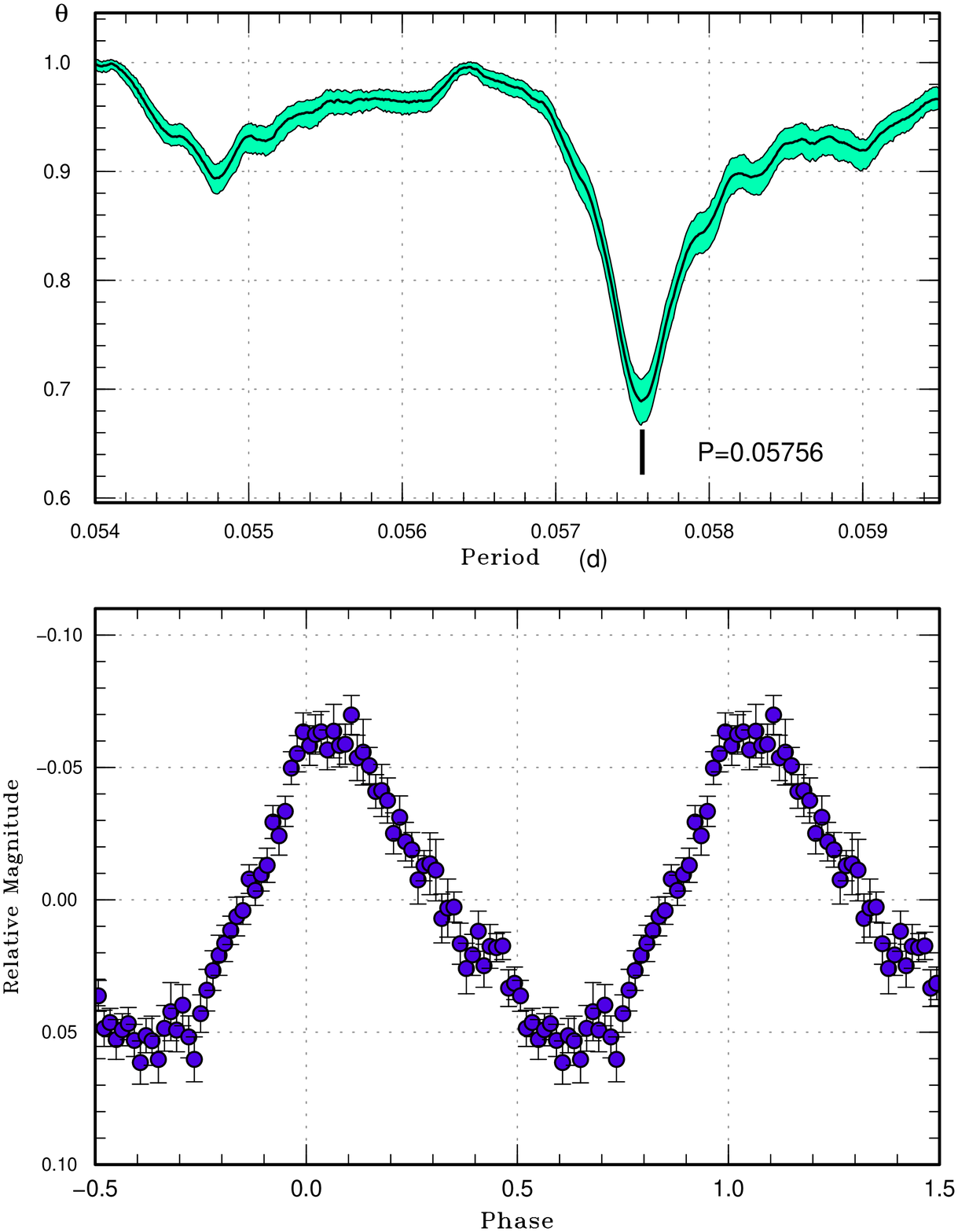}
  \end{center}
  \caption{Ordinary superhumps in MASTER J181953 (2013).
     (Upper): PDM analysis.
     (Lower): Phase-averaged profile.}
  \label{fig:j1819shpdm}
\end{figure}

   This object (hereafter MASTER J181953) was discovered by
MASTER network \citep{shu13j1819atel5196} 
on 2013 July 5 at an unfiltered CCD magnitude
of 13.9.  The quiescent counterpart has a magnitude of 21.6.
The large outburst amplitude was suggestive of a WZ Sge-type 
dwarf nova.  Subsequent observations detected likely early
superhumps (vsnet-alert 15929, 15940; figure \ref{fig:j1819eshpdm}).
Ordinary superhump then appeared (vsnet-alert 15950, 15956, 15971, 
15977, 15987, 15994, 16014, 16023; figure \ref{fig:j1819shpdm}).
The object started fading rapidly
14~d after the initial detection (vsnet-alert 16052).

\begin{figure}
  \begin{center}
    \FigureFile(88mm,70mm){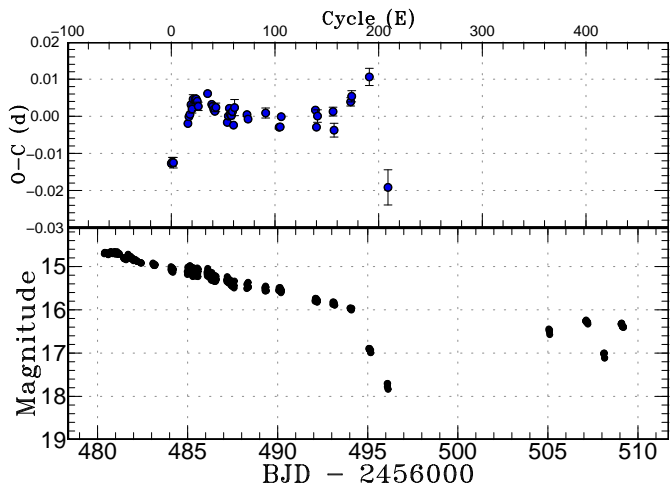}
  \end{center}
  \caption{$O-C$ diagram of superhumps in MASTER J181953 (2013).
     (Upper): $O-C$ diagram.  A period of 0.057549~d
     was used to draw this figure.
     (Lower): Light curve.  The observations were binned to 0.012~d.}
  \label{fig:j1819humpall}
\end{figure}

   The period of early superhump was determined to be 0.05684(2)~d
(figure \ref{fig:j1819eshpdm}).
The times of superhump maxima are listed in table
\ref{tab:j181953oc2013}.  There were clear stage A and B superhumps
(figure \ref{fig:j1819humpall}).
In determining $P_{\rm dot}$ of stage B superhumps, we excluded
the part of the rapid decline ($E \ge 173$), as in ASASSN-13ax.
The last epoch $E=209$ appeared
to show a phase reversal, whose origin in WZ Sge-type dwarf novae
is still unclear.
Using the period of stage A superhump ($E \le 21$), we could
obtain $\epsilon^*$=0.0259(3).  This value corresponds to
$q$=0.069(1).

   The object showed an oscillating type long rebrightening
similar to WZ Sge (2001) and OT J012059.6$+$325545 \citep{Pdot3}
(figure \ref{fig:j1819humpall}).

\begin{table}
\caption{Superhump maxima of MASTER J181953 (2013)}\label{tab:j181953oc2013}
\begin{center}
\begin{tabular}{ccccc}
\hline
$E$ & max\commenta & error & $O-C$\commentb & $N$\commentc \\
\hline
0 & 56484.0826 & 0.0008 & $-$0.0127 & 167 \\
1 & 56484.1405 & 0.0012 & $-$0.0123 & 199 \\
2 & 56484.1979 & 0.0014 & $-$0.0125 & 86 \\
16 & 56485.0141 & 0.0005 & $-$0.0019 & 115 \\
17 & 56485.0735 & 0.0004 & $-$0.0001 & 165 \\
18 & 56485.1317 & 0.0003 & 0.0006 & 275 \\
19 & 56485.1919 & 0.0004 & 0.0032 & 183 \\
20 & 56485.2482 & 0.0011 & 0.0020 & 92 \\
21 & 56485.3084 & 0.0012 & 0.0046 & 49 \\
23 & 56485.4234 & 0.0006 & 0.0045 & 64 \\
24 & 56485.4811 & 0.0005 & 0.0047 & 64 \\
25 & 56485.5381 & 0.0004 & 0.0041 & 64 \\
26 & 56485.5942 & 0.0011 & 0.0027 & 38 \\
35 & 56486.1156 & 0.0005 & 0.0062 & 122 \\
39 & 56486.3429 & 0.0006 & 0.0032 & 61 \\
40 & 56486.4000 & 0.0005 & 0.0027 & 62 \\
41 & 56486.4566 & 0.0005 & 0.0019 & 58 \\
42 & 56486.5137 & 0.0005 & 0.0014 & 61 \\
43 & 56486.5723 & 0.0012 & 0.0024 & 55 \\
54 & 56487.2013 & 0.0004 & $-$0.0016 & 81 \\
55 & 56487.2606 & 0.0008 & 0.0001 & 65 \\
56 & 56487.3202 & 0.0008 & 0.0022 & 63 \\
57 & 56487.3761 & 0.0012 & 0.0005 & 50 \\
58 & 56487.4334 & 0.0006 & 0.0003 & 50 \\
59 & 56487.4920 & 0.0005 & 0.0013 & 61 \\
60 & 56487.5459 & 0.0009 & $-$0.0023 & 54 \\
61 & 56487.6081 & 0.0022 & 0.0024 & 30 \\
73 & 56488.2968 & 0.0007 & 0.0005 & 54 \\
74 & 56488.3532 & 0.0009 & $-$0.0007 & 67 \\
91 & 56489.3331 & 0.0013 & 0.0009 & 63 \\
104 & 56490.0774 & 0.0008 & $-$0.0029 & 64 \\
105 & 56490.1351 & 0.0005 & $-$0.0028 & 65 \\
106 & 56490.1954 & 0.0008 & $-$0.0001 & 26 \\
139 & 56492.0963 & 0.0007 & 0.0017 & 58 \\
140 & 56492.1492 & 0.0010 & $-$0.0029 & 53 \\
141 & 56492.2098 & 0.0017 & 0.0001 & 28 \\
156 & 56493.0742 & 0.0012 & 0.0013 & 53 \\
157 & 56493.1267 & 0.0019 & $-$0.0037 & 55 \\
173 & 56494.0552 & 0.0011 & 0.0039 & 60 \\
174 & 56494.1142 & 0.0015 & 0.0054 & 35 \\
191 & 56495.0978 & 0.0023 & 0.0107 & 42 \\
209 & 56496.1039 & 0.0047 & $-$0.0191 & 38 \\
\hline
  \multicolumn{5}{l}{\commenta BJD$-$2400000.} \\
  \multicolumn{5}{l}{\commentb Against max $= 2456484.0953 + 0.057549 E$.} \\
  \multicolumn{5}{l}{\commentc Number of points used to determine the maximum.} \\
\end{tabular}
\end{center}
\end{table}

\subsection{MASTER OT J212624.16$+$253827.2}\label{obj:j212624}

\begin{figure}
  \begin{center}
    \FigureFile(88mm,110mm){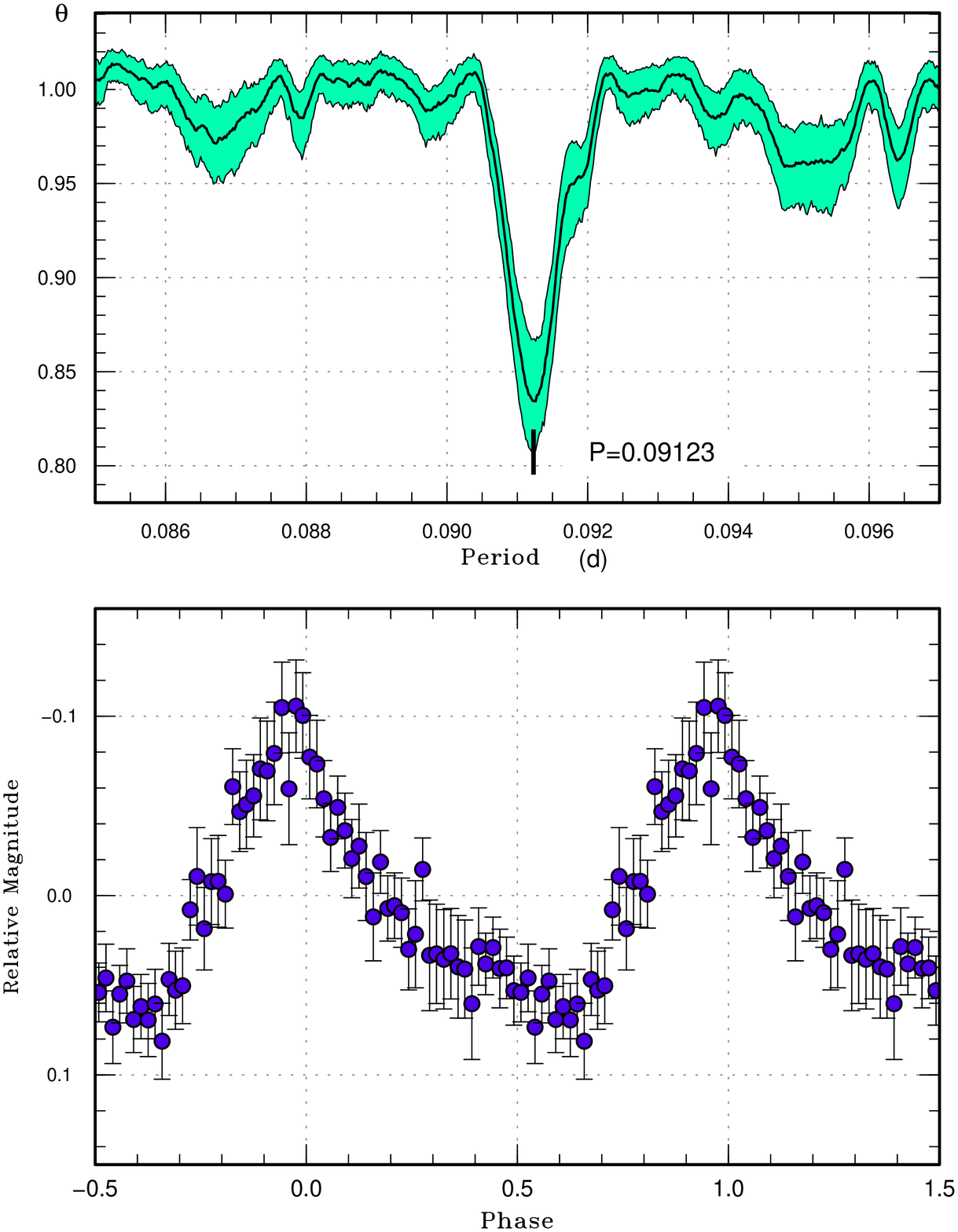}
  \end{center}
  \caption{Superhumps in MASTER J212624 (2013).
     (Upper): PDM analysis.
     (Lower): Phase-averaged profile.}
  \label{fig:j212624shpdm}
\end{figure}

   This object (hereafter MASTER J212624) was discovered by
MASTER network \citep{den13j2126atel5111} 
on 2013 June 6 at an unfiltered CCD magnitude
of 14.1.  The object underwent an even brighter (13.8 mag)
outburst in 2012 December.  The SU UMa-type nature was
suggested from the large outburst amplitude.
Subsequent observations confirmed the presence of superhumps
(vsnet-alert 15813, 15821, 15829, 15842; figure \ref{fig:j212624shpdm}).
The times of superhump maxima are listed in table
\ref{tab:j212624oc2013}.  The observation recorded mostly
stage B superhumps with a large positive $P_{\rm dot}$ of
$+29(4) \times 10^{-5}$.  It was likely only the later part
of the superoutburst was recorded; this is consistent with
the fainter recorded maximum than in 2012 December.
Such a large positive $P_{\rm dot}$
is rather unusual for such a long-$P_{\rm SH}$ object.
There have been a small number of similar objects
(GX Cas: \cite{Pdot3}; OT J145921.8$+$354806: \cite{Pdot4})
showing a large positive $P_{\rm dot}$.

\begin{table}
\caption{Superhump maxima of MASTER J212624 (2013)}\label{tab:j212624oc2013}
\begin{center}
\begin{tabular}{ccccc}
\hline
$E$ & max\commenta & error & $O-C$\commentb & $N$\commentc \\
\hline
0 & 56451.5930 & 0.0008 & 0.0018 & 75 \\
9 & 56452.4186 & 0.0004 & 0.0054 & 64 \\
10 & 56452.5092 & 0.0006 & 0.0047 & 65 \\
20 & 56453.4195 & 0.0006 & 0.0016 & 64 \\
21 & 56453.5100 & 0.0005 & 0.0008 & 54 \\
31 & 56454.4169 & 0.0076 & $-$0.0056 & 24 \\
32 & 56454.5099 & 0.0007 & $-$0.0040 & 55 \\
42 & 56455.4222 & 0.0005 & $-$0.0051 & 65 \\
43 & 56455.5109 & 0.0006 & $-$0.0077 & 48 \\
53 & 56456.4268 & 0.0007 & $-$0.0051 & 95 \\
54 & 56456.5232 & 0.0010 & $-$0.0001 & 108 \\
64 & 56457.4418 & 0.0011 & 0.0052 & 134 \\
65 & 56457.5297 & 0.0012 & 0.0018 & 45 \\
75 & 56458.4332 & 0.0017 & $-$0.0081 & 90 \\
76 & 56458.5422 & 0.0020 & 0.0095 & 51 \\
86 & 56459.4508 & 0.0019 & 0.0048 & 44 \\
\hline
  \multicolumn{5}{l}{\commenta BJD$-$2400000.} \\
  \multicolumn{5}{l}{\commentb Against max $= 2456451.5911 + 0.091335 E$.} \\
  \multicolumn{5}{l}{\commentc Number of points used to determine the maximum.} \\
\end{tabular}
\end{center}
\end{table}

\subsection{OT J112619.4$+$084651}\label{obj:j112619}

\begin{figure}
  \begin{center}
    \FigureFile(88mm,110mm){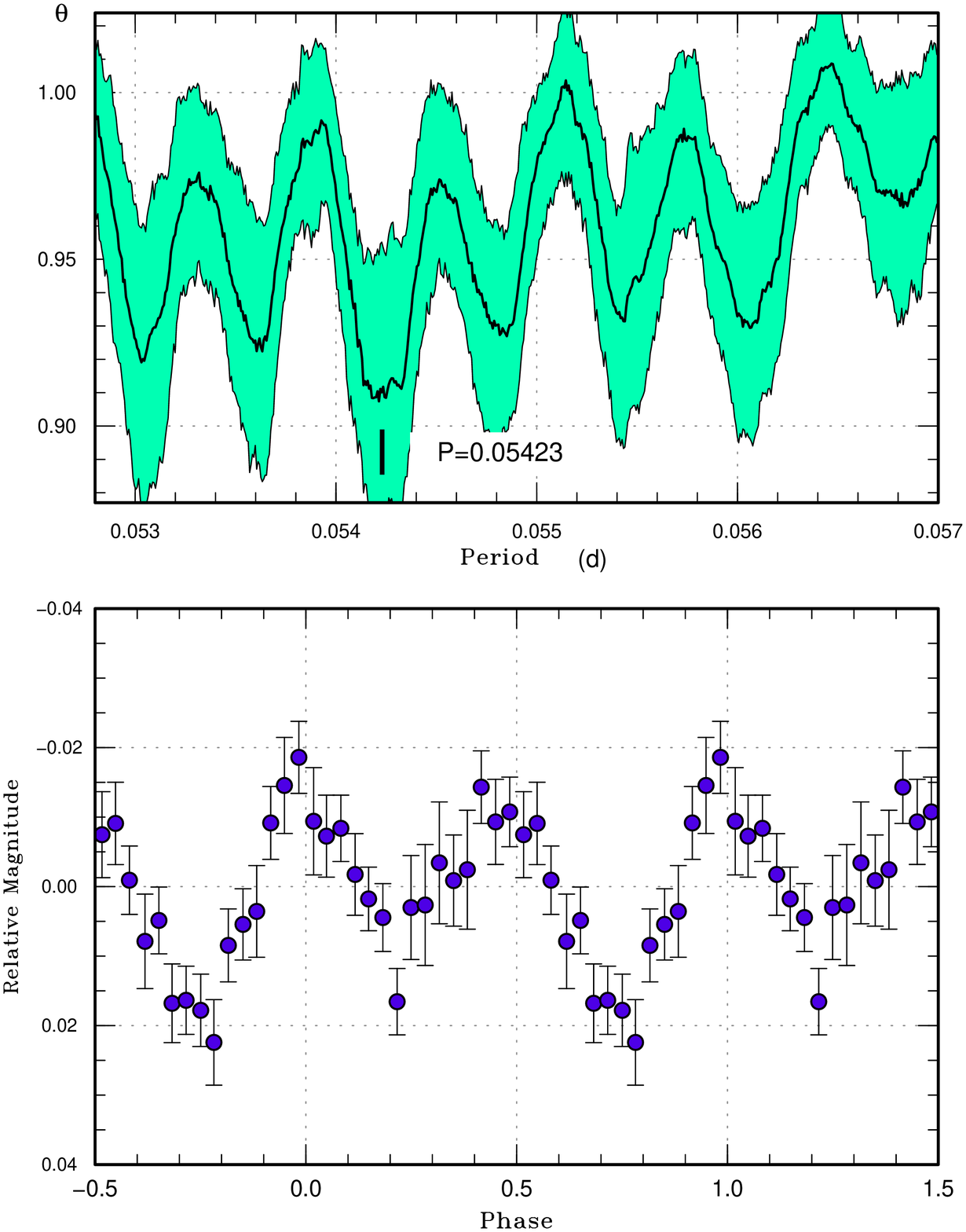}
  \end{center}
  \caption{Early superhumps in OT J112619 (2013). (Upper): PDM analysis.y
     (Lower): Phase-averaged profile.}
  \label{fig:j112619eshpdm}
\end{figure}

\begin{figure}
  \begin{center}
    \FigureFile(88mm,110mm){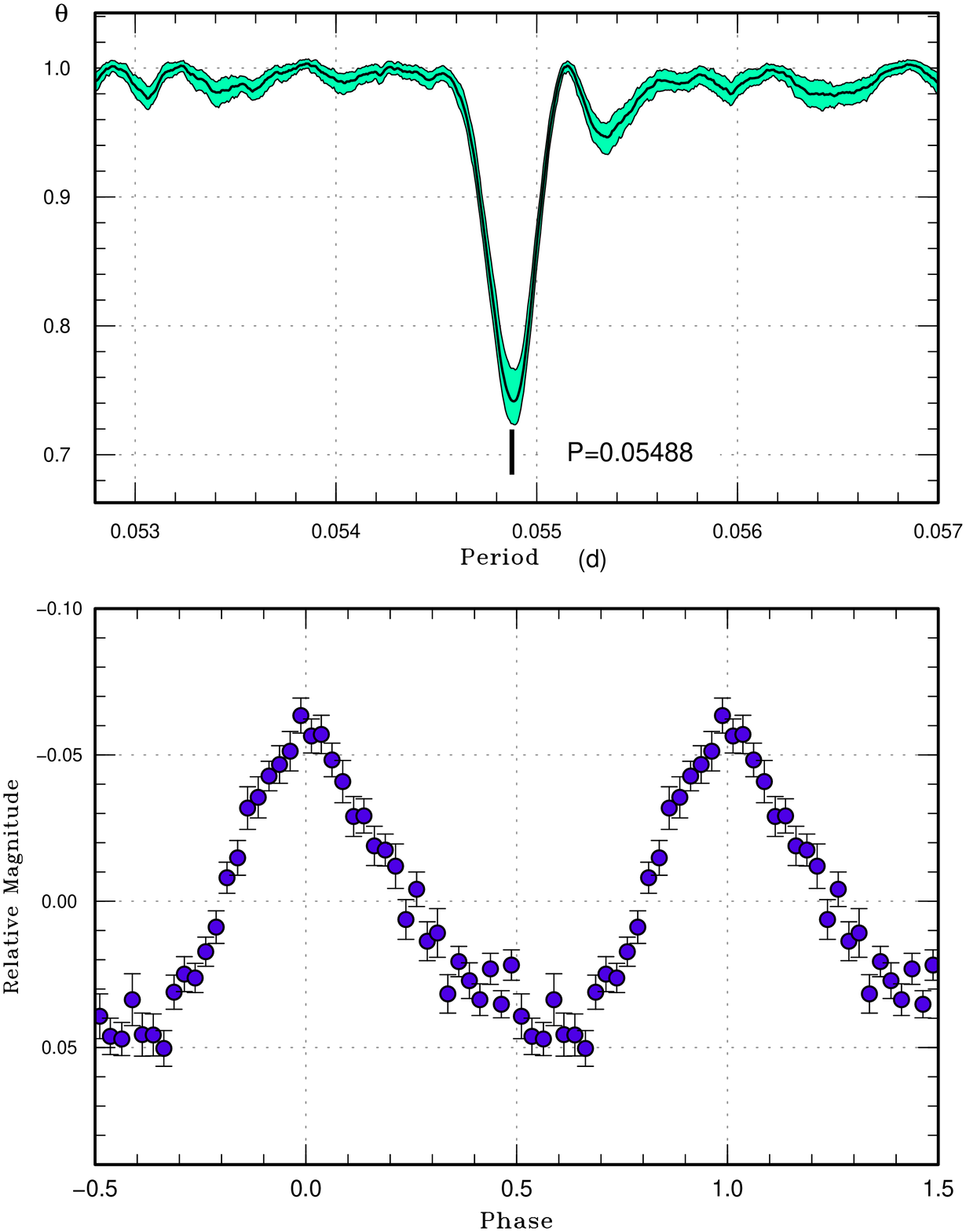}
  \end{center}
  \caption{Superhumps in OT J112619 (2013). (Upper): PDM analysis.
     (Lower): Phase-averaged profile.}
  \label{fig:j112619shpdm}
\end{figure}

\begin{figure}
  \begin{center}
    \FigureFile(88mm,70mm){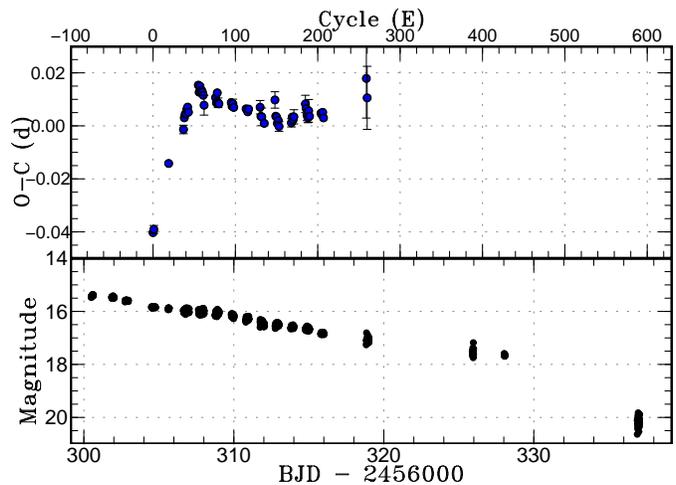}
  \end{center}
  \caption{$O-C$ diagram of superhumps in OT J112619 (2013).
     (Upper): $O-C$ diagram.  A period of 0.05492~d
     was used to draw this figure.
     (Lower): Light curve.  The observations were binned to 0.0055~d.}
  \label{fig:j112619humpall}
\end{figure}

   This is a dwarf nova discovered by CRTS
(=CSS130106:112619$+$084651, hereafter OT J112619) on 2013 January 6
at a magnitude of 14.8.  There was no past outburst in the CRTS data.
The quiescent counterpart is very faint ($g$=21.8), and the outburst
amplitude immediately suggested a WZ Sge-type dwarf nova
(vsnet-alert 15246).  Early observation detected double-wave
early superhumps (vsnet-alert 15249, 15252, 15264; figure
\ref{fig:j112619eshpdm}).  The object started to show ordinary
superhumps on January 12 (figure \ref{fig:j112619shpdm}).
The times of superhump maxima are listed in table
\ref{tab:j112619oc2013}.  The $O-C$ diagram very clearly
shows stage A ($E \le 43$) and stage B ($55 \le E \le 260$)
(figure \ref{fig:j112619humpall}).
The value of $\varepsilon^*$ for stage A superhumps was 0.0317(6),
which corresponds to $q$=0.086(2).

\begin{table}
\caption{Superhump maxima of OT J112619 (2013)}\label{tab:j112619oc2013}
\begin{center}
\begin{tabular}{ccccc}
\hline
$E$ & max\commenta & error & $O-C$\commentb & $N$\commentc \\
\hline
0 & 56304.5700 & 0.0015 & $-$0.0402 & 39 \\
1 & 56304.6261 & 0.0015 & $-$0.0391 & 26 \\
19 & 56305.6396 & 0.0011 & $-$0.0149 & 59 \\
37 & 56306.6410 & 0.0016 & $-$0.0028 & 36 \\
38 & 56306.7003 & 0.0007 & 0.0016 & 50 \\
39 & 56306.7566 & 0.0009 & 0.0029 & 59 \\
40 & 56306.8117 & 0.0009 & 0.0030 & 24 \\
41 & 56306.8682 & 0.0009 & 0.0046 & 25 \\
42 & 56306.9240 & 0.0007 & 0.0054 & 23 \\
43 & 56306.9770 & 0.0007 & 0.0035 & 28 \\
55 & 56307.6463 & 0.0009 & 0.0132 & 35 \\
56 & 56307.6985 & 0.0006 & 0.0105 & 51 \\
57 & 56307.7557 & 0.0006 & 0.0127 & 59 \\
58 & 56307.8083 & 0.0008 & 0.0103 & 25 \\
59 & 56307.8640 & 0.0010 & 0.0111 & 24 \\
60 & 56307.9182 & 0.0009 & 0.0103 & 22 \\
61 & 56307.9720 & 0.0009 & 0.0091 & 25 \\
62 & 56308.0231 & 0.0037 & 0.0053 & 12 \\
76 & 56308.7949 & 0.0010 & 0.0077 & 71 \\
77 & 56308.8478 & 0.0014 & 0.0056 & 79 \\
78 & 56308.9065 & 0.0010 & 0.0094 & 68 \\
79 & 56308.9577 & 0.0018 & 0.0056 & 30 \\
80 & 56309.0122 & 0.0015 & 0.0052 & 20 \\
95 & 56309.8365 & 0.0005 & 0.0050 & 57 \\
96 & 56309.8900 & 0.0007 & 0.0036 & 38 \\
97 & 56309.9458 & 0.0018 & 0.0045 & 26 \\
98 & 56309.9994 & 0.0014 & 0.0031 & 25 \\
113 & 56310.8227 & 0.0007 & 0.0019 & 77 \\
115 & 56310.9314 & 0.0004 & 0.0007 & 67 \\
116 & 56310.9873 & 0.0005 & 0.0017 & 75 \\
130 & 56311.7569 & 0.0025 & 0.0019 & 22 \\
131 & 56311.8084 & 0.0011 & $-$0.0016 & 23 \\
132 & 56311.8631 & 0.0033 & $-$0.0018 & 17 \\
135 & 56312.0255 & 0.0011 & $-$0.0044 & 40 \\
148 & 56312.7482 & 0.0031 & 0.0039 & 12 \\
149 & 56312.7970 & 0.0012 & $-$0.0023 & 43 \\
150 & 56312.8516 & 0.0008 & $-$0.0026 & 97 \\
151 & 56312.9041 & 0.0005 & $-$0.0051 & 92 \\
152 & 56312.9602 & 0.0006 & $-$0.0040 & 103 \\
153 & 56313.0128 & 0.0018 & $-$0.0063 & 52 \\
168 & 56313.8380 & 0.0016 & $-$0.0055 & 49 \\
169 & 56313.8951 & 0.0010 & $-$0.0034 & 57 \\
170 & 56313.9489 & 0.0008 & $-$0.0045 & 67 \\
171 & 56314.0050 & 0.0027 & $-$0.0034 & 41 \\
185 & 56314.7788 & 0.0033 & 0.0009 & 32 \\
186 & 56314.8319 & 0.0015 & $-$0.0009 & 38 \\
187 & 56314.8844 & 0.0021 & $-$0.0034 & 30 \\
188 & 56314.9383 & 0.0019 & $-$0.0044 & 40 \\
189 & 56314.9958 & 0.0043 & $-$0.0019 & 41 \\
190 & 56315.0487 & 0.0013 & $-$0.0039 & 11 \\
204 & 56315.8188 & 0.0009 & $-$0.0033 & 116 \\
205 & 56315.8731 & 0.0009 & $-$0.0040 & 80 \\
206 & 56315.9289 & 0.0012 & $-$0.0031 & 86 \\
207 & 56315.9817 & 0.0006 & $-$0.0052 & 114 \\
259 & 56318.8525 & 0.0150 & 0.0076 & 25 \\
260 & 56318.9001 & 0.0119 & 0.0002 & 24 \\
\hline
  \multicolumn{5}{l}{\commenta BJD$-$2400000.} \\
  \multicolumn{5}{l}{\commentb Against max $= 2456304.6103 + 0.054960 E$.} \\
  \multicolumn{5}{l}{\commentc Number of points used to determine the maximum.} \\
\end{tabular}
\end{center}
\end{table}

\subsection{OT J191443.6$+$605214}\label{obj:j191443}

   The 2008 superoutburst of this object, which was discovered
by K. Itagaki \citep{yam08j1914cbet1535}, was studied
in \citep{Pdot}.  The 2012 outburst was detected by
E. Muyllaert (baavss-alert 2989).  Subsequent observations
detected superhumps (vsnet-alert 14864, 14873).  The times
of superhump maxima are listed in table \ref{tab:j191443oc2012}.
In table \ref{tab:perlist}, we listed a period obtained
by the PDM method.

\begin{table}
\caption{Superhump maxima of OT J191443 (2012)}\label{tab:j191443oc2012}
\begin{center}
\begin{tabular}{ccccc}
\hline
$E$ & max\commenta & error & $O-C$\commentb & $N$\commentc \\
\hline
0 & 56159.3779 & 0.0003 & $-$0.0003 & 44 \\
1 & 56159.4498 & 0.0002 & 0.0003 & 75 \\
13 & 56160.3042 & 0.0004 & $-$0.0000 & 70 \\
\hline
  \multicolumn{5}{l}{\commenta BJD$-$2400000.} \\
  \multicolumn{5}{l}{\commentb Against max $= 2456159.3782 + 0.071234 E$.} \\
  \multicolumn{5}{l}{\commentc Number of points used to determine the maximum.} \\
\end{tabular}
\end{center}
\end{table}

\subsection{OT J205146.3$-$035828}\label{obj:j205146}

   This is a dwarf nova discovered by CRTS
(=CSS121004:205146$-$035827, hereafter OT J205146) on 2012 October 4
at a magnitude of 14.1.  There was no past outburst in the CRTS data.
Subsequent observations recorded superhumps (vsnet-alert 14987, 14996)

The times of superhump maxima are listed in table
\ref{tab:j205146oc2012}.  Although the scatters were large,
especially during the later stage, stages B and C can be
recognized.  The pattern of period variation suggests that the
object more resemble ordinary SU UMa-type dwarf novae rather than
extreme WZ Sge-type dwarf novae.  $E=250$ was excluded in
determining the period of stage C superhumps in table \ref{tab:perlist}.

\begin{figure}
  \begin{center}
    \FigureFile(88mm,110mm){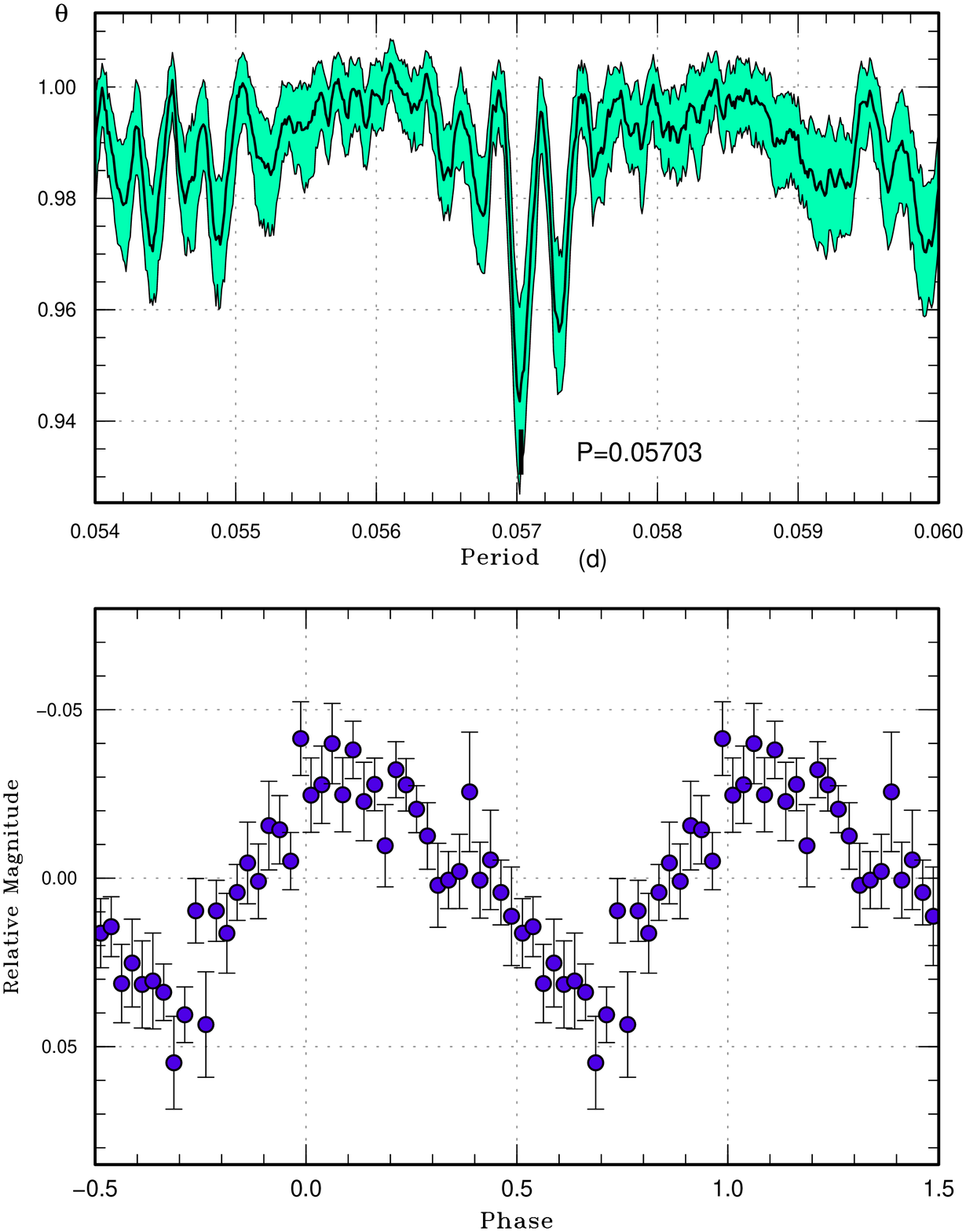}
  \end{center}
  \caption{Superhumps in OT J205146 (2012). (Upper): PDM analysis.
     (Lower): Phase-averaged profile.}
  \label{fig:j205146shpdm}
\end{figure}

\begin{table}
\caption{Superhump maxima of OT J205146}\label{tab:j205146oc2012}
\begin{center}
\begin{tabular}{ccccc}
\hline
$E$ & max\commenta & error & $O-C$\commentb & $N$\commentc \\
\hline
0 & 56207.9950 & 0.0005 & $-$0.0048 & 116 \\
1 & 56208.0512 & 0.0006 & $-$0.0057 & 88 \\
22 & 56209.2503 & 0.0003 & $-$0.0041 & 92 \\
23 & 56209.3068 & 0.0003 & $-$0.0046 & 95 \\
24 & 56209.3661 & 0.0009 & $-$0.0024 & 61 \\
57 & 56211.2476 & 0.0028 & $-$0.0028 & 16 \\
58 & 56211.3102 & 0.0009 & 0.0028 & 27 \\
59 & 56211.3636 & 0.0022 & $-$0.0009 & 28 \\
88 & 56213.0293 & 0.0025 & 0.0110 & 117 \\
92 & 56213.2612 & 0.0008 & 0.0149 & 61 \\
93 & 56213.3205 & 0.0011 & 0.0171 & 38 \\
126 & 56215.1870 & 0.0035 & 0.0017 & 36 \\
127 & 56215.2440 & 0.0029 & 0.0017 & 51 \\
139 & 56215.9418 & 0.0038 & 0.0151 & 114 \\
162 & 56217.2272 & 0.0014 & $-$0.0112 & 28 \\
163 & 56217.2814 & 0.0020 & $-$0.0140 & 28 \\
180 & 56218.2549 & 0.0013 & $-$0.0099 & 43 \\
215 & 56220.2514 & 0.0020 & $-$0.0094 & 43 \\
216 & 56220.3066 & 0.0022 & $-$0.0112 & 44 \\
250 & 56222.2734 & 0.0030 & 0.0166 & 10 \\
\hline
  \multicolumn{5}{l}{\commenta BJD$-$2400000.} \\
  \multicolumn{5}{l}{\commentb Against max $= 2456207.9998 + 0.057028 E$.} \\
  \multicolumn{5}{l}{\commentc Number of points used to determine the maximum.} \\
\end{tabular}
\end{center}
\end{table}

\subsection{OT J220641.1$+$301436}\label{obj:j220641}

\begin{figure}
  \begin{center}
    \FigureFile(88mm,110mm){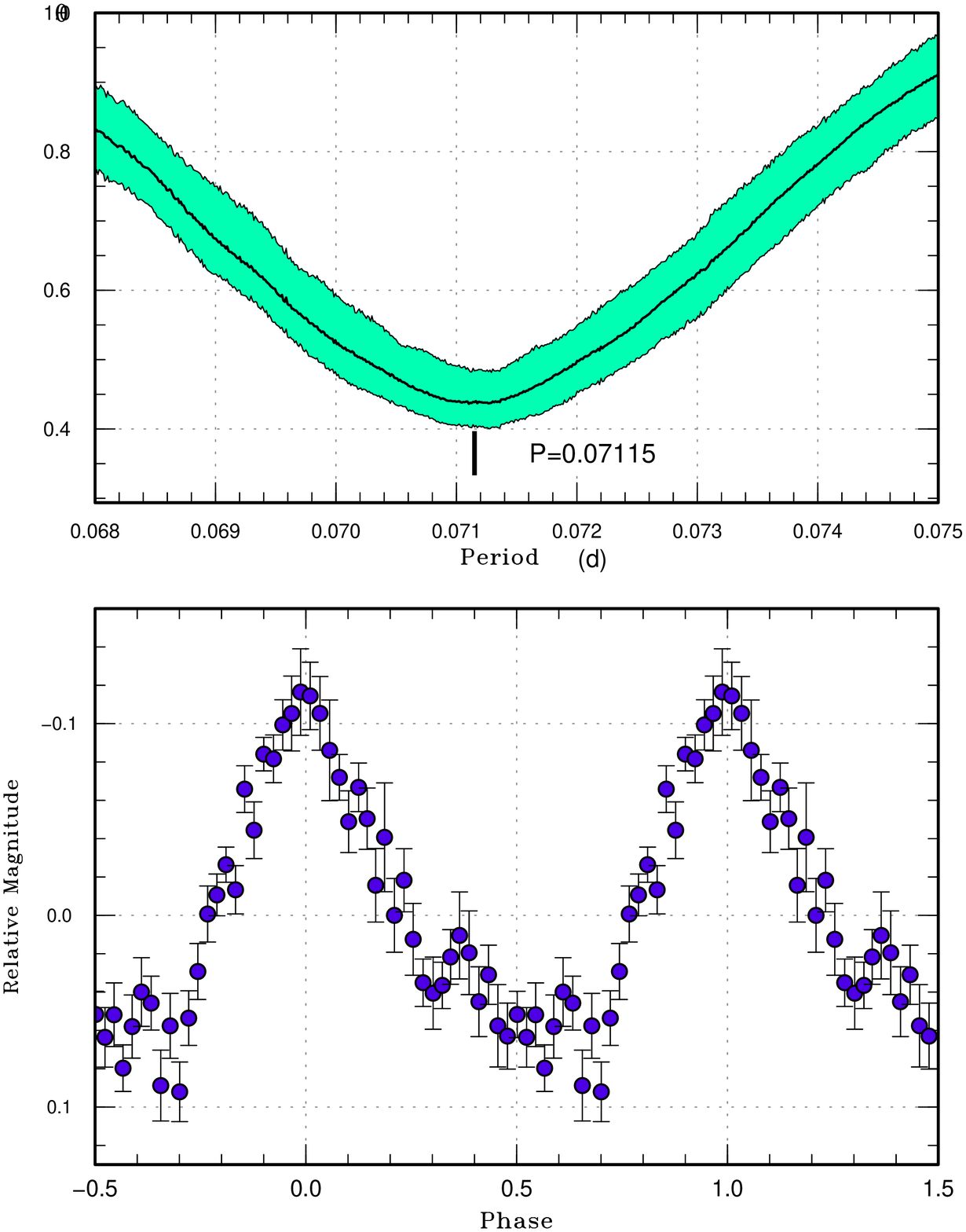}
  \end{center}
  \caption{Superhumps in OT J220641 (2012). (Upper): PDM analysis.
     (Lower): Phase-averaged profile.}
  \label{fig:j220641shpdm}
\end{figure}

   This object was detected by CRTS (=CSS110921:220641$+$301436,
hereafter OT J220641) on 2011 September 21.  A new outburst
was detected by MASTER network (vsnet-alert 15029).
The object has a faint ($g=23.2$) SDSS counterpart and the
amplitude suggested a superoutburst (vsnet-alert 15033).
Subsequent observations detected superhumps
(vsnet-alert 15036, 15038; figure \ref{fig:j220641shpdm}).
The times of superhump maxima are listed in table
\ref{tab:j220641oc2012}.  The superhump period in table
\ref{tab:perlist} was determined by the PDM method.

\begin{table}
\caption{Superhump maxima of OT J220641 (2012)}\label{tab:j220641oc2012}
\begin{center}
\begin{tabular}{ccccc}
\hline
$E$ & max\commenta & error & $O-C$\commentb & $N$\commentc \\
\hline
0 & 56223.2038 & 0.0007 & 0.0003 & 47 \\
1 & 56223.2730 & 0.0009 & $-$0.0015 & 45 \\
6 & 56223.6321 & 0.0005 & 0.0028 & 69 \\
7 & 56223.7020 & 0.0007 & 0.0017 & 68 \\
8 & 56223.7686 & 0.0009 & $-$0.0026 & 68 \\
9 & 56223.8415 & 0.0028 & $-$0.0007 & 35 \\
\hline
  \multicolumn{5}{l}{\commenta BJD$-$2400000.} \\
  \multicolumn{5}{l}{\commentb Against max $= 2456223.2035 + 0.070966 E$.} \\
  \multicolumn{5}{l}{\commentc Number of points used to determine the maximum.} \\
\end{tabular}
\end{center}
\end{table}

\subsection{OT J232727.2$+$085539}\label{obj:j232727}

\begin{figure}
  \begin{center}
    \FigureFile(88mm,110mm){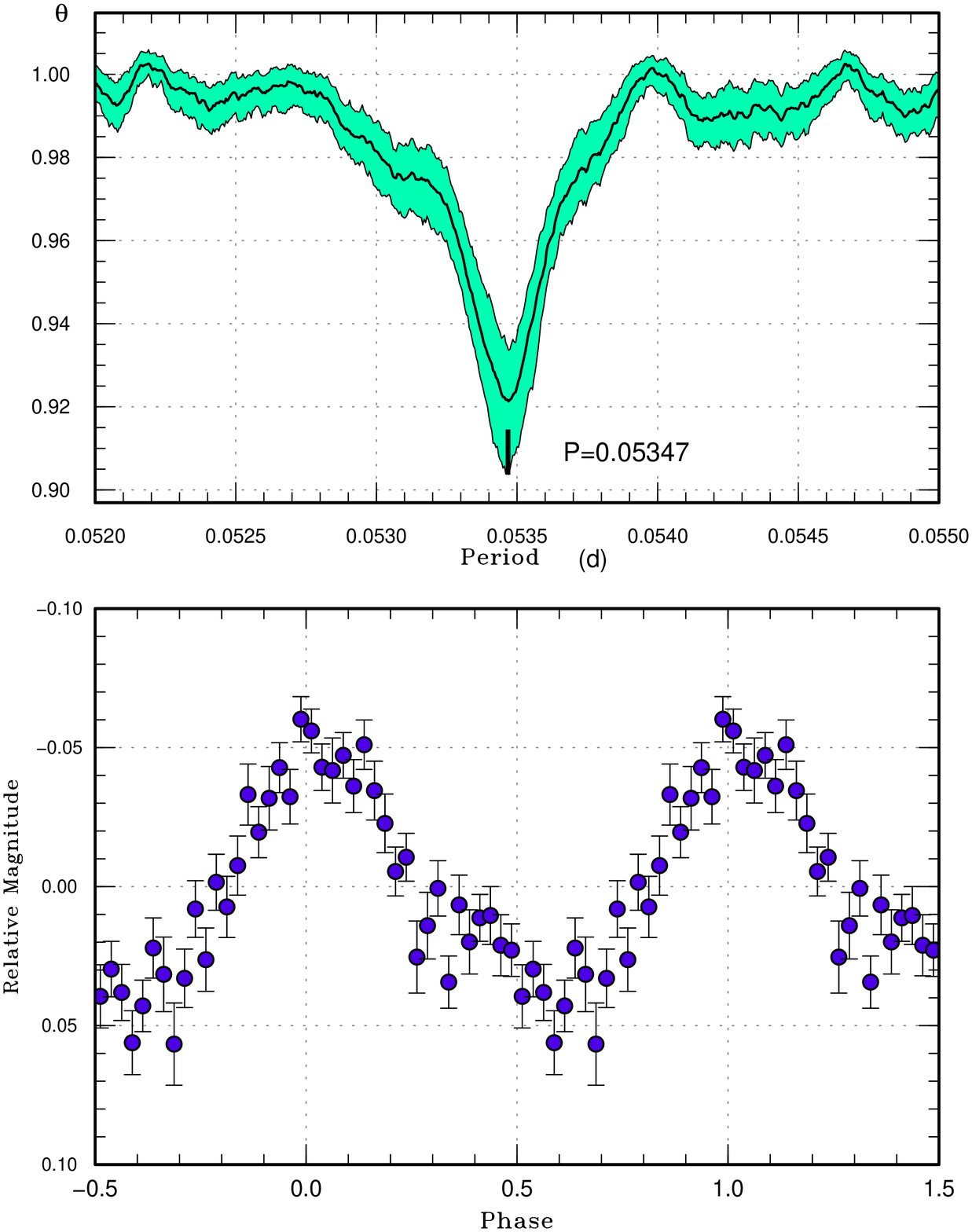}
  \end{center}
  \caption{Superhumps in OT J232727 (2012). (Upper): PDM analysis.
     (Lower): Phase-averaged profile.}
  \label{fig:j232727shpdm}
\end{figure}

\begin{figure}
  \begin{center}
    \FigureFile(88mm,110mm){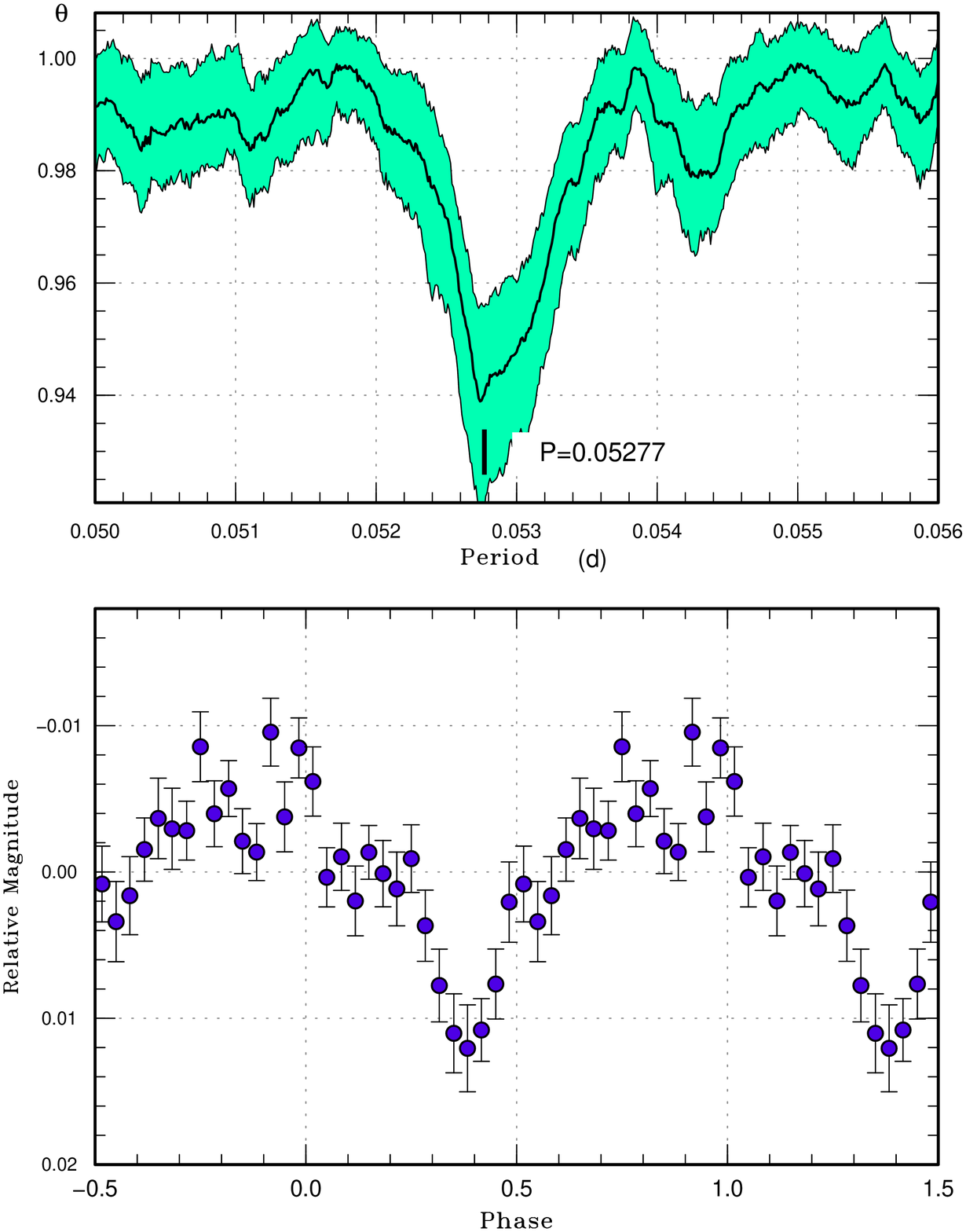}
  \end{center}
  \caption{Early superhumps in OT J232727 (2012). (Upper): PDM analysis.
     (Lower): Phase-averaged profile.}
  \label{fig:j232727eshpdm}
\end{figure}

   This object (=PNV J23272715$+$0855391, hereafter OT J232727)
was discovered by K. Itagaki at a magnitude of 13.9 on
2012 September 13.568 UT 
\citep{ita12j2327cbet3228}.\footnote{
   See also $<$http://www.cbat.eps.harvard.edu/\\unconf/followups/J23272715+0855391.html$>$.
}
Although the object was initially reported as a possible nova,
the presence of a blue quiescent counterpart in SDSS and
a UV object in the GALEX catalog already suggested
a WZ Sge-type dwarf nova (vsnet-alert 14921).
Shortly after the discovery announcement, some indication
of short-period variation was reported (vsnet-alert 14925).
Although the period was difficult to determine due to the small
amplitude and large airmass degrading the quality of photometry,
a very short-$P_{\rm orb}$ was already inferred
(vsnet-alert 14931, 14937).  The object started to show ordinary
superhumps (vsnet-alert 14938, 14939, 14948, 14952;
figure \ref{fig:j232727shpdm}).
Since the comparison star was much redder than the variable
and some observations were done at high airmasses, 
we corrected observations by using a second-order atmospheric 
extinction whose coefficients were experimentally determined.
The period of early superhumps up to BJD 2456188.5 was
0.05277(2)~d (figure \ref{fig:j232727eshpdm}),
which is almost the same value as the expected $P_{\rm orb}$ of
V1265 Tau \citep{sha07j0329}, the record holder of
the shortest $P_{\rm orb}$ of (hydrogen-rich, ordinary)
SU UMa-type dwarf nova.

   The times of superhump maxima are listed in table
\ref{tab:j232727oc2012}.  Stages A and B were very clearly
detected.  The $P_{\rm dot}$ for stage B superhumps was
small [$+4.0(1.1) \times 10^{-5}$], which is usual for
a very short-$P_{\rm orb}$ object.
The period of stage A superhumps was measured when the superhump
was growing ($E \le 29$).  The resultant value of
$\varepsilon^*=$0.0303(5) corresponds to $q=$0.082(2).
For an averaged white dwarf mass of 0.75 M$_{\odot}$
(\cite{lit08eclCV}; \cite{sav11CVeclmass}), the secondary
has a mass of 0.06 M$_{\odot}$.  The object is located
at the exact position of the period minimum expected by
the experimentally modified model of \citet{kni06CVsecondary}.

   No post-superoutburst rebrightening was reported, although
it may have been missed since the object became more difficult
to observe in the later season of this year.

\begin{table}
\caption{Superhump maxima of OT J232727 (2012)}\label{tab:j232727oc2012}
\begin{center}
\begin{tabular}{ccccc}
\hline
$E$ & max\commenta & error & $O-C$\commentb & $N$\commentc \\
\hline
0 & 56192.5540 & 0.0013 & $-$0.0263 & 20 \\
15 & 56193.3710 & 0.0007 & $-$0.0114 & 104 \\
16 & 56193.4258 & 0.0006 & $-$0.0101 & 103 \\
28 & 56194.0773 & 0.0003 & $-$0.0002 & 57 \\
29 & 56194.1329 & 0.0003 & 0.0019 & 48 \\
32 & 56194.2960 & 0.0006 & 0.0045 & 49 \\
33 & 56194.3484 & 0.0003 & 0.0035 & 54 \\
34 & 56194.4023 & 0.0003 & 0.0039 & 54 \\
35 & 56194.4537 & 0.0016 & 0.0018 & 23 \\
37 & 56194.5644 & 0.0009 & 0.0055 & 29 \\
40 & 56194.7256 & 0.0004 & 0.0063 & 92 \\
41 & 56194.7794 & 0.0008 & 0.0066 & 89 \\
65 & 56196.0635 & 0.0027 & 0.0075 & 68 \\
66 & 56196.1137 & 0.0012 & 0.0042 & 109 \\
67 & 56196.1649 & 0.0014 & 0.0019 & 106 \\
68 & 56196.2208 & 0.0020 & 0.0043 & 83 \\
72 & 56196.4341 & 0.0011 & 0.0037 & 26 \\
73 & 56196.4917 & 0.0024 & 0.0079 & 27 \\
74 & 56196.5408 & 0.0019 & 0.0035 & 29 \\
85 & 56197.1257 & 0.0011 & 0.0001 & 110 \\
87 & 56197.2339 & 0.0028 & 0.0014 & 19 \\
88 & 56197.2833 & 0.0008 & $-$0.0026 & 41 \\
89 & 56197.3410 & 0.0005 & 0.0015 & 111 \\
90 & 56197.3943 & 0.0005 & 0.0014 & 89 \\
91 & 56197.4454 & 0.0005 & $-$0.0009 & 84 \\
92 & 56197.4990 & 0.0003 & $-$0.0009 & 60 \\
103 & 56198.0931 & 0.0023 & 0.0050 & 90 \\
104 & 56198.1377 & 0.0036 & $-$0.0038 & 111 \\
105 & 56198.1929 & 0.0018 & $-$0.0021 & 101 \\
108 & 56198.3532 & 0.0042 & $-$0.0023 & 30 \\
182 & 56202.3055 & 0.0137 & $-$0.0069 & 21 \\
183 & 56202.3641 & 0.0007 & $-$0.0018 & 48 \\
184 & 56202.4190 & 0.0012 & $-$0.0004 & 44 \\
185 & 56202.4756 & 0.0014 & 0.0028 & 46 \\
186 & 56202.5192 & 0.0017 & $-$0.0071 & 46 \\
187 & 56202.5740 & 0.0010 & $-$0.0058 & 46 \\
201 & 56203.3306 & 0.0013 & 0.0023 & 40 \\
202 & 56203.3806 & 0.0018 & $-$0.0013 & 42 \\
216 & 56204.1326 & 0.0046 & 0.0021 & 93 \\
217 & 56204.1842 & 0.0040 & 0.0003 & 51 \\
\hline
  \multicolumn{5}{l}{\commenta BJD$-$2400000.} \\
  \multicolumn{5}{l}{\commentb Against max $= 2456192.5803 + 0.053473 E$.} \\
  \multicolumn{5}{l}{\commentc Number of points used to determine the maximum.} \\
\end{tabular}
\end{center}
\end{table}

\subsection{PNV J06270375$+$3952504}\label{obj:j062703}

\begin{figure}
  \begin{center}
    \FigureFile(88mm,110mm){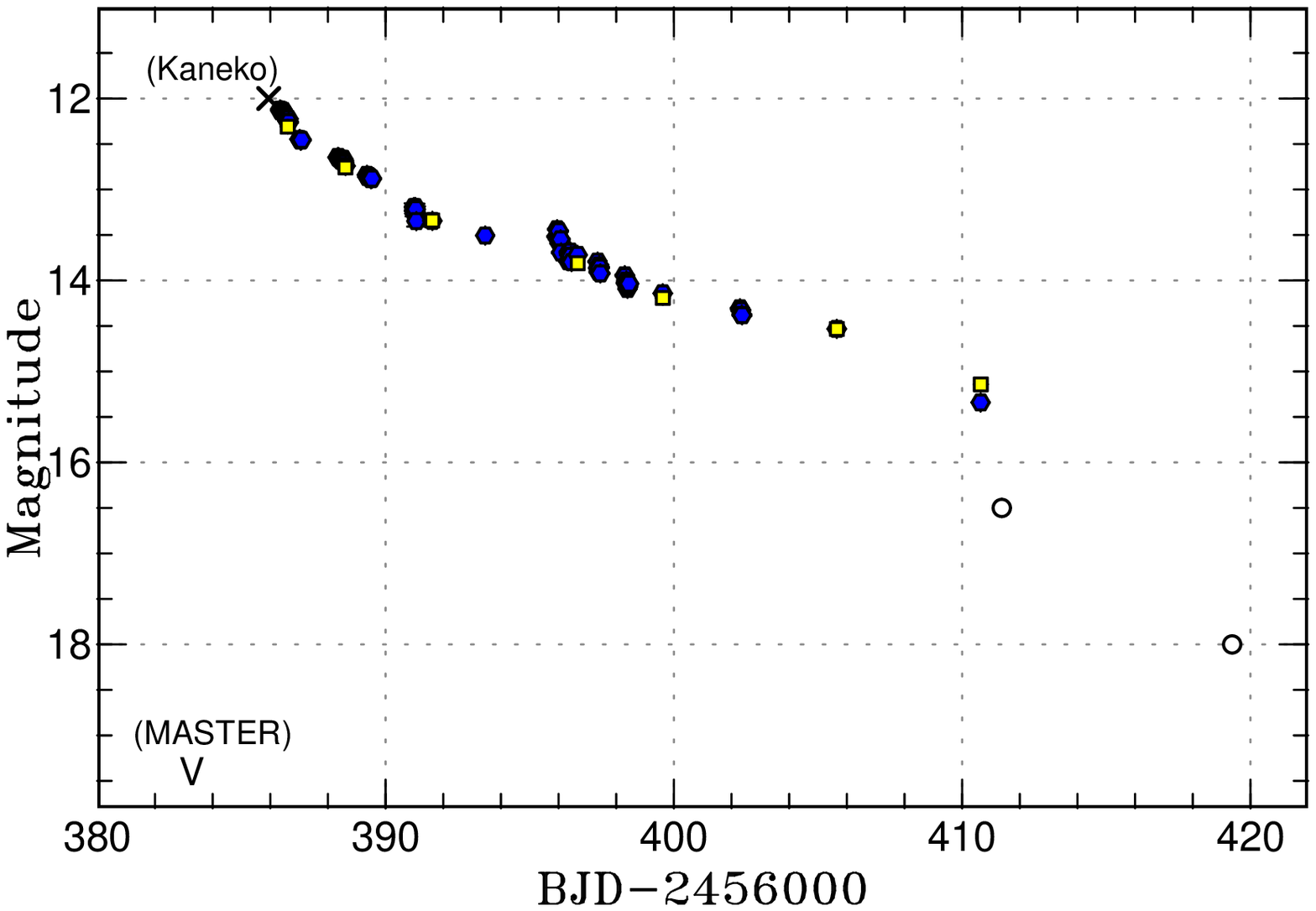}
  \end{center}
  \caption{Light curve of PNV J062703 (2013).  Filled circles,
  filled squares and open circles represent our CCD observations
  (binned to 0.02~d), AAVSO $R$-band observations and unfiltered
  snapshot photometry by E. Morillon, respectively.
  The cross represent the discovery observation and the ``V'' sign
  represents the upper limit by MASTER.}
  \label{fig:j0627lc}
\end{figure}

\begin{figure}
  \begin{center}
    \FigureFile(88mm,110mm){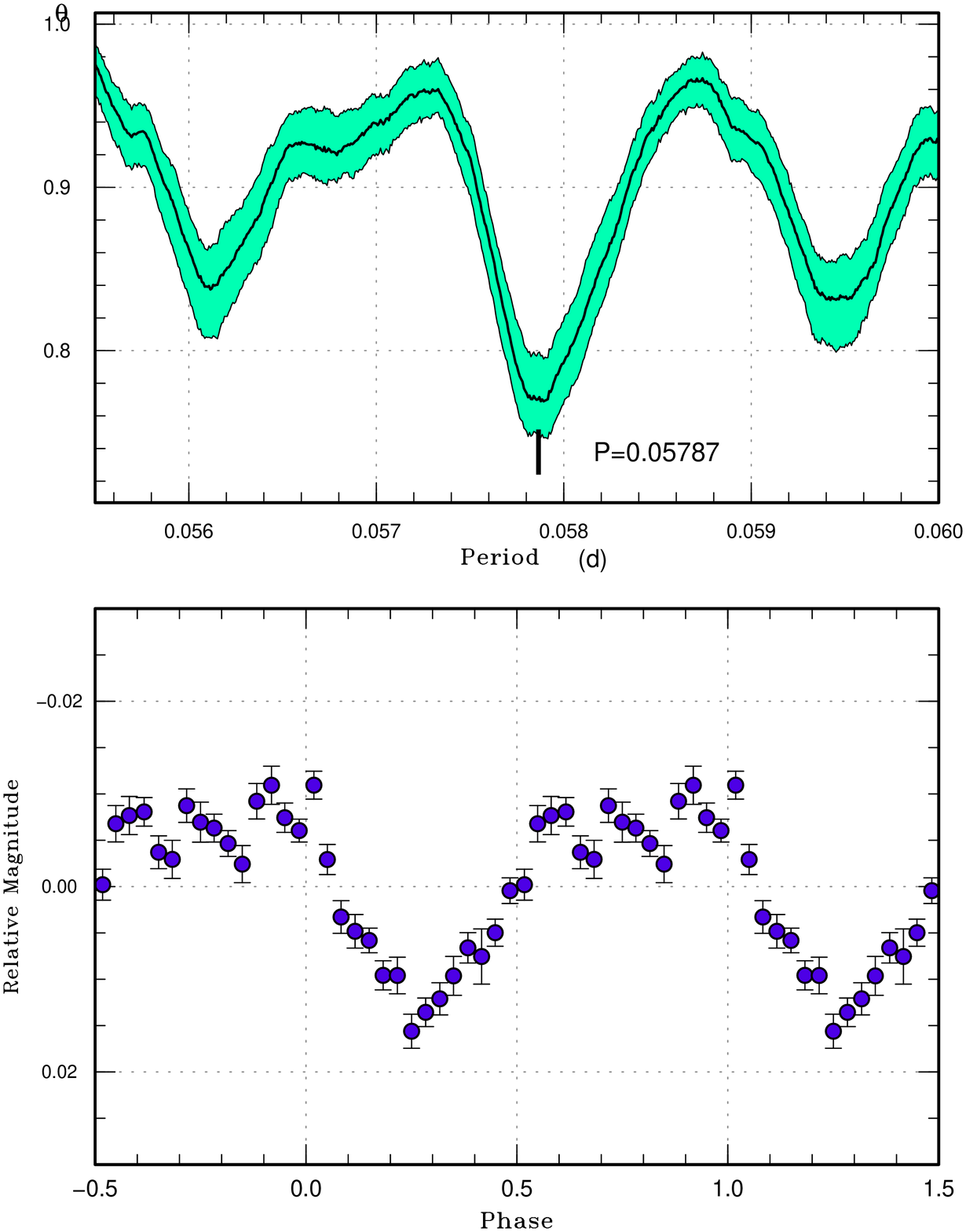}
  \end{center}
  \caption{Early superhumps in PNV J062703 (2013). (Upper): PDM analysis.
     (Lower): Phase-averaged profile.}
  \label{fig:j062703eshpdm}
\end{figure}

\begin{figure}
  \begin{center}
    \FigureFile(88mm,110mm){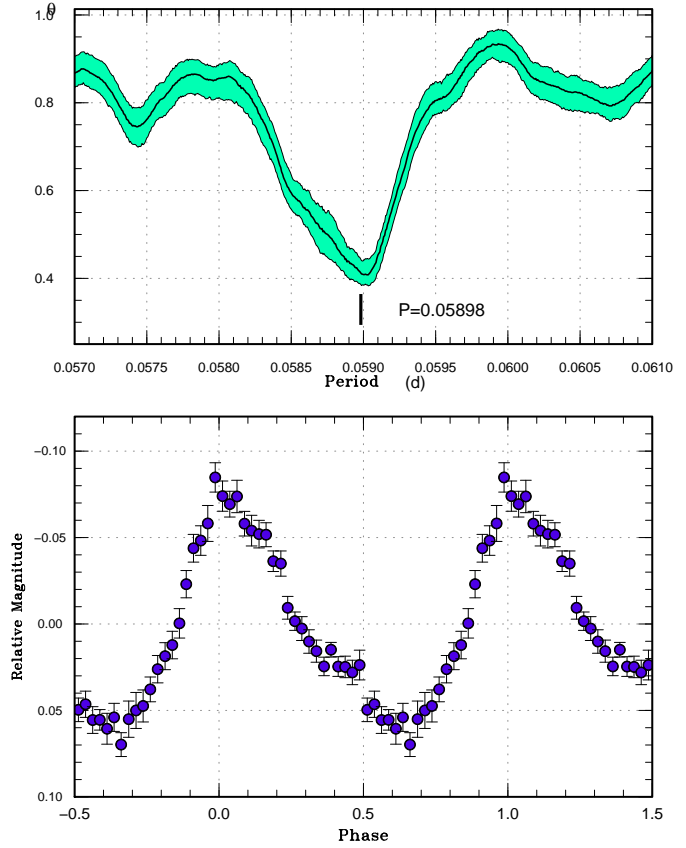}
  \end{center}
  \caption{Ordinary superhumps in PNV J062703 (2013).
     The data after BJD 2456394 were used in the analysis.
     (Upper): PDM analysis.
     (Lower): Phase-averaged profile.}
  \label{fig:j062703shpdm}
\end{figure}

   This object (hereafter PNV J062703)
is a transient initially reported as a possible nova by 
S. Kaneko.\footnote{
$<$http://www.cbat.eps.harvard.edu/unconf/followups/\\J06270375+3952504.html$>$.
}  This object was detected in outburst on 2013 April 3.445 UT
at an unfiltered CCD magnitude of 12.0.  According to the
observation of the MASTER network, the object was not in outburst
on March 31 (figure \ref{fig:j0627lc}).
There is a 20th mag star in the USNO catalog, and
the object was suspected to be a WZ Sge-type dwarf nova.
Subsequent observations detected double-wave early superhumps
[vsnet-alert 15581, 15592, 15594, 15607; figure
\ref{fig:j062703eshpdm}; period 0.05787(2)~d], 
qualifying the WZ Sge-type classification.
On April 8, the object started to show ordinary superhumps
(vsnet-alert 15599) and they later evolved (vsnet-alert
15609, 15616; figure \ref{fig:j062703shpdm}).
The times of superhump maxima are listed in table
\ref{tab:j062703oc2013}.  Although the observations on April 8
likely detected stage A superhumps, we could not convincingly
measure the times of maxima, and are not listed in the table.

   Considering the non-detection by MASTER, the duration of
early superhumps was less than 8~d, which is shorter than
those in well-observed WZ Sge-type dwarf novae.
The object apparently started rapid fading 25~d after
the discovery (figure \ref{fig:j0627lc}).  The late phase
of the outburst was not well observed due to the proximity
to the Sun.

\begin{table}
\caption{Superhump maxima of PNV J062703 (2013)}\label{tab:j062703oc2013}
\begin{center}
\begin{tabular}{ccccc}
\hline
$E$ & max\commenta & error & $O-C$\commentb & $N$\commentc \\
\hline
0 & 56395.9469 & 0.0003 & 0.0030 & 97 \\
1 & 56396.0050 & 0.0003 & 0.0021 & 97 \\
2 & 56396.0630 & 0.0004 & 0.0010 & 98 \\
7 & 56396.3582 & 0.0004 & 0.0012 & 67 \\
8 & 56396.4165 & 0.0003 & 0.0004 & 73 \\
24 & 56397.3561 & 0.0009 & $-$0.0045 & 48 \\
25 & 56397.4190 & 0.0007 & $-$0.0006 & 73 \\
40 & 56398.3043 & 0.0005 & $-$0.0007 & 49 \\
41 & 56398.3611 & 0.0003 & $-$0.0028 & 59 \\
42 & 56398.4205 & 0.0010 & $-$0.0025 & 38 \\
108 & 56402.3220 & 0.0016 & 0.0033 & 56 \\
\hline
  \multicolumn{5}{l}{\commenta BJD$-$2400000.} \\
  \multicolumn{5}{l}{\commentb Against max $= 2456395.9439 + 0.059026 E$.} \\
  \multicolumn{5}{l}{\commentc Number of points used to determine the maximum.} \\
\end{tabular}
\end{center}
\end{table}

\subsection{SDSS J075107.50$+$300628.4}\label{obj:j075107}

\begin{figure}
  \begin{center}
    \FigureFile(88mm,110mm){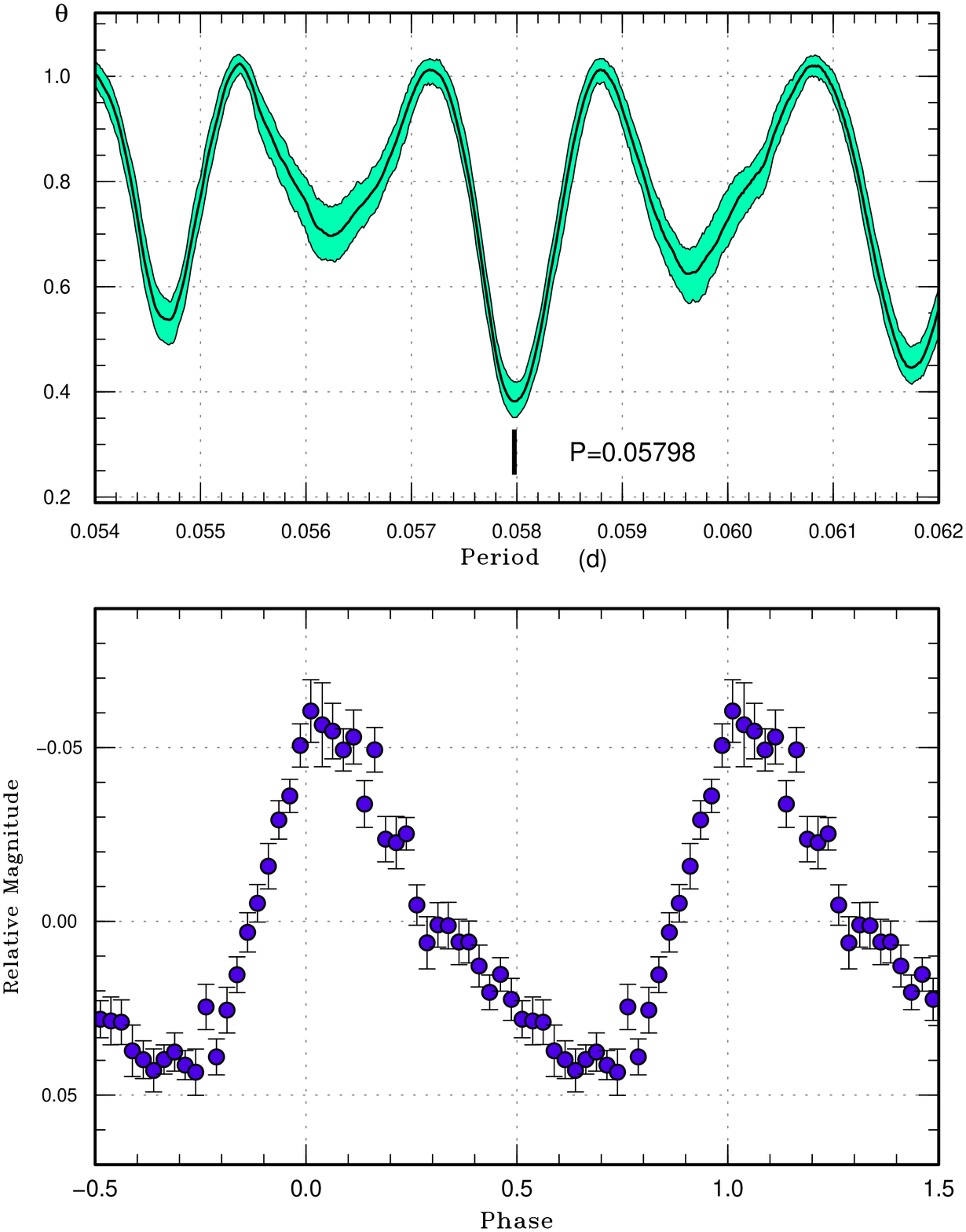}
  \end{center}
  \caption{Superhumps in SDSS J075107 (2013). (Upper): PDM analysis.
     (Lower): Phase-averaged profile.}
  \label{fig:j075107shpdm}
\end{figure}

   This object (hereafter SDSS J075107) was selected as a dwarf nova
by \citet{wil10newCVs}.  The SDSS colors suggested an orbital period
of 0.067~d \citep{kat12DNSDSS}.  E. Muyllaert detected an outburst
on 2013 February 12 (cvnet-outburst 5248).  Subsequent observations
detected superhumps (vsnet-alert 15395, 15408; figure
\ref{fig:j075107shpdm}).
The times of superhump maxima are listed in table
\ref{tab:j075107oc2013}.  It is not known whether stage B or C
was recorded.

\begin{table}
\caption{Superhump maxima of SDSS J075107 (2013)}\label{tab:j075107oc2013}
\begin{center}
\begin{tabular}{ccccc}
\hline
$E$ & max\commenta & error & $O-C$\commentb & $N$\commentc \\
\hline
0 & 56338.4881 & 0.0035 & $-$0.0009 & 29 \\
1 & 56338.5471 & 0.0005 & 0.0001 & 63 \\
2 & 56338.6055 & 0.0005 & 0.0005 & 58 \\
32 & 56340.3460 & 0.0069 & 0.0016 & 27 \\
33 & 56340.4017 & 0.0008 & $-$0.0006 & 60 \\
34 & 56340.4606 & 0.0005 & 0.0003 & 50 \\
35 & 56340.5174 & 0.0007 & $-$0.0008 & 53 \\
36 & 56340.5765 & 0.0008 & 0.0002 & 56 \\
50 & 56341.3887 & 0.0008 & 0.0007 & 59 \\
51 & 56341.4448 & 0.0014 & $-$0.0011 & 33 \\
\hline
  \multicolumn{5}{l}{\commenta BJD$-$2400000.} \\
  \multicolumn{5}{l}{\commentb Against max $= 2456338.4890 + 0.057980 E$.} \\
  \multicolumn{5}{l}{\commentc Number of points used to determine the maximum.} \\
\end{tabular}
\end{center}
\end{table}

\subsection{SDSS J080033.86$+$192416.5}\label{obj:j080033}

\begin{figure}
  \begin{center}
    \FigureFile(88mm,110mm){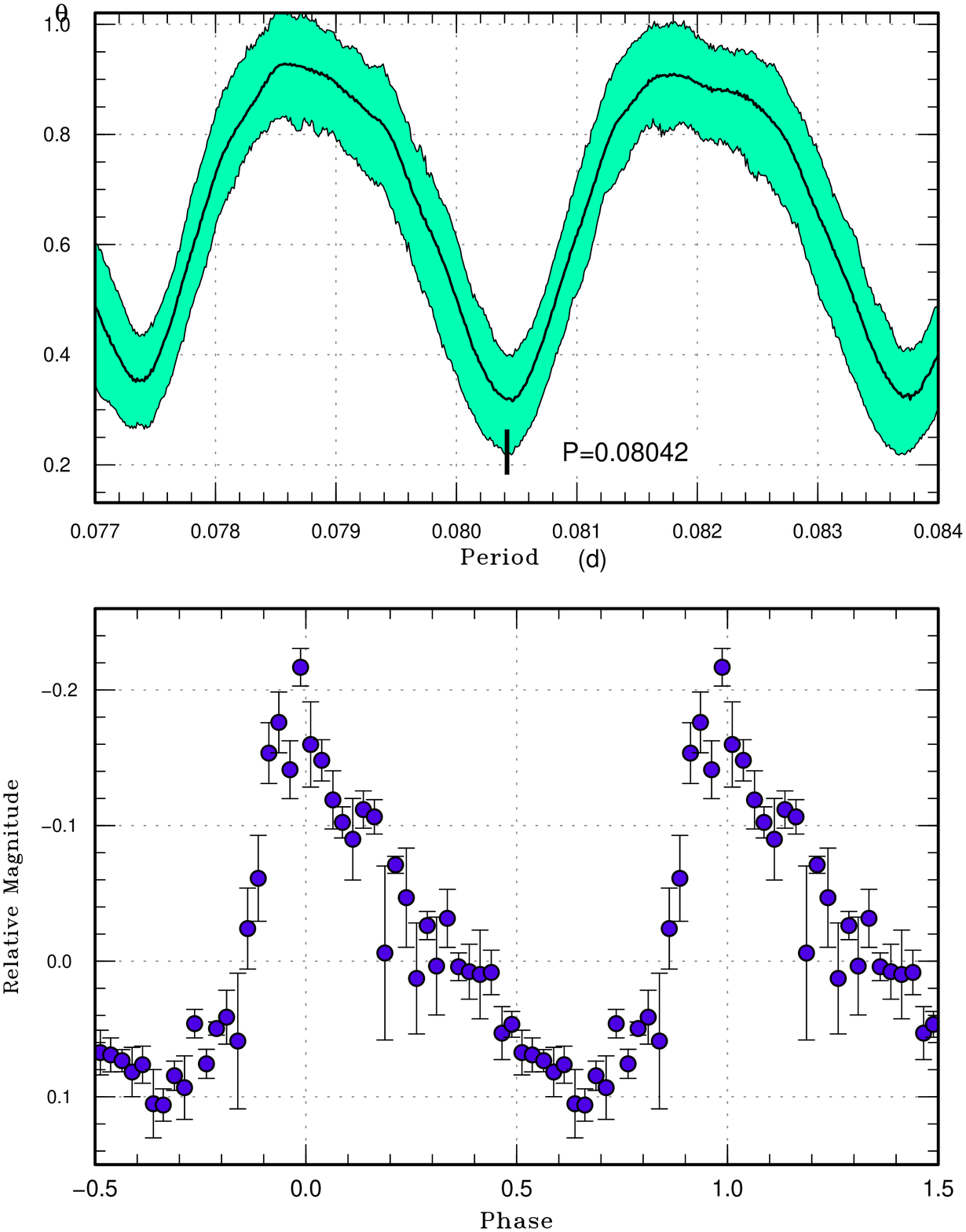}
  \end{center}
  \caption{Superhumps in SDSS J080033 (2012). (Upper): PDM analysis.
     (Lower): Phase-averaged profile.}
  \label{fig:j080033shpdm}
\end{figure}

   This object is a dwarf nova identified by \citet{wil10newCVs}
(hereafter SDSS J080033).
The SDSS color suggested an orbital period of 0.065--0.074~d
\citep{kat12DNSDSS}.  The MASTER network detected a bright outburst 
on 2012 October 19 (vsnet-alert 15021).  Subsequent observations
detected superhumps (vsnet-alert 15028, 15040; figure
\ref{fig:j080033shpdm}).  SDSS J080033 is a long-$P_{\rm SH}$
object with large superhump amplitudes. 
The times of superhump maxima are listed in table
\ref{tab:j080033oc2012}.  The superhump period in table
\ref{tab:perlist} was determined by the PDM method.

   According to the CRTS data, the object showed very frequent
outbursts and the duty cycle of the outburst (magnitude brighter
than 17.5) was 0.23.  The object must be a long-$P_{\rm orb}$,
active SU UMa-type dwarf nova like YZ Cnc.

\begin{table}
\caption{Superhump maxima of SDSS J080033 (2012)}\label{tab:j080033oc2012}
\begin{center}
\begin{tabular}{ccccc}
\hline
$E$ & max\commenta & error & $O-C$\commentb & $N$\commentc \\
\hline
0 & 56221.5897 & 0.0004 & $-$0.0026 & 84 \\
1 & 56221.6756 & 0.0018 & 0.0028 & 21 \\
25 & 56223.6045 & 0.0019 & $-$0.0005 & 43 \\
26 & 56223.6858 & 0.0044 & 0.0004 & 17 \\
\hline
  \multicolumn{5}{l}{\commenta BJD$-$2400000.} \\
  \multicolumn{5}{l}{\commentb Against max $= 2456221.5923 + 0.080507 E$.} \\
  \multicolumn{5}{l}{\commentc Number of points used to determine the maximum.} \\
\end{tabular}
\end{center}
\end{table}

\subsection{SDSS J162520.29$+$120308.7}\label{obj:j162520}

\begin{figure}
  \begin{center}
    \FigureFile(88mm,110mm){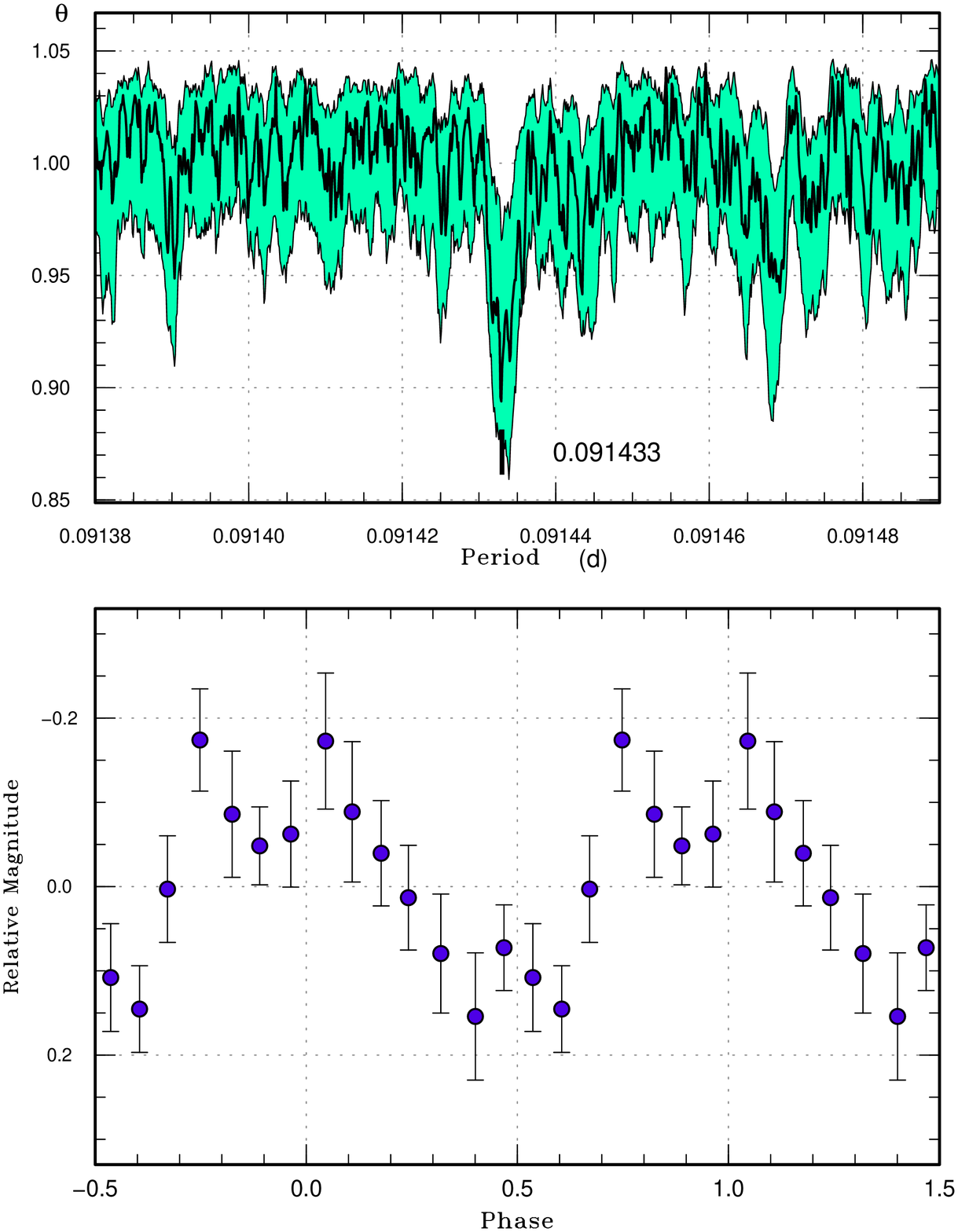}
  \end{center}
  \caption{Orbital modulation in SDSS J162520 in quiescence.
     (Upper): PDM analysis.
     (Lower): Phase-averaged profile.}
  \label{fig:j1625porbpdm}
\end{figure}

   This object (hereafter SDSS J162520) in the period gap
was studied in \citet{Pdot}, \citet{ole11j1625}.
Although we do not have new data, we discuss on this object again
since we have been able to identify the orbital period
in the CRTS data.  The resultant orbital period was
0.091433(1)~d (figure \ref{fig:j1625porbpdm}).
This value is slightly (2$\sigma$) shorter than 0.09111(15)~d
obtained by radial-velocity study \citet{ole11j1625}.
We consider that this period is more likely because
any candidate period within the error of \citet{ole11j1625}
did not yield a smooth orbital light curve
(in figure \ref{fig:j1625porbpdm}, only a small segment
of the period search is shown because the signal is very
narrow due to the long baseline of the CRTS data).
The $\varepsilon^*$ for stage A superhumps \citet{Pdot2}
amounts to 7.2(2)\%.  The $q$ estimated using the relation
in \citet{kat13qfromstageA} is 0.23(1).
This object showed a post-superoutburst rebrightening 
(\cite{Pdot2}; \cite{ole11j1625}), rather unusual for
a long-$P_{\rm orb}$ object.  We could detect superhumps
in the faint state between the main superoutburst and
the rebrightening.  The period was 0.09579(5)~d,
corresponding to $\varepsilon^*$=4.54(5)\%.  Using the
relation in \citet{kat13qfromstageA}, \citet{kat13j1222},
this precession rate can be translated to a disk radius
of 0.36(1)$A$, since the pressure effect can be ignored
in a cold disk (\cite{osa13v344lyrv1504cyg};
\cite{kat13qfromstageA}).  This value is a usual one
for the post-superoutburst disk \citep{kat13qfromstageA}.

   Although \citet{ole11j1625} considered the double-wave
hump signal during the early stage of the outburst as
early superhumps, we consider it unlikely because
ordinary superhumps in the growing stage are known to
sometimes appear doubly humped, and the object was
quickly fading (0.37 mag d$^{-1}$), which is apparently
characteristic of a precursor outburst and not of
the early superhump stage.  The growing superhump
signal during the fading branch of a precursor is
also very commonly seen (\cite{osa13v1504cygKepler};
\cite{osa13v344lyrv1504cyg}).  The short period
detected by \citet{ole11j1625} was probably due to 
the large error in period analysis for a very short 
(0.22~d) segment.  We therefore see no reason
to assume early superhumps present in this object,
and the claim by \citet{ole11j1625} regarding the origin
of the early superhumps is not justified.

\subsection{SSS J122221.7$-$311523}\label{obj:j122221}

\begin{figure}
  \begin{center}
    \FigureFile(75mm,90mm){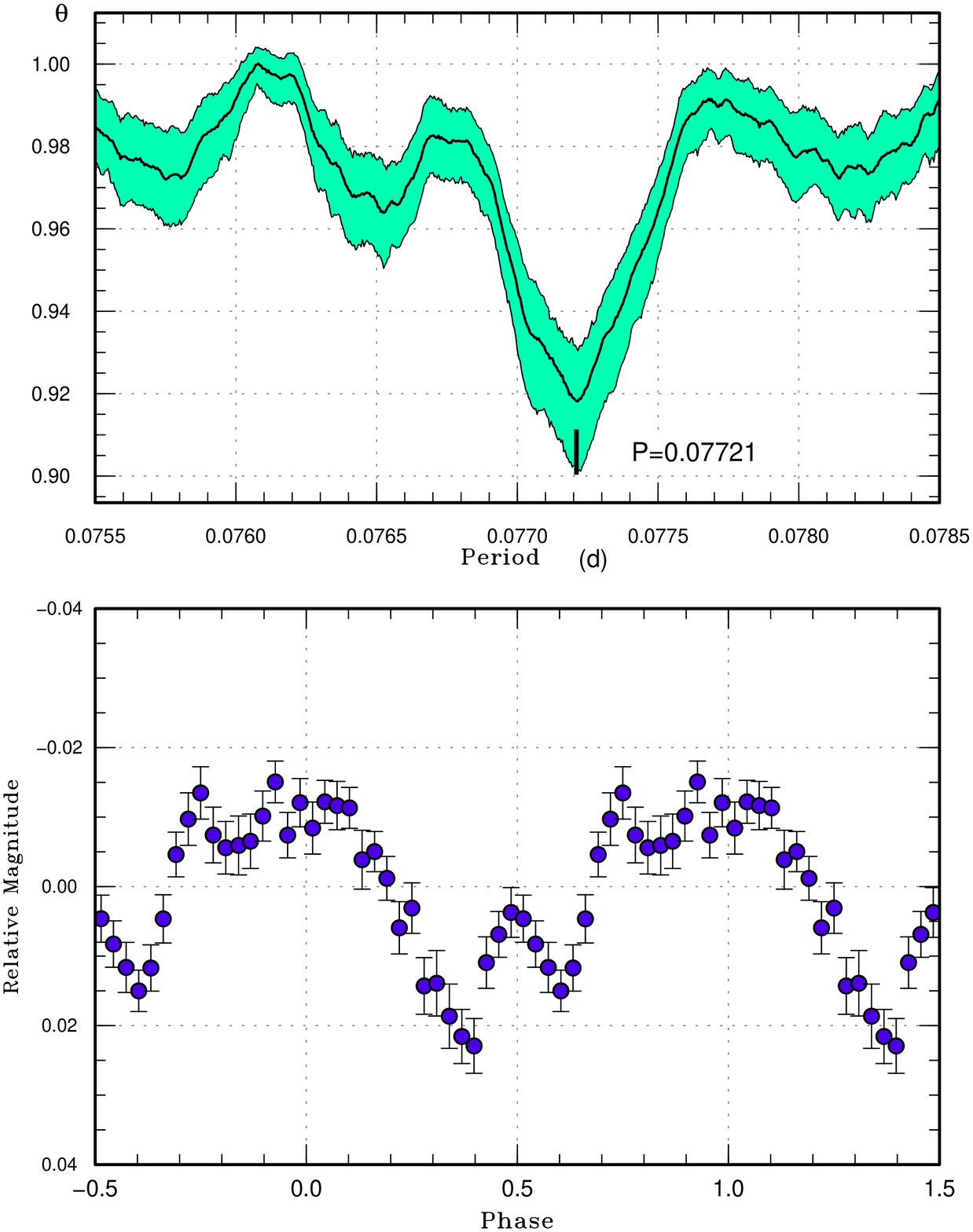}
  \end{center}
  \caption{Stage A superhumps in SSS J122221.
  (Upper): PDM analysis.
  (Lower): Phase-averaged profile.}
  \label{fig:j1222ashpdm}
\end{figure}

   The times of superhump maxima of (SSS J122221.7$-$311523,
hereafter SSS J122221) used in \citet{kat13j1222} are listed
in table \ref{tab:j122221oc2013}.   The mean profile of stage A 
superhumps is also given (figure \ref{fig:j1222ashpdm}).

\begin{table}
\caption{Superhump maxima of SSS J122221 (2013)}\label{tab:j122221oc2013}
\begin{center}
\begin{tabular}{ccccc}
\hline
$E$ & max\commenta & error & $O-C$\commentb & $N$\commentc \\
\hline
0 & 56304.7444 & 0.0026 & $-$0.0639 & 17 \\
1 & 56304.8131 & 0.0019 & $-$0.0718 & 26 \\
7 & 56305.2810 & 0.0006 & $-$0.0641 & 124 \\
13 & 56305.7377 & 0.0020 & $-$0.0674 & 19 \\
27 & 56306.8266 & 0.0018 & $-$0.0521 & 28 \\
33 & 56307.2834 & 0.0007 & $-$0.0553 & 136 \\
39 & 56307.7508 & 0.0019 & $-$0.0480 & 23 \\
40 & 56307.8341 & 0.0015 & $-$0.0414 & 33 \\
53 & 56308.8362 & 0.0012 & $-$0.0362 & 38 \\
58 & 56309.2329 & 0.0019 & $-$0.0228 & 133 \\
59 & 56309.2946 & 0.0023 & $-$0.0378 & 110 \\
79 & 56310.8469 & 0.0013 & $-$0.0191 & 42 \\
136 & 56315.2502 & 0.0010 & 0.0134 & 107 \\
152 & 56316.4853 & 0.0005 & 0.0216 & 207 \\
156 & 56316.7909 & 0.0018 & 0.0205 & 30 \\
157 & 56316.8619 & 0.0006 & 0.0149 & 31 \\
161 & 56317.1689 & 0.0095 & 0.0151 & 94 \\
162 & 56317.2512 & 0.0016 & 0.0207 & 136 \\
165 & 56317.4788 & 0.0003 & 0.0183 & 269 \\
166 & 56317.5448 & 0.0006 & 0.0076 & 108 \\
168 & 56317.7064 & 0.0015 & 0.0159 & 21 \\
169 & 56317.7778 & 0.0010 & 0.0105 & 44 \\
170 & 56317.8654 & 0.0011 & 0.0215 & 34 \\
174 & 56318.1698 & 0.0036 & 0.0192 & 101 \\
175 & 56318.2541 & 0.0029 & 0.0268 & 132 \\
181 & 56318.7087 & 0.0021 & 0.0214 & 20 \\
182 & 56318.7815 & 0.0020 & 0.0175 & 45 \\
183 & 56318.8606 & 0.0010 & 0.0199 & 37 \\
187 & 56319.1724 & 0.0013 & 0.0250 & 116 \\
188 & 56319.2395 & 0.0013 & 0.0154 & 132 \\
190 & 56319.4006 & 0.0037 & 0.0231 & 92 \\
191 & 56319.4762 & 0.0004 & 0.0220 & 329 \\
192 & 56319.5529 & 0.0005 & 0.0220 & 329 \\
194 & 56319.7122 & 0.0058 & 0.0280 & 19 \\
195 & 56319.7794 & 0.0018 & 0.0184 & 17 \\
203 & 56320.4049 & 0.0004 & 0.0306 & 232 \\
204 & 56320.4761 & 0.0004 & 0.0251 & 330 \\
205 & 56320.5504 & 0.0004 & 0.0227 & 329 \\
213 & 56321.1655 & 0.0013 & 0.0243 & 89 \\
214 & 56321.2345 & 0.0011 & 0.0167 & 137 \\
216 & 56321.3938 & 0.0062 & 0.0226 & 112 \\
217 & 56321.4715 & 0.0004 & 0.0237 & 328 \\
218 & 56321.5476 & 0.0003 & 0.0231 & 329 \\
219 & 56321.6261 & 0.0007 & 0.0249 & 204 \\
240 & 56323.2273 & 0.0006 & 0.0158 & 126 \\
246 & 56323.6925 & 0.0011 & 0.0209 & 16 \\
247 & 56323.7681 & 0.0026 & 0.0198 & 40 \\
248 & 56323.8379 & 0.0011 & 0.0129 & 36 \\
252 & 56324.1601 & 0.0025 & 0.0284 & 81 \\
253 & 56324.2285 & 0.0008 & 0.0201 & 136 \\
256 & 56324.4544 & 0.0004 & 0.0160 & 328 \\
257 & 56324.5283 & 0.0003 & 0.0132 & 329 \\
258 & 56324.6079 & 0.0012 & 0.0161 & 252 \\
269 & 56325.4505 & 0.0003 & 0.0153 & 319 \\
\hline
  \multicolumn{5}{l}{\commenta BJD$-$2400000.} \\
  \multicolumn{5}{l}{\commentb Against max $= 2456304.8083 + 0.076680 E$.} \\
  \multicolumn{5}{l}{\commentc Number of points used to determine the maximum.} \\
\end{tabular}
\end{center}
\end{table}

\addtocounter{table}{-1}
\begin{table}
\caption{Superhump maxima of SSS J122221 (2013) (continued)}
\begin{center}
\begin{tabular}{ccccc}
\hline
$E$ & max\commenta & error & $O-C$\commentb & $N$\commentc \\
\hline
270 & 56325.5272 & 0.0005 & 0.0153 & 329 \\
271 & 56325.6027 & 0.0004 & 0.0141 & 286 \\
272 & 56325.6836 & 0.0014 & 0.0184 & 15 \\
273 & 56325.7548 & 0.0014 & 0.0129 & 69 \\
274 & 56325.8328 & 0.0007 & 0.0142 & 76 \\
278 & 56326.1472 & 0.0040 & 0.0218 & 54 \\
279 & 56326.2158 & 0.0017 & 0.0138 & 96 \\
286 & 56326.7450 & 0.0010 & 0.0062 & 29 \\
287 & 56326.8277 & 0.0015 & 0.0123 & 32 \\
288 & 56326.9011 & 0.0046 & 0.0089 & 9 \\
298 & 56327.6675 & 0.0030 & 0.0085 & 10 \\
299 & 56327.7395 & 0.0010 & 0.0038 & 23 \\
300 & 56327.8174 & 0.0008 & 0.0051 & 30 \\
306 & 56328.2771 & 0.0024 & 0.0047 & 63 \\
321 & 56329.4335 & 0.0005 & 0.0109 & 224 \\
322 & 56329.5036 & 0.0006 & 0.0044 & 329 \\
323 & 56329.5744 & 0.0004 & $-$0.0016 & 329 \\
333 & 56330.3544 & 0.0011 & 0.0116 & 219 \\
334 & 56330.4201 & 0.0004 & 0.0006 & 325 \\
335 & 56330.4974 & 0.0003 & 0.0013 & 329 \\
336 & 56330.5701 & 0.0004 & $-$0.0027 & 293 \\
347 & 56331.4130 & 0.0004 & $-$0.0032 & 329 \\
348 & 56331.4921 & 0.0003 & $-$0.0009 & 329 \\
349 & 56331.5633 & 0.0003 & $-$0.0063 & 329 \\
360 & 56332.4074 & 0.0007 & $-$0.0057 & 328 \\
361 & 56332.4874 & 0.0004 & $-$0.0024 & 328 \\
362 & 56332.5593 & 0.0004 & $-$0.0072 & 324 \\
416 & 56336.6777 & 0.0014 & $-$0.0295 & 28 \\
417 & 56336.7533 & 0.0015 & $-$0.0305 & 28 \\
418 & 56336.8308 & 0.0015 & $-$0.0298 & 26 \\
442 & 56338.6806 & 0.0031 & $-$0.0203 & 26 \\
443 & 56338.7549 & 0.0011 & $-$0.0226 & 26 \\
455 & 56339.6873 & 0.0016 & $-$0.0104 & 25 \\
456 & 56339.7563 & 0.0022 & $-$0.0181 & 25 \\
457 & 56339.8489 & 0.0009 & $-$0.0022 & 27 \\
468 & 56340.6872 & 0.0012 & $-$0.0074 & 26 \\
469 & 56340.7597 & 0.0021 & $-$0.0115 & 25 \\
470 & 56340.8351 & 0.0008 & $-$0.0128 & 27 \\
483 & 56341.8333 & 0.0020 & $-$0.0114 & 26 \\
484 & 56341.9082 & 0.0019 & $-$0.0133 & 7 \\
494 & 56342.6694 & 0.0010 & $-$0.0188 & 23 \\
496 & 56342.8266 & 0.0015 & $-$0.0150 & 28 \\
507 & 56343.6651 & 0.0022 & $-$0.0200 & 26 \\
508 & 56343.7494 & 0.0043 & $-$0.0123 & 25 \\
509 & 56343.8154 & 0.0012 & $-$0.0230 & 27 \\
510 & 56343.9100 & 0.0140 & $-$0.0051 & 11 \\
520 & 56344.6745 & 0.0018 & $-$0.0075 & 25 \\
521 & 56344.7422 & 0.0030 & $-$0.0164 & 25 \\
522 & 56344.8264 & 0.0085 & $-$0.0089 & 14 \\
523 & 56344.9001 & 0.0058 & $-$0.0119 & 18 \\
535 & 56345.8158 & 0.0027 & $-$0.0164 & 28 \\
536 & 56345.8957 & 0.0043 & $-$0.0131 & 21 \\
549 & 56346.8933 & 0.0017 & $-$0.0123 & 24 \\
562 & 56347.8933 & 0.0035 & $-$0.0092 & 23 \\
\hline
  \multicolumn{5}{l}{\commenta BJD$-$2400000.} \\
  \multicolumn{5}{l}{\commentb Against max $= 2456304.8083 + 0.076680 E$.} \\
  \multicolumn{5}{l}{\commentc Number of points used to determine the maximum.} \\
\end{tabular}
\end{center}
\end{table}

\subsection{SSS J224739.7$-$362253}\label{obj:j224739}

\begin{figure}
  \begin{center}
    \FigureFile(88mm,70mm){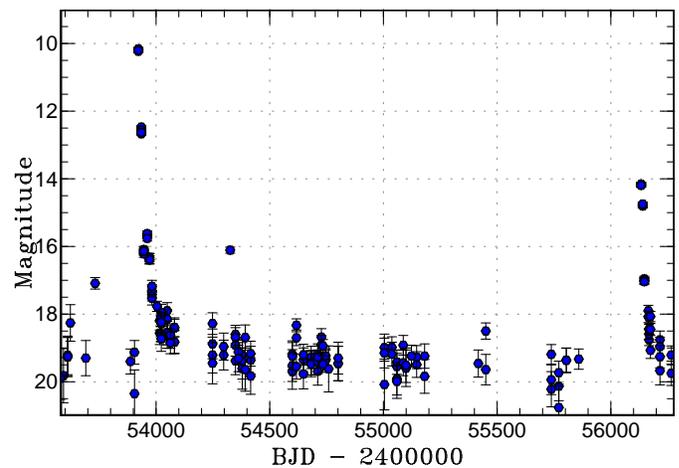}
  \end{center}
  \caption{Long term curve of SSS J224739.
    The data are from CRTS SSS observations.}
  \label{fig:j224739long}
\end{figure}

\begin{figure}
  \begin{center}
    \FigureFile(88mm,70mm){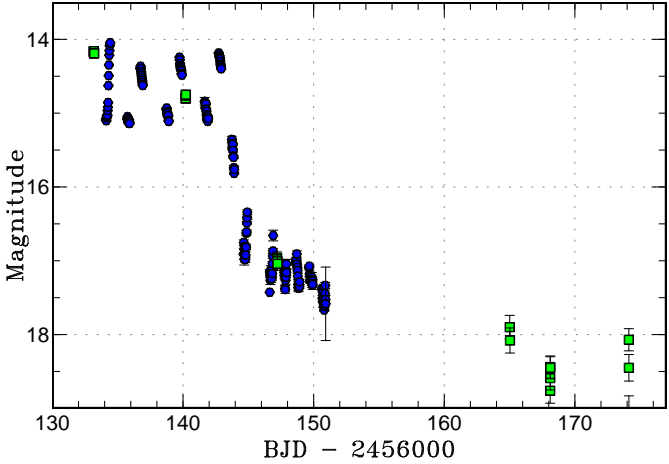}
  \end{center}
  \caption{Overall light curve of SSS J224739 (2012).
    The time-series data (filled circles) were binned to 0.02~d.
    The filled squares are CRTS SSS observations.}
  \label{fig:j224739reb}
\end{figure}

\begin{figure}
  \begin{center}
    \FigureFile(88mm,110mm){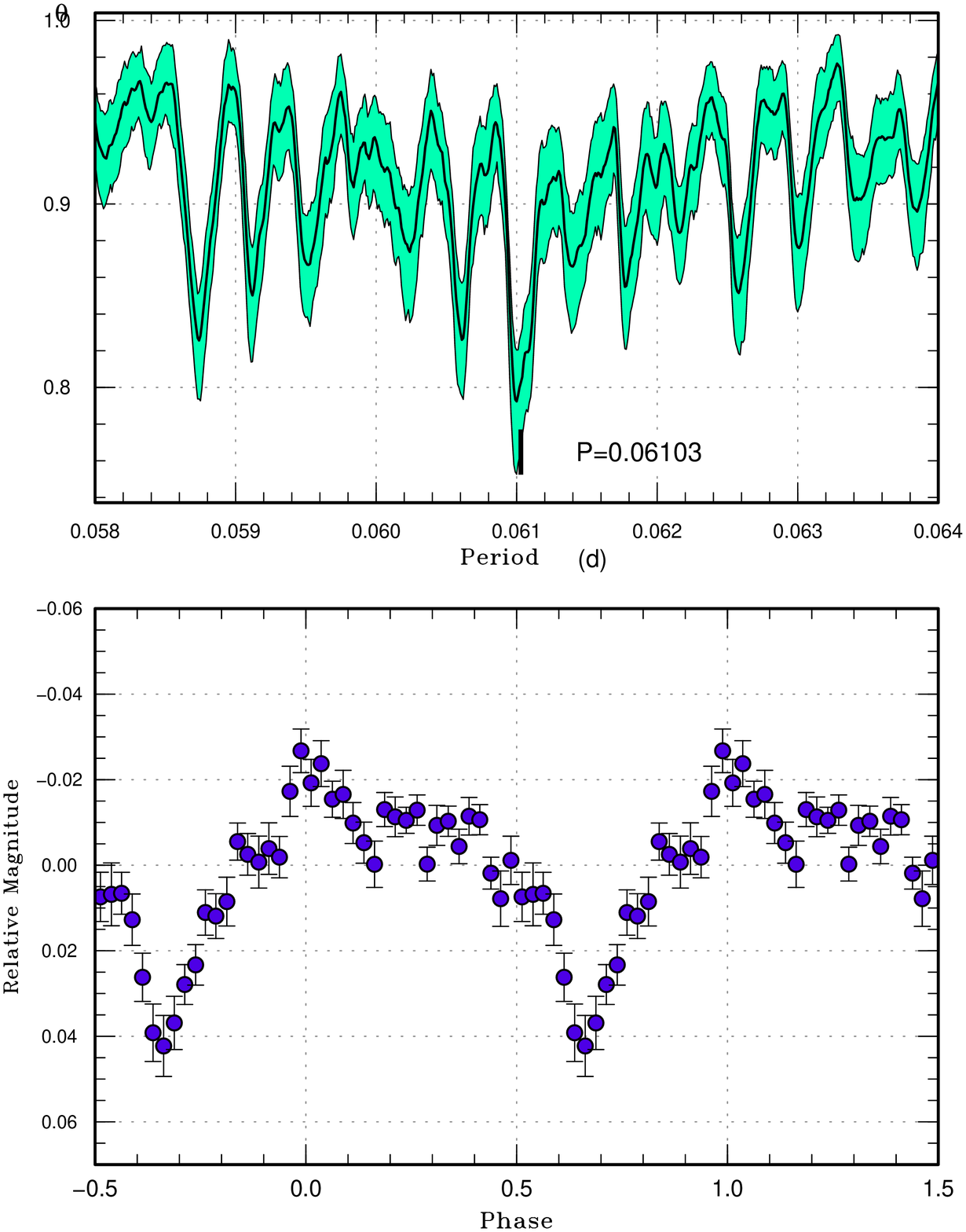}
  \end{center}
  \caption{Superhumps in SSS J224739 (2012). (Upper): PDM analysis.
     (Lower): Phase-averaged profile.}
  \label{fig:j224739shpdm}
\end{figure}

   SSS J224739.7$-$362253 is a dwarf nova discovered by CRTS SSS
(=SSS120724:224740$-$362254, hereafter SSS J224739) on 2012 July 24
at a magnitude of 14.2.  The CRTS data indicated that the object
underwent a brighter (11.0 mag) outburst in 2006 July
(vsnet-alert 14791, figure \ref{fig:j224739long}).
Subsequent observations recorded large-amplitude variations with
a scale of days, and short-term ($\sim$0.06 d) variations
resembling superhumps (vsnet-alert 14797, 14804, 14805, 14808).
Later it became apparent that these variations were post-superoutburst
repetitive rebrightening phase of a WZ Sge-type dwarf nova
(vsnet-alert 14821, figure \ref{fig:j224739reb}).
The case is similar to EL UMa \citep{Pdot2},
for which only multiple rebrightening phase was recorded.
The amplitudes of these rebrightenings were rather small
($\sim$1 mag), suggesting that the outburst was an intermediate
between WZ Sge-type (type-A rebrightening) and discrete
multiple rebrightenings (type-B).
The mean superhump period during the post-superoutburst stage
was 0.06103(1)~d (PDM method, figure \ref{fig:j224739shpdm}).
The times of superhump maxima are listed in table
\ref{tab:j224739oc2012}.  A PDM analysis of the observations
after the rebrightenings yielded a period of 0.06101(3)~d,
which is almost identical to the superhump period
during the rebrightening phase.  No signal of the orbital
modulation was detected either in our time-series data
and the CRTS data.

   The relatively short interval between two superoutbursts
(2006 and 2012) suggests that superoutbursts are more frequent
in this object than in other WZ Sge-type dwarf novae.
The case is similar to EZ Lyn (\cite{she07j0804};
\cite{pav07j0804}; \cite{kat09j0804}; \cite{Pdot3})
which underwent superoutbursts in 2006 and 2010 and also
showed multiple post-superoutburst rebrightenings.
Since the object is one of the brightest (10 mag) WZ Sge-type
dwarf novae, further detailed observations are encouraged.
The object was also detected as a GALEX source
[NUV magnitude 19.67(9)].

\begin{table}
\caption{Superhump maxima of SSS J224739}\label{tab:j224739oc2012}
\begin{center}
\begin{tabular}{ccccc}
\hline
$E$ & max\commenta & error & $O-C$\commentb & $N$\commentc \\
\hline
0 & 56134.1020 & 0.0063 & $-$0.0013 & 84 \\
1 & 56134.1628 & 0.0009 & $-$0.0016 & 109 \\
2 & 56134.2225 & 0.0016 & $-$0.0028 & 125 \\
3 & 56134.2891 & 0.0008 & 0.0028 & 125 \\
4 & 56134.3427 & 0.0005 & $-$0.0047 & 125 \\
5 & 56134.4004 & 0.0009 & $-$0.0081 & 124 \\
27 & 56135.7549 & 0.0016 & 0.0039 & 15 \\
28 & 56135.8173 & 0.0020 & 0.0052 & 16 \\
29 & 56135.8775 & 0.0010 & 0.0045 & 22 \\
46 & 56136.9167 & 0.0116 & 0.0061 & 17 \\
78 & 56138.8692 & 0.0018 & 0.0057 & 22 \\
79 & 56138.9152 & 0.0042 & $-$0.0092 & 14 \\
125 & 56141.7416 & 0.0038 & 0.0100 & 22 \\
126 & 56141.7870 & 0.0051 & $-$0.0057 & 26 \\
127 & 56141.8580 & 0.0023 & 0.0042 & 26 \\
128 & 56141.9145 & 0.0031 & $-$0.0003 & 21 \\
143 & 56142.8272 & 0.0045 & $-$0.0029 & 26 \\
159 & 56143.8033 & 0.0016 & $-$0.0033 & 27 \\
160 & 56143.8654 & 0.0023 & $-$0.0022 & 27 \\
161 & 56143.9284 & 0.0064 & $-$0.0003 & 13 \\
\hline
  \multicolumn{5}{l}{\commenta BJD$-$2400000.} \\
  \multicolumn{5}{l}{\commentb Against max $= 2456134.1033 + 0.061027 E$.} \\
  \multicolumn{5}{l}{\commentc Number of points used to determine the maximum.} \\
\end{tabular}
\end{center}
\end{table}

\subsection{TCP J15375685$-$2440136}\label{obj:j153756}

\begin{figure}
  \begin{center}
    \FigureFile(88mm,110mm){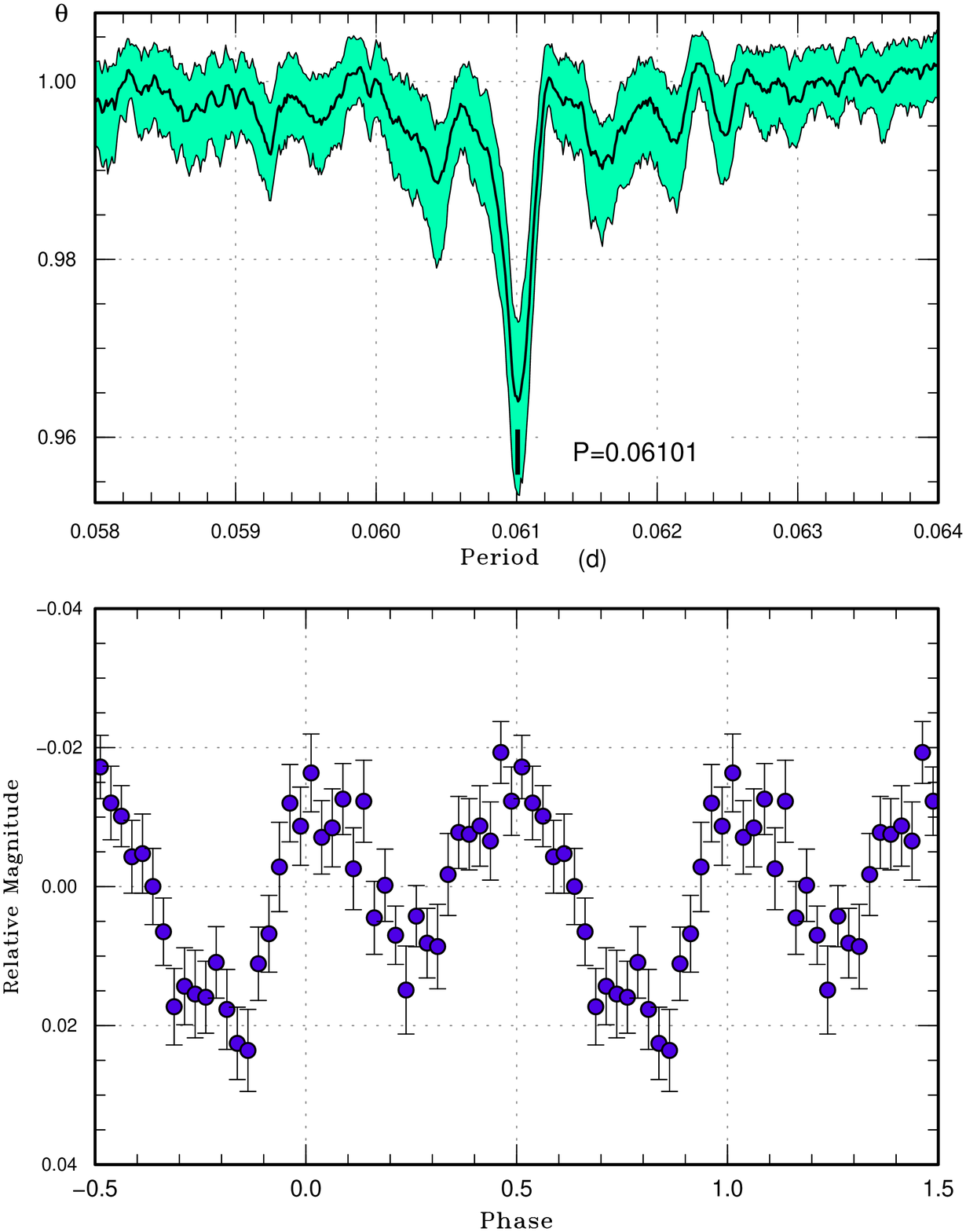}
  \end{center}
  \caption{Early superhumps in TCP J153756 (2013). (Upper): PDM analysis.
     (Lower): Phase-averaged profile.}
  \label{fig:j153756eshpdm}
\end{figure}

\begin{figure}
  \begin{center}
    \FigureFile(88mm,110mm){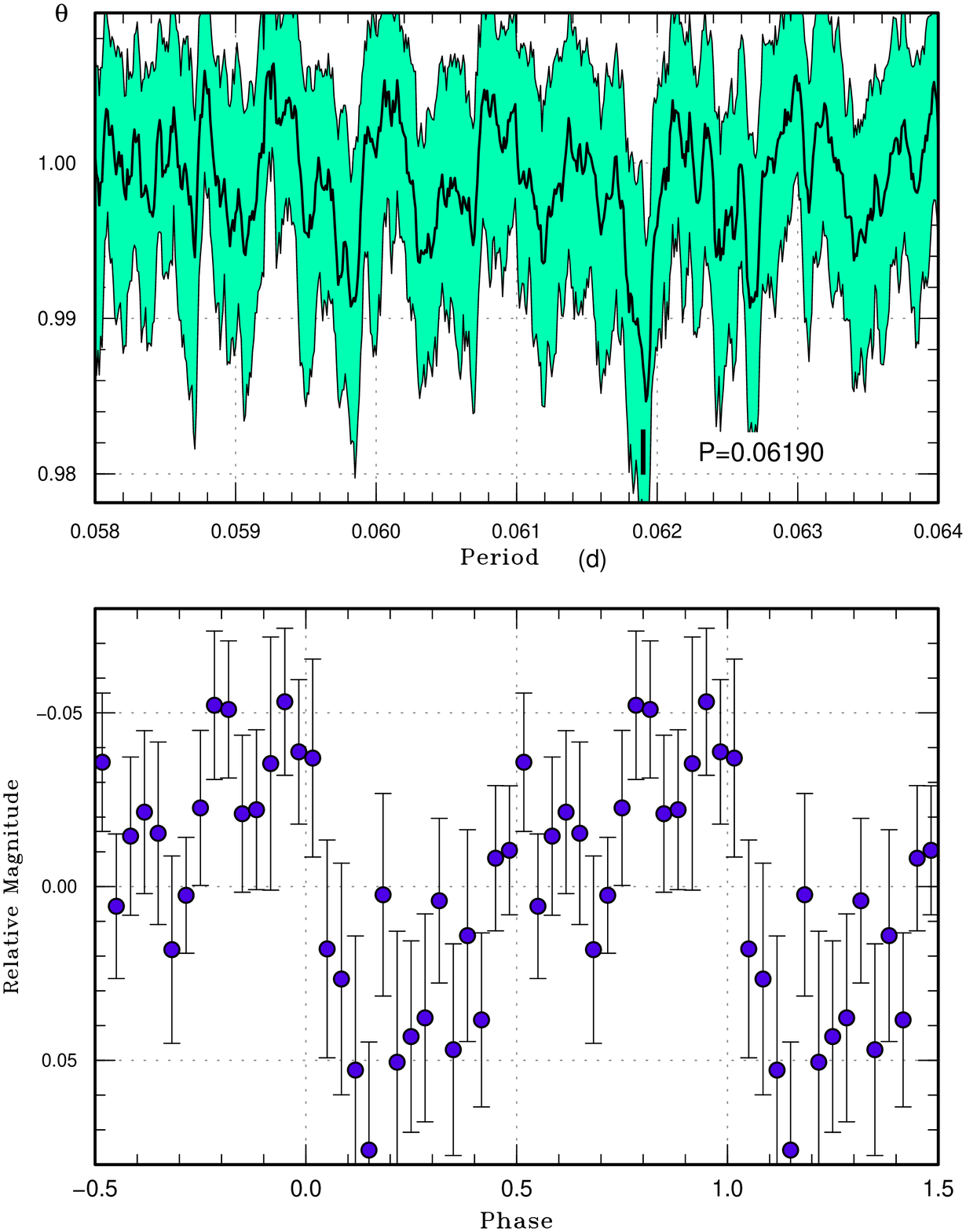}
  \end{center}
  \caption{Possible superhumps in TCP J153756 (2013). (Upper): PDM analysis.
     (Lower): Phase-averaged profile.}
  \label{fig:j153756shpdm}
\end{figure}

   This is a transient object (hereafter TCP J153756)\footnote{
$<$http://www.cbat.eps.harvard.edu/unconf/\\followups/J15375685-2440136.html$>$.
} discovered by K. Itagaki at a magnitude of 13.6 on 
2013 February 8.779 UT.
There was a 21.7 mag quiescent counterpart and the large
outburst amplitude suggested a WZ Sge-type dwarf nova
(vsnet-alert 15366).  Double-wave early superhumps were
immediately observed (vsnet-alert 15368, 15382, 15390, 15416;
figure \ref{fig:j153756eshpdm}).
After this, a possible signal of ordinary superhumps were
detected (vsnet-alert 15443; figure \ref{fig:j153756shpdm}).
Individual times of superhump maxima could not be determined.
The detected signal [0.06190(2)~d] was 1.5\% longer than
the period of early superhumps [0.061007(14)~d].
This fractional superhump excess is too small for stage A
superhumps, and it was likely that stage A evolution was missed
during the 2-d gap in the observation.  The period is longer
than typical WZ Sge-type dwarf novae, and is rather typical
for a WZ Sge-type dwarf novae with multiple rebrightenings
\citep{nak13j2112j2037}.  Observations for the post-superoutburst
stage were not available and it was not known whether this object
underwent rebrightenings.

\subsection{TCP J17521907$+$5001155}\label{obj:j175219}

\begin{figure}
  \begin{center}
    \FigureFile(88mm,110mm){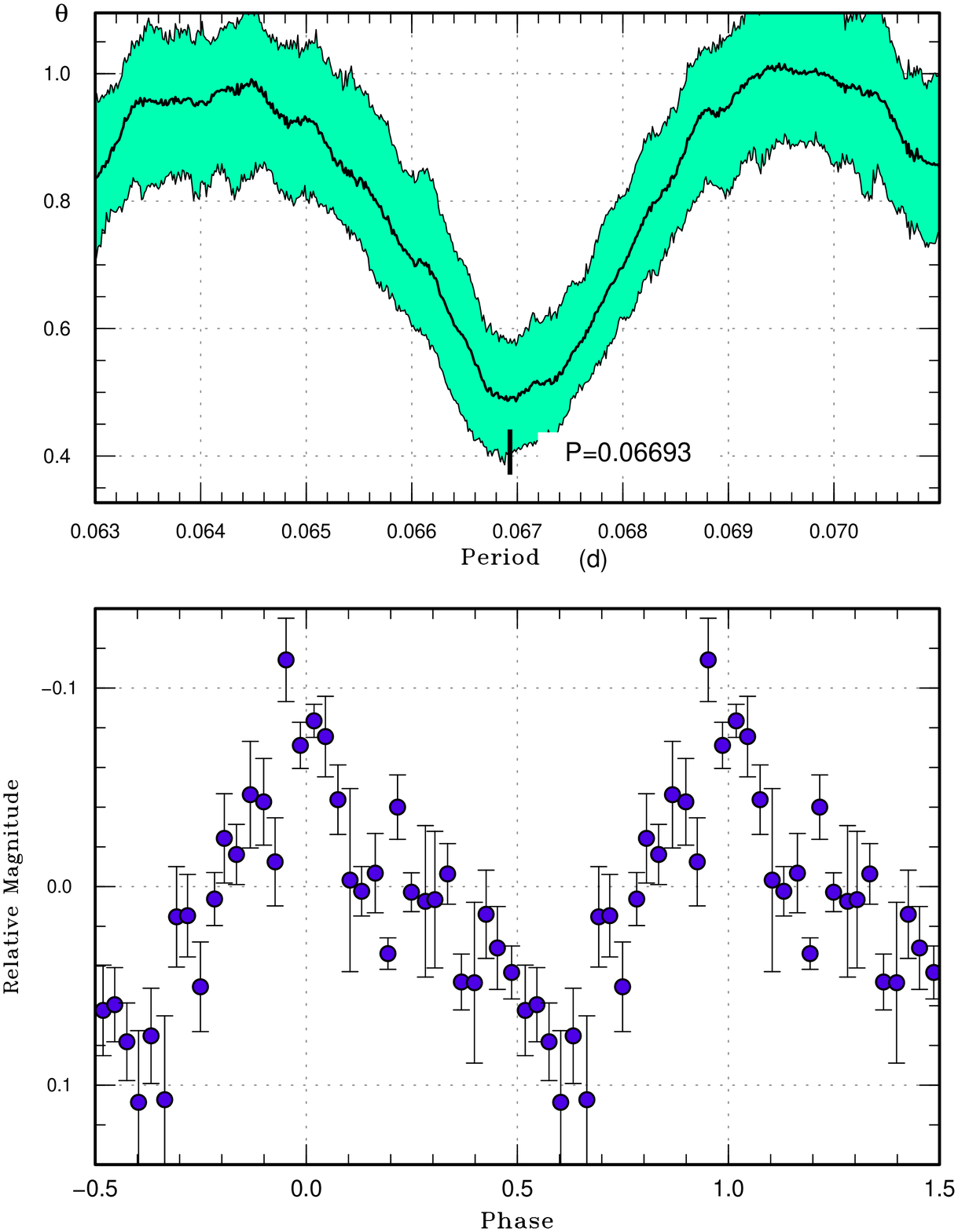}
  \end{center}
  \caption{Superhumps in TCP J175219 (2012). (Upper): PDM analysis.
     (Lower): Phase-averaged profile.}
  \label{fig:j175219shpdm}
\end{figure}

   This is a transient object (hereafter TCP J175219)\footnote{
$<$http://www.cbat.eps.harvard.edu/unconf/\\followups/J17521907+5001155.html$>$.
} discovered by H. Mikuz on 2012 August 17.  Subsequent
observations confirmed the presence of superhumps
(vsnet-alert 14888, 14893; figure \ref{fig:j175219shpdm}).
The times of superhump maxima are listed in table \ref{tab:j175219oc2012}.
The period adopted in table \ref{tab:perlist} was by
the PDM analysis.

\begin{table}
\caption{Superhump maxima of TCP J175219 (2012)}\label{tab:j175219oc2012}
\begin{center}
\begin{tabular}{ccccc}
\hline
$E$ & max\commenta & error & $O-C$\commentb & $N$\commentc \\
\hline
0 & 56162.5043 & 0.0020 & $-$0.0001 & 39 \\
12 & 56163.3113 & 0.0011 & 0.0007 & 33 \\
13 & 56163.3772 & 0.0013 & $-$0.0006 & 37 \\
\hline
  \multicolumn{5}{l}{\commenta BJD$-$2400000.} \\
  \multicolumn{5}{l}{\commentb Against max $= 2456162.5044 + 0.067186 E$.} \\
  \multicolumn{5}{l}{\commentc Number of points used to determine the maximum.} \\
\end{tabular}
\end{center}
\end{table}

\section{General Discussion}\label{sec:generaldiscuss}

   We report in this section general statistical properties
of the sample together with the earlier sample as in \citet{Pdot4}.

\subsection{Period Derivatives during Stage B}\label{sec:stagebpdot}

   As in \citet{Pdot4}, we have determined period derivatives
of new superoutbursts during stage B.  All the objects with
$P_{\rm orb} < 0.086$~d followed the trend reported in
\citet{Pdot4}.  Although AQ Eri (2012) showed a large $P_{\rm dot}$,
this was due to the limited coverage and the resultant error
was large.  We found three additional objects with large positive 
$P_{\rm dot}$ despite the long $P_{\rm orb}$:
V444 Peg, CSS J203937 and MASTER J212624.  It has become more
evident that large positive-$P_{\rm dot}$ objects are relatively
common among long-$P_{\rm orb}$ systems than has been thought.

\begin{figure*}
  \begin{center}
    \FigureFile(160mm,110mm){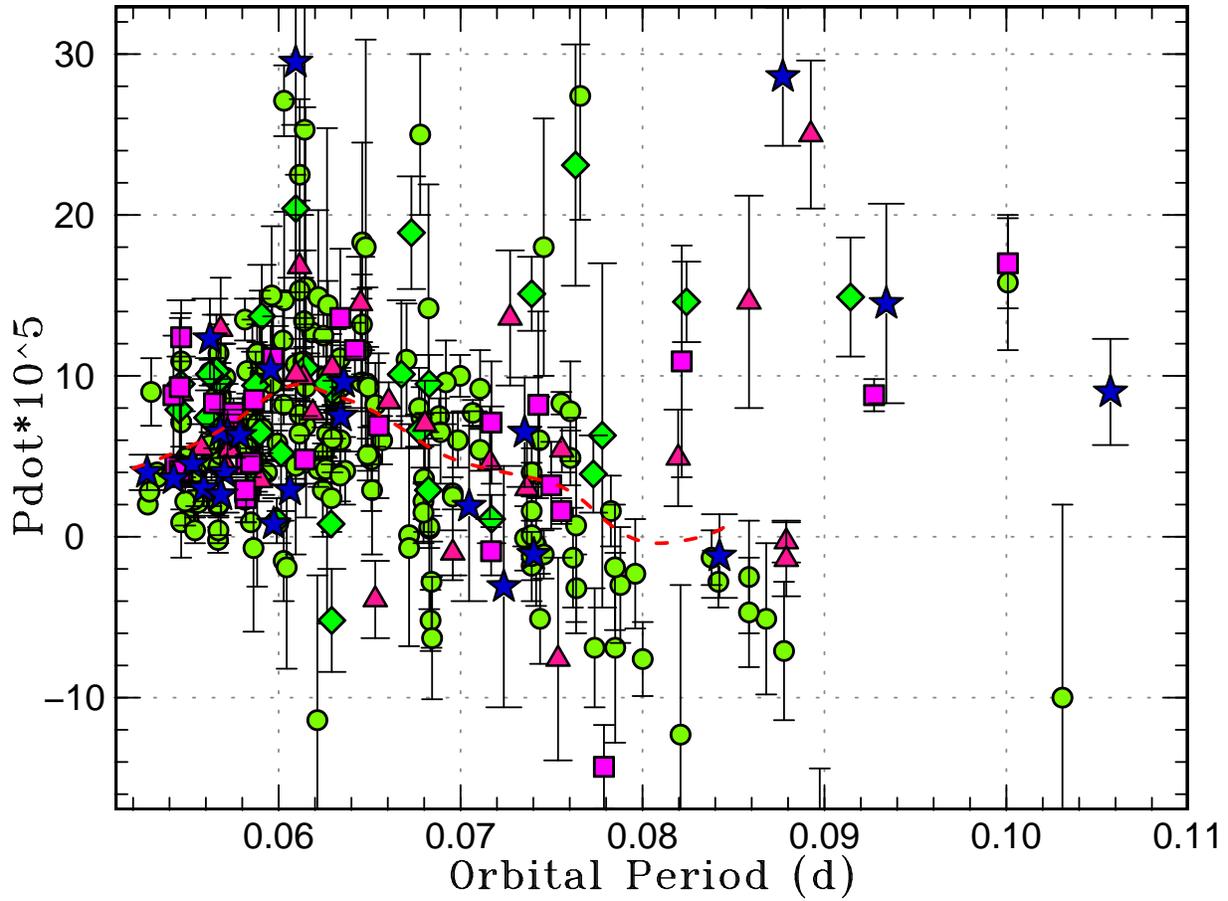}
  \end{center}
  \caption{$P_{\rm dot}$ for stage B versus $P_{\rm orb}$.
  Filled circles, filled diamonds, filled triangles, filled squares
  and filled stars represent samples in 
  \citet{Pdot}, \citet{Pdot2}, \citet{Pdot3},
  \citet{Pdot4} and this paper, respectively.
  The curve represents the spline-smoothed global trend.
  }
  \label{fig:pdotporb5}
\end{figure*}

   \citet{pol11OGLEDNe} reported a large negative ($-2\times 10^{-3}$)
period variation of superhumps in OGLE-BLG-DN-001,
an SU UMa-type dwarf nova in the period gap.  Their data,
however, covered the growing stage of superhumps and
this large negative $P_{\rm dot}$ is most likely a result
of a stage A--B transition.  Since only the initial part
of stage B was observed, more data are needed to determine
the true $P_{\rm dot}$.  The frequency of the outbursts 
in this object looks much lower than in MN Dra and NY Ser
(the objects \cite{pol11OGLEDNe} compared), and it would be
interesting to see how $P_{\rm dot}$ depends on the outburst
frequency in long-$P_{\rm orb}$ systems.
The object, however, very much resembles SDSS J162520
(subsection \ref{obj:j162520}), which showed a post-superoutburst
rebrightening and only two outbursts in the CRTS data.

\subsection{Periods of Stage A Superhumps}

\begin{figure*}
  \begin{center}
    \FigureFile(160mm,110mm){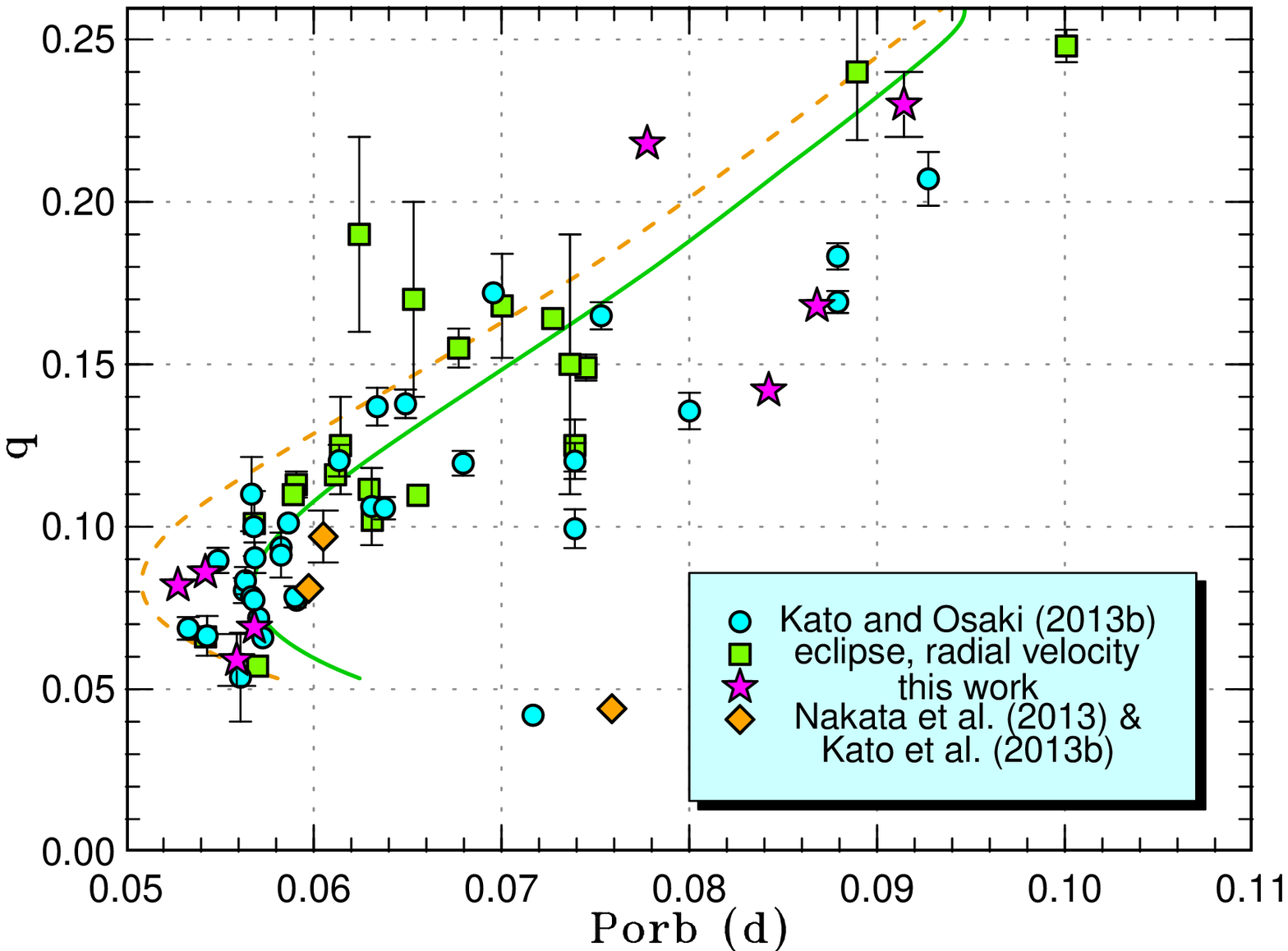}
  \end{center}
  \caption{Mass ratio versus orbital period.
  The dashed and solid curves represent the standard and optimal
  evolutionary tracks in \citet{kni11CVdonor}, respectively.
  The filled circles, filled squares, filled stars, filled diamonds
  represent $q$ values from \citet{kat13qfromstageA}
  [improved $q$ for V1504 Cyg in subsection 4.2 of
  \citet{kat13qfromstageA} is adopted here],
  known $q$ values from quiescent eclipses or radial-velocity
  study (see \cite{kat13qfromstageA} for the data source),
  $q$ estimated in this work and two publications
  \citet{nak13j2112j2037} and \citet{kat13j1222}, respectively.}
  \label{fig:qall2}
\end{figure*}

   It has recently been shown that stage A superhumps can be
directly used in estimating the binary $q$
(\cite{osa13v344lyrv1504cyg}; \cite{kat13qfromstageA}).
Stage A superhumps recorded in the present study are listed
in table \ref{tab:pera}.  The $q$ values can be determined
if the orbital period is known.  The $q$ values for
such objects are discussed section \ref{sec:individual}
individually.  A summary of newly determined $q$ values
is shown in table \ref{tab:newqstageA}.
The best examples are MASTER J211258 and
MASTER J203749, WZ Sge-type dwarf novae with multiple 
rebrightenings, whose $q$ values were estimated using
stage A superhumps \citep{nak13j2112j2037}.
\citet{nak13j2112j2037} suggested that WZ Sge-type dwarf novae
with multiple rebrightenings do not have strikingly smaller
$q$ than other SU UMa-type dwarf novae with similar
orbital periods, and that these objects are not good
candidate for period bouncers.  This suggestion needs
to be tested by further data and observations.
A updated summary of $q$ estimates is shown in
figure \ref{fig:qall2}.
The objects in this table but without known orbital periods
will be prime targets to measure the orbital periods, 
which will lead to an addition of a number of new $q$ estimates 
from stage A superhumps.

\begin{table}
\caption{New estimates for the binary mass ratio from stage A superhumps}\label{tab:newqstageA}
\begin{center}
\begin{tabular}{ccc}
\hline
Object & $\varepsilon^*$ (stage A) & $q$ from stage A \\
\hline
YZ Cnc & 0.0559(12) & 0.168(5) \\
V503 Cyg & 0.0688(12) & 0.218(5) \\
TY PsA & 0.0486(12) & 0.142(4) \\
MASTER J094759 & 0.023(3) & 0.060(8) \\
MASTER J181953 & 0.0259(3) & 0.069(1) \\
OT J112619 & 0.0317(6) & 0.086(2) \\
OT J232727 & 0.0303(5) & 0.082(2) \\
SDSS J162520 & 0.072(2) & 0.23(1) \\
\hline
\end{tabular}
\end{center}
\end{table}

\begin{table}
\caption{Superhump Periods during Stage A}\label{tab:pera}
\begin{center}
\begin{tabular}{cccc}
\hline
Object & Year & period (d) & err \\
\hline
YZ Cnc & 2011 & 0.09194 & 0.00044 \\
V503 Cyg & 2012 & 0.08350 & 0.00011 \\
V660 Her & 2012 & 0.07826 & -- \\
V660 Her & 2013 & 0.07826 & -- \\
V1227 Her & 2012a & 0.06442 & -- \\
V1227 Her & 2012b & 0.06442 & -- \\
V1227 Her & 2013 & 0.06442 & -- \\
TY PsA & 2012 & 0.08854 & 0.00011 \\
AW Sge & 2012 & 0.07743 & 0.00014 \\
ASAS SN-13ax & 2013 & 0.05712 & 0.00002 \\
CSS J102842 & 2013 & 0.03849 & 0.00000 \\
CSS J174033 & 2013 & 0.04640 & 0.00006 \\
MASTER J081110 & 2012 & 0.05876 & 0.00008 \\
MASTER J094759 & 2013 & 0.05717 & 0.00020 \\
MASTER J181953 & 2013 & 0.05835 & 0.00004 \\
MASTER J203749 & 2012 & 0.06271 & -- \\
MASTER J211258 & 2012 & 0.06158 & 0.00005 \\
OT J112619 & 2013 & 0.05601 & 0.00004 \\
OT J232727 & 2012 & 0.05442 & 0.00003 \\
SSS J122221 & 2013 & 0.07721 & 0.00011 \\
\hline
\end{tabular}
\end{center}
\end{table}

\subsection{WZ Sge-Type Stars}\label{sec:wzsgestat}

   New WZ Sge-type dwarf novae and candidates are listed in
table \ref{tab:wztab}.  It is noteworthy that both MASTER
network and amateur astronomers (Itagaki and Kaneko)
were very productive in detecting new WZ Sge-type dwarf novae.

\begin{figure*}
  \begin{center}
    \FigureFile(140mm,100mm){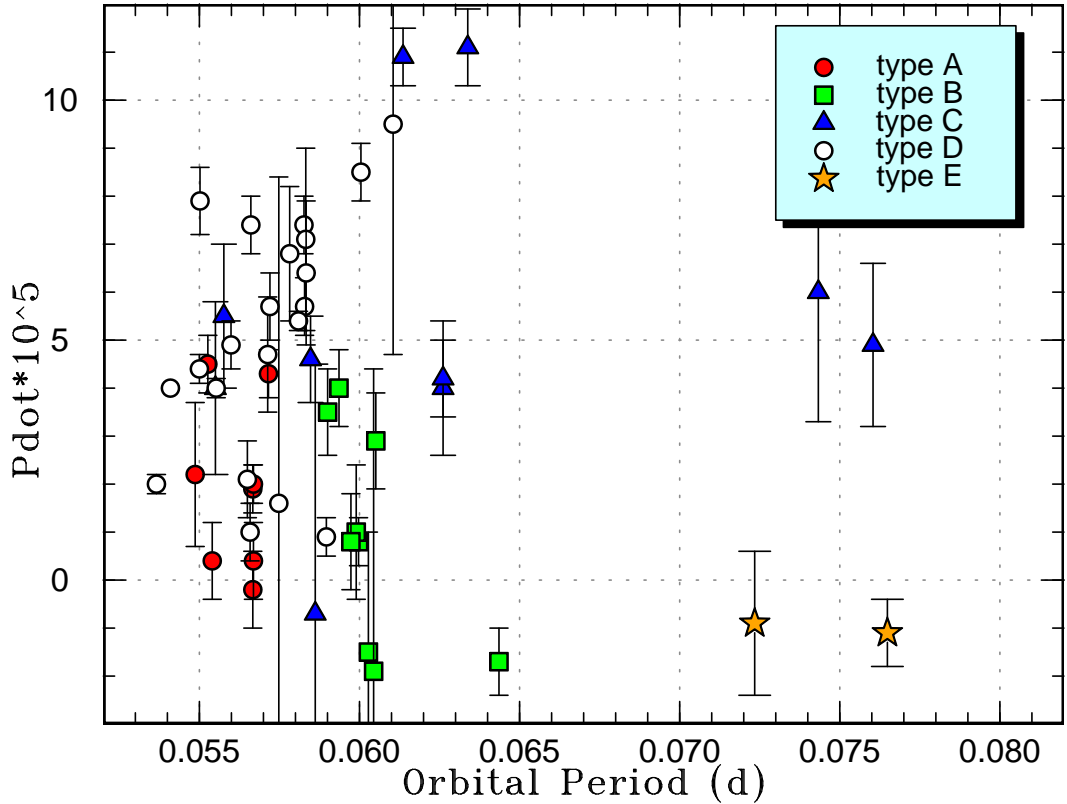}
  \end{center}
  \caption{$P_{\rm dot}$ versus $P_{\rm orb}$ for WZ Sge-type
  dwarf novae.  Symbols represent the type (cf. \cite{Pdot}) of outburst:
  type-A (filled circles), type-B (filled squares),
  type-C (filled triangles), type-D (open circles) and
  newly introduced type-E (filled stars) which show double
  superoutburst as shown in \citet{kat13j1222}.
  }
  \label{fig:wzpdottype5}
\end{figure*}

   Figure \ref{fig:wzpdottype5} shows the updated relation
between $P_{\rm dot}$ and $P_{\rm orb}$ and its relation to
the type of post-superoutburst rebrightening phenomenon.
We used an updated $P_{\rm dot}$ of $+3.5(0.9) \times 10^{-5}$
($31 \le E \le 143$) for stage B superhumps in EZ Lyn (2010)
since the original identification of stage B in \citet{Pdot3}
included the radid fading stage when the period sharply
increased due to the decrease of the pressure effect
\citep{nak13j2112j2037}.

   We here use the types of superoutburst in terms of
rebrightenings as introduced in
\citet{ima06tss0222} and \citet{Pdot}: type-A outburst
(long-duration rebrightening), type-B outburst
(multiple discrete rebrightenings), type-C outburst
(single rebrightening) and type-D outburst (no rebrightening)
(see e.g. figure 35 in \cite{Pdot}). 
In this figure, we newly introduced type-E which show double
superoutburst (superoutburst with early superhumps and
another superoutburst with ordinary superhumps)
as in \citet{kat13j1222}.  OT J184228.1$+$483742 was reclassified
to type-E following the discussion in \citet{kat13j1222}.
These objects showing type-E superoutbursts are good
candidate for period bouncers (\cite{Pdot4}; \cite{kat13j1222}).

\begin{table*}
\caption{Parameters of WZ Sge-type superoutbursts.}\label{tab:wztab}
\begin{center}
\begin{tabular}{cccccccccccc}
\hline
Object & Year & $P_{\rm SH}$ & $P_{\rm orb}$ & $P_{\rm dot}$\commenta & err\commenta & $\epsilon$ & Type\commentb & $N_{\rm reb}$\commentc & delay\commentd & Max & Min \\
\hline
GR Ori & 2013 & 0.058333 & -- & 6.4 & 1.5 & -- & D & 0 & 9 & 13.0 & 22.4 \\
ASAS SN-13ax & 2013 & 0.056155 & -- & 4.5 & 0.6 & -- & A & 1(2) & $\geq$7 & ]13.5 & 21.2 \\
CSS J174033 & 2013 & 0.045548 & 0.045048 & 1.6 & 0.1 & 0.011 & A & 1 & $\geq$7 & ]14.0 & 20.1 \\
MASTER J081110 & 2012 & 0.058147 & -- & 4.5 & 0.3 & -- & -- & -- & -- & ]14.1 & 22.1 \\
MASTER J094759 & 2013 & 0.056121 & 0.05588 & 3.0 & 1.1 & 0.004 & -- & -- & $\geq$2 & ]13.6 & 20.4 \\
MASTER J165236 & 2013 & 0.084732 & -- & -- & -- & -- & -- & -- & -- & ]14.8 & 21.9 \\
MASTER J181953 & 2013 & 0.057519 & 0.05684 & 2.6 & 1.1 & 0.012 & A & $ 1(\geq$3) & $\geq$3 & ]13.9 & 21.6 \\
MASTER J203749 & 2012 & 0.061307 & 0.06051 & 2.9 & 1.0 & 0.013 & B & $\geq$4 & $\geq$3 & ]14.1 & 21.3 \\
MASTER J211258 & 2012 & 0.060227 & 0.059732 & 0.8 & 1.0 & 0.008 & B & 8 & 12 & 14.1 & 21.3 \\
OT J112619 & 2013 & 0.054886 & 0.05423 & 3.6 & 0.4 & 0.012 & -- & -- & $\geq$6 & ]14.8 & 21.8 \\
OT J232727 & 2012 & 0.053438 & 0.05277 & 4.0 & 1.1 & 0.013 & -- & -- & $\geq$11 & ]13.9 & 21.8 \\
PNV J062703 & 2013 & 0.059026 & 0.05787 & 6.3 & 1.3 & 0.020 & -- & -- & $\geq$4 & ]12.0 & 21.0 \\
SSS J122221 & 2013 & 0.076486 & -- & $-$1.1 & 0.7 & -- & E & 0 & -- & ]11.8 & 18.7 \\
TCP J153756 & 2013 & 0.061899 & 0.06101 & -- & -- & 0.015 & -- & -- & $\geq$4 & ]13.6 & 21.7 \\
\hline
  \multicolumn{12}{l}{\commenta Unit $10^{-5}$.} \\
  \multicolumn{12}{l}{\commentb A: long-lasting rebrightening; B: multiple rebegitehnings; C: single rebrightening; D: no rebrightening.} \\
  \multicolumn{12}{l}{\commentc Number of rebrightenings.} \\
  \multicolumn{12}{l}{\commentd Days before ordinary superhumps appeared.} \\
\end{tabular}
\end{center}
\end{table*}

\begin{table*}
\caption{Rate of slow decline in SU UMa-type dwarf novae}\label{tab:suumarate}
\begin{center}
\begin{tabular}{cccccccc}
\hline
Object & Year & Rate\commenta & Error\commenta & Start\commentb & End\commentb & Mean $P_{\rm SH}$ (stage B)\commentc & Source\commentd \\
\hline
FO And & 2010 & 0.106 & 0.001 & 55487.3 & 55494.7 & 0.07451 & 3 \\
KV And & 1994 & 0.135 & 0.003 & 49578.2 & 49583.3 & 0.07460 & 1 \\
KV And & 2002 & 0.079 & 0.001 & 52584.1 & 52591.2 & 0.07450 & 1 \\
V402 And & 2005 & 0.182 & 0.007 & 53671.1 & 53673.7 & 0.06323 & 1 \\
V402 And & 2008 & 0.130 & 0.002 & 54755.0 & 54761.2 & 0.06353 & 1 \\
DH Aql & 2002 & 0.099 & 0.001 & 52483.1 & 52493.2 & 0.08002 & 1 \\
V1141 Aql & 2003 & 0.131 & 0.007 & 52824.1 & 52827.5 & 0.06296 & 1 \\
VY Aqr & 2008 & 0.113 & 0.001 & 54648.1 & 54656.3 & 0.06466 & 1 \\
EG Aqr & 2006 & 0.140 & 0.001 & 54050.2 & 54057.0 & 0.07896 & 1 \\
EG Aqr & 2008 & 0.143 & 0.001 & 54802.9 & 54808.0 & 0.07876 & 1 \\
QV Aqr & 2011 & 0.173 & 0.004 & 55834.4 & 55836.5 & 0.07531 & 4 \\
QZ Aqr & 2008 & 0.127 & 0.005 & 54819.8 & 54821.9 & 0.06463 & 1 \\
BF Ara & 2002 & 0.101 & 0.001 & 52505.0 & 52510.4 & 0.08789 & 1 \\
V663 Ara & 2004 & 0.110 & 0.002 & 53195.4 & 53199.5 & 0.07642 & 1 \\
V877 Ara & 2002 & 0.117 & 0.001 & 52435.0 & 52443.3 & 0.08393 & 1 \\
BG Ari & 2009 & 0.076 & 0.001 & 55086.5 & 55099.6 & 0.08510 & 2 \\
TT Boo & 2004 & 0.073 & 0.001 & 53161.0 & 53174.8 & 0.07809 & 1 \\
TT Boo & 2010 & 0.104 & 0.001 & 55311.1 & 55318.3 & 0.07812 & 3 \\
TT Boo & 2012 & 0.066 & 0.001 & 56016.4 & 56021.3 & 0.07808 & 4 \\
NN Cam & 2007 & 0.077 & 0.001 & 54364.5 & 54367.6 & 0.07429 & 1 \\
NN Cam & 2009 & 0.073 & 0.001 & 55142.3 & 55147.8 & 0.07426 & 2 \\
NN Cam & 2011 & 0.108 & 0.001 & 55905.0 & 55913.1 & 0.07420 & 4 \\
V342 Cam & 2008 & 0.149 & 0.001 & 54532.2 & 54538.8 & 0.07840 & 1 \\
V342 Cam & 2010 & 0.142 & 0.001 & 55444.1 & 55451.3 & 0.07846 & 2 \\
V391 Cam & 2005 & 0.156 & 0.001 & 53450.0 & 53455.4 & 0.05716 & 1 \\
V391 Cam & 2008 & 0.161 & 0.001 & 54477.9 & 54481.0 & 0.05713 & 1 \\
SY Cap & 2011 & 0.085 & 0.003 & 55803.1 & 55805.1 & 0.06375 & 4 \\
GX Cas & 1999 & 0.080 & 0.001 & 51472.1 & 51477.3 & 0.09352 & 1 \\
GX Cas & 2010 & 0.116 & 0.001 & 55502.0 & 55505.3 & 0.09300 & 3 \\
KP Cas & 2008 & 0.131 & 0.001 & 54767.0 & 54771.0 & 0.08553 & 1 \\
V359 Cen & 2002 & 0.146 & 0.002 & 52425.8 & 52430.8 & 0.08121 & 1 \\
V1040 Cen & 2002 & 0.171 & 0.001 & 52366.2 & 52372.7 & 0.06218 & 1 \\
WX Cet & 1989 & 0.120 & 0.001 & 47683.6 & 47695.7 & 0.05962 & 1 \\
WX Cet & 1998 & 0.120 & 0.001 & 51128.9 & 51138.2 & 0.05962 & 1 \\
WX Cet & 2001 & 0.162 & 0.002 & 52092.3 & 52097.9 & 0.05955 & 1 \\
WX Cet & 2004 & 0.102 & 0.001 & 53347.9 & 53358.0 & 0.05953 & 1 \\
RX Cha & 2009 & 0.091 & 0.005 & 54858.0 & 54860.1 & 0.08492 & 1 \\
BZ Cir & 2004 & 0.111 & 0.001 & 53184.2 & 53193.5 & 0.07661 & 1 \\
PU CMa & 2005 & 0.152 & 0.001 & 53401.9 & 53407.1 & 0.05801 & 1 \\
PU CMa & 2009 & 0.221 & 0.001 & 55159.1 & 55162.4 & 0.05809 & 2 \\
AQ CMi & 2010 & 0.117 & 0.001 & 55296.0 & 55303.0 & 0.06618 & 2 \\
YZ Cnc & 2007 & 0.105 & 0.001 & 54144.0 & 54152.0 & 0.09031 & 1 \\
AK Cnc & 1992 & 0.092 & 0.002 & 48639.1 & 48644.3 & 0.06751 & 1 \\
CC Cnc & 2001 & 0.117 & 0.001 & 52226.3 & 52233.4 & 0.07589 & 1 \\
GZ Cnc & 2010 & 0.096 & 0.001 & 55269.1 & 55273.2 & 0.09277 & 1 \\
GZ Cnc & 2013 & 0.151 & 0.002 & 56331.9 & 56335.1 & 0.09284 & 5 \\
KK Cnc & 2007 & 0.118 & 0.001 & 54424.9 & 54430.3 & 0.06105 & 1 \\
GO Com & 2003 & 0.145 & 0.001 & 52795.1 & 52801.4 & 0.06308 & 1 \\
GO Com & 2005 & 0.188 & 0.001 & 53484.0 & 53488.3 & 0.06305 & 1 \\
GO Com & 2010 & 0.170 & 0.002 & 55287.6 & 55291.1 & 0.06307 & 2 \\
GO Com & 2012 & 0.139 & 0.002 & 55984.3 & 55987.7 & 0.06302 & 4 \\
V728 CrA & 2003 & 0.141 & 0.001 & 52821.3 & 52825.6 & 0.08238 & 1 \\
VW CrB & 2003 & 0.109 & 0.001 & 52849.4 & 52855.5 & 0.07292 & 1 \\
VW CrB & 2006 & 0.085 & 0.004 & 53842.7 & 53846.7 & 0.07268 & 1 \\
\hline
  \multicolumn{8}{l}{\commenta Unit: mag d$^{-1}$.}\\
  \multicolumn{8}{l}{\commentb BJD$-$2400000.}\\
  \multicolumn{8}{l}{\commentc Unit: d.}\\
  \multicolumn{8}{l}{\commentd 1: \citet{Pdot}, 2: \citet{Pdot2}, 3: \citet{Pdot3}, 4: \citet{Pdot4}, 5: this work.}\\
\end{tabular}
\end{center}
\end{table*}

\addtocounter{table}{-1}
\begin{table*}
\caption{Rate of slow decline in SU UMa-type dwarf novae (continued)}
\begin{center}
\begin{tabular}{cccccccc}
\hline
Object & Year & Rate\commenta & Error\commenta & Start\commentb & End\commentb & Mean $P_{\rm SH}$ (stage B)\commentc & Source\commentd \\
\hline
TU Crt & 2001 & 0.121 & 0.002 & 52010.0 & 52020.0 & 0.08518 & 1 \\
TV Crv & 2001 & 0.123 & 0.001 & 51961.1 & 51968.2 & 0.06500 & 1 \\
TV Crv & 2004 & 0.136 & 0.002 & 53162.5 & 53168.7 & 0.06509 & 1 \\
TV Crv & 2009 & 0.138 & 0.002 & 55183.2 & 55189.4 & 0.06506 & 2 \\
V337 Cyg & 2006 & 0.167 & 0.002 & 53886.4 & 53888.6 & 0.07000 & 1 \\
V337 Cyg & 2010 & 0.152 & 0.001 & 55421.4 & 55425.2 & 0.07033 & 2 \\
V503 Cyg & 2002 & 0.153 & 0.002 & 52478.2 & 52483.3 & 0.08139 & 1 \\
V503 Cyg & 2008 & 0.090 & 0.001 & 54822.9 & 54828.9 & 0.08177 & 1 \\
V503 Cyg & 2011 & 0.097 & 0.001 & 55744.4 & 55751.6 & 0.08131 & 4 \\
V503 Cyg & 2011b & 0.113 & 0.001 & 55831.2 & 55840.5 & 0.08124 & 4 \\
V503 Cyg & 2012 & 0.069 & 0.001 & 56093.7 & 56099.0 & 0.08123 & 5 \\
V630 Cyg & 2008 & 0.122 & 0.002 & 54692.0 & 54698.2 & 0.07918 & 1 \\
V632 Cyg & 2008 & 0.112 & 0.001 & 54782.9 & 54791.0 & 0.06583 & 1 \\
V1028 Cyg & 1995 & 0.124 & 0.001 & 49928.1 & 49938.3 & 0.06175 & 1 \\
V1028 Cyg & 1999 & 0.129 & 0.002 & 51430.1 & 51435.3 & 0.06170 & 1 \\
V1028 Cyg & 2002 & 0.142 & 0.002 & 52618.9 & 52624.0 & 0.06177 & 1 \\
V1113 Cyg & 1994 & 0.120 & 0.004 & 49600.0 & 49603.1 & 0.07906 & 1 \\
V1113 Cyg & 2008 & 0.135 & 0.002 & 54757.3 & 54760.0 & 0.07905 & 1 \\
V1454 Cyg & 2006 & 0.099 & 0.001 & 54068.9 & 54080.9 & 0.06102 & 1 \\
V1454 Cyg & 2009 & 0.129 & 0.002 & 55058.1 & 55063.6 & 0.05765 & 2 \\
V1504 Cyg & 2007 & 0.129 & 0.002 & 54326.4 & 54333.6 & 0.07232 & 3 \\
HO Del & 1994 & 0.124 & 0.002 & 49591.0 & 49595.2 & 0.06456 & 1 \\
HO Del & 2008 & 0.127 & 0.001 & 54684.4 & 54690.8 & 0.06436 & 1 \\
BC Dor & 2003 & 0.025 & 0.001 & 52958.0 & 52968.1 & 0.06847 & 1 \\
CP Dra & 2003 & 0.130 & 0.002 & 52648.1 & 52653.2 & 0.08370 & 1 \\
CP Dra & 2009 & 0.199 & 0.001 & 54917.4 & 54922.5 & 0.08382 & 1 \\
DM Dra & 2003 & 0.152 & 0.002 & 52706.2 & 52710.4 & 0.07571 & 1 \\
KV Dra & 2002 & 0.169 & 0.003 & 52519.5 & 52523.4 & 0.06030 & 1 \\
KV Dra & 2004 & 0.155 & 0.001 & 53120.0 & 53127.2 & 0.06045 & 1 \\
KV Dra & 2005 & 0.171 & 0.002 & 53465.2 & 53468.3 & 0.06034 & 1 \\
AQ Eri & 2006 & 0.083 & 0.002 & 54070.1 & 54076.2 & 0.06168 & 1 \\
AQ Eri & 2008 & 0.101 & 0.001 & 54830.0 & 54836.1 & 0.06236 & 1 \\
AQ Eri & 2010 & 0.160 & 0.001 & 55202.0 & 55206.2 & 0.06237 & 2 \\
AQ Eri & 2012 & 0.087 & 0.001 & 56213.3 & 56217.2 & 0.06240 & 5 \\
KY Eri & 2004 & 0.172 & 0.003 & 53198.5 & 53204.9 & 0.06864 & 1 \\
AX For & 2005 & 0.116 & 0.001 & 53555.6 & 53563.7 & 0.08120 & 1 \\
UV Gem & 2003 & 0.087 & 0.001 & 52647.9 & 52653.6 & 0.09355 & 1 \\
AW Gem & 1995 & 0.076 & 0.002 & 50001.2 & 50005.4 & 0.07983 & 1 \\
AW Gem & 2008 & 0.111 & 0.001 & 54565.0 & 54572.1 & 0.07899 & 1 \\
AW Gem & 2010 & 0.122 & 0.002 & 55257.3 & 55261.6 & 0.07906 & 2 \\
AW Gem & 2011 & 0.136 & 0.001 & 55572.1 & 55579.3 & 0.07938 & 3 \\
CI Gem & 2005 & 0.094 & 0.003 & 53476.3 & 53485.7 & 0.11931 & 1 \\
V660 Her & 2012 & 0.119 & 0.003 & 56177.4 & 56179.5 & 0.08089 & 5 \\
V844 Her & 1997 & 0.091 & 0.001 & 50592.4 & 50602.5 & 0.05601 & 1 \\
V844 Her & 1999 & 0.107 & 0.002 & 51451.9 & 51457.0 & 0.05591 & 1 \\
V844 Her & 2006 & 0.151 & 0.001 & 53854.0 & 53860.3 & 0.05587 & 1 \\
V844 Her & 2008 & 0.134 & 0.001 & 54577.1 & 54584.5 & 0.05593 & 1 \\
V844 Her & 2009 & 0.136 & 0.001 & 54887.5 & 54892.7 & 0.05592 & 2 \\
V844 Her & 2010b & 0.085 & 0.002 & 55496.3 & 55499.3 & 0.05611 & 3 \\
V844 Her & 2012 & 0.155 & 0.001 & 56052.0 & 56057.3 & 0.05590 & 4 \\
V1227 Her & 2012b & 0.131 & 0.002 & 56183.6 & 56187.8 & 0.06508 & 5 \\
RU Hor & 2003 & 0.104 & 0.001 & 52912.4 & 52917.4 & 0.07095 & 1 \\
RU Hor & 2008 & 0.128 & 0.001 & 54686.5 & 54688.7 & 0.07103 & 1 \\
CT Hya & 1999 & 0.127 & 0.002 & 51224.9 & 51231.2 & 0.06642 & 1 \\
CT Hya & 2000 & 0.169 & 0.002 & 51880.1 & 51887.4 & 0.06639 & 1 \\
CT Hya & 2002a & 0.099 & 0.003 & 52317.1 & 52321.3 & 0.06638 & 1 \\
\hline
  \multicolumn{8}{l}{\commenta Unit: mag d$^{-1}$.}\\
  \multicolumn{8}{l}{\commentb BJD$-$2400000.}\\
  \multicolumn{8}{l}{\commentc Unit: d.}\\
  \multicolumn{8}{l}{\commentd 1: \citet{Pdot}, 2: \citet{Pdot2}, 3: \citet{Pdot3}, 4: \citet{Pdot4}, 5: this work.}\\
\end{tabular}
\end{center}
\end{table*}

\addtocounter{table}{-1}
\begin{table*}
\caption{Rate of slow decline in SU UMa-type dwarf novae (continued)}
\begin{center}
\begin{tabular}{cccccccc}
\hline
Object & Year & Rate\commenta & Error\commenta & Start\commentb & End\commentb & Mean $P_{\rm SH}$ (stage B)\commentc & Source\commentd \\
\hline
CT Hya & 2002b & 0.107 & 0.002 & 52592.1 & 52601.4 & 0.06641 & 1 \\
CT Hya & 2009 & 0.161 & 0.003 & 54848.0 & 54852.0 & 0.06663 & 1 \\
MM Hya & 1998 & 0.073 & 0.005 & 50882.3 & 50886.3 & 0.05896 & 1 \\
MM Hya & 2011 & 0.145 & 0.001 & 55661.0 & 55665.2 & 0.05885 & 3 \\
MM Hya & 2012 & 0.117 & 0.001 & 55993.6 & 56000.8 & 0.05887 & 4 \\
V498 Hya & 2008 & 0.156 & 0.001 & 54491.0 & 54497.2 & 0.06047 & 1 \\
VW Hyi & 2011 & 0.095 & 0.001 & 55895.5 & 55901.8 & 0.07691 & 4 \\
RZ LMi & 2012 & 0.084 & 0.002 & 55985.5 & 55989.7 & 0.05944 & 4 \\
RZ LMi & 2012b & 0.064 & 0.001 & 56013.4 & 56017.8 & 0.05947 & 4 \\
RZ LMi & 2012c & 0.076 & 0.001 & 56031.6 & 56036.8 & 0.05941 & 4 \\
SX LMi & 1994 & 0.160 & 0.004 & 49702.2 & 49705.3 & 0.06948 & 1 \\
SX LMi & 2001 & 0.121 & 0.002 & 51938.3 & 51944.3 & 0.06914 & 1 \\
SX LMi & 2002 & 0.144 & 0.003 & 52300.3 & 52307.3 & 0.06934 & 1 \\
BK Lyn & 2012b & 0.059 & 0.001 & 56023.3 & 56030.7 & 0.07851 & 4 \\
FV Lyn & 2007 & 0.076 & 0.001 & 54161.0 & 54172.2 & 0.06977 & 1 \\
AY Lyr & 2008 & 0.076 & 0.002 & 54754.9 & 54758.1 & 0.07623 & 1 \\
AY Lyr & 2009 & 0.085 & 0.001 & 54963.0 & 54970.3 & 0.07616 & 1 \\
DM Lyr & 1996 & 0.147 & 0.003 & 50280.0 & 50282.2 & 0.06709 & 1 \\
DM Lyr & 1997 & 0.095 & 0.003 & 50509.3 & 50513.4 & 0.06721 & 1 \\
DM Lyr & 2002 & 0.188 & 0.013 & 52583.9 & 52587.0 & 0.06723 & 1 \\
V344 Lyr & 1993 & 0.099 & 0.002 & 49133.1 & 49140.3 & 0.09135 & 1 \\
V419 Lyr & 1999 & 0.065 & 0.002 & 51415.1 & 51422.2 & 0.09015 & 1 \\
V419 Lyr & 2006 & 0.109 & 0.001 & 53935.4 & 53941.5 & 0.09006 & 1 \\
V585 Lyr & 2003 & 0.152 & 0.001 & 52900.3 & 52906.9 & 0.06036 & 1 \\
FQ Mon & 2004 & 0.090 & 0.001 & 53068.9 & 53081.1 & 0.07335 & 1 \\
FQ Mon & 2006 & 0.105 & 0.002 & 53759.0 & 53764.2 & 0.07392 & 1 \\
FQ Mon & 2007 & 0.083 & 0.001 & 54466.1 & 54472.2 & 0.07335 & 1 \\
AB Nor & 2002 & 0.066 & 0.003 & 52519.2 & 52522.0 & 0.07962 & 1 \\
AB Nor & 2013 & 0.107 & 0.001 & 56436.2 & 56441.7 & 0.07976 & 5 \\
DT Oct & 2003 & 0.137 & 0.001 & 52645.9 & 52652.2 & 0.07476 & 1 \\
V699 Oph & 2003 & 0.086 & 0.001 & 52824.2 & 52829.1 & 0.07033 & 1 \\
V699 Oph & 2008 & 0.070 & 0.001 & 54618.1 & 54624.2 & 0.07013 & 1 \\
V2527 Oph & 2004 & 0.093 & 0.001 & 53211.9 & 53219.1 & 0.07205 & 1 \\
V2527 Oph & 2008 & 0.103 & 0.001 & 54711.0 & 54720.0 & 0.07194 & 1 \\
V1159 Ori & 2002 & 0.076 & 0.001 & 52605.2 & 52610.3 & 0.06414 & 1 \\
EF Peg & 1991 & 0.079 & 0.001 & 48547.9 & 48555.1 & 0.08693 & 1 \\
EF Peg & 1997 & 0.058 & 0.001 & 50758.0 & 50764.1 & 0.08704 & 1 \\
EF Peg & 2009 & 0.045 & 0.001 & 55187.9 & 55195.0 & 0.08735 & 2 \\
V364 Peg & 2004 & 0.173 & 0.004 & 53329.2 & 53331.7 & 0.08534 & 1 \\
V368 Peg & 2000 & 0.105 & 0.001 & 51786.0 & 51792.3 & 0.07038 & 1 \\
V368 Peg & 2005 & 0.135 & 0.001 & 53621.0 & 53628.1 & 0.07038 & 1 \\
V368 Peg & 2009 & 0.144 & 0.001 & 55101.5 & 55112.9 & 0.07036 & 2 \\
V444 Peg & 2008 & 0.132 & 0.001 & 54778.0 & 54783.3 & 0.09945 & 1 \\
V444 Peg & 2012 & 0.212 & 0.001 & 56193.9 & 56199.6 & 0.09764 & 5 \\
UV Per & 2000 & 0.124 & 0.001 & 51904.5 & 51913.1 & 0.06663 & 1 \\
UV Per & 2003 & 0.110 & 0.001 & 52949.3 & 52957.4 & 0.06667 & 1 \\
UV Per & 2010 & 0.122 & 0.003 & 55203.4 & 55205.6 & 0.06671 & 2 \\
PV Per & 2008 & 0.019 & 0.002 & 54745.4 & 54753.2 & 0.08080 & 1 \\
QY Per & 1999 & 0.123 & 0.001 & 51545.1 & 51552.3 & 0.07861 & 1 \\
QY Per & 2005 & 0.124 & 0.001 & 53667.0 & 53676.3 & 0.07861 & 1 \\
TY PsA & 2008 & 0.135 & 0.001 & 54798.9 & 54805.0 & 0.08799 & 1 \\
TY PsA & 2012 & 0.118 & 0.001 & 56163.1 & 56171.1 & 0.08781 & 5 \\
TY Psc & 2005 & 0.067 & 0.002 & 53614.2 & 53616.3 & 0.07034 & 1 \\
TY Psc & 2008 & 0.097 & 0.001 & 54754.9 & 54759.5 & 0.07066 & 1 \\
GV Psc & 2011 & 0.109 & 0.002 & 55852.3 & 55854.9 & 0.09431 & 4 \\
VZ Pyx & 2008 & 0.084 & 0.001 & 54790.2 & 54796.4 & 0.07605 & 1 \\
DT Pyx & 2005 & 0.082 & 0.002 & 53448.0 & 53450.1 & 0.06289 & 1 \\
\hline
  \multicolumn{8}{l}{\commenta Unit: mag d$^{-1}$.}\\
  \multicolumn{8}{l}{\commentb BJD$-$2400000.}\\
  \multicolumn{8}{l}{\commentc Unit: d.}\\
  \multicolumn{8}{l}{\commentd 1: \citet{Pdot}, 2: \citet{Pdot2}, 3: \citet{Pdot3}, 4: \citet{Pdot4}, 5: this work.}\\
\end{tabular}
\end{center}
\end{table*}

\addtocounter{table}{-1}
\begin{table*}
\caption{Rate of slow decline in SU UMa-type dwarf novae (continued)}
\begin{center}
\begin{tabular}{cccccccc}
\hline
Object & Year & Rate\commenta & Error\commenta & Start\commentb & End\commentb & Mean $P_{\rm SH}$ (stage B)\commentc & Source\commentd \\
\hline
DV Sco & 2004 & 0.135 & 0.002 & 53272.2 & 53277.3 & 0.09978 & 1 \\
QW Ser & 2002 & 0.060 & 0.001 & 52427.2 & 52435.2 & 0.07703 & 1 \\
V493 Ser & 2007 & 0.114 & 0.001 & 54312.0 & 54324.1 & 0.08296 & 1 \\
RZ Sge & 1994 & 0.094 & 0.002 & 49576.0 & 49580.1 & 0.07057 & 1 \\
RZ Sge & 1996 & 0.139 & 0.002 & 50305.1 & 50308.3 & 0.07064 & 1 \\
RZ Sge & 2002 & 0.096 & 0.001 & 52549.0 & 52558.1 & 0.07044 & 1 \\
AW Sge & 2012 & 0.108 & 0.001 & 56127.1 & 56132.9 & 0.07473 & 5 \\
V551 Sgr & 2003 & 0.143 & 0.001 & 52903.9 & 52910.4 & 0.06760 & 1 \\
V4140 Sgr & 2004 & 0.083 & 0.001 & 53269.2 & 53286.7 & 0.06351 & 1 \\
V701 Tau & 1995 & 0.088 & 0.001 & 50078.0 & 50090.1 & 0.06908 & 1 \\
V1208 Tau & 2000 & 0.106 & 0.004 & 51580.7 & 51586.2 & 0.07050 & 1 \\
V1208 Tau & 2002 & 0.120 & 0.001 & 52635.1 & 52642.1 & 0.07054 & 1 \\
V1212 Tau & 2011 & 0.101 & 0.001 & 55589.6 & 55601.4 & 0.07011 & 3 \\
V1265 Tau & 2006 & 0.085 & 0.001 & 54035.4 & 54043.5 & 0.05341 & 1 \\
EK TrA & 2009 & 0.124 & 0.001 & 55028.9 & 55036.1 & 0.06483 & 2 \\
FL TrA & 2005 & 0.136 & 0.001 & 53581.2 & 53585.4 & 0.05985 & 1 \\
SU UMa & 1999 & 0.113 & 0.001 & 51189.9 & 51197.1 & 0.07909 & 1 \\
SU UMa & 2010 & 0.128 & 0.002 & 55220.5 & 55223.6 & 0.07907 & 2 \\
SW UMa & 1996 & 0.134 & 0.001 & 50191.3 & 50198.0 & 0.05819 & 1 \\
SW UMa & 1997 & 0.119 & 0.001 & 50744.5 & 50752.7 & 0.05828 & 1 \\
SW UMa & 2000 & 0.133 & 0.001 & 51589.9 & 51601.4 & 0.05826 & 1 \\
SW UMa & 2002 & 0.103 & 0.001 & 52572.2 & 52581.3 & 0.05832 & 1 \\
SW UMa & 2006 & 0.113 & 0.001 & 53997.2 & 54009.3 & 0.05821 & 1 \\
SW UMa & 2010 & 0.155 & 0.001 & 55538.2 & 55545.4 & 0.05821 & 3 \\
BC UMa & 2000 & 0.131 & 0.001 & 51638.3 & 51647.1 & 0.06456 & 1 \\
BC UMa & 2003 & 0.125 & 0.001 & 52674.0 & 52682.4 & 0.06457 & 1 \\
BC UMa & 2009 & 0.129 & 0.001 & 55108.9 & 55118.7 & 0.06455 & 2 \\
BZ UMa & 2007 & 0.119 & 0.001 & 54205.3 & 54213.6 & 0.07018 & 1 \\
CI UMa & 2003 & 0.100 & 0.001 & 52739.3 & 52747.5 & 0.06269 & 1 \\
CI UMa & 2011 & 0.211 & 0.002 & 55660.4 & 55663.6 & 0.06269 & 3 \\
CI UMa & 2013 & 0.222 & 0.001 & 56385.3 & 56396.3 & 0.06238 & 5 \\
CY UMa & 1998 & 0.110 & 0.001 & 50882.4 & 50891.6 & 0.07246 & 1 \\
CY UMa & 1999 & 0.142 & 0.001 & 51222.9 & 51230.0 & 0.07222 & 1 \\
CY UMa & 2009 & 0.146 & 0.001 & 54917.9 & 54924.6 & 0.07222 & 1 \\
DI UMa & 2007 & 0.077 & 0.002 & 54206.3 & 54211.5 & 0.05531 & 4 \\
KS UMa & 2003 & 0.125 & 0.001 & 52691.0 & 52696.7 & 0.07018 & 1 \\
KS UMa & 2007 & 0.129 & 0.001 & 54150.1 & 54155.3 & 0.07026 & 1 \\
MR UMa & 2002 & 0.102 & 0.001 & 52342.0 & 52347.4 & 0.06516 & 1 \\
MR UMa & 2003 & 0.123 & 0.001 & 52712.0 & 52717.5 & 0.06514 & 1 \\
SS UMi & 2012 & 0.159 & 0.002 & 56008.4 & 56012.0 & 0.07036 & 4 \\
CU Vel & 2002 & 0.107 & 0.001 & 52620.2 & 52630.2 & 0.08094 & 1 \\
HS Vir & 1996 & 0.119 & 0.001 & 50155.2 & 50161.3 & 0.08006 & 1 \\
HS Vir & 2008 & 0.119 & 0.001 & 54619.0 & 54624.0 & 0.08003 & 1 \\
QZ Vir & 1993 & 0.134 & 0.001 & 48993.2 & 49000.3 & 0.06035 & 1 \\
QZ Vir & 2007 & 0.135 & 0.001 & 54111.1 & 54113.4 & 0.06048 & 1 \\
QZ Vir & 2008 & 0.159 & 0.001 & 54470.2 & 54473.4 & 0.06044 & 1 \\
QZ Vir & 2009 & 0.133 & 0.001 & 54857.1 & 54860.4 & 0.06038 & 1 \\
RX Vol & 2003 & 0.135 & 0.001 & 52764.0 & 52770.3 & 0.06136 & 1 \\
TY Vul & 2010 & 0.122 & 0.001 & 55371.4 & 55375.9 & 0.08046 & 2 \\
DO Vul & 2008 & 0.073 & 0.003 & 54671.0 & 54674.1 & 0.05820 & 1 \\
NSV 04838 & 2007 & 0.098 & 0.001 & 54143.1 & 54151.2 & 0.06992 & 1 \\
NSV 14652 & 2004 & 0.111 & 0.002 & 53251.4 & 53256.4 & 0.08151 & 1 \\
1RXS J231935.0$+$364705 & 2011 & 0.158 & 0.001 & 55835.4 & 55843.3 & 0.06599 & 4 \\
ASAS J224349$+$0809.5 & 2009 & 0.097 & 0.001 & 55113.0 & 55121.8 & 0.06981 & 2 \\
CSS J015051.7$+$332621 & 2012 & 0.156 & 0.001 & 56211.6 & 56217.1 & 0.07271 & 5 \\
\hline
  \multicolumn{8}{l}{\commenta Unit: mag d$^{-1}$.}\\
  \multicolumn{8}{l}{\commentb BJD$-$2400000.}\\
  \multicolumn{8}{l}{\commentc Unit: d.}\\
  \multicolumn{8}{l}{\commentd 1: \citet{Pdot}, 2: \citet{Pdot2}, 3: \citet{Pdot3}, 4: \citet{Pdot4}, 5: this work.}\\
\end{tabular}
\end{center}
\end{table*}

\addtocounter{table}{-1}
\begin{table*}
\caption{Rate of slow decline in SU UMa-type dwarf novae (continued)}
\begin{center}
\begin{tabular}{cccccccc}
\hline
Object & Year & Rate\commenta & Error\commenta & Start\commentb & End\commentb & Mean $P_{\rm SH}$ (stage B)\commentc & Source\commentd \\
\hline
CSS J105835.1$+$054703 & 2012 & 0.104 & 0.001 & 56268.2 & 56275.0 & 0.05788 & 5 \\
CSS J203937.7$-$042907 & 2012 & 0.087 & 0.001 & 56155.6 & 56163.7 & 0.11121 & 5 \\
Lanning 420 & 2010 & 0.119 & 0.001 & 55438.6 & 55444.7 & 0.06159 & 2 \\
MASTER OT J042609.34$+$354144.8 & 2012 & 0.188 & 0.003 & 56202.6 & 56205.3 & 0.06756 & 5 \\
MASTER OT J054317.95$+$093114.8 & 2012 & 0.101 & 0.001 & 56205.8 & 56218.9 & 0.07595 & 5 \\
MASTER OT J064725.70$+$491543.9 & 2013 & 0.134 & 0.001 & 56362.6 & 56366.1 & 0.06777 & 5 \\
MASTER OT J081110.46$+$660008.5 & 2012 & 0.081 & 0.001 & 56225.3 & 56234.0 & 0.05815 & 5 \\
MisV 1446 & 2012 & 0.171 & 0.002 & 55936.5 & 55940.1 & 0.07807 & 4 \\
SDSS J033449.86$-$071047.8 & 2009 & 0.122 & 0.001 & 54856.0 & 54860.1 & 0.07477 & 1 \\
SDSS J073208.11$+$413008.7 & 2010 & 0.082 & 0.001 & 55199.6 & 55210.0 & 0.07995 & 2 \\
SDSS J074640.62$+$173412.8 & 2009 & 0.127 & 0.002 & 54874.9 & 54884.2 & 0.06679 & 1 \\
SDSS J075107.50$+$300628.4 & 2013 & 0.108 & 0.002 & 56338.5 & 56340.7 & 0.05798 & 5 \\
SDSS J080303.90$+$251627.0 & 2011 & 0.096 & 0.004 & 55654.3 & 55657.1 & 0.09195 & 4 \\
SDSS J080306.99$+$284855.8 & 2011 & 0.096 & 0.004 & 55654.3 & 55657.1 & 0.07510 & 3 \\
SDSS J081207.63$+$131824.4 & 2008 & 0.142 & 0.002 & 54754.2 & 54759.3 & 0.07756 & 3 \\
SDSS J081207.63$+$131824.4 & 2011 & 0.133 & 0.001 & 55645.5 & 55653.0 & 0.07789 & 3 \\
SDSS J083931.35$+$282824.0 & 2010 & 0.139 & 0.002 & 55295.3 & 55298.5 & 0.07852 & 2 \\
SDSS J125023.85$+$665525.5 & 2008 & 0.169 & 0.004 & 54491.9 & 54496.9 & 0.06033 & 2 \\
SDSS J125023.85$+$665525.5 & 2011 & 0.174 & 0.003 & 55665.0 & 55668.7 & 0.06026 & 3 \\
SDSS J161027.61$+$090738.4 & 2009 & 0.113 & 0.001 & 55040.4 & 55050.5 & 0.05782 & 2 \\
SDSS J162520.29$+$120308.7 & 2010 & 0.154 & 0.001 & 55386.4 & 55389.8 & 0.09605 & 2 \\
SDSS J162718.39$+$120435.0 & 2008 & 0.103 & 0.001 & 54620.6 & 54627.6 & 0.10974 & 1 \\
SDSSp J173008.38$+$624754.7 & 2001 & 0.134 & 0.003 & 52205.9 & 52209.1 & 0.07941 & 1 \\
OT J014150.4$+$090822 & 2010 & 0.124 & 0.002 & 55527.9 & 55532.2 & 0.06249 & 3 \\
OT J040659.8$+$005244 & 2008 & 0.074 & 0.002 & 54687.3 & 54690.3 & 0.07995 & 1 \\
OT J041350.0$+$094515 & 2011 & 0.090 & 0.001 & 55587.6 & 55591.8 & 0.05483 & 3 \\
OT J043112.5$-$031452 & 2011 & 0.129 & 0.002 & 55575.2 & 55580.9 & 0.06758 & 3 \\
OT J050617.4$+$354738 & 2009 & 0.121 & 0.002 & 55162.1 & 55171.6 & 0.06932 & 2 \\
OT J055718$+$683226 & 2006 & 0.082 & 0.001 & 54087.1 & 54097.3 & 0.05351 & 1 \\
OT J055721.8$-$363055 & 2011 & 0.142 & 0.003 & 55926.5 & 55932.2 & 0.05976 & 4 \\
OT J064608.2$+$403305 & 2011 & 0.126 & 0.004 & 55923.4 & 55925.0 & 0.06110 & 4 \\
OT J064804.5$+$414702 & 2011 & 0.148 & 0.003 & 55590.6 & 55593.5 & 0.06632 & 3 \\
OT J075414.5$+$313216 & 2011 & 0.130 & 0.003 & 55666.3 & 55668.5 & 0.06308 & 3 \\
OT J081418.9$-$005022 & 2008 & 0.051 & 0.002 & 54765.6 & 54770.3 & 0.07652 & 1 \\
OT J094854.0$+$014911 & 2012 & 0.139 & 0.001 & 56002.3 & 56007.9 & 0.05750 & 4 \\
OT J102616.0$+$192045 & 2010 & 0.117 & 0.001 & 55534.5 & 55541.4 & 0.08283 & 3 \\
OT J102637.0$+$475426 & 2010 & 0.144 & 0.001 & 55270.5 & 55274.6 & 0.06873 & 2 \\
OT J130030.3$+$115101 & 2008 & 0.142 & 0.002 & 54653.0 & 54660.1 & 0.06439 & 1 \\
OT J132900.9$-$365859 & 2011 & 0.109 & 0.002 & 55656.4 & 55660.7 & 0.07100 & 3 \\
OT J144011.0$+$494734 & 2009 & 0.046 & 0.001 & 54985.4 & 54991.9 & 0.06462 & 2 \\
OT J144252.0$-$225040 & 2012 & 0.120 & 0.001 & 56035.7 & 56043.7 & 0.06513 & 4 \\
OT J144341.9$-$175550 & 2009 & 0.119 & 0.001 & 54942.1 & 54953.3 & 0.07218 & 1 \\
OT J145921.8$+$354806 & 2011 & 0.107 & 0.005 & 55728.4 & 55731.5 & 0.08511 & 4 \\
OT J163120.9$+$103134 & 2008 & 0.174 & 0.001 & 54592.3 & 54597.6 & 0.06413 & 1 \\
OT J191443.6$+$605214 & 2008 & 0.086 & 0.001 & 54743.1 & 54753.1 & 0.07135 & 1 \\
OT J210950.5$+$134840 & 2011 & 0.129 & 0.001 & 55707.8 & 55718.3 & 0.06005 & 4 \\
OT J214738.4$+$244553 & 2011 & 0.099 & 0.001 & 55839.2 & 55845.8 & 0.09715 & 4 \\
OT J215818.5$+$241925 & 2011 & 0.180 & 0.001 & 55863.2 & 55866.0 & 0.06740 & 4 \\
OT J221232.0$+$160140 & 2011 & 0.110 & 0.001 & 55922.2 & 55931.3 & 0.09032 & 4 \\
\hline
  \multicolumn{8}{l}{\commenta Unit: mag d$^{-1}$.}\\
  \multicolumn{8}{l}{\commentb BJD$-$2400000.}\\
  \multicolumn{8}{l}{\commentc Unit: d.}\\
  \multicolumn{8}{l}{\commentd 1: \citet{Pdot}, 2: \citet{Pdot2}, 3: \citet{Pdot3}, 4: \citet{Pdot4}, 5: this work.}\\
\end{tabular}
\end{center}
\end{table*}

\begin{table*}
\caption{Rate of slow decline in WZ Sge-type dwarf novae}\label{tab:wzsgerate}
\begin{center}
\begin{tabular}{ccccccccc}
\hline
Object & Year & Rate\commenta & Error\commenta & Start\commentb & End\commentb & Mean $P_{\rm SH}$ (stage B)\commentc & Type\commentd & Source\commente \\
\hline
LL And & 1993 & 0.035 & 0.001 & 49330.9 & 49334.1 & 0.05690 & 1 & 1 \\
LL And & 2004 & 0.071 & 0.002 & 53151.2 & 53161.3 & 0.05658 & 1 & 1 \\
V455 And & 2007 & 0.133 & 0.001 & 54355.3 & 54366.3 & 0.05713 & 1 & 1 \\
V466 And & 2008 & 0.092 & 0.001 & 54722.2 & 54734.6 & 0.05720 & 1 & 1 \\
V500 And & 2008 & 0.085 & 0.001 & 54810.9 & 54821.0 & 0.05689 & 1 & 1 \\
V572 And & 2005 & 0.123 & 0.001 & 53695.0 & 53704.2 & 0.05559 & 1 & 1 \\
SV Ari & 2011 & 0.074 & 0.001 & 55776.6 & 55796.9 & 0.05552 & 1 & 4 \\
UZ Boo & 2003 & 0.087 & 0.001 & 52981.3 & 52989.4 & 0.06192 & 2 & 1 \\
DY CMi & 2008 & 0.104 & 0.001 & 54486.6 & 54494.6 & 0.06074 & 2 & 1 \\
EG Cnc & 1996 & 0.090 & 0.001 & 50425.2 & 50429.4 & 0.06034 & 2 & 1 \\
AL Com & 1995 & 0.090 & 0.001 & 49824.9 & 49836.2 & 0.05723 & 1 & 1 \\
V1251 Cyg & 2008 & 0.121 & 0.001 & 54764.2 & 54772.1 & 0.07597 & 1 & 1 \\
V2176 Cyg & 1997 & 0.031 & 0.002 & 50704.5 & 50710.7 & 0.05624 & 1 & 1 \\
VX For & 2009 & 0.121 & 0.001 & 55092.0 & 55097.3 & 0.06133 & 2 & 1 \\
PR Her & 2011 & 0.102 & 0.002 & 55900.2 & 55905.6 & 0.05502 & 1 & 4 \\
V592 Her & 2010 & 0.108 & 0.001 & 55413.7 & 55424.1 & 0.05661 & 1 & 2 \\
RZ Leo & 2000 & 0.126 & 0.001 & 51903.0 & 51913.4 & 0.07866 & 1 & 1 \\
RZ Leo & 2006 & 0.123 & 0.001 & 53886.0 & 53896.0 & 0.07843 & 1 & 1 \\
IK Leo & 2006 & 0.048 & 0.001 & 54060.8 & 54065.7 & 0.05631 & 1 & 1 \\
GW Lib & 2007 & 0.088 & 0.001 & 54212.3 & 54223.2 & 0.05409 & 1 & 1 \\
QZ Lib & 2004 & 0.135 & 0.001 & 53043.2 & 53048.9 & 0.06460 & 1 & 1 \\
EZ Lyn & 2010 & 0.135 & 0.001 & 55464.9 & 55471.0 & 0.05963 & 2 & 3 \\
V358 Lyr & 2008 & 0.049 & 0.001 & 54793.4 & 54815.0 & 0.05563 & 1 & 1 \\
V453 Nor & 2005 & 0.122 & 0.001 & 53533.9 & 53543.2 & 0.06497 & 1 & 1 \\
GR Ori & 2013 & 0.110 & 0.001 & 56343.4 & 56352.0 & 0.05833 & 1 & 5 \\
FL Psc & 2004 & 0.139 & 0.001 & 53263.8 & 53271.0 & 0.05709 & 1 & 1 \\
BW Scl & 2011 & 0.100 & 0.001 & 55865.0 & 55877.0 & 0.05500 & 1 & 4 \\
WZ Sge & 2001 & 0.102 & 0.001 & 52131.2 & 52138.2 & 0.05720 & 1 & 1 \\
CT Tri & 2008 & 0.066 & 0.001 & 54773.4 & 54793.0 & 0.05366 & 1 & 1 \\
V355 UMa & 2011 & 0.094 & 0.001 & 55604.4 & 55615.8 & 0.05809 & 1 & 3 \\
HV Vir & 1992 & 0.114 & 0.002 & 48744.0 & 48749.2 & 0.05828 & 1 & 1 \\
HV Vir & 2002 & 0.130 & 0.001 & 52281.0 & 52290.0 & 0.05827 & 1 & 1 \\
HV Vir & 2008 & 0.133 & 0.001 & 54517.1 & 54526.3 & 0.05832 & 1 & 1 \\
V498 Vul & 2005 & 0.106 & 0.003 & 53606.4 & 53612.1 & 0.05992 & 1 & 1 \\
ASAS J102522$-$1542.4 & 2006 & 0.139 & 0.001 & 53762.9 & 53773.0 & 0.06337 & 1 & 1 \\
1RXS J023238.8$-$371812 & 2007 & 0.141 & 0.001 & 54379.1 & 54383.1 & 0.06617 & 2 & 1 \\
MASTER OT J203749.39$+$552210.3 & 2012 & 0.052 & 0.001 & 56228.3 & 56233.7 & 0.06131 & 2 & 6 \\
MASTER OT J211258.65$+$242145.4 & 2012 & 0.127 & 0.001 & 56119.2 & 56124.9 & 0.06023 & 2 & 6 \\ 
MisV 1443 & 2011 & 0.115 & 0.001 & 55571.4 & 55581.5 & 0.05672 & 1 & 3 \\
SDSS J220553.98$+$115553.7 & 2011 & 0.178 & 0.002 & 55702.2 & 55708.0 & 0.05815 & 1 & 4 \\
OT J111217.4$-$353829 & 2007 & 0.050 & 0.001 & 54469.2 & 54493.3 & 0.05896 & 1 & 1 \\
OT J112619.4$+$084651 & 2013 & 0.104 & 0.001 & 56308.7 & 56319.0 & 0.05489 & 1 & 5 \\
OT J213806.6$+$261957 & 2010 & 0.128 & 0.001 & 55332.8 & 55339.9 & 0.05502 & 1 & 2 \\
OT J012059.6$+$325545 & 2010 & 0.089 & 0.001 & 55545.0 & 55553.4 & 0.05783 & 1 & 3 \\
OT J184228.1$+$483742 & 2011 & 0.045 & 0.001 & 55841.2 & 55851.5 & 0.07234 & 3 & 4 \\
OT J232727.2$+$085539 & 2012 & 0.092 & 0.001 & 56192.3 & 56205.1 & 0.05344 & 1 & 5 \\
PNV J06270375$+$3952504 & 2013 & 0.102 & 0.002 & 56391.0 & 56402.4 & 0.05892 & 1 & 5 \\
PNV J19150199$+$0719471 & 2013 & 0.069 & 0.001 & 56455.5 & 56464.9 & 0.05821 & 1 & 8 \\
SSS J122221.7$-$311523 & 2013 & 0.020 & 0.001 & 56312.3 & 56328.3 & 0.07649 & 3 & 7 \\
\hline
  \multicolumn{8}{l}{\commenta Unit: mag d$^{-1}$.}\\
  \multicolumn{8}{l}{\commentb BJD$-$2400000.}\\
  \multicolumn{8}{l}{\commentc Unit: d.}\\
  \multicolumn{8}{l}{\commentd 1: no or one rebrightening, 2: multiple rebrightenings, 3: double superoutburst.}\\
  \multicolumn{8}{l}{\commente 1: \citet{Pdot}, 2: \citet{Pdot2}, 3: \citet{Pdot3}, 4: \citet{Pdot4}, 5: this work,}\\
  \multicolumn{8}{l}{6: \citet{nak13j2112j2037}, 7: \citet{kat13j1222}, 8. Nakata et al. in prep.}\\
\end{tabular}
\end{center}
\end{table*}

\begin{table*}
\caption{Rate of slow decline in systems with evolved secondaries}\label{tab:evolvedrate}
\begin{center}
\begin{tabular}{cccccccc}
\hline
Object & Year & Rate\commenta & Error\commenta & Start\commentb & End\commentb & Mean $P_{\rm SH}$ (stage B)\commentc & Source\commentd \\
\hline
V485 Cen & 2001 & 0.130 & 0.003 & 51999.8 & 52005.2 & 0.04207 & 1 \\
V485 Cen & 2013 & 0.169 & 0.003 & 56399.8 & 56402.9 & 0.04214 & 3 \\
GZ Cet & 2003 & 0.073 & 0.001 & 52996.9 & 53007.1 & 0.05677 & 1 \\
EI Psc & 2001 & 0.249 & 0.001 & 52218.0 & 52221.1 & 0.04635 & 1 \\
EI Psc & 2009 & 0.266 & 0.015 & 54993.6 & 54995.7 & 0.04635 & 1 \\
CSS J102842.8$-$081930 & 2009 & 0.118 & 0.002 & 54923.0 & 54925.3 & 0.03814 & 1 \\
CSS J102842.8$-$081930 & 2010 & 0.109 & 0.002 & 55544.7 & 55551.3 & 0.03815 & 3 \\
CSS J102842.8$-$081930 & 2012 & 0.090 & 0.001 & 55958.6 & 55963.9 & 0.03817 & 2 \\
CSS J102842.8$-$081930 & 2013 & 0.116 & 0.001 & 56394.3 & 56400.5 & 0.03820 & 3 \\
SBS 1108$+$574 & 2012 & 0.074 & 0.001 & 56040.6 & 56050.7 & 0.03912 & 2 \\
\hline
  \multicolumn{8}{l}{\commenta Unit: mag d$^{-1}$.}\\
  \multicolumn{8}{l}{\commentb BJD$-$2400000.}\\
  \multicolumn{8}{l}{\commentc Unit: d.}\\
  \multicolumn{8}{l}{\commentd 1: \citet{Pdot}, 2: \citet{Pdot4}, 3: this work.}\\
\end{tabular}
\end{center}
\end{table*}

\begin{table*}
\caption{Rate of slow decline in peculiar systems}\label{tab:pecrate}
\begin{center}
\begin{tabular}{ccccccc}
\hline
Object & Year & Rate\commenta & Error\commenta & Start\commentb & End\commentb & Mean $P_{\rm SH}$ (stage B)\commentc \\
\hline
CC Scl & 2011 & 0.145 & 0.004 & 55870.5 & 55873.8 & 0.06001 \\
MASTER OT J072948.66$+$593824.4 & 2012 & 0.154 & 0.002 & 55976.3 & 55978.7 & 0.06625 \\
OT J173516.9$+$154708 & 2011 & 0.038 & 0.001 & 55739.4 & 55742.2 & 0.05827 \\
\hline
  \multicolumn{7}{l}{\commenta Unit: mag d$^{-1}$.}\\
  \multicolumn{7}{l}{\commentb BJD$-$2400000.}\\
  \multicolumn{7}{l}{\commentc Unit: d.}\\
\end{tabular}
\end{center}
\end{table*}

\section{Fading Rate during Superoutbursts}\label{sec:fadingrate}

\subsection{Background}

\begin{figure*}
  \begin{center}
    \FigureFile(160mm,110mm){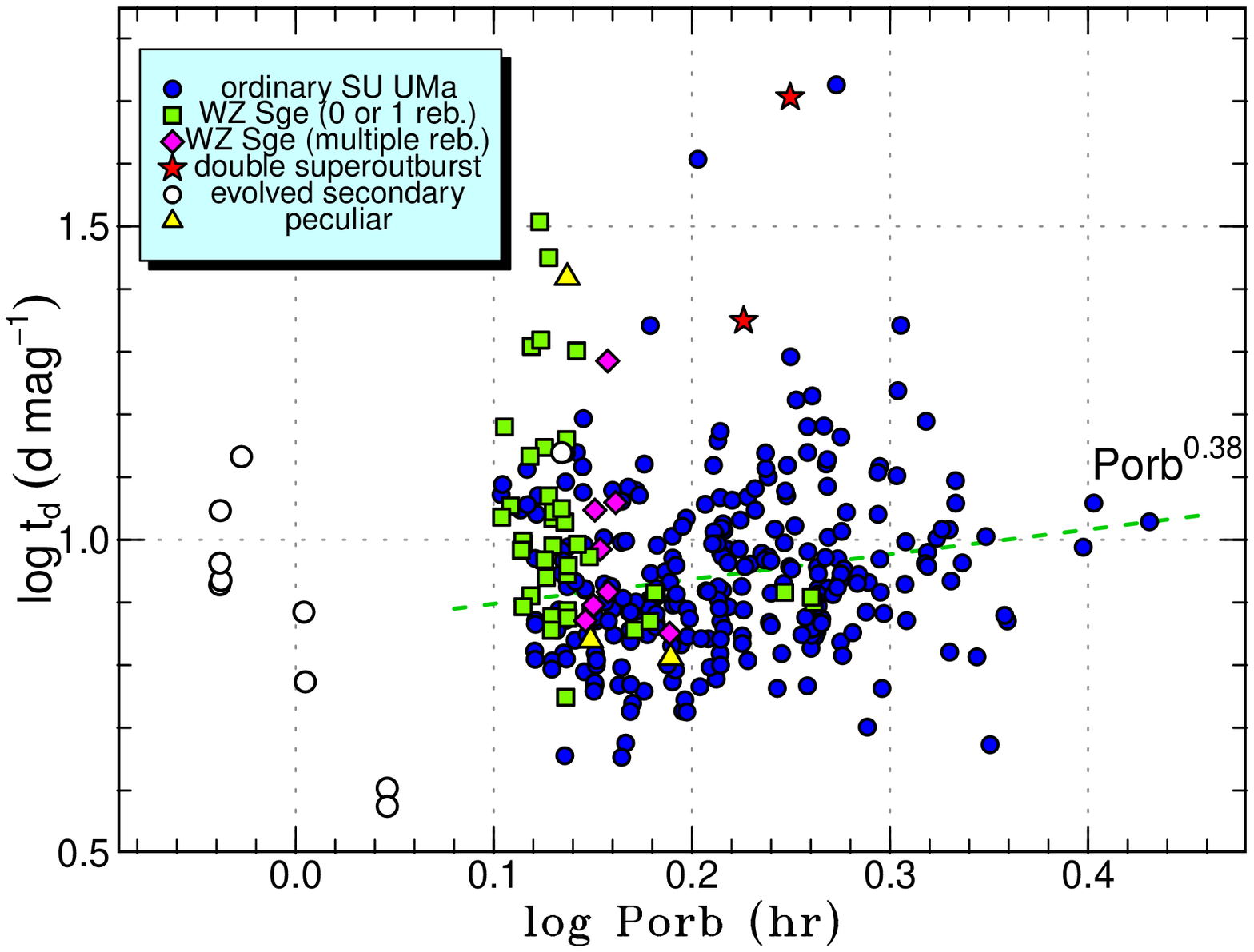}
  \end{center}
  \caption{Dependence of slow fading rate of superoutburst
  on the orbital period.  The orbital period was estimated
  from the period of stage B superhumps using the relation
  in \citet{kat12DNSDSS}.  Filled circles, filled squares,
  filled diamonds, filled stars, open circles and 
  filled triangles represent ordinary SU UMa-type dwarf novae, 
  WZ Sge-type dwarf novae with zero or one rebrightening, 
  WZ Sge-type dwarf novae with multiple rebrightenings and 
  objects with double superoutbursts
  (candidate period bouncers), systems with evolved secondaries,
  and objects with unusually low-amplitude superhumps, respectively.}
  \label{fig:sodecline}
\end{figure*}

   During the superoutburst of SU UMa-type dwarf novae,
there exists an almost exponential, slow decline phase.
\citet{osa89suuma} was the first to derive the time scale
of this slow fading.
\begin{equation}
t_{\rm d} \simeq 8.14\; {\rm d}\; R_{\rm d,10}^{0.4} \alpha_{0.3}^{-0.7},
\label{equ:declinerateosaki}
\end{equation}
where $R_{\rm d,10}$ is the disk radius in a unit of 10$^{10}$ cm
and $\alpha_{0.3} = \alpha_{\rm hot}/0.3$, respectively.
[Note that there are two different definitions of $\alpha$
(cf. subsection 4.3 in \cite{osa96review}) and $\alpha$
here corresponds to $\alpha$=0.2 in \citet{sma84DI} or \citet{war95suuma}].
\citet{war95suuma}, \citet{war95book} discussed the dependence
of this relation to system parameters, and obtained the following
relation:
\begin{equation}
t_{\rm d}\sim t_\nu = 17\; {\rm d}\; \alpha_{\rm hot, -1}^{-4/5} P_{\rm h}^{1/4} m_1^{1/6},
\label{equ:visdecay}
\end{equation}
where $\alpha_{\rm hot, -1} = \alpha_{\rm hot}/0.1$ and
$m1 = M_1/M_{\odot}$ \citep{can10v344lyr}.

   \citet{can10v344lyr} studied the Kepler light curves of
V344 Lyr and V1504 Cyg and compared the slow fading rate during
the superoutburst with other SU UMa-type dwarf novae.
\citet{can10v344lyr} suggested from Kepler observations
that the fading rate has a much stronger dependence on 
the orbital period than equations (\ref{equ:declinerateosaki})
(\ref{equ:visdecay}).

   \citet{can10v344lyr} postulated that this different
could arise from the strong dependence of the viscosity
in quiescence ($\alpha_{\rm cold}$), that is, a smaller
$\alpha_{\rm cold}$ gives rise to a larger surface density
at the start of the superoutburst and hence a steeper
viscous decay.

\subsection{Our Sample}

   We examined this study by a more homogeneous survey of 
fading rates during the superoutburst of SU UMa-type dwarf novae, 
since there was only a limited number of (not necessarily
homogeneous) samples listed in \citet{can10v344lyr}.
We used the superoutburst data in \citet{Pdot}, \citet{Pdot2},
\citet{Pdot3}, \citet{Pdot4} and this paper.
We did not attempt a complete survey, but chose observations 
with high quality.  We also did not attempt to survey
the data in the literature which may be potentially useful.
We also excluded deeply eclipsing systems since it is difficult
to remove the effect of the eclipses and since these objects
usually show a beat phenomenon.
Note that the data ``with high quality'' do not necessarily
correspond to the data giving high-quality times of superhump
maxima, and the qualification of the data does not agree to
the data classification in the table of outburst observation
(such as table \ref{tab:perlist}).

   We examined each light curve of a superoutburst (after
correcting the difference between different observers)
visually and identified the linear part to obtain the fading rate.
Some objects tend to brighten when stage C superhumps
appear (see also \cite{kat03hodel}) and we neglected this part.

   We here conveniently divided the objects into two
categories: ordinary SU UMa-type dwarf novae and WZ Sge-type
dwarf novae including candidate period bouncers.
Since there is no clearly dividing line between ordinary SU UMa-type 
dwarf novae and WZ Sge-type, we adopted the category
in \citet{Pdot}: the objects that show early superhumps,
a manifestation of the 2:1 resonance, and/or the objects with
large ($\sim$8 mag) outburst amplitudes and long (several
to tens of years) supercycle.  The objects with frequent
normal outbursts are not included in WZ Sge-type dwarf novae
even if the amplitude satisfies this condition.

   The results are listed in tables \ref{tab:suumarate} and
\ref{tab:wzsgerate}.  Unless new variable star designations
were given, we used the names of the objects given in
the references for reader's convenience.
Note that the errors given in the tables only refer to
random errors, and the real errors should be larger 
due to systematic errors arising from a number of factors,
such as poor zero-point calibration, difference between observers, 
variable weather condition, poor flat-fielding, atmospheric extinction 
and selection of the segment used for calculation.
A comparison between different superoutbursts
of the same object suggests an order of 0.01--0.02 mag d$^{-1}$
as the typical systematic error.

\subsection{Comparison with the Previous Study}

   Although \citet{can10v344lyr} suggested a much steeper
dependence of $P_{\rm hr}$, our data on SU UMa-type dwarf novae
do not show a significant deviation from equations
(\ref{equ:declinerateosaki}) (\ref{equ:visdecay}) 
(figure \ref{fig:sodecline}).
The $P_{\rm orb}$-dependence in $t_{\rm d} \propto P_{\rm orb}^\psi$
was estimated to be $\psi=$0.38(0.13), consistent with
the predicted value $\psi=0.25$ by a 1$\sigma$ level.
For at least ordinary SU UMa-type
dwarf novae, we do not need a special explanation for 
the deviation from this relation.

   The case of the difference between V344 Lyr and V1504 Cyg
in \citet{can10v344lyr} appears to reflect
the intrinsic difference between different objects,
sinfec there appears to be a sufficient scatter among
SU UMa-type dwarf novae in our sample.
Is is possibly related to the condition that V1504 Cyg 
($P_{\rm orb}$=1.67 hr) has an anomalously high outburst
frequency, and hence has a higher mass-transfer rate than in
other SU UMa-type dwarf nova with similar $P_{\rm orb}$.

   It is not clear whether the long outburst
of U Gem \citet{can10v344lyr} can be compared to these
results, partly because that the magnitudes of
the comparison stars for U Gem used in the 1980s were
different from the modern $V$ magnitudes (e.g. 9.0 mag
for the modern $V$=9.51 star HD 64813)
and the variation in the brighter part of the outburst was 
systematically underestimated.  Furthermore, the VSOLJ record 
suggests that there were two peaks in the light curve with
a intervening temporary fading (to 10 mag around 1985 October 18--19)
between them.  This fading was recorded by a very skillful
observer (H. Narumi).\footnote{
   H. Narumi is one of the world top observers in the history
   of visual variable star observation \citep{kiy12japanVS},
   and is renowned for his very accurate observation.
}
While two AAVSO observers recorded the
same fading (10.5 mag on 1985 October 18), other observers
tended to give brighter magnitudes as before, which was
likely a psychological bias.
We thus regard this outburst of U Gem was composed of
two different parts and it is inadequate to 
estimate the linear fading rate.

   Although there was also a suggestion that this outburst
was a superoutburst with superhumps \citep{sma04ugemSH},
\citet{sch07ugemSH} questioned the statistical significance
of the superhump detection.  We also doubt the detection
because the resultant period was 0.20~d, which was
one fifth of the sampling interval of the same longitude
and it could be easily produced by systematic errors.
The poor quality of the comparison star sequence at that time
and the lack of suitable comparison stars between 9.3 and 10.2
(magnitudes used at the time of the 1985 observation)
made the detection of subtle variations such as superhumps
by human eyes particularly difficult.  There was likely 
a psychological bias as stated above, and visual observations
under such adverse conditions should not be weighed too much.\footnote{
   One of the authors (TK) actually visually observed U Gem
   during this outburst, and realized the difficulty in
   estimating this object without suitable comparison stars.
}

\subsection{WZ Sge-Type Objects}

   The other problem in \citet{can10v344lyr} is that they
used the rapidly fading (not linearly fading) part of
the WZ Sge-type outbursts.  This part of the outburst
is different from those of ordinary SU UMa-type dwarf novae
in that it reflects the viscous depletion process
of the high-mass disk \citep{osa95wzsge} and also
this phase corresponds to early superhumps, which are
different from ordinary superhumps.  This part of WZ Sge-type
outbursts therefore cannot be directly compared to
the linear decline rate in SU UMa-type dwarf novae.
Our results in the linear part of the outbursts in WZ Sge-type
dwarf novae are in good agreement with the $P_{\rm orb}^{1/4}$
dependence for objects with longer ($P_{\rm orb} \geq 0.07$~d),
and these objects (notably RZ Leo and V1251 Cyg) appear
to have the same properties with the ordinary SU UMa-type
dwarf novae.  The fading rates for the systems with shorter
$P_{\rm orb}$ appear to be smaller than what are expected
for the $P_{\rm orb}^{1/4}$ dependence.  We interpret that
this is an effect of a smaller degree of removal of 
the angular momentum caused by the tidal instability as explained
in the following subsection.  Our conclusion is that
the linear fading part of WZ Sge-type dwarf novae are
essentially the same as in those in ordinary SU UMa-type
dwarf novae and there is no special need for a steep
dependence of the quiescent viscosity on the orbital period
(nor the linear fading rate cannot reflect the quiescent
viscosity), though a subtle deviation exists in WZ Sge-type 
dwarf novae with small $q$.

\subsection{Candidate Period Bouncers and
            Objects with Multiple Rebrightenings}

   Among the studied objects, OT J184228.1$+$483742
(\cite{Pdot4}; \cite{kat13j1842})
and SSSJ122221.7$-$311523 \citep{kat13j1222} have remarkably
small slow fading rates compared to the objects with similar
orbital periods.  These objects are suggested to be candidate
period bouncers, whose very low $q$ was inferred from the dynamical
precession rates of the stage A superhumps (\cite{kat13qfromstageA};
\cite{kat13j1222}).

   We first consider the effect of the disk radius
on the fading rate.  Since there is a relation
$t_{\rm d} \propto R_{\rm d}^{0.4}$
[equation (\ref{equ:declinerateosaki})], a larger disk would produce
a slower fading.  The binary separation $A$ is proportional to
$M_1^{1/3}(1+q)^{1/3}P_{\rm orb}^{2/3}$ (cf. equation 2.1b
in \cite{war95book}).  The radius of the 3:1
resonance also depends on $q$ as in the equation:
\begin{equation}
\label{equ:radius31}
r_{3:1}=3^{(-2/3)}(1+q)^{-1/3}.
\end{equation} 
Combining these effects, the fading rate is 3.6\% smaller in
$q=0.05$ case than in $q=0.2$ case.  This value is clearly
insufficient to explain the observation.

   Although the mass of the primary also matters with
a dependence of $m_1^{0.13}$, the known primary masses of
short-period systems are confined to a relatively narrow
region ($M_1$=0.83 $M_\odot$ with an intrinsic scatter of 
0.07 $M_\odot$ for systems with $P_{\rm orb} \le 0.066$~d,
\cite{sav11CVeclmass}), and it is unlikely the main reason.

   We then consider the effect of the disk viscosity
in the hot state.  Since the theoretical fading rate
[equations (\ref{equ:declinerateosaki})
(\ref{equ:visdecay})] is correlated to $\alpha_{\rm hot}$,
we consider that $\alpha_{\rm hot}$ is smaller in these
candidate period bouncer.  As shown in \citet{Pdot4},
\citet{kat13j1222}, the amplitudes of ordinary superhumps
in these systems were small compared to those in ordinary
SU UMa-type dwarf novae.  This likely reflects the very small
tidal effect on the disk resulting from low $q$.  The small
tidal effect produces a smaller degree of the removal of
the angular momentum (such as enhanced viscous dissipation in the
compressed region, cf. \cite{woo11v344lyr})
due to the development of the tidal instability, and the net
$\alpha_{\rm hot}$ is expected to smaller than in higher-$q$
systems.  The unusually slow fading rates may become
a discriminative signature for candidate period bouncers.
A recently discovered WZ Sge-type dwarf nova
OT J075418.7$+$381225 (=CSS130131:075419$+$381225), which
has an 8-mag amplitude despite the long (0.0716~d)
superhump period (Nakata et al. in preparation),
showed a very slow fading rate [0.0189(3) mag d$^{-1}$],
and is also a good candidate for a period bouncer.
OT J230425.8+062546 \citep{Pdot3} also showed a slow fading 
rate [0.0340(4) mag d$^{-1}$ for the entire plateau, 
0.061(1) mag d$^{-1}$ if we restrict the portion 
(BJD after 24455573.5) after the change in the superhump period]
despite its relatively long (0.0663--0.0672~d) superhump period.
This object also may be a candidate period bouncer.

   Most of the objects with multiple rebrightenings 
(UZ Boo, EG Cnc, DY CMi, VX For, EZ Lyn,
MASTER OT J211258.65$+$242145.4, 1RXS J023238.8$-$371812)
showed the fading rates similar to those of ordinary
SU UMa-type dwarf novae with the corresponding superhump
periods.  We should note, however, that we selected the
linear part of the light curve as in ordinary SU UMa-type
dwarf novae.  At least one of the object with multiple 
rebrightenings (VX For) appeared to show a slower fading part
before this linear part, and the inclusion of the slower part
would produce a smaller fading rate as those of candidate
period bouncers.  We did not include this slower fading part
because this object was observed in high air-mass and
the mean magnitudes were not very well calibrated, which
may have produced an artificial trend.
The only two WZ Sge-type objects 
OT J111217.4$-$353829 and MASTER OT J203749.39$+$552210.3
showed smaller fading rates.  The former is a WZ Sge-type 
dwarf nova without a detected rebrightening \citep{Pdot} and 
the latter showed multiple rebrightenings.  The present result
seems to favor the recent suggestion that object showing
multiple rebrightenings are not considered to be
good candidates for period bouncers \citep{nak13j2112j2037}.

\subsection{Systems with Evolved Secondaries}

   The results for the systems with secondaries having an evolved
core are summarized in table \ref{tab:evolvedrate}.
Since equation (\ref{equ:declinerateosaki}) only weakly depends
on $q$, the decline rate for these ultrashort-$P_{\rm orb}$
objects are generally expected to follow the relation of
a smooth extension of ordinary SU UMa-type dwarf novae.
Figure \ref{fig:sodecline} appears to confirm this expectation.
Several objects need more explanation.
Although EI Psc showed larger decline rates than expected,
both superoutbursts (2001, 2009) were observed in their
late stage (cf. \cite{uem02j2329letter}; \cite{ski02j2329};
\cite{Pdot}), these measurements may not reflect the proper 
slowly fading part.  Future observation of the full superoutburst
is waited.  SBS 1108$+$574 has a light secondary and shows
infrequent outbursts (\cite{Pdot4}; \cite{car13sbs1108};
\cite{lit13sbs1108}), resembling a WZ Sge-type dwarf nova.
The relatively slow decline rate of this object may be
explained as in the same way as WZ Sge-type dwarf novae.

\subsection{Systems with Unusually Low-amplitude Superhumps}

   In \citet{Pdot4}, we reported three systems with unusually
low-amplitude superhumps.  These systems apparently showed
closely separated two periods, one of which is likely
the superhump period.   The results for these systems are
listed in table \ref{tab:pecrate} (we tentatively adopted
the longer period as the superhump period).  Both CC Scl and
OT J072948.66$+$593824.4 showed usual decline rates comparable
to the systems with similar $P_{\rm SH}$.  OT J173516.9$+$154708,
however, showed a very slow decline rate.  Although only
a limited segment of the light curve was recorded, this
decline rate might suggest a candidate period bouncer.

\section{Comment on Dwarf Novae in the OGLE Data}\label{sec:OGLEDNe}

   Quite recently, \citet{mro13OGLEDN2} reported detection of
a number of dwarf novae in the OGLE-III data.  They reported
a number of ``SU UMa-type dwarf novae'' having supercycles
in the range of 20--90~d, and claimed that there is no gap
in the distribution of supercycles between (ordinary) SU UMa-type
dwarf novae and ER UMa stars.  Since this conclusion is against
our knowledge based on which we compiled the present series of papers
starting with \citet{Pdot}, we here examine the validity of
their claim.  We also checked superhump periods reported in
\citet{mro13OGLEDN2} and tried to interpret the data 
referring to the superhump stages by our definition.
The results are not included in table \ref{tab:perlist}.
We used the public available electronic data for these objects.

\subsection{OGLE-GD-DN-001} 

   As stated in \citet{mro13OGLEDN2}, this object showed
four post-superoutburst rebrightenings.  In the light curve,
there was a jump around BJD 2454177--2454178.  This jump most 
likely corresponds to the growth of ordinary superhumps
(cf. \cite{nak13j2112j2037}).  After this we detected
superhumps with a mean period of 0.06067(2)~d
by the profile fitting using MCMC method \citep{Pdot2}. 
Although early superhumps were expected before this jump,
we could not detect a convincing signal.  The duration of
this phase was only 6~d, which is much shorter than those
of ordinary WZ Sge-type dwarf novae (see a discussion
of the implication in \cite{nak13j2112j2037}).

\subsection{OGLE-GD-DN-007}

   We obtained the mean superhumps period of 0.08083(2)~d
using the 2007 and 2008 superoutbursts, assuming the common
period and independent phases between them.
An interesting point is that there was a precursor 10~d
preceding the 2008 superoutburst.  We analyzed the data
between this precursor and the main superoutburst and obtained
a possible period of 0.08201(6)~d.  This long period is
consistent with the expected period of stage A superhumps,
and the growing stage of the superhump took place during
the quiescence state between the precursor and the main
superoutburst.  This finding strengthens the TTI model
in which the superhumps play a central role in triggering
a superoutburst (cf. \cite{osa13v1504cygv344lyrpaper3}).
We detected a possible signal of the orbital
period at 0.0782183(4)~d, which needs future verification.

\subsection{OGLE-GD-DN-008}

   We obtained a mean superhump period of 0.08400(1)~d during
the 2007 superoutburst.  This outburst had a precursor which was
followed by a small dip.  During this dip, the superhump
was not yet evident.  This phenomenon exactly reproduces
the Kepler observations of V344 Lyr and V1504 Cyg
(\cite{osa13v1504cygKepler}; \cite{osa13v344lyrv1504cyg})
and again strengthens the TTI model.  In quiescence, this
objects showed fairly strong orbital humps with a mean
amplitude of 0.23 mag.  The orbital period is 0.080919(1)~d.
The resultant $\varepsilon$=3.7\% is typical for this
orbital period.

\subsection{OGLE-GD-DN-014} 

   This object showed two post-superoutburst rebrightenings.
During the plateau phase, superhumps with a mean period of
0.08921(2)~d.  This period is much longer than the most of
WZ Sge-type dwarf novae with multiple rebrightenings
(cf. \cite{nak13j2112j2037}).  The object may be similar to
QZ Ser (Ohshima et al. in prep.), which has and orbital
period of 0.083161~d with an undermassive, evolved secondary
and showed two post-superoutburst rebrightenings.
OGLE-GD-DN-014 may also have an unusual secondary, and would
be worth a further study.

\subsection{OGLE-GD-DN-039} 

   We obtained the mean superhumps period of 0.08347(1)~d
using three recorded superoutbursts.  The duration of the
superoutburst (13.0~d) comprises 16\% of the supercycle (81.3~d),
which is much smaller than in ER UMa (46\%) in 1995
\citep{kat95eruma}.  OGLE-GD-DN-039 resembles BF Ara
which has a supercycle 83.4~d (\cite{kat01bfara}; \cite{kat03bfara})
and a duration of the superoutburst (11--17~d).
\citet{kat03bfara} suggested that BF Ara and SS UMi
(\cite{kat98ssumi}; \cite{kat00ssumi}) are ordinary SU UMa-type
dwarf novae with the shortest supercycle, which are on
a smooth extension of the already known objects.
OGLE-GD-DN-039 apparently fits to this category.

\subsection{Objects with Short Supercycles}

\begin{figure}
  \begin{center}
    \FigureFile(88mm,65mm){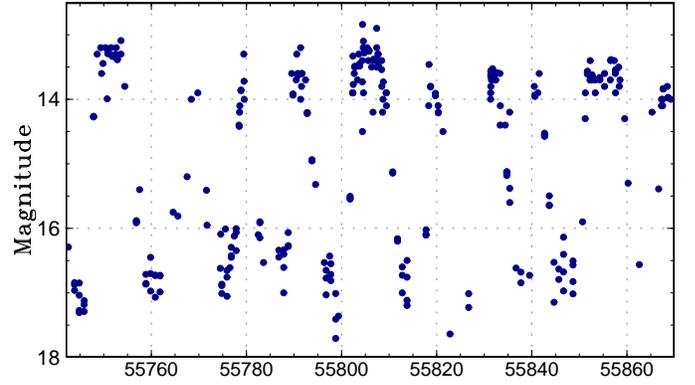}
  \end{center}
  \caption{Light curve of CY Lyr from AAVSO observations.
  Long outbursts sometimes appear quasi-cyclically.}
  \label{fig:cylyr}
\end{figure}

   \citet{mro13OGLEDN2} reported four objects with very short
(24.5--86.8~d) supercycles (OGLE-GD-DN-003, OGLE-GD-DN-004,
OGLE-GD-DN-009 and OGLE-GD-DN-036) and claimed that they fill 
the gap of distribution of the supercycle between ordinary SU UMa-type
dwarf novae and ER UMa stars.  We examined the OGLE light curves
of these objects and could not find any evidence of superhumps
during the long outbursts of these objects.  Furthermore,
the light curves of these long outbursts often did not have
the linear segment which is present in superoutbursts
(cf. section \ref{sec:fadingrate}).  The duration of the long
outbursts are shorter (10--14~d) than in those of typical
superoutbursts.  \citet{mro13OGLEDN2} detected a period of
0.1310(3)~d in OGLE-GD-DN-009, which is a typical value for
an SS Cyg-type dwarf nova.  Although we cannot completely exclude
the possibility that some of these objects resemble the unusual
object RZ LMi (\cite{rob95eruma}; \cite{nog95rzlmi}; 
\cite{ole08rzlmi}), which shows low superhump amplitudes
and has a short duration of the superoutburst, we conclude
that these OGLE objects are SS Cyg-type dwarf novae.
If it is the case, the discussion on the distribution
of the supercycle in \citet{mro13OGLEDN2} is pointless.
In figure \ref{fig:cylyr}, one can see how well an SS Cyg-type
dwarf nova [CY Lyr, $P_{\rm orb}$=0.207584(7)~d,
\cite{tho98cylyrtwtrivwvul}] mimics a quasi-supercycle
of 50--60~d.  Another dwarf nova VW Vul [$P_{\rm orb}$=0.16870(7)~d]
was also long been suspected to be an SU UMa-type dwarf nova
based on its ``superoutburst''-like behavior
(e.g. \cite{sha85cmdelv380orivwvul}; \cite{rob87swumaQPO}).

   \citet{pat13bklyn} recently proposed an evolutionary scenario
of ER UMa stars are transitional objects during the cooling
phase of post-eruption classical novae [the idea was not new
and it was already proposed in \citet{kat95eruma}].
Following \citet{pat13bklyn}, the post-eruption classical novae
below the period gap spend $\sim 10^3$~yr in a state like
BK Lyn (borderline novalike and ER UMa-type object) and
$\sim 10^4$~yr in a state like ER UMa, and longer time
with a further reduced mass-transfer rate.  This picture predicts
ER UMa stars are ten times more numerous than BK Lyn-like stars,
and intermediate objects between ER UMa stars and ordinary
SU UMa-type dwarf novae are 10--100 times more numerous
than ER UMa stars.  Apparently, it is not what is actually
observed [see figure 14 in \citet{mro13OGLEDN2} after omitting 
the four objects discussed here].  This tendency is even
clearer on the mass-transfer rate versus $P_{\rm orb}$ plane
(figure 9 in \cite{pat13bklyn}).  While the longer-$P_{\rm orb}$
region ($P_{\rm orb} \geq$ 0.06--0.07~d), ER UMa stars and
ordinary SU UMa-type dwarf novae look like to form a continuum,\footnote{
   Note that \citet{pat13bklyn} changed the original classification
   of ER UMa stars and included SU UMa-type dwarf novae with
   short supercycles.
} the gap between ER UMa stars and ordinary SU UMa-type dwarf novae
still remains in the short-$P_{\rm orb}$ ($P_{\rm orb} \leq$ 0.06~d)
region.  These two problems (number density and the lack of intermediate
objects in the short-$P_{\rm orb}$ region) have not yet been
solved even with the OGLE data, and the picture may not be
not as simple as in \citet{pat13bklyn}.

\section{Topics on Some Objects}\label{sec:topics}

   In this section, we provide new phenomena and findings for
the objects we treated in the earlier series of this work.
These subsections provide supplementary information to earlier
studies, mainly reporting on new phenomena occurring in the objects
we reported earlier.  Subsection \ref{sec:lassowzsgehtcas} is
provided to illustrate the result of Lasso two-dimensional
power spectral analysis in order to help readers interpreting
the Lasso two-dimensional power spectral analysis by comparing with 
our familiar objects WZ Sge and HT Cas, since this method is 
first introduced for the ground-based data in this paper.
Subsection \ref{sec:kvuma} is a new application of
the method in \citet{kat13qfromstageA}.
Since the sources of the data were shown (with some addition
for BK Lyn) in earlier papers or the collective data
(such as from the AAVSO) were used, these objects are not
included in tables \ref{tab:outobs} and \ref{tab:perlist}.

\subsection{BK Lyncis Returned to Novalike State}

\begin{figure}
  \begin{center}
    \FigureFile(88mm,95mm){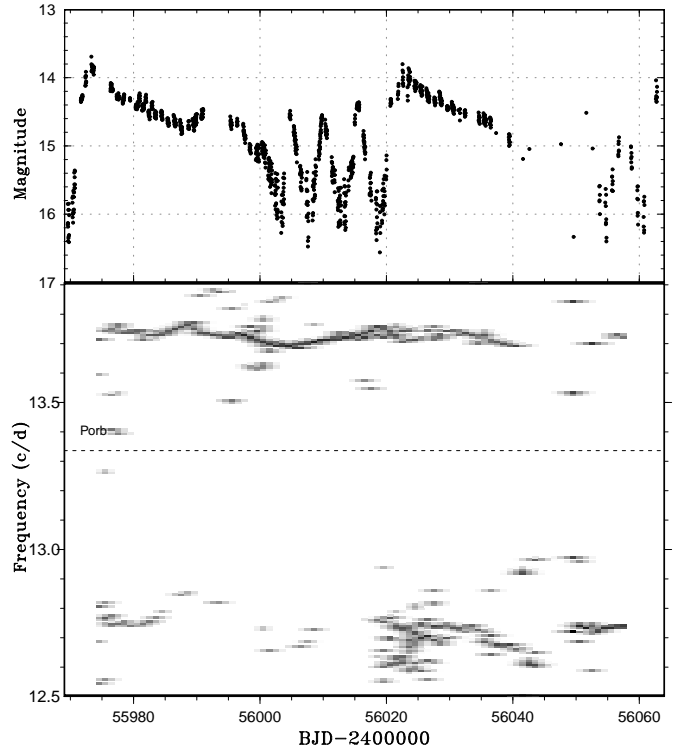}
  \end{center}
  \caption{Two-dimensional Lasso period analysis of BK Lyn (2012).
     (Upper): Light curve (binned to 0.02~d).
     (Lower): Two-dimensional Lasso analysis (10~d window,
     1~d shift and $\log \lambda=-7.8$).
     The signal of negative superhumps was continuously
     present with a systematic variation of the frequency
     in relation to the supercycle phase.   Positive superhumps
     only appeared during the early phase of the superoutbursts.
     Note that the density of the spectrum varies according
     to the density of observations.  No spectrum was detected
     when sufficiently long observation runs were not obtained.}
  \label{fig:bklyn2012lasso}
\end{figure}

   Since we have developed a new technique of two dimensional
power spectrum using Lasso after the publication of \citet{Pdot4}, 
we present a power spectrum of the corresponding data.
The data include those obtained by E. de Miguel in addition
to the ones in \citet{Pdot4}.  \citet{osa13v1504cygv344lyrpaper3}
also present the result of period analysis using PDM,
upon the question raised by \citet{sma13SHnature}.
The continuous presence of negative superhumps in the 2012
data can be very clearly seen (figure \ref{fig:bklyn2012lasso}).
Positive superhumps were only present during the early phase
of the superoutbursts just as in ER UMa (\cite{ohs12eruma};
Ohshima et al. in prep.).  The frequency of the negative
superhumps systematically varied: the frequency reached
the minimum just after the superoutburst and increases
towards the end of the supercycle.  Since the bin used to
draw this figure (10~d) is longer than the interval
of normal outbursts in contrast to V1504 Cyg or V344 Lyr
\citep{osa13v344lyrv1504cyg}, the frequency variation 
associated with the ignition of normal outbursts was not
resolved in this figure.  This short-term variation
can be better recognized with the PDM analysis, and is
presented in \citet{osa13v1504cygv344lyrpaper3}.

\begin{figure}
  \begin{center}
    \FigureFile(88mm,95mm){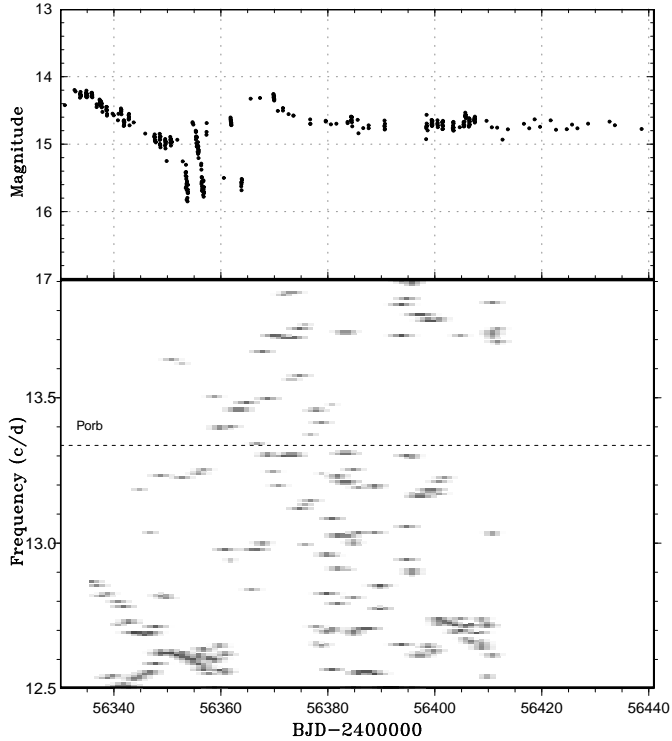}
  \end{center}
  \caption{Two-dimensional Lasso period analysis of BK Lyn (2013).
     (Upper): Light curve (binned to 0.02~d).
     (Lower): Two-dimensional Lasso analysis (10~d window,
     1~d shift and $\log \lambda=-7.8$).
     The signal of negative superhumps was not detected and 
     positive superhumps were the prevalent signal.
     The object entered a standstill (novalike state)
     after the second superoutburst.}
  \label{fig:bklyn2013lasso}
\end{figure}

   On the other hand, the object did not show a strong signal
of negative superhumps in 2013 February--March (including a
superoutburst in BJD 2456333--2456352) (figure \ref{fig:bklyn2013lasso}).
Following the next outburst (this outburst appears to be
superoutburst, but its maximum was much fainter than the maxima
of other superoutbursts; no time-resolved photometry is available
other than on BJD 2456369, when unidentified 0.04~d semi-periodic
variations with amplitudes of 0.1 mag were recorded) on BJD 2456365, 
the object entered a standstill or a novalike state.  
This transition was quite similar to that of a Z Cam-type dwarf nova
as well as in the model calculation of a higher mass-transfer
rate with system parameters of ER UMa \citep{osa95eruma}.
The interval between the superoutbursts, however, did not increase
as in \citet{osa95eruma} just before the object entered
the novalike state.
This transition to a novalike state has confirmed that
the object can undergo a state transition between dwarf nova
state and novalike state (as in Z Cam-type dwarf novae)
and the transition from the novalike state to the dwarf nova-type
state in 2004--2012 cannot be directly attributed to
secular evolution in a cooling postnova state as proposed
in \citet{pat13bklyn}.  As suggested in \citet{Pdot4},
the time-scale (several years) of this transition appears to be 
too short compared to the proposed duration ($\sim$1900~yr) of 
the post-nova state.  This transition may be either driven
by a varying mass-transfer rate or by the disappearance
of the disk tilt.  Since the interval between the superoutbursts
did not show variation as expected from the increased
mass-transfer, the latter possibility might deserve consideration.
In the presence of a disk tilt, the accretion stream hits
the inner part of the disk, and it is expected to provide
less material in the outer part of the disk compared to
a condition without a tilt.  In this condition, the mass-transfer
rate in the outer part of the disk may be insufficient to
maintain the hot (novalike) state. 
It would be very interesting how the disk tilt can govern the disk
dynamics and it needs to be investigated in more detail.

\subsection{Variation in Supercycles in CR Bootis}\label{sec:crboocycle}

   In \citet{Pdot4}, we reported that CR Boo in 2012 showed
a period variation of superhumps very similar to those of 
hydrogen-rich SU UMa-type dwarf novae.  In the meantime,
\citet{hon13crboo} quite recently published the RoboScope
light curve of CR Boo in 1990--2012.  It has become apparent
that this object has at least two distinct states: 
(a) fainter quiescence and regular superoutbursts 
(1990 December--1991 July, 1994 May--1995 January, 
1997 May--1999 March, likely 1999
July and 2007 November--2012 August), which corresponds to
the ``ER UMa-like'' state reported by \citet{kat00crboo},
(b) brighter quiescence with frequent outbursts
(1992 December--1994 April, 1995 February--1997 April,
1999 March--2007 June).  The behavior in the latter state
was somewhat variable from regular outbursts, regular short
superoutbursts to almost standstill (in 1996 May-July
and 1997 April).  If short superoutbursts were present
in the latter state, the intervals of these outbursts
tended to be 10--20~d \citep{kat01crboo}.  Note that
\citet{hon13crboo} used a criterion of superoutbursts different
from ours; we distinguished superoutbursts by morphology
rather than the magnitude, i.e. based on the presence
of a slowly fading segment (cf. section \ref{sec:fadingrate}).
We must also note that there is no guarantee that superhumps
were excited in these short superoutbursts, but we regard
it likely considering the presence of superhumps in
similar states \citep{pat97crboo}.

\begin{figure}
  \begin{center}
    \FigureFile(88mm,70mm){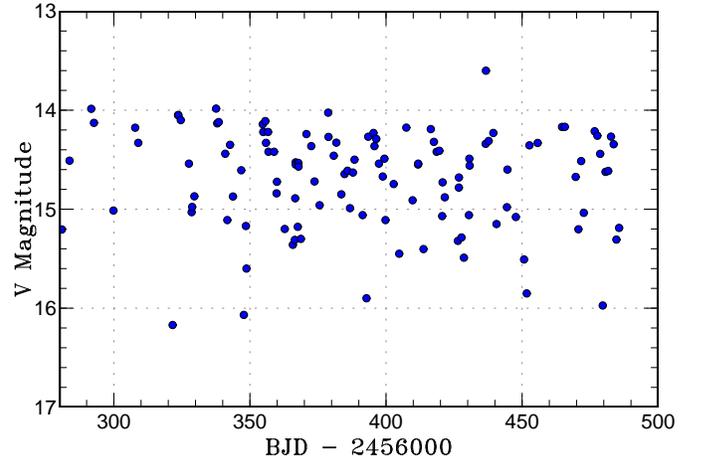}
  \end{center}
  \caption{AAVSO light curve of CR Boo (2013).  The $\sim$46~d
  supercycle seen in the 2012 data disappeared.  Instead, the object
  showed frequent outbursts with a typical interval of 10--20~d.
  }
  \label{fig:crboo2013lc}
\end{figure}

   Our 2012 observation in \citet{Pdot4} happened to record 
the most typical ``ER UMa-like'' state of this object.
\citet{ram12amcvnLC} also recorded this state in 2009--2011.
The situation was very different in the season of 
2012 December--2013 July.  In this season, the object
showed bright quiescence and frequent outburst (the second
state of those described above).  The cycle lengths of
superoutbursts were $\sim$15~d with some variation
(figure \ref{fig:crboo2013lc}).
It looks like that this object switches between these two
states and the maximum duration of each state is several years.
The reason why the object shows these distinct states is
not yet known.

\subsection{Unusual Superoutbursts in NY Serpentis}\label{sec:nyser}

\begin{figure}
  \begin{center}
    \FigureFile(88mm,100mm){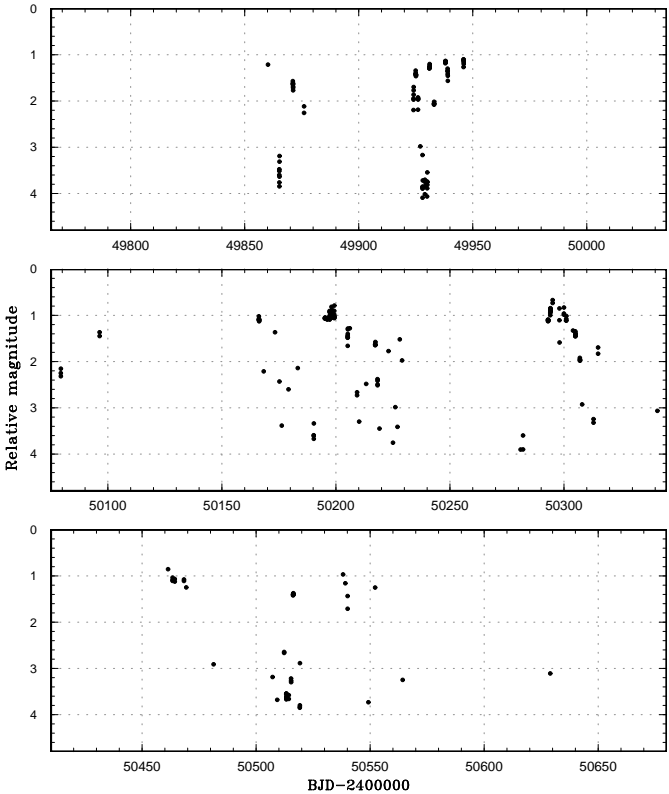}
  \end{center}
  \caption{Light curve of NY Ser in the 1990s.  The data were
  taken from \citet{nog98nyser} and were binned to 0.02~d.
  The magnitudes were relative to the comparison star
  in \citet{nog98nyser}.}
  \label{fig:nyser1990s}
\end{figure}

\begin{figure}
  \begin{center}
    \FigureFile(88mm,100mm){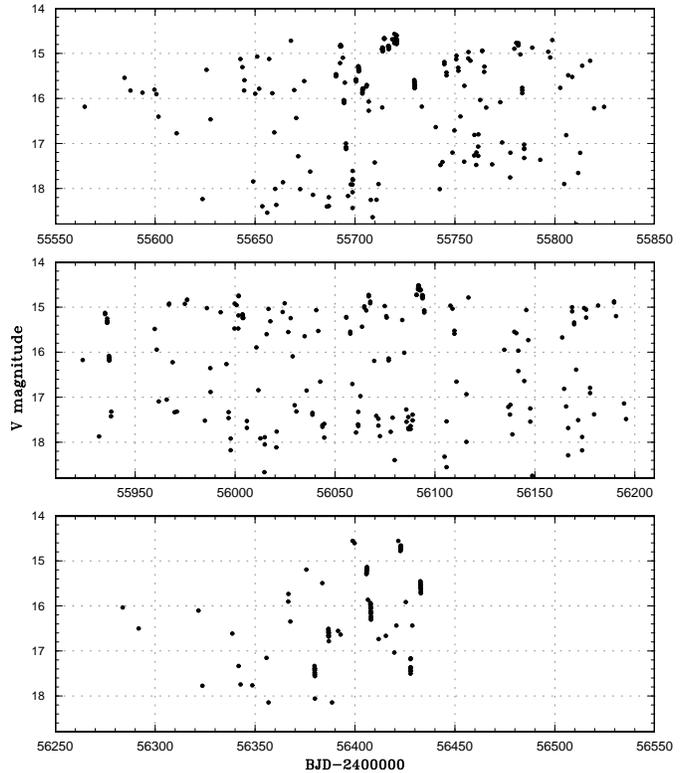}
  \end{center}
  \caption{Light curve of NY Ser in 2011--2013.  The data were
  taken from the AAVSO database and were binned to 0.02~d.}
  \label{fig:nyser2010s}
\end{figure}

   In the late 1990s, NY Ser was known to show superoutbursts
separated by 85--100~d and normal outbursts with recurrence
times of 6--9~d (\cite{nog98nyser}; \cite{iid95nyser},
figure \ref{fig:nyser1990s}).  We, however, have noticed that
this pattern disappeared recently (this tendency may be even
traced back to 2009--2010) in the AAVSO data 
(figure \ref{fig:nyser2010s}).

   In 2011, there was a distinct superoutburst
around BJD 2455714--2455730.  This outburst has been confirmed
to be a genuine superoutburst by the presence of superhumps
on BJD 2455719--2455720.  The outburst immediately preceding 
this superoutburst (BJD 2455700) was unusual in that it showed 
slower decline than in normal outbursts.  Although there may 
have been superhumps, the limited data hindered the secure
detection.  Around BJD 2455779--2455783, there was an outburst
of an intermediate duration ($\sim$4~d), 65~d after the superoutburst.

   In 2012, long superoutbursts were not apparent
(we cannot exclude a long superoutburst around 
BJD 2456116--2456126), and only normal outbursts and outbursts 
with intermediate durations (around BJD 2456000, 2456024, 2456064, 
2456091, 2456108) were frequently recorded.  The tendency
appears to be the same in 2013.

   The intervals of these outbursts with intermediate durations
were 17--40~d, which is much shorter than the supercycles
recorded in \citet{nog98nyser}.  There appear to be at least
two different states in NY Ser, as in CR Boo (subsection
\ref{sec:crboocycle}).  The quiescent brightness or the outburst
amplitudes, however, were not so different between these states
unlike CR Boo.  Observations of superhumps during such
states might shed light on the mechanism of the change
in the outburst behavior.

\subsection{Two-dimensional Lasso Period Analysis of WZ Sagittae and
            HT Cassiopeiae}\label{sec:lassowzsgehtcas}

\begin{figure}
  \begin{center}
    \FigureFile(88mm,100mm){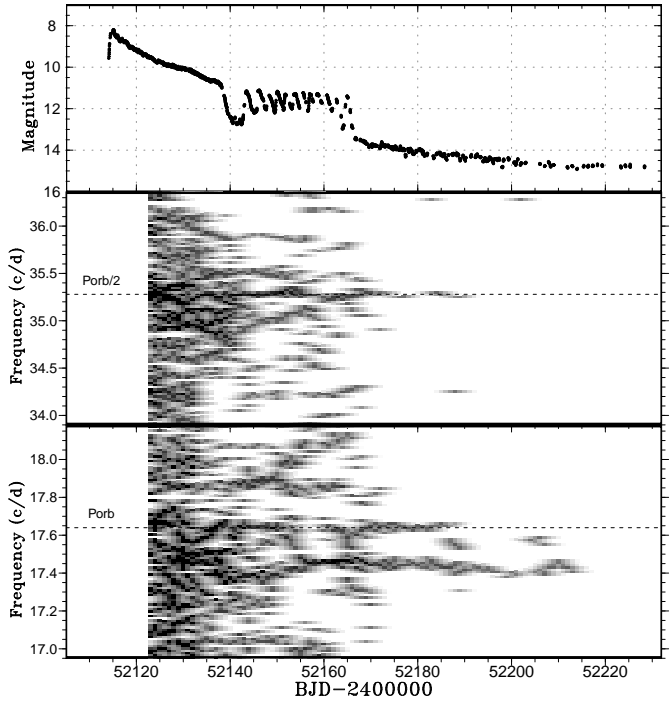}
  \end{center}
  \caption{Lasso analysis of WZ Sge (2001).
  (Upper:) Light curve.  The data were taken from \citet{Pdot}
  and were binned to 0.02~d.
  (Middle:) First harmonics of the superhump and orbital signals.
  (Lower:) Fundamental of the superhump and orbital signal.
  The orbital signal was present both in the fundamental and
  the first harmonic.  The signal of positive superhumps with
  variable frequency was recorded during the superoutburst
  plateau, rebrightening phase and post-superoutburst stage.
  Possible signals of negative superhumps were present during
  the dip phase and rebrightening phases near the frequencies
  17.85~c/d and 35.4~c/d, as suggested by \citet{pat02wzsge}.
  There was also possibly a signal around 35.9 c/d. 
  $\log \lambda=-3.8$ was used.  The width of 
  the sliding window and the time step used are 10~d and 1~d,
  respectively.
  }
  \label{fig:wzsge2001lasso}
\end{figure}

   Since we have successfully obtained a two-dimensional
Lasso spectrum for the long cadence Kepler data of
V585 Lyr \citep{kat13j1939v585lyrv516lyr} using two spectral
windows, we applied the same technique to the 2001 superoutburst
of WZ Sge using the data in \citet{Pdot}.  This application
was motivated by the report of the possible detection of 
transient negative superhumps in \citet{pat02wzsge} and
the possible detection of negative superhumps in FL Psc
during the post-superoutburst stage \citep{Pdot3}.
The result is shown in figure \ref{fig:wzsge2001lasso}.
Unlike Kepler data, ground-based observations have uneven
coverage with occasional gaps.  One should remind that 
the detected frequencies may not be real signals.
We used a long window size of 10~d to resolve different
signals, rather than trying to follow the rapid frequency
variation (note that the initial part has a dirty spectrum
because this part contained the segments of early superhumps
and ordinary superhumps together).
The orbital signal was present both in the fundamental and
the first harmonic.  The signal of positive superhumps with
variable frequency was recorded during the superoutburst
plateau, rebrightening phase and post-superoutburst stage.
Possible signals of negative superhumps were present during
the dip phase and rebrightening phases near the frequencies
17.85~c/d and 35.4~c/d, as suggested by \citet{pat02wzsge}.
There was also a possible signal around 35.9~c/d.
We consider the detection of 17.85~c/d likely, because
this frequency did not match any side lobe of the frequencies
of the orbital and (positive) superhumps, although there
might remain a possibility strong variation of the orbital
signal could produce such a signal as described in
\citet{pat02wzsge}.  Our result seems to strengthen
the detection of these signals in \citet{pat02wzsge}.
If these signals are indeed negative superhumps and arise
from the disk tilt, such a tilt could affect the behavior
of the disk (\cite{ohs12eruma}; \cite{osa13v1504cygKepler}).
It is interesting that the signal of possible negative superhumps
only appeared after the dip and during rebrightenings.
Although we have no concrete explanation, the repeated 
small-amplitude rebrightenings in WZ Sge may be somehow 
related to the disk tilt.

\begin{figure}
  \begin{center}
    \FigureFile(88mm,100mm){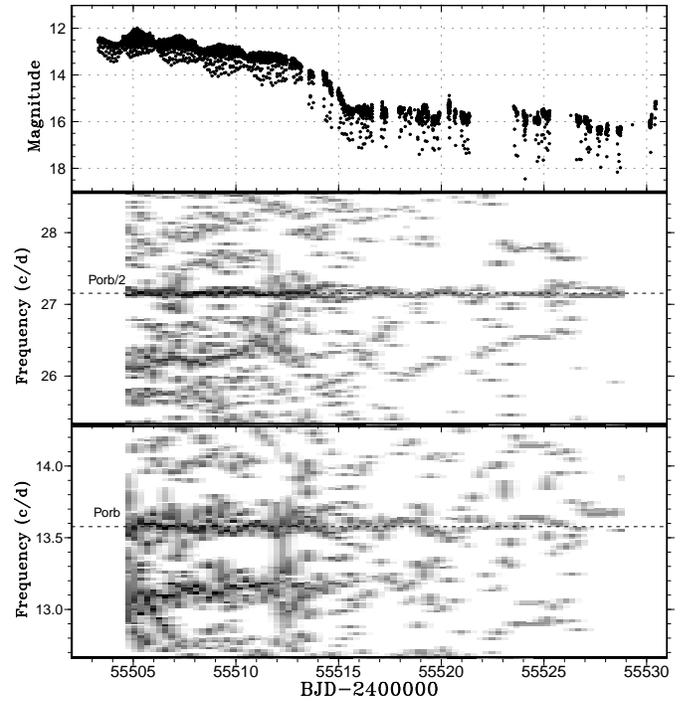}
  \end{center}
  \caption{Lasso analysis of HT Cas (2010).
  (Upper:) Light curve.  The data were taken from \citet{Pdot}
  and were binned to 0.003~d.
  (Middle:) First harmonics of the superhump and orbital signals.
  (Lower:) Fundamental of the superhump and orbital signal.
  The orbital signal was present both in the fundamental and
  the first harmonic.  The signal of positive superhumps with
  variable frequency was recorded during the superoutburst
  plateau and and the post-superoutburst stage.
  The decay of the superhump signal was quicker than in WZ Sge.
  $\log \lambda=-5.3$ was used.  The width of 
  the sliding window and the time step used are 3~d and 0.3~d,
  respectively.
  }
  \label{fig:htcas2010lasso}
\end{figure}

   Figure \ref{fig:htcas2010lasso} presents a Lasso
two-dimensional analysis of the 2010 superoutburst of HT Cas
\citep{Pdot3}.  No strong negative superhumps were detected.
There was possibly an increase of the superhump frequency
when the system started to fade more quickly (BJD 2455513).
This increase in the superhump frequency was probably a result
of the decrease in the pressure effect.  At other times,
the superhump frequency rather monotonously decreased, reflecting
the decreasing disk radius.

\subsection{Late-stage Superhumps in KV Ursae Majoris}\label{sec:kvuma}

   We may apply the analysis of precession frequency is
low-mass X-ray binaries (LMXBs).  As shown in \citet{pea06SH},
the disk in LMXBs appear to have pressure effects different
from CVs, possibly due to the higher temperature.
We may avoid the problem of the unknown pressure effect
by using superhumps after the outburst.  There is such
an observation in KV UMa = XTE J1118$+$480 \citep{zur02j1118},
reporting superhumps near quiescence having $\epsilon^*$=0.0033.
If we assume that the disk radius is KV UMa after the outburst
is similar to those of SU UMa-type dwarf novae \citep{kat13qfromstageA},
we can expect a radius in a range of 0.30--0.38$A$.
The mass ratio is estimated to be $q$=0.023 for 0.30$A$ and
$q=0.014$ for 0.38$A$.  Although these values are slightly smaller 
than spectroscopically determined mass ratios, e.g. 0.037(7)
\citet{oro01kvumaatel67}, 0.027(9) \citep{gon08kvuma}, 
0.024(9) or 0.0435(100) \citep{cal09kvuma}, the value for
the radius 0.30$A$ is marginally in agreement with 
the spectroscopic value considering the intrinsic uncertainty
in analysis of the spectroscopic data.  If a typical disk
radius for post-outburst LMXBs is established, the superhump
period would become a promising tool in determining the mass ratio,
and hence the mass of the black hole or neutron star in LMXBs.

\section{Summary}\label{sec:summary}

   Continuing the project described by \citet{Pdot}, we collected
times of superhump maxima for SU UMa-type dwarf novae 
mainly observed during the 2012--2013 season.  Most of
the short-$P_{\rm orb}$ objects showed period variations
consistent with the trend reported up to \citet{Pdot4}.
We found three objects (V444 Peg, CSS J203937 and MASTER J212624)
having strongly positive $P_{\rm dot}$ despite the long
$P_{\rm orb}$.  It appears that $P_{\rm dot}$-objects are
more numerous in the long-$P_{\rm orb}$ region than had been
considered.  V444 Peg also showed a post-superoutburst
rebrightening.

   We studied ten new WZ Sge-type dwarf novae and updated
the relation between $P_{\rm orb}$ and the rebrightening type.
Although many of them were neither discovered early nor sufficiently
observed to determine the duration of early superhump stage
and the presence of rebrightenings, a greatly increased number
of WZ Sge-type dwarf novae illustrates the ability of modern
transient surveys, especially the MASTER survey.

   By using the period of growing stage (stage A)
superhumps, we obtained mass ratios for seven objects and
updated the $P_{\rm orb}-q$ relation.  This result strengthened
the two conclusions in \citet{kat13qfromstageA} that 
most of WZ Sge-type dwarf novae have secondaries close to the border 
of the lower main-sequence and brown dwarfs, and that most of 
the objects have not yet reached the evolutionary stage of 
period bouncers.  Combined with the $P_{\rm orb}-q$ diagram,
our result seems to support the minimum period at around
0.054--0.055~d in ordinary hydrogen-rich CVs.

   We made a pilot survey of the decline rate of slowly fading
part of SU UMa-type and WZ Sge-type outbursts.
the decline time-scale was found to generally follow the expected
$P_{\rm orb}^{1/4}$ and WZ Sge-type outbursts also generally 
follow this trend.  There is no need for introducing a steep
dependence of the quiescent viscosity parameter $\alpha_{\rm cold}$
on $P_{\rm orb}$ as suggested by \citet{can10v344lyr}.
There are, however, some objects which show significantly slower
decline rates, and we consider these objects good candidates
for the period bouncers.  MASTER OT J165236.22$+$460513.2
showed an outburst with a large amplitude and with long-period
($\sim$0.084~d) superhumps.  This object might be a period bouncer.
We also suggested OT J173516.9$+$154708 as a candidate
period bouncer.

   In addition to these main results, we studied:

\begin{itemize}

\item We re-examined dwarf novae discovered in \citet{mro13OGLEDN2}
and indicated that their claim of detections of a number of
SU UMa-type dwarf novae with short supercycles is unfounded.

\item BK Lyn, which was recently found to show a transition
from a novalike object to an ER UMa-type dwarf nova, again
returned to the novalike state in 2013.  This observation
indicates that the transition to the ER UMa-type state does
not immediately reflect the secular decrease of the mass-transfer
rate as proposed by \citet{pat13bklyn}.  Instead, the supercycle
length did not significantly vary before the transition to
the novalike state, indicating that the mass-transfer rate
did not change greatly.  The signal of negative superhumps
showed dramatic disappearance in 2013, and this may be related
to the transition to the novalike state.

\item CR Boo stopped showing ``ER UMa-like'' state with
$\sim$46~d supercycle in the 2012--2013 season.  The object
instead showed outbursts with intermediate durations with
smaller outburst amplitudes.

\item NY Ser showed frequent occurrence of outbursts with
intermediate durations (likely faint superoutbursts) in
2012--2013.  The shortest interval was even 17~d.

\item We applied least absolute shrinkage and selection operator (Lasso) 
power spectral analysis, which has been proven to be very effective 
in analyzing the Kepler data, to ground-based photometry.
We detected possible negative superhumps in TY PsA and
confirmed the result of possible detection of negative superhumps
in WZ Sge (2001) by \citet{pat02wzsge}.  The frequency variation of
positive superhumps was also well visualized in WZ Sge and HT Cas
(2010).

\item We studied whether our interpretation of the precession
rate of positive superhumps can be applied to black-hole binaries.
An application to KV UMa resulted $q$=0.023 assuming a disk radius
of 0.30$A$ after the outburst.  This method would become
a promising tool in determining the mass ratios if a typical
disk radius for post-outburst black-hole binaries is established.

\end{itemize}

\medskip

This work was supported by the Grant-in-Aid
``Initiative for High-Dimensional Data-Driven Science through Deepening
of Sparse Modeling'' from the Ministry of Education, Culture, Sports, 
Science and Technology (MEXT) of Japan.
The authors are grateful to observers of VSNET Collaboration and
VSOLJ observers who supplied vital data.
We acknowledge with thanks the variable star
observations from the AAVSO International Database contributed by
observers worldwide and used in this research.
This work is deeply indebted to outburst detections and announcement
by a number of variable star observers worldwide, including participants of
CVNET and BAA VSS alert.
We thank Prof. Yoji Osaki for his comments on the discussion
on the decay rate of superoutbursts and the interpretation of BK Lyn.
The CCD operation of the Bronberg Observatory is partly sponsored by
the Center for Backyard Astrophysics.
The CCD operation by Peter Nelson is on loan from the AAVSO,
funded by the Curry Foundation.
We are grateful to the Catalina Real-time Transient Survey
team for making their real-time
detection of transient objects available to the public.

\end{document}